\documentclass[11pt, onecolumn]{book}
\usepackage{graphicx}
\usepackage{amssymb}
\usepackage{color}
\usepackage{rotating}

\newcommand{\one}{\hat{\mbox{1\hspace{-0.097cm}l}}}
\newcommand{\dd}{\mbox{d}}
\newcommand{\half}{\frac{1}{2}}
\newcommand{\nn}{\nonumber} 
\newcommand{\Eqref}[1]{Eq.~(\ref{#1})}
\newcommand{\ket}[1]{\left|{#1}\right\rangle\hspace{-2pt}}
\newcommand{\bra}[1]{\hspace{-1pt}\left\langle{#1}\right|}
\newcommand{\scalar}[2]{\left.\hspace{-3pt}\left\langle{#1}\right|{#2}\right\rangle\hspace{-2pt}}
\newcommand{\tfrac}[2]{\mbox{$\frac{{#1}}{{#2}}$}}

\begin{document}

\frontmatter

\title{\huge \hspace{-6mm} Open Quantum System Dynamics from a \\ \hspace{-6mm}Measurement Perspective: Applications \\ \hspace{-6mm} to Coherent Particle Transport and to\\\hspace{-6mm} Quantum~Brownian Motion} \author{\\ \Large\hspace{-6mm} \textbf{Ingo Kamleitner }\\ \hspace{-3mm}Dipl.\ Phys. \vspace{3mm}\\ \hspace{-6mm}\emph{Centre for Quantum Computer Technology,}\\ \hspace{-6mm}\emph{Department of Physics, Macquarie University,}\\ \hspace{-6mm}\emph{Sydney, New South Wales 2109, Australia.}} \date{\hspace{-6mm}{\Large 19/03/2010} \\ \vspace{6.6cm}{\hspace{-6mm}This thesis is presented for the degree of Doctor of Philosophy in Physics}}

\maketitle
\tableofcontents

\chapter[Abstract]{\huge Abstract}

We employ the theoretical framework of positive operator valued measures, to study Markovian open quantum systems. In particular, we discuss how a quantum system influences its environment, or parts thereof. Using the theory of indirect measurements, we then draw conclusions about the information we could hypothetically obtain about the system by observing the environment. Although the environment is not actually observed, we can use these results to describe the change of the quantum system due to its interaction with the environment. We apply this technique to two different problems.

In the first part, we study the coherently driven dynamics of a particle on a rail of quantum dots. This particle can tunnel between adjacent quantum dots and the tunnel rates can be controlled externally. We employ an adiabatic scheme similar to stimulated Raman adiabatic passage in quantum optics, to transfer the particle between different quantum dots. 

We compare two fundamentally different sources of decoherence. The first is the frequent but weak measurements of the position of the particle by nearby quantum point contacts. This Markovian effect destroys the position coherence required for a good particle transport fidelity.  Second, we couple the particle to two-level fluctuators, which are randomly located near the quantum dot rail. These result in non-Markovian dephasing, which reduces the transport fidelity in a quite different way.

In the second and larger part of this thesis, we study the dynamics of a free quantum particle, which experiences random collisions with gas particles. Previous studies on this topic, which applied scattering theory to momentum eigenstates, found controversial results and were not conclusive. Therefore, we present a supplementary approach, where we start by solving the time dependent  Schr\"odinger equation for two colliding particles, each described by a Gaussian wave function.

Next, we use our results about a single collision, to develop a rigorous measurement interpretation of the collision process, in which the colliding gas particle performs a simultaneous momentum and position measurement on the tracer particle. This in turn leads to the collisional transformation of the tracer particle's density operator. We then derive a master equation by using the correct collision statistics. Finally, we study the collisional decoherence process in terms of the Wigner function.

Because our approach is more complex than the typical scattering theory approach, we restrict ourselves to one spatial dimension. Nevertheless, we find some interesting new insight, including that the previously celebrated quantum contribution to position diffusion is not real, but a consequence of the Markovian approximation. Further, we discover that the leading decoherence process is due to phase averaging, rather than induced by the information transfer between the Brownian particle and the gas.

\chapter[Statement of Candidate]{\huge Statement of Candidate}

I certify that the work in this thesis entitled ``\textbf{Open Quantum System Dynamics from a Measurement Perspective: Applications to Coherent Particle Transport and to Quantum Brownian Motion}" has not previously been submitted for a degree nor has it been submitted as part of requirements for a degree to any other university or institution other than Macquarie University.  \vspace{7mm}\\
 I also certify that the thesis is an original piece of research and it has been written by me. Any help and 
assistance that I have received in my research work and the preparation of the thesis itself have been 
appropriately acknowledged.   \vspace{7mm}\\ 
In addition, I certify that all information sources and literature used are indicated in the thesis. 
\vspace{3cm}\\
\textbf{Ingo Kamleitner (41036107)}\\
\textbf{19/03/2010}

\chapter[Acknowledgments]{\huge Acknowledgments}

Special acknowledgment goes to my primary PhD supervisor, Dr.\ James Cresser, for his time and energy, used in many ways to help me at all stages during my thesis, as well as for his positive feedback and encouragement. I also want to thank my associate supervisor, Prof.\ Jason Twamley; his reliability in support and advice was very much appreciated. Also my fellow PhD students deserve gratitude, not only for distracting me with frequent coffee breaks, but more generally for a good working atmosphere.

I am also grateful to Macquarie University and the Centre for Quantum Computing Technology for their financial support, without which this thesis would never have started.

Last but not least, I want to thank my family. In particular, my parents Barbara and Ewald Kamleitner, for their constant support throughout my life, even when it meant that I could not see them as frequent as I would have liked to. Also my sister Mia and brother Nico deserve acknowledgment, for making every effort to maintain a great brothers and sisters relationship, despite the large distance between us. Further thanks goes to my wife, Mary Chiu, for her strong belief in me.

\chapter[List of Publications]{\huge List of Publications}

\begin{enumerate}
\item I. Kamleitner, J. Cresser, and J. Twamley, \emph{Adiabatic information transport in the presence of decoherence}, Phys. Rev. A \textbf{77}, 032331 (2008).
\item I. Kamleitner and J. Cresser, \emph{Quantum position diffusion and its implications for the quantum linear Boltzmann equation}, Phys. Rev. A \textbf{81}, 012107 (2010).
\end{enumerate}

\chapter[Acronyms and Notations]{\huge Acronyms and Notations}

Throughout the thesis, we used the following acronyms:\vspace{5mm}
\linespread{1.3}
\begin{table}[htdp]\caption{\textbf{Acronyms}}
\begin{center}
\begin{tabular}{c|c}
	CTAP & coherent tunneling by adiabatic passage  \\
	POVM & positive operator valued measure \\
	QBM & quantum Brownian motion  \\
	QD & quantum dot  \\
	QLBE & quantum linear Boltzmann equation \\
	QPC & quantum point contact  \\
	QPD & quantum contribution to position diffusion \\
	STIRAP & stimulated Raman adiabatic passage \\
	TLS & two level system
\end{tabular}
\end{center}
\label{fdgs}
\end{table}%

\linespread{1.3}

We used hats for all operators acting in Hilbert space, except for the density operator, the free evolution operator, and the Hamiltonian. We further used calligraphic fonts for super operators and tildes for continuous measurement outcomes. The index $g$ refers to a gas particle. Below is a list of the notations, which where used in more than one chapter.

\linespread{1}

\begin{table}[htdp]\caption{\textbf{Notations}}
\begin{center}
\begin{tabular}{@{}c@{ }|p{5cm}||}
	$a$ & distance between QD rail and QPC rail \\
	$\alpha$ & measurement sensitivity (part I) \\
	$\alpha$ & mass ratio $m_g/m$ (part II) \\
	$\hat A$ & measurement Kraus operator \\
	$\chi_n$ & coupling constant to $n$-th TLS \\
	$d$ & distance between neighboring QDs \\
	d & differential \\
	$D$ & decoherence rate \\
	$D_{pp}$ & momentum diffusion constant  \\
	$D_{xx}$ & position diffusion constant  \\
	$\hat D (x,p)$ & Glauber-Sudarshan displacement operator \\ 
	$\delta$ & coarse graining time \\
	$E$ & energy \\
	$\gamma$ & friction constan \\
	$\hat \Gamma_\delta(x_g,p_g)$ & phase space projection operator \\
	$\hbar$ & reduced Planck constant $h/(2\pi)$ \\
	$H$ & Hamiltonian \\
	$\mathcal H$ & Hilbert space \\
	$\imath$ & imaginary unit \\
	$k_B$ & Boltzmann constant \\
	$\kappa$ & see \Eqref{gamma} \\
	$\mathcal{L}$ & Liouville super operator \\
	$m$ & mass \\
	$\mu_T(p_g)$ & Maxwell-Boltzmann distribution\\
	$n_g$ & gas particle density \\
	$\hat O$ & general operator \\
	$\widehat O$ & general observable \\
\end{tabular}
\begin{tabular}{c@{ }|p{5cm}@{}}
	$\Omega$ & tunneling rate times $\hbar$ \\
	$p$ & momentum \\
	$\bar p$ & momentum after collision \\
	$\hat p$ & momentum operator \\	
	$\tilde p$ & momentum measurement result \\
	$\hat \pi$ & effect operator \\
	$\ket\psi$ / $\ket{\Psi}$ & state vector of a single / composite system \\
	$q$ & momentum transfer \\
	$r_{ij}$ & distance between $i$-th QD and $j$-th QPC \\
	$\rho(x,p)$ & classical phase space probability distribution \\
	$\rho$ / $\varrho$ & density operator of a single / composite system \\
	$R$ & measurement rate \\\vspace{-2.4mm}
	$\hat R,\,\hat R(p_g),$ & \raisebox{-1.7ex}{collision rate operators} \\
	$\hat R_\delta(x_g,p_g)$ &  \\\vspace{-4mm}
	&\\
	$\sigma(\mathbf p)$ & total scattering cross section \\
	$\hat \sigma$ & $\hat\pi(\tilde x=0,\tilde p=0)$ \\
	$\hat \sigma_{z,n}$ & Pauli $z$-operator for $n$-th TLS \\
	$\hat S$ & scattering operator \\
	$t_c$ & collision time \\
	$T$ & temperature \\
	$\hat T$ & transition operator \\
	$U(t)$ & free evolution operator \\
	$v$ & velocity \\
	$V$ & volume \\
	$W$ & position variance of Gaussian wave packet \\
	$W_\rho$(x,p) & Wigner function for the state $\rho$\\
	$x,\,\bar x,\, \hat x,\,\tilde x$ & analogous to $p$'s \\
\end{tabular}
\end{center}
\label{fdgp}
\end{table}%

\mainmatter

\chapter{General Introduction}

All physical systems, whether governed by the laws of classical or quantum physics, are open systems, that is, they interact to some extent with their surrounding environment. For quantum systems, this interaction leads to the process known as decoherence, that has the effect of destroying the peculiar quantum feature of such systems, which is that quantum systems can behave as if they are simultaneously in a number of different, distinct states. This is both, a benefit and a curse: decoherence is believed to be responsible for the `emergence' of the classical laws of physics from their underlying quantum form, but it also makes it difficult to exploit these same quantum features in, for instance, the  development of quantum computers. 

One of the most fundamental differences between classical and quantum physics are their respective descriptions and interpretations of a measurement process. The classical one is rather trivial, in that a measurement changes our knowledge about the state of the measured system, but does not necessarily affect the state of the system itself. In fact, a measurement just means using our senses to gain information about a system, as e.g. our eyes for its position. If we can not use our senses directly, either because they lack in precision, or because we do not have the required sense for a physical observable (i.e.\ magnetic field), we employ a classical measurement apparatus. This magnifies the desired observable and / or translates it into another observable for which we have a sense, as the position of a pointer, or the click of a Geiger-M\"uller counter. We say, the position of the pointer becomes correlated to the observable.

A quantum mechanical measurement process also translates and magnifies a physical observable, but in general, it also irreversibly changes the state of the measured system. The reason is that in quantum mechanics, the building up of correlations between two systems (observed system and measurement apparatus), necessarily changes the states of both. 

If we are interested in the evolution of the state of a quantum system, we should therefore know when a measurement on it is performed. Of course, from the systems point of view, a measurement is performed each time it becomes correlated with any other system, regardless whether or not the latter can be observed by one of the human senses. We conclude that a quantum system is also measured by its environment~\cite{Zurek}.

This poses the following questions: Can we use the very well developed theory of quantum measurements, to describe a general open quantum system? If so, is this new approach advantageous to the many other means of studying open quantum systems?  Rather than trying to answer these very general questions, in this thesis, we will apply the above reasoning to two special problems, which are both of general interest to the physics community. In both instances, we find that our measurement approach presents us with the necessary insight for a good interpretation of our results.

A challenge in our approach is certainly to work out the details of the measurements, which are carried out by the environment. But once that is achieved, in some instances, it is straightforward to write down a master equation, which governs the dynamics of the quantum system of interest. A further nice property of our approach is, that in addition to the mathematical master equation, it naturally presents a physical interpretation thereof. Especially in the second part of our work, this leads us to a number of conclusions, which were not found in previous studies.

We wish to note that the idea of a measurement interpretation of the influence of an environment is not new. Let us quote Joos and Zeh in their influential paper about collisional decoherence~\cite{Joos}: \emph{The destruction of interference terms is often considered as caused in a classical way by an `uncontrollable influence' of the environment on the system of interest. In fact, this interpretation seems to date back to Heisenberg~\cite{Heisen}. But the opposite is true: the system disturbs the environment, thereby dislocalizing the phases. If the system is initially in a superposition $\sum_nc_n\ket{\psi_n},$ it may influence the environment as if being measured by it.} 

As this point of view is now widespread, especially in regard to collisional decoherence and quantum Brownian motion, it is surprising that there is no rigorous attempt in the literature, to work out the precise details of such an environmental measurement. In the second part of this thesis, we present these details for the very topic of collisional decoherence. Interestingly, these measurements turn out to be too ``weak" to account for a leading contribution to decoherence. Instead, we find that the main cause of decoherence is indeed the `uncontrolled influence' of the random gas particle momentum.

\chapter{General Concepts in Quantum Mechanics\label{quantmech}}

In this chapter we introduce the basic concepts used throughout this thesis as well as establish the notation. There are many sources for the material presented here: the lecture notes of Hornberger~\cite{lectures} provide a good introduction, with a more thorough development to be found in Breuer and Petruccione~\cite{Breuer}.

\section{Density operators and composite quantum systems}

One of the basic tenets of quantum mechanics is that a closed quantum mechanical system $S$ is described by a \emph{state vector}, which is a normalized element of a Hilbert space $\mathcal H_S$. We will use Dirac's \emph{bra-ket} notation when talking about state vectors $\ket\psi_S$, where the subscript refers to the quantum mechanical system of interest. The generally time dependent state vector describing the system $S$ evolves according to the Schr\"odinger equation
	\begin{eqnarray}
		\frac\dd{\dd t} \ket{\psi(t)}_S &=& \frac1{\imath\hbar} H_S(t)\ket{\psi(t)}_S,  \label{SGL}
	\end{eqnarray}
where $H_S$ is the Hamiltonian of the system $S$.

An alternative approach to quantum mechanics is the use of a \emph{state operator} $\rho_S(t)$ to describe $S$. This operator is known as the \emph{density operator}, or \emph{density matrix} when talking about its components with respect to some basis of $\mathcal H_S$. In the most simple case, when a ket $\ket{\psi(t)}_S$ can be assigned to the system, then the corresponding density operator is the projector onto this ket, i.e. $\rho_S(t)=\ket{\psi(t)}_S\bra{\psi(t)}$. It follows from \Eqref{SGL}, that the density operator then evolves according to the von Neumann equation
	\begin{eqnarray}
		\frac\dd{\dd t} \rho_S(t) &=& \frac1{\imath\hbar} [H_S(t),\rho_S(t)] ,  \label{vNGL}
	\end{eqnarray}
where $[\cdot ,\cdot ]$ denotes the commutator.

One of the major advantages of the density operator formalism over the state vector formalism is that it includes the description of statistical mixtures. Such mixtures are commonly needed if one is not certain about the state of the quantum system, but only knows the probability $p_j$ for the system being in the state $\ket{\phi_j}\,$. Then, one can use the density operator
	\begin{eqnarray}
		\rho_S &=& \sum_j p_j \ket{\phi_j}_S\bra{\phi_j}. \label{decomposition}
	\end{eqnarray}
which will again evolve according to the von Neumann equation~(\ref{vNGL}), if the system $S$ is a closed one.

Some basic properties of the density operator follow directly from the probability interpretation of $p_j$ in \Eqref{decomposition}, in particular its positivity, i.e. ${}_S\bra{\phi}\rho_S\ket\phi_S\ge 0$ for all $\ket\phi_S\in\mathcal H_S$, and its normalization Tr$\rho_S=1$. If the state of the system is known with certainty, that is, if the density operator is a projector, then the system is said to be in a \emph{pure state}, whereas otherwise we say the system is in a \emph{mixed state}.

It is important to note that for a given density operator $\rho_S$, the decomposition \Eqref{decomposition} is not unique. For example, a system being in the state $\ket{\phi_1}_S$ or $\ket{\phi_2}_S$ with respective probabilities 1/2 has the same density operator as a system being in the state $(\ket{\phi_1}_S+\ket{\phi_2}_S)/\sqrt2$ or $(\ket{\phi_1}_S-\ket{\phi_2}_S)/\sqrt2$ with the same probabilities.  This imposes the question whether the density operator formalism is sufficient to describe all physical situations. The answer is a definite yes if one is interested in the prediction of probabilities of measurement outcomes only, as is usually the case in any physical theory. More precisely, it can be shown that no measurement can distinguish between two physical situations which are described by the same density operator. 

The most general density operator can always be written in terms of an orthonormal basis $\ket i_S$ as
	\begin{eqnarray}
		\rho_S &=& \sum_{i,i'} c_{ii'} \ket i_S\bra{i'} , \qquad c_{ii}\ge 0, \qquad \sum_ic_{ii}=1,
	\end{eqnarray}
where $c_{ii'}={}_S\bra i\rho_S\ket i'_S$. As $\rho_S$ is positive and therefore Hermitian, one can choose the eigenvectors of $\rho_S$ as orthonormal basis to diagonalize the density matrix
	\begin{eqnarray}
		\rho_S &=& \sum_i p_i \ket i_S\bra i , \qquad p_i\ge 0, \qquad \sum_ip_i=1,
	\end{eqnarray}
where $p_i$ is the eigenvalue of $\rho_S$ corresponding to the eigenvector $\ket i_S$. 

The concept of density operators is most appreciated when the system $S$ of interest is part of a larger, composite quantum systems $S+T$. If $\ket i_S$ and $\ket j_T$ are respective orthonormal bases of $\mathcal H_S$ and $\mathcal H_T$, then $\ket i_S\otimes \ket j_T$ is a basis for the product Hilbert space $\mathcal H=\mathcal H_S\otimes\mathcal H_T$ of the combined system. The symbol $\otimes$ denotes the direct product, and will frequently be omitted for shorter notation. Any pure state of the combined system can now be written either as state vector $\ket\Psi_{S+T}=\sum_{ij} \gamma_{ij} \ket i_S\ket j_T$ or as density matrix $\varrho_{S+T}=\sum_{ii'jj'} c_{ii'jj'} \ket i_S\bra{i'}\otimes\ket j_T\bra {j'}$, and the two descriptions are related by $c_{ii'jj'}=\gamma_{ij}\gamma^*_{i'j'}$. If the state vector of the composite system can be written as a direct product $\ket\Psi_{S+T}=\ket\phi_S\otimes\ket\varphi_T$, then the composite system is said to be in a \emph{product state}. If this is not the case, then the subsystems $S$ and $T$ are called \emph{entangled} and this is the general situation if the subsystems are interacting with each other (or did interact in the past).

If there is entanglement between $S$ and $T$, then it is not possible to assign a state vector to any of the subsystems. On the other hand, the density operator formalism tells us how the \emph{reduced density operators} for each subsystem is obtained in terms of the partial trace of the density operator of the composite system:
	\begin{eqnarray}
		\rho_S &=& \mbox{Tr}_T(\varrho_{S+T}) 
		\;=\; \sum_j{}_T\bra j \varrho_{S+T}\ket j_T 
		\;=\; \sum_{ii} c_{ii'} \ket i_S\bra{i'} \qquad \label{partialtrace}
	\end{eqnarray}
with $c_{ii'}=\sum_jc_{ii'jj}$. Note that if the two systems are entangled, the reduced density operator of each subsystem is never in a pure state even if the composite system is. 

It is this property of the possibility to describe subsystems of larger systems, which makes the density operator formalism so suitable for the study of open quantum systems, where the system of interest is only a small part of a much larger system often incorporating an environment with infinite degrees of freedom.

An important property of density matrices is $\mbox{Tr} \rho_S^2 \leq 1$ where the equal sign holds if and only if $\rho_S$ describes a pure state. Therefore Tr$\rho_S^2$ is often used as a measure of how close the system is to a pure state and is commonly referred to as the \emph{purity}. If $S$ is a subsystem of $S+T$ and the combined state is pure, then Tr$\rho_S^2$ can also serve as a measure of entanglement between $S$ and $T$. Whether a composite system described by a mixed state $\varrho_{S+T}$ is entangled or not is not as easily seen. One can show that a system $S+T$ is not entangled if and only if its density operator is \emph{separable}, that is if it can be written as $\varrho_{S+T}=\sum_m\gamma_m\rho_{S,m}\otimes\rho_{T,m}$ where $\rho_{S,m}$ and $\rho_{T,m}$ are density matrices of the respective subsystem and $\gamma_m\geq 0$.
	
	Unitary operations $\ket\psi_S\to\ket{\psi'}_S=U_S\ket\psi_S$ which are well known from the state vector formalism of quantum mechanics are easily generalized to density operator transformations by $\rho_S\to\rho_S'=U_S\rho_SU_S^\dag$. But there is a larger class of ``allowed" linear transformations of density operators. Any linear map which maps operators on operators is called a \emph{super operator} and will be denoted with calligraphic fonts. A super operator $\mathcal T_S:\; \rho_S\to\rho_S'=\mathcal T_S(\rho_S)$  is called \emph{positive} if $\mathcal T_S(\rho_S)$ is positive for all positive $\rho_S$. It is called \emph{completely positive} if $\mathcal T_S\otimes\one$ is positive for all identity operators $\one$ in any dimension. An ``allowed" linear map $\mathcal T_S$ is trace preserving and completely positive. Trace preserving and positivity are required for the transformed density operator to be a valid density operator. Complete positivity is less intuitive at first glance, but is required because there is always another system $T$ which does not interact with $S$ (e.g. a molecule on the moon). The combined system is then transformed by $\mathcal T_S\otimes \one_T$ which again should be a positive operation therefore demanding complete positivity for $\mathcal T_S$. Any such transformation can be written in terms of \emph{Kraus operators}
	\begin{eqnarray}
		\mathcal T_S(\rho_S) &=& \sum_m \hat K_m\rho_S \hat K_m^\dag
	\end{eqnarray}
with $\sum_m\hat K_m\hat K_m^\dag=\one_S$. 

Remember that any linear operator $\hat O$ is uniquely specified by its action on a complete basis $\ket i$, i.e. knowing $\hat O\ket i$ for all $i$ is equivalent to knowing $\hat O$. Similarly, $\mathcal T_S (\ket i\bra{i'})$ for all $i,i'$ uniquely determines the super operator $\mathcal T_S$. Note that the action of $\mathcal T_S$ on all diagonals $\ket i\bra{i}$ alone is not sufficient to uniquely determine $\mathcal T_S$.

\section{Decoherence\label{decointro}}

A diagonal density operator $\rho_S=\sum_ip_i\ket i_S\bra i$ is sometimes called ``classical", because it can be interpreted as having a state $\ket i_S$ with the classical probability $p_i$. If there are ``quantum" superpositions between states $\ket i_S$ and $\ket{i'\neq i}_S$ involved, then the density operator will also include off-diagonals, which are termed \emph{coherences}. The coherences therefore reflect quantum behavior.

The reduction of coherences is called \emph{decoherence}\footnote{The nomenclature to this topic varies vastly in the literature. A good overview is given in \cite{lectures}.}, and leads to a classical like behavior of a quantum system. This process typically results from the coupling to another system, which can be either classical or quantum in nature. If  studying quantum properties of a system, one aims to reduce decoherence effects as much as experimentally feasible. For this reason, decoherence is a major obstacle in quantum information science, where superpositions of many ``quantum bits" have to be maintained over a sufficiently long period of time. On the other hand, decoherence is necessary for any measurement process in quantum mechanics, and is believed to be responsible for a classical macroscopic world.

Of course, whether a density operator is diagonal or not, depends on the basis $\ket i_S$. Therefore, when studying decoherence, one has to single out a basis of the Hilbert space. Often, the microscopic decoherence mechanism points out an appropriate bases in which decoherence is most pronounced. In this \emph{pointer basis}, all coherences often vanish completely over time. One should mention that many problems regarding pointer bases, are not resolved yet \cite{lectures}, and the concept remains somewhat vague, except for some special cases.

In the most simple situation, when the diagonals are constants in time, the off-diagonals often follow exponential decay
	\begin{eqnarray}
		{}_S\bra i\rho_S(t)\ket i'_S &=& e^{-D_{ii'}t}\,{}_S\bra i\rho_S(0)\ket i'_S.
	\end{eqnarray}
The decoherence rate $D_{ii'}$ is often called \emph{dephasing rate} in processes which do only affect the off-diagonals. If dephasing is due to multiplication of coherences by a random and unknown phase, then we will also call it \emph{phase averaging}.

A somewhat different definition of decoherence is the loss of purity. This notion is independent of any designated basis, but instead depends on the initial state of the system.

\section{The von Neumann-L\"uders measurement\label{lueder}}

The von Neumann-L\"uders measurement \cite{Neumann,Luders} is easily generalized to density operators. As in the state vector formalism, one assigns an Hermitian operator
	\begin{eqnarray}
		\widehat O=\sum_m r_m\ket m_S\bra m
	\end{eqnarray}
to the observable to be measured. Here, $r_m$ are the possible measurement outcomes which will be found with the probability
	\begin{eqnarray}
		P(r_m)&=&{}_S\bra m \rho_S\ket m_S.
	\end{eqnarray}
The  expectation value of an observable measured in a von Neumann-L\"uders measurement is easily obtained via the trace operation
	\begin{eqnarray}
		\langle \widehat O \rangle &=& \sum_m  P(r_m) r_m \nn\\
		&=& \mbox{Tr}(\widehat O\rho_S).
	\end{eqnarray}

After a measurement result $r_m$ is found, the system is known to be in the measured state $\ket m_S\bra m$. That is, if a second measurement of the same observable is performed, the measurement outcome will be $r_m$ with certainty. This change of the system's state into the measured one is known as the \emph{projection postulate} of quantum mechanics. The set of states $\{\ket m\}$ are called the \emph{measurement basis} of the measurement.

\section{Generalized measurements\label{genmeas}}

For both parts of this work the concept of generalized measurements \cite{lectures,Breuer,Busch} will be of importance. We will refer to this as the theory of positive operator valued measures (POVM). It generalizes the von Neumann-L\"uders measurement to a larger class of measurements. The von Neumann-L\"uders measurements can be considered as the limit of a \emph{strong} measurement in that it ensures that the measured system is in the measured state after the measurement. POVMs also include measurements where the measured system will generally not end up in the measured state, but might more or less differ from this state. Consecutive measurements on the same system will then generally give different results, but there will be some correlation between probabilities of consecutive measurement results (if there were no correlation at all, the name ``measurement" would barely be justified). If the correlation between consecutive measurements are small, the measurement is called a \emph{weak} one. These typically disturb the system only slightly, but do not reveal much information about the measured system.

If a measurement is performed on a system described by a general density operator $\rho_S$, then for every possible measurement outcome $r_m$ there exists a positive operator, the so-called \emph{effect operator} $\hat \pi_m$ such that the probability of this measurement outcome is 
	\begin{eqnarray}
		P(r_m) &=& \mbox{Tr}(\rho_S\hat \pi_m). 
	\end{eqnarray}
The normalization of probabilities requires $\sum_m\hat \pi_m=\one_S$. The state of the system changes due to the measurement with outcome $r_m$ according to
	\begin{eqnarray}
		\rho_S' &=& \frac{\sum_n \hat A_{mn}\rho_S\hat A^\dag_{mn}}{\mbox{Tr}(\rho_S\hat \pi_m)}  \label{C1.9}
	\end{eqnarray}
where $\sum_n \hat A_{mn}^\dag \hat A_{mn}=\hat \pi_m$. It follows that the density operator after a non-readout measurement, that is after a measurement is performed but without knowing the outcome, is
	\begin{eqnarray}
		\rho_S' &=& \sum_{mn} \hat A_{mn}\rho_S\hat A^\dag_{mn}.
	\end{eqnarray}
The operators $\hat A_{mn}$ will be called \emph{Kraus operators} according to their more general use outside of measurement theory (see previous section). A special case which is often encountered is if for every effect operator there is only one Kraus operator $\hat \pi_m=\hat A_m^\dag \hat A_m$. In this case, the measurement is called \emph{efficient} and the transformation $\rho_S\to \hat A_m\rho_S\hat A^\dag_m/\mbox{Tr}(\hat A_m\rho_S\hat A^\dag_m)$ maps pure states to pure states. In some sense, these measurements can be considered as optimal, because no unnecessary decoherence is introduced during the measurement process.

The generalized measurements include the von Neumann-L\"uders measurements as the special case $\hat \pi_{m}=\hat A_m=\hat P_m$ where $\hat P_m$ are projection operators. Interestingly, one can also derive POVMs from indirect von Neumann-L\"uders measurements. For this, one couples the system $S$ to some ancilla system $M$ which is often called \emph{measurement apparatus} or \emph{meter} in the literature (although it can only be considered as the``first step" of a measurement apparatus). After some coupling time, which is usually assumed to be short compared to all other time scales, the ancilla's  state depends on the initial state of $S$. Performing a von Neumann-L\"uders measurement on the ancilla will then reveal information about the state of $S$. The effect and Kraus operators acting on $S$ to describe this indirect measurement may depend on the initial state of $M$, the coupling Hamiltonian $H_{MS}(t)$, and on the type of measurement performed on $M$. 

Because it is possible to show that all POVMs can in principle be derived in such a way~\cite{Busch}, one is tempted to consider the von Neumann-L\"uders measurement as the fundamental one. This would however be misleading as often strong von Neumann-L\"uders measurements are actually the result of a large number of  weak POVMs. 

One should mention here that there is still no satisfactory microscopic derivation of a quantum mechanical measurement process. Especially the projection postulate or its generalization~\Eqref{C1.9} has to be used without further justification~\cite{Busch}.

There are two distinct causes for the possibility of an indirect measurement to reveal only little information about the system. First the coupling of $M$ to the system $S$ might be a weak one in a sense that the combined state of $M$ and $S$ is only little influenced by the coupling. Then $M$ and $S$ are only weakly correlated and even the best measurement on $M$ gives only little information about $S$. Such a measurement influences the system only little, and is therefore called a \emph{weak measurement}. In this case the effect operator is far away from being a projection operator, as are the Kraus operators. Such a weak measurement can also be an efficient one.

If the coupling between $M$ and $S$ is optimal in that $M$ and $S$ are perfectly correlated in the desired measurement basis of $S$, it is still possible that a measurement on $M$ gives only little information about $S$. This can happen either because the von Neumann-L\"uders measurement on $M$ has a different measurement basis which is not the basis of the correlations, or because the measurement on $M$ is a weak POVM itself. In both cases, the effect operators acting on $S$ indicate little information gain, whereas the Kraus operators indicate strong disturbance on $S$ and can even be projectors. Such a measurement will influence the system considerably despite giving only limited information about it. These sort of measurements will never be efficient ones. In some sense, on a microscopic description a strong measurement is performed, but then followed by an imprecise readout of the meter.

\section{Open quantum systems}

\begin{figure}
		\centering
		\includegraphics[width=10cm]{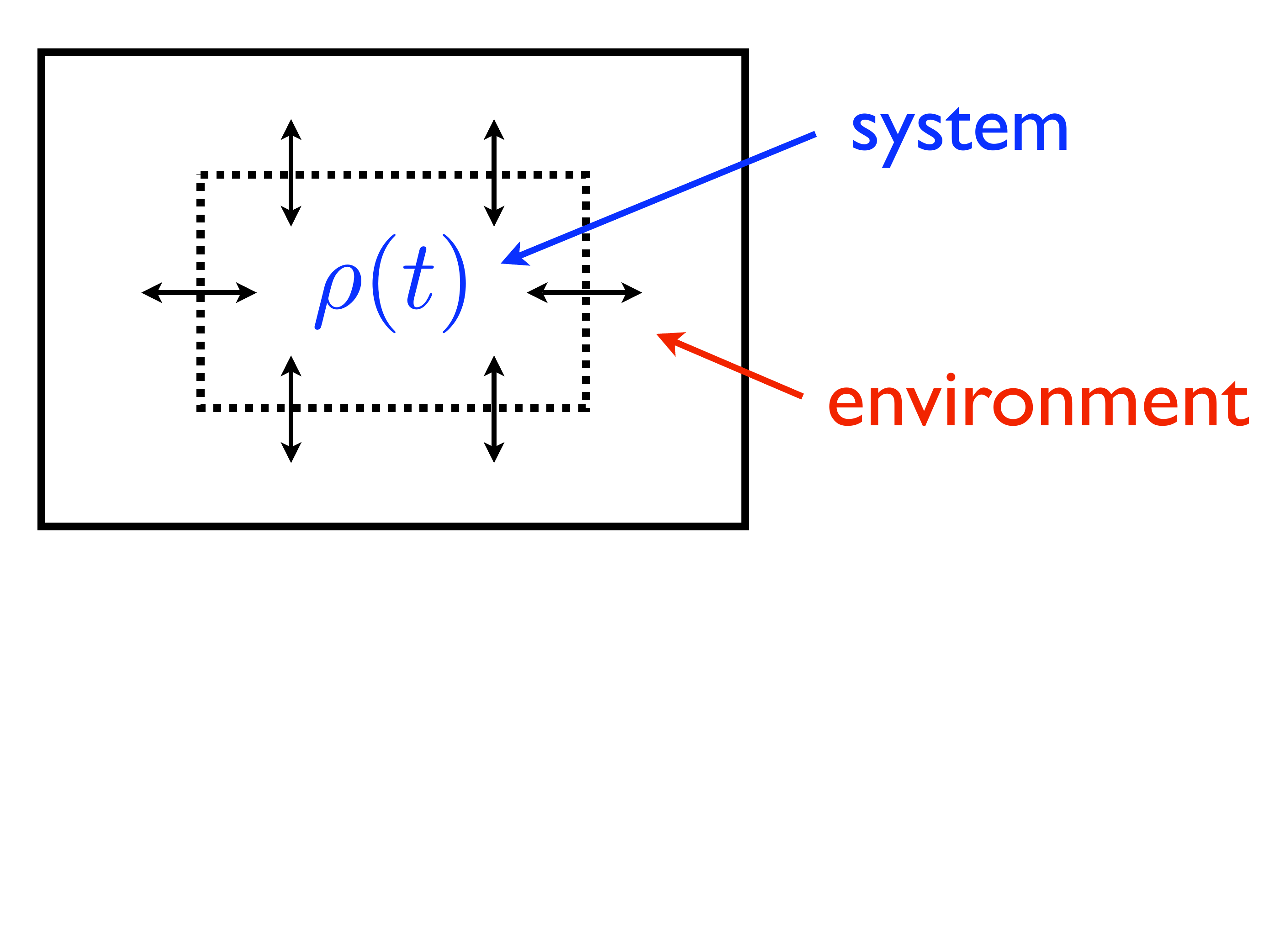}\vspace{-3cm}
		\caption{\small The system $S$ interacts with the environment. The environment's state then changes depending on the system's state. Therefore, at some later time the system will see an environment which depends on its own past. This process is called back action, and leads to memory effects. \label{enviro}}
	\end{figure}

If the quantum system $S$ of interest is coupled to an environment $E$, then we call it an \emph{open quantum system}. The environment is usually a quantum system with a large number of degrees of freedom, such as the electromagnetic field in a vacuum or the phonons in a solid state matrix. The combined system is then a closed system (see Fig.~\ref{enviro}) evolving according to the von Neumann equation
	\begin{eqnarray}
		\frac\dd{\dd t} \varrho_{S+ E}(t) &=& \frac1{\imath\hbar} [H_{S+ E}(t),\varrho_{S+ E}(t)] .
	\end{eqnarray}
Once this equation is solved one could trace out the environment to find the density operator $\rho_S(t)$. However, because of the many degrees of freedom of the environment, the solution of the von Neumann equation is generally not feasible. Instead one usually tries to find a first order differential equation for the reduced density operator $\rho_S(t)=\mbox{Tr}_{E}\varrho_{S+E}$, which is called the \emph{master equation}. In open systems (classical as well as quantum), the change of the state during an infinitesimal time interval $(t, t+\dd t)$ does not only depend on the system's state at time $t$, but also on its past. That is because the state of the environment at time $t$ depends on the history of the state of the interacting system $S$, and the evolution of $S$ in turn depends on the state of $E$. This process is called \emph{back action}, and the open system is then said to behave \emph{non-Markovian}, or the environment is said to exhibit \emph{memory}. The master equation can be shown, by, for instance, the Zwanzig-Nakajima projection operator technique, to take the following form
	\begin{eqnarray}
		\frac\dd{\dd t} \rho_{S}(t) &=& \int_0^t \dd t' \mathcal K_S(t-t')[\rho_S(t')],
	\end{eqnarray}
where the super operator $\mathcal K_S(t-t')$ is the \emph{memory kernel}.

The Markovian approximation $\mathcal K_S(t-t')[\rho_S(t')] = \delta(t-t')\mathcal L_S[\rho_S(t')]$ results in a master equation which is local in time. It is generally valid on a coarse grained time scale on which correlations between environment and system vanish. It can be shown that under quite general conditions the \emph{Liouvillian} $\mathcal L_S$ has to take the following Lindblad form~\cite{Lindblad}
	\begin{eqnarray}
		\frac\dd{\dd t} \rho_{S}(t) &=& \mathcal L_S[\rho_S(t)] \nn\\
		&=& \frac1{\imath\hbar}[H,\rho_S(t)] + \sum_m \left( \hat L_m\rho_S(t)\hat L^\dag_m -\half \hat L_m^\dag \hat L_m\rho_S(t)-\half\rho_S(t)\hat L_m^\dag \hat L_m \right), \qquad\quad \label{C1.13}
	\end{eqnarray}
in order to ensure completely positive and trace preserving dynamics. The Hamiltonian consists of the system Hamiltonian $H_S$ plus a Lamb-shift term $H_{LS}$ which is a coherent modification due to the coupling to the environment. The non-coherent effects of the environment are described by the \emph{Lindblad operators} $\hat L_m$.

It should be noted here that it is often very difficult to derive a Markovian master equation from the underlying microscopic physics. But once this is achieved it provides a clear physical picture in terms of quantum trajectories~\cite{Breuer}, and there are many methods of solving such equations. One should however not forget, that any Markovian description is an approximate one. It involves one or several of the following assumptions~\cite{Breuer}: 
	\begin{enumerate}
		\item weak coupling to the environment,
		\item high temperature of the environment,
		\item low density of the environment,
		\item instantaneous change of state at random times (e.g. collisions, measurements, ...).
	\end{enumerate}
Furthermore, any Markovian master equation is only valid on a coarse grained time scale, not resolving times small compared to correlation time between system and environment.

An example which will be used in this thesis is if the environment can be described as a measurement apparatus, which performs measurements at random times at a rate $R$~\cite{MME}. In this case the Lindblad operators are the Kraus operators from the theory of POVM, multiplied with $\sqrt R$. It is then clear that the time scale on which the master equation is  valid is the time a single measurement takes.

Most work about open quantum systems in the literature is concerned with Markovian systems described by a master equation of Lindblad form. This is on the one hand because in many physical situations the aforementioned assumptions are good approximations, and on the other hand because general non-Markovian master equations are much harder to solve. We will encounter non-Markovian behavior in the first part of this work, but we limited ourselves to a very small ``environment" which enables us to solve the full dynamics of $S+E$.

\part{\textbf{Coherently Driven Adiabatic Transport}}

\chapter{Introduction}

In the last two decades, the field of quantum computing has drawn the attention not only of a large number of physicists, but also mathematicians, computer scientists, engineers and to a smaller extent even the general public. At first glance this might be surprising, as we are not anywhere near of being able to build a useful machine for quantum information processing, but in fact there are many reasons for the excitement about the topic of quantum computing. 

Quantum mechanics these days is very well understood and was used for the development of a broad range of items which influence our lives. Just to name a few, there are lasers and all its applications; improvements in semiconductors which are used in LEDs, solar cells or classical computers; nuclear magnetic resonance essential for magnetic imaging in medicine; nuclear fission as an energy resource as well as for radiation needed for medical purposes; and many more. Yet it is very hard to imagine what consequences a quantum computer would have in our daily life. The situation might be compared with the introduction of electronics during the last, say six decades. Electromagnetism was well understood at that time and its applications already changed the world. But yet a new revolution was to come with electronics, the processing of information by means of electromagnetism.

Quantum physicists are very excited about the idea of quantum computing because it is believed that only such a machine could be capable of simulating many particle quantum systems. Mathematicians are fascinated by the challenge a quantum computer imposes on the Church-Turing thesis \cite{Turing}, which was long believed to be true without ever being proven. For computer scientists a whole new range of possible algorithms, so-called quantum algorithms open up. Finally, engineers are closest to the technical challenge of building such a machine.

One of the main challenges to a quantum computer is decoherence, a process which leads to classical like behavior of a quantum system. It is either caused by imperfect control of physical parameters, or by coupling to an environment. As an environment can be considered to change physical parameters in an non-controllable way, these two causes of decoherence behave very similarly and can generally be described within the same language. A problem lies in the fact, that systems which naturally couple only very weakly to any environment (e.g. nuclear spins), also couple weakly to measurement apparatuses, creating a new challenge in the readout of a computational result. 

In this part of our work, we are concerned with transfer of quantum information as it would be necessary to transport information in any computational device. Especially, if a two-qubit operation between distant qubits is needed (as will be in any realistic computation), the two qubits have to get close together to interact. One could also imagine that during the processing, quantum information has to move from some storage area to a processing area. 

In particular, we are concerned with information stored on a single quantum system. To be specific, we will use an electron and discuss the encoding of information into its spin (spin qubit), as well as into its location (charge qubit), and it will turn out that the encoding into the position is somewhat disadvantaged over the spin qubit. But our work equivalently applies to ions, where the information can be encoded in its energy levels, or other systems. This electron is allowed to tunnel between discrete positions, which we assume to be quantum dots (QD) fixed in space, but again one could use ion traps or similar.
\begin{figure}[htbp]
\begin{center}
	\includegraphics[width=12cm, height=4cm]{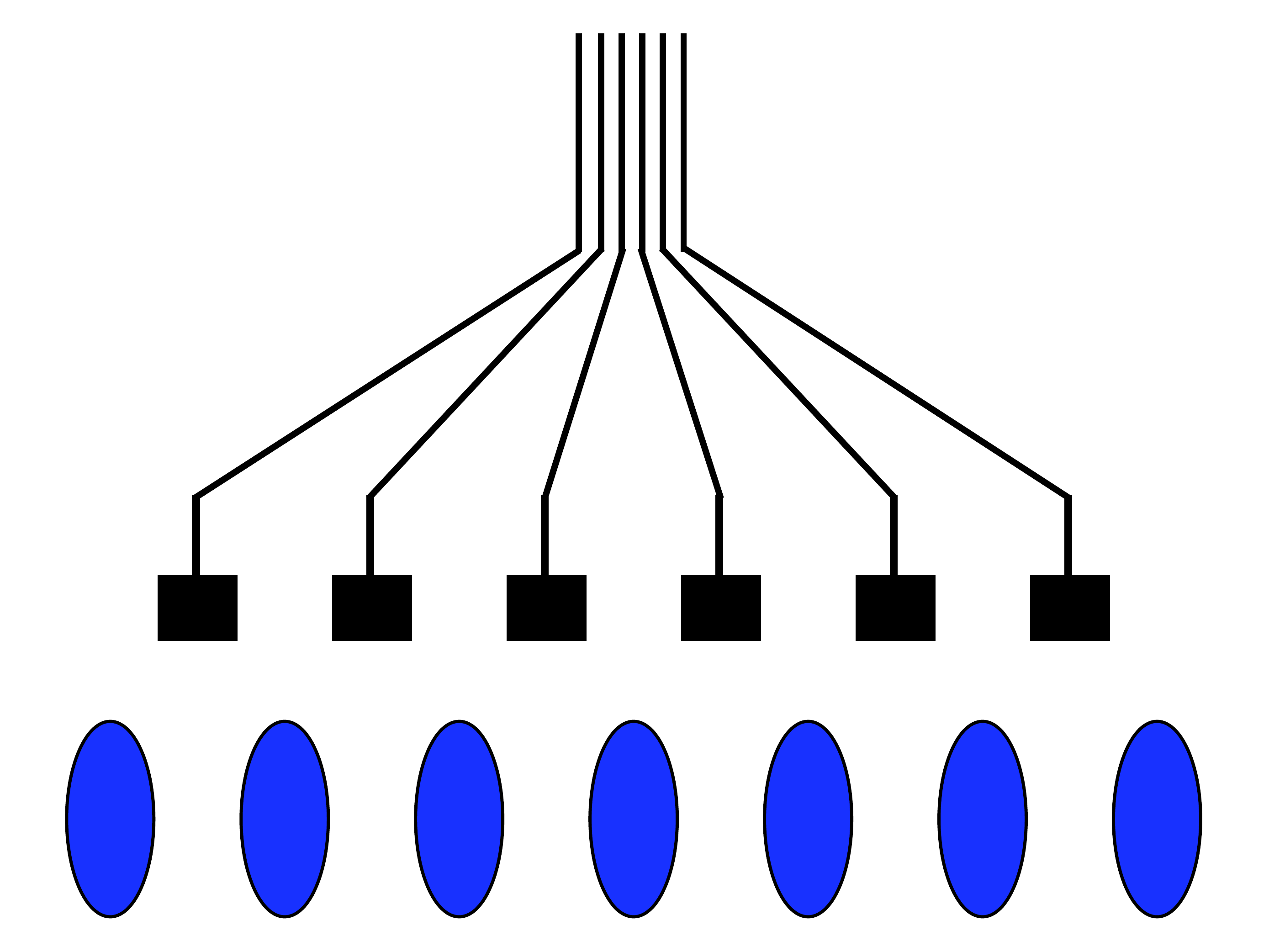}
	\caption{\small The electron carrying the quantum information can be located on any of the quantum dots (blue circles) of the quantum dot rail. It can move from one dot to another by tunneling, which can be controlled externally by applying voltages to the gates between them.}
	\label{figure1}
\end{center}
\end{figure}

We suppose that the tunneling rates between the QDs can be controlled externally, in our specific example by voltages applied to external gates~\cite{main} as is shown in Fig~\ref{figure1}. We will use a particular transport concept called Coherent Tunneling by Adiabatic Passage (CTAP)~\cite{main}, which in a different application is well know in quantum optics as Stimulated Raman Adiabatic Passage (STIRAP)~\cite{QO}. Indeed, it was recently shown~\cite{Rahman,Jong} by a Schr\"odinger-wave description with experimentally feasible parameters, that CTAP can be applied to a system where the QDs are replaced by phosphorus ions embedded in a silicon matrix. This latter system is a promising candidate for the implementation of quantum computing~\cite{Kane}.

As an adiabatic process, CTAP does not require a very precise control over the tunneling rates. All what is required is the possibility to suppress tunneling and to switch it on slowly, but the precise time when tunneling is switched on as well as the tunneling rate need only qualitative control. This advantage however comes with a downside in that adiabatic processes are typically slow, therefore giving the environment more time to destroy quantum behavior. 

We include two fundamentally different types of environmental decoherence in our study, and investigate their influence on the transfer fidelity. First, we assume that there are measurement apparatuses near the chain of QDs to measure the position of an electron. This could be a realistic situation as we might need these apparatuses for readout of quantum information, but more generally there are many types of Markovian environments (including dephasing), which have very much the same effect as such a measurement apparatus. Therefore our model for Markovian decoherence is quite realistic in many practical circumstances.

Second, we include a non-Markovian decoherence effect by coupling the electron's position to a two level system (TLS). This could model a two level fluctuator often found in solid state physics~\cite{TLF}. However, our non-Markovian model is very much simplified because we do not take into account the coupling of the TLSs to their environment. The reason is that full non-Markovian open quantum systems are very complicated. Furthermore, we are more interested in qualitative differences of Markovian and non-Markovian effects, and already with our simple model we find these to be quite pronounced. 

One should mention that there are many other types of decoherence sources, depending on the physical system CTAP is applied to. E.g.\ for spin qubits in a solid state matrix, the nuclear spins of nearby atoms lead to decoherence~\cite{nucspin}. Although a ${}^{28}$Si matrix does not have nuclear spins, a small fraction of ${}^{29}$Si is unavoidable. Other defects might also have nuclear spins. For charge qubits, fluctuating electrical fields are a prominent decoherence source~\cite{Stern}. Furthermore, $1/f$-noise is of importance in most solid state applications~\cite{oneoverf}. Of course, this work can not take into account all possible decoherence sources, but our conclusions from the case studies of a Markovian and a non-Markovian decoherence process should be valid at least qualitatively for similar processes.

\section{Quantum dots and quantum point contacts\label{things}}

Although our work should be usable for many physical systems as indicated in the introduction, we used the specific example of electrons on quantum dots. Similarly, we could model any Markovian dephasing source, but as a specific example we model dephasing due to measurements performed with quantum point contacts. We therefore devote this section of the introduction to say a few words about quantum dots and quantum point contacts (QPC), as well as two-level fluctuators, which also lead to decoherence. 

A \textbf{quantum dot}~\cite{QD} is mostly made of semiconducting material, but is so small (between one and a hundred nanometer), that it is often called a zero dimensional quantum system. An electron in a quantum dot is so much confined, that it can move in no direction (therefore zero dimensional). The quantum dot acts as a potential well for the electron, which can only inhabit the discrete energy levels obtained by solving the time independent Schr\"odinger equation. The smaller the dot, the larger is the difference between neighboring energy levels. The spacing between energy levels is further increased by semiconductor materials which result in a low effective mass of the electrons, typically about ten percent of the free electron mass.

The quantum properties of a quantum dot are most apparent when the difference between energies of different levels are comparable (or even large) to the thermal energy of the electron. Therefore, in quantum dot physics, one uses either small temperatures or extremely small quantum dots. As the optical properties of any semiconductor depend on the energy level structure, one can influence the optical properties of quantum dots by engineering their size.

It is also possible to add an additional electron to a quantum dot. The energy levels of this additional electron is also determined by the potential well of the quantum dot, that is, by the dots size. If the quantum dot is small enough and if the temperature is sufficiently low, then it is possible to ensure that the electron only occupies the ground state. This is the realm we will use.

If two QDs are located close to each other, they can act as a double well, and if the barrier between the wells is small enough, electrons can tunnel between the two QDs. The tunneling rate depends crucially on the distance between the quantum dots, but as electrons couple strongly to electric fields, the tunneling rate can also be controlled externally. For this purpose one can place electrodes between the two QDs and by adjusting voltages at these electrodes (often called gates), the electric field and hence the tunneling rate can be varied.

A \textbf{quantum point contact}~\cite{QPC1} is a narrow constriction between two wide electrically conducting regions, of a width comparable to the electronic wavelength (nano- to micrometer). At low temperatures and voltages, the conductance of a QPC is quantized according to $G=NG_Q$. Here, $N$ is a natural number which depends on the properties of the QPC, and $G_Q=2e^2/h$ is the conductance quantum. 

At finite temperatures, there is a steep, but continuous transition between the quantized conductance plateaus, at which $G$ becomes extremely sensitive to its electrostatic environment~\cite{QPC1,QPC}. Therefore, if the QPC conductance is set between such a transition, it can act as a very sensitive charge detector, able to detect single electrons on a QD in its vicinity.

\textbf{Two-level fluctuators}~\cite{TLF} are often found in solid state systems and can be a major source of decoherence. It is not exactly clear what these are, and there might be different physical reasons for their existence for different systems. What is known is that they behave like a two level system, randomly fluctuating between the two states. Their energy splitting is thought to have a broad distribution and the coupling between the TLS and the quantum system of interest also seems to be quite random and might differ between consecutive experiments.

\chapter{Adiabatic Dark State Population Transfer\label{stirap}}

The coherent transport of quantum information is an essential element in any scalable architecture for a quantum information processor. Much attention has been focused on dark state adiabatic passage for coherent state transport. Originally studied in the context of quantum optics \cite{overview} where it is called stimulated Raman adiabatic passage (STIRAP), dark state transport uses the existence of a ``dark state'', which is a zero-energy eigenstate of a driven quantum system. By manipulating the driving of the system, one can sculpt this dark state to coherently transport quantum states using STIRAP-like procedures. This intra-atomic dark-state transport has been demonstrated experimentally \cite{Goto2006}. However, the method has more recently been applied to  spatial transport of quantum information (which we specifically denote CTAP - coherent tunneling by adiabatic passage following \cite{main})  in a variety of physical systems, including chains of neutral atoms \cite{Eck}, quantum dots \cite{main, Hohenster2006, Petrosyan2006}, superconductors \cite{Siewert2006}, and photons in nearby waveguides \cite{Longhi2007}.  It has also been proposed as a crucial element in the scale up to large quantum processors \cite{Hollenberg}. The method possesses two very crucial benefits over other quantum transport methods: since the transport is via a zero energy state the quantum state acquires no dynamical phase, and due to the adiabatic theorem, the process is very robust to a wide range of system variations. 

\section{Adiabatic theorem\label{adia}}

This section is devoted to the \emph{adiabatic theorem}, because the adiabatic state transfer described in the following section relies heavily on it. This theorem (often called \emph{adiabatic approximation}) \cite{Messiah} is an approximation used in many fields of quantum mechanics, typically if the coherent dynamics induced by a time dependent Hamiltonian involves two different time scales. Its application requires that the time dependence of the Hamiltonian $H(t)$ is slow compared to internal time scales, defined by frequencies $\omega_{mn}(t)=({}E_n(t)-{}E_m(t))/\hbar$, where ${}E_j(t)$ are the eigenvalues of the Hamiltonian. The adiabatic theorem then states, that the time dependance of the Hamiltonian will not induce transitions between different instantaneous eigenstates of the Hamiltonian.

This situation is commonly achieved when the Hamiltonian can be controlled externally. For example one can slowly change the phase or magnitude of a laser field, which determines the Hamiltonian of an atom; or slowly change the direction of a magnetic field to control particles with a magnetic moment. There are also important applications to systems which are not controlled externally, as for example in molecular physics, where the Hamiltonian describing the electronic structure depends on the configuration of the nucleuses within the molecule. Because the dynamics of the heavy nucleuses are slow compared to the electronic dynamics, the adiabatic theorem can be applied for the latter. In particular, the adiabatic theorem is a requirement for the use of the Born-Oppenheimer approximation in molecular physics~\cite{Messiah}.

The theorem can be stated in a more precise way:\\
\textbf{Adiabatic Theorem:} \begin{minipage}[t]{10.6cm} Let $H(t)$ be non-degenerate with $H(t)\ket{n(t)}={}E_n(t)\ket{n(t)}$.\\ Assume $|\scalar{m(t)}{\dot n(t)}|\ll |{}E_m(t)-{}E_n(t)|/\hbar\quad\forall m\neq n$.\\ If the initial state of the system is an eigenstate $\ket{\psi(0)}=\ket{n(0)}$, \\ then it will approximately stay in the corresponding eigenstate $\ket{\psi(t)}\approx e^{\imath\phi(t)}\ket{n(t)}$ during the entire evolution.\\
\end{minipage}

Here $\ket{\dot n(t)}$ denotes the time derivative of the time dependent energy eigenstate $\ket{n(t)}$. We will not go into the proof of the adiabatic theorem as it is quite technical and can be found in standard text books~\cite{Messiah}. Instead, we will discuss some properties which are important for this chapter in more detail. First the reader should note, that the assumption in the second line of the theorem is not required for all $n$, but only for the one corresponding to the initial state. 

We also want to say a few words about the scaling of errors due to the adiabatic approximation. Under quite general conditions, the undesired populations $p_m(t)=|\scalar{m(t)}{\psi(t)}|$ of states $\ket{m(t)}\neq\ket{n(t)}$ satisfy\footnote{This is easily seen from the proof of the adiabatic theorem as given in \cite{Messiah} and is essentially due to the fact that the integral of a product of a fast oscillating function ($e^{i \omega_{mn}t}$) with a slowly varying function ($|\scalar{m(t)}{\dot n(t)}|$) vanishes in first order.\label{poia}}
	\begin{eqnarray}
		p_m(t)&\lesssim&\mbox{max}_{0<t'<t} \left(\frac{\hbar|\scalar{m(t')}{\dot n(t')}|}{|{}E_m(t')-{}E_n(t')|}\right)^{2}.\label{3.1}
	\end{eqnarray}
This implies that if we want to change an initial state $\ket{\psi(0)}=\ket{n(0)}$ adiabatically to some final state $\ket{\psi(t_T)}\propto\ket{n(t_T)},$ the undesired populations $p_m$ for $m\neq n$ decrease with the square of the transfer time $t_T$ (because $\ket{\dot n(t)}\propto t_T^{-1}$)
	\begin{eqnarray}
		p_m(t_T)&\propto& \left(\frac\hbar{t_T|{}E_m-{}E_n|}\right)^2.\label{3.2}
	\end{eqnarray}

There are very rare situations when (\ref{3.1}) and (\ref{3.2}) do not hold~\cite{Wu}, in particular if $|\scalar{m(t)}{\dot n(t)}|$ oscillates with the frequency $\omega_{mn}$. But we will not encounter this situation in the following. 

More important for our work is that the relation~(\ref{3.2}) can be significantly improved, if some higher derivations of $\ket{n(t)}$ are small. In our case, all higher derivations are small as our tunable parameters are described by Gaussian curves. Then, the final populations $p_m(t_T)$ decrease exponentially with $t_T$~\cite{Davis}, which means that exceptional fidelities can be achieved with reasonable short transfer times. However, this statement is only correct once the change of the state is completed (and $\dot H(t>t_T)=0$), while the populations $p_m$ during the process (for the same reason as in footnote~\ref{poia}) still follow relation~(\ref{3.1}). We will see such a behavior later in our studies, when small populations in undesired states $\ket{m(t)}$ appear during the state change, which almost magically disappear once the change of state is completed.

\section{Coherent population transfer\label{transport}}

We consider a system $A$ described by an $N$-dimensional Hilbert space where $N$ is an odd number, and we assume that we can tune the coupling between neighboring states. The Hamiltonian for such a system is
	\begin{equation}
		\hspace{3mm}H_A = \left( \begin{array}{ccccccc}
			  0 & \Omega_1&0& \cdots& 0&0 &0 \\ \vspace{0mm}
			  \Omega_1& 0 &\Omega_{2}  & &0& 0&0 \\ 
			  0 & \Omega_{2} & 0 & &0&0&0 \\
			  \vdots&  & &\ddots  &&&\vdots \\
			  0&0&0&&0&\!\Omega_{N-2}\!&0 \\
			  0 &0&0 && \!\Omega_{N-2}\! &0& \!\Omega_{N-1}  \! \\
			  0 & 0 &0&\cdots&0&  \!\Omega_{N-1}  \!& 0  \\
		\end{array} \right) \label{C14}
	\end{equation}
and by appropriate tuning of the couplings $\Omega_n$, we want to adiabatically change the state of the system from $\ket{1}_A$ to $\ket{N}_A$.  The couplings $\Omega_1=\Omega_P$ and $\Omega_{N-1}=\Omega_S$ are often called pump and Stokes pulse, respectively. 
	
In the quantum optical process of STIRAP (often three-level atoms are used), the different levels are energy levels of an atom, and $\Omega_n$ is the Rabi coupling induced by a laser field with frequency $\nu_n$. One can derive this form of the Hamiltonian~\cite{QO} if the laser field has frequency $\nu_n=|\epsilon_n-\epsilon_{n-1}|/\hbar$, where $\epsilon_n$ are atomic energy levels; that is, if there is no detuning. To do so, one has to use the interaction picture and perform the rotating wave approximation. If there is detuning, one has to include entries on the diagonals \cite{multilevel}, and we will return to this situation in chapter~\ref{full}.

In this work, the different levels correspond to the discrete positions of an electron, which is allowed to tunnel between $N$ quantum dots. We aim to transport the electron coherently from the first quantum dot to the last one. The spin of the electron is disregarded, which is justified if the tunneling rates $\Omega_n$ do not depend on it. Furthermore, we assume that the electron populates only the ground states of each quantum dot. Quite realistically, the electron can only tunnel between neighboring quantum dots leading us to the structure of the Hamiltonian above. We assume that we can tune the tunneling rates within $0<\Omega_j<\Omega_{max}$ by external gates, as shown in Fig.~\ref{figure1}.  In fact, during the transport process  we will only change the tunneling rates $\Omega_1(t)$ and $\Omega_{N-1}(t)$ between the first two dots and the last two dots, respectively, and keep $\Omega_n=\Omega_{max}$ for $n=2,3,\cdots,N-2$. The ket $\ket{n}_A$ denotes the state of the electron when it is found with certainty on the $n$-th quantum dot. 

Of importance for this process is the zero energy eigenstate (un-normalized)~\cite{main} of the Hamiltonian \Eqref{C14}
	\begin{eqnarray}
		\ket{\psi_0}_A &=& \cos \Theta\ket 1_A +(-1)^{(N-1)/2}\sin\Theta \ket N_A - X\!\!\sum_{j=2}^{(N-1)/2}(-1)^j \ket{2j-1}_A \qquad \label{C3.4}
	\end{eqnarray}
with
	\begin{equation}
		\Theta = \arctan\frac{\Omega_1}{\Omega_{N-1}}, \qquad \quad X = \frac{\Omega_1\Omega_{N-1}}{\Omega_{max}\sqrt{\Omega_1^2+\Omega_{N-1}^2}}.
	\end{equation}
It is easily seen that $\ket{\psi_0}_A=\ket{1}_A$ if $\Omega_1=0$ and $\Omega_{N-1}\neq 0$, and $\ket{\psi_0}_A=\ket{N}_A$ if $\Omega_1\neq 0$ and $\Omega_{N-1}= 0$. Therefore we can use this eigenstate for the adiabatic electron transport by performing the following steps:
	\begin{enumerate}
		\item While keeping $\Omega_{1}=0$ one switches $\Omega_{n}=\Omega_{max}$ for all $n=2,\cdots,N-1$. This ensures that the Hamiltonian is non-degenerate \footnote{The Hamiltonian is then non-degenerate only if $N$ is an odd number. This is the reason for assuming odd $N$ in the first place.}.
		\item Then $\Omega_1$ is increased and $\Omega_{N-1}$ is decreased until it vanishes. This process has to be done slowly to satisfy adiabaticity. It is this step where the electron moves from the first to the last quantum dot of the rail.
		\item Finally all couplings can be set to zero.
	\end{enumerate}
	
In the following, we will refer to these steps  as step one, step two and step three. We emphasize that the electron moves exclusively in the second step, and therefore, the first and  the third steps can be done arbitrarily fast (up to experimental limitations). Hence, we set $t=t_0$ at the beginning of step two and $t=t_{1}$ at the end of step two, and call $t_T=t_1-t_0$ the transfer time. 

During the entire process, the electron will be in the state $\ket{\psi_0(t)}$ (up to non-adiabatic corrections). It is interesting to note that this state does not populate the even numbered QDs, which is very counter intuitive in the following sense: Despite the fact that the electron can only tunnel between neighboring QDs, it can move from one end of the chain to the other end, without ever populating the even numbered QDs between them. 

This paradox can be understood if one has a look at the non-adiabatic corrections. These give  amplitudes on the even numbered QDs which are proportional to the inverse of the transfer time $t_T$, therefore validating the intuitive picture that the wave function passes through all QDs. The reason we say that even numbered QDs do not get populated in the adiabatic limit is that population is the squared absolute value of the amplitude, therefore scaling like $t_T^{-2}$. This gives \emph{population} $\times$ \emph{transfer time} $\propto t_T^{-1}$, resulting in negligible population of the even numbered QDs in the limit of long transfer times. 

This described paradox has important implications for STIRAP, as even numbered levels are usually excited atomic levels exposed to spontaneous decay. This decay is successfully avoided in STIRAP experiments because these levels do not get populated significantly. For our work, this paradox is interesting to mention, but has no implications as all QDs are equally subject to decoherence and there is no advantage of populating some more than others. 

The real advantage of CTAP compared to non-adiabatic population transfers is that the tunneling rates as well as the timing of their switching do not need to be very precise (see \cite{Eck} for a detailed study).  This is crucial in many experiments because these parameters are often hard to control. The only requirement is that tunneling can be suppressed completely when needed, and an approximate control of tunneling rates and their timing. As mentioned earlier, the drawback of CTAP is a relatively long transport time, which is limited by the adiabatic theorem and usually is about an order of magnitude longer compared to diabatic population transfer schemes.

\begin{figure}[htbp]\vspace{0cm}
	{\begin{rotate}{270} \includegraphics[width=18cm]{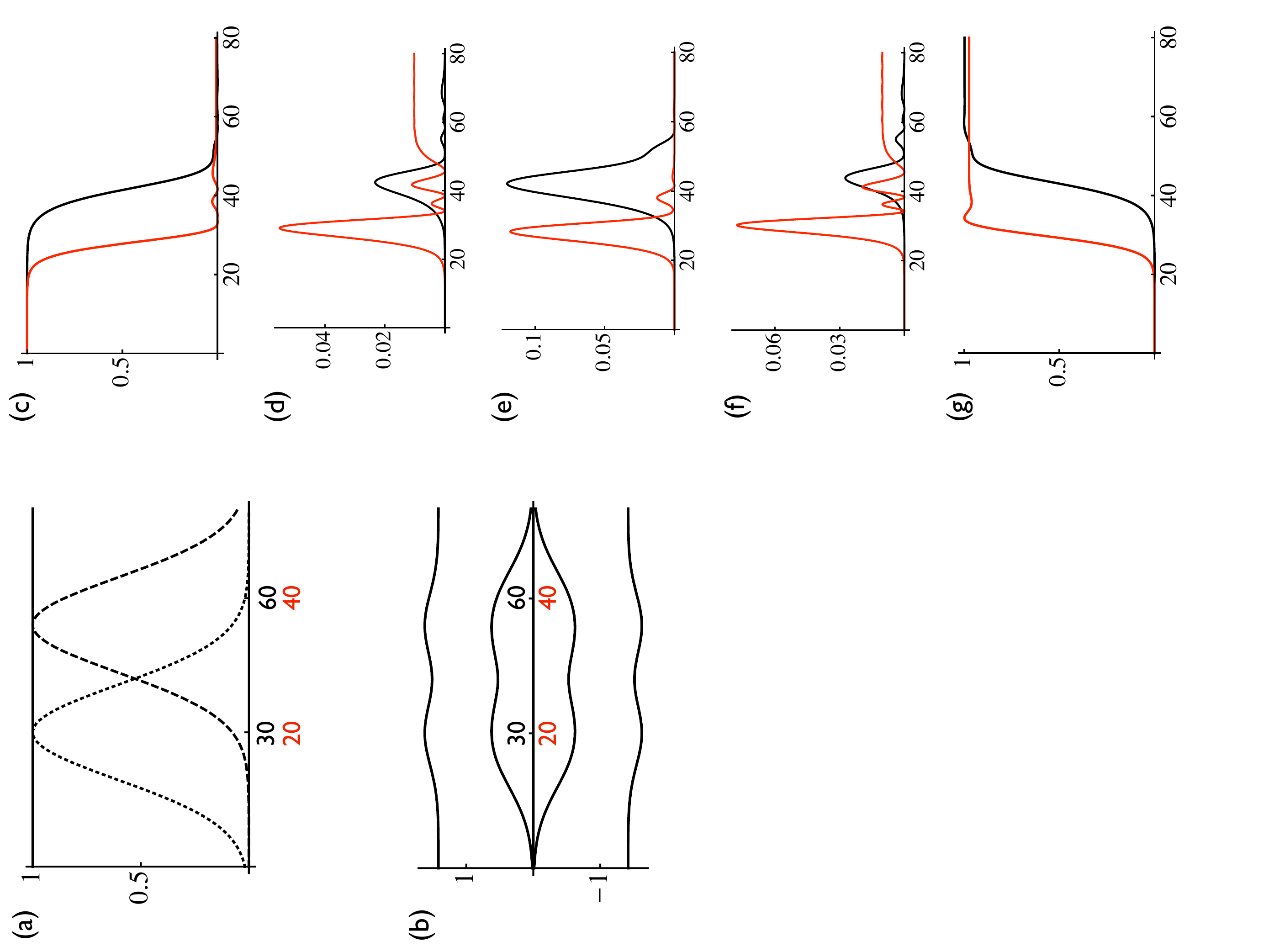} \end{rotate}}\vspace{17cm}
	\caption{\small (a): Pump pulse $\Omega_1/\Omega_{max}$ (dashed), Stokes pulse $\Omega_4/\Omega_{max}$ (dotted), and intermediate couplings $\Omega_{2,3}/\Omega_{max}$ (solid). (b): Eigenenergies of the Hamiltonian in units of $\Omega_{max}$. (c) - (g): Populations on the different quantum dots, starting with the first dot. Red lines correspond to faster transport, black lines to slow transport. Time is in units of $\hbar/\Omega_{max}$. Note the exponential increase in fidelity with transfer time.}
	\label{figure2}
\end{figure}

We conclude this section with an example of CTAP on five quantum dots. The tunnel rates
	\begin{eqnarray}
		\Omega_1 &=& \Omega_{max}\exp\left\{ -\frac{(t-0.9T)^2}{(T/4)^2}   \right\} \nn\\ 
		\Omega_4 &=& \Omega_{max}\exp\left\{ -\frac{(t-0.5T)^2}{(T/4)^2}   \right\} \nn\\
		\Omega_{2,3} &=& \Omega_{max}
	\end{eqnarray}
are plotted over time in Fig.~\ref{figure2}~(a), where the red and black marks on the time axis correspond to $T=40\hbar/\Omega_{max}$ and $T=60\hbar/\Omega_{max}$, respectively. Note that due to the Gaussian shape of these rates, there is no clear distinction between step one, two, and three. As $t_{T}$ is the time when all pulses are switched on, we have approximately $t_{T}=20$ for the red curves and  $t_{T}=30$ for the black ones. The energy levels are shown in Fig.~\ref{figure2}~(b) and show the required energy spacing during step two. The populations on the respective dots are shown in Fig.~\ref{figure2}~(c)~-~(g). Comparing the red with the black lines, it is clearly seen how a relatively small increase in the transfer time results in a dramatic increase of the fidelity (from 0.973 to 0.998). In the limit of slow populations transfer there would be zero populations on the second and fourth quantum dot. For finite transfer times we find populations on these dots due to non-adiabatic corrections. During the transfer these scale according to proportionality (\ref{3.2}), but most of it vanishes towards the end of the transfer (scaling exponentially in $t_{T}$). This effect, which was explained at the end of the previous section, is responsible for the for extremely good fidelities (with reasonable times) in the coherent transport of an electron.

\newpage

\chapter{Modeling of Decoherence\label{model}}

An important question, particularly with regard to the use of CTAP within large scale quantum computer architectures,  is to determine the effects of decoherence on the transport. The effects of dephasing and spontaneous emission has previously been examined in the case of STIRAP in a three level atom in a $\Lambda$ configuration \cite{Ivanov2004, Ivanov2005}. In that work, a master equation was postulated and its effects  on the population transfer studied. 

We examine the effects of two types of physically-motivated decoherence sources effecting the CTAP transport in a quantum dot (QD) chain. We first study the effects of delocalised non-readout measurements on the systems making up the QD chain. In particular we imagine quantum point contacts (QPC) close to each QD to measure the electric charge on the respective QD (see Fig.~\ref{figure3}). These QPCs however, are non-local measurement devices in that their charge sensitivity falls off continuously with distance. Such devices or similar will be required to either modulate or readout the quantum information in a real device. In large scale quantum processors one will routinely wish to have quantum information in a superposition of two ``distant'' spatial locations\footnote{If the qubit is encoded in the location itself, then such superpositions arise naturally. But even if the qubit is encoded on a different degree of freedom (i.e. spin), such superpositions appear during quantum information transport.}. It is known that from numerous studies of  cat-states in quantum Brownian motion - a single harmonic oscillator coupled to a bath of harmonic oscillators - the rate of decoherence suffered by the cat grows quadratically with the spatial separation of the two superposition states of the ``cat'' \cite{Joos}. We find that such an effect is also present in our case, i.e. the decoherence rate of a ``cat-state'' on the CTAP chain increases with cat-separation, but surprisingly we find that this decoherence rate saturates beyond a critical cat spatial separation.  This is a positive result for the CTAP transport protocol, and is essentially due to the rapid spatial fall off of the measurement sensitivity of the QPCs. 

This first model is an example of Markovian decoherence. We also include a second non-Markovian dephasing source, and consider that each quantum dot interacts with a nearby two level system (TLS). This could model two-level fluctuators in a solid state CTAP scheme.  Interestingly, we find that qubit transport still seems relatively robust in the presence of these combined Markovian and non-Markovian decoherence sources. More worryingly, however, we find that this non-Markovian dephasing slightly entangles the two level systems with the transported qubit. Surprisingly, this effect does not seriously detract from the transport of an electron isolated on one QD, but causes serious degradation if the qubit transported is in a superposition state. As the latter situation will be the typical case in a large scale quantum processor, the present analysis might indicate a much lower density of TLSs will be required when using CTAP in a large scale quantum computer, if operated with charge qubits. The problem is avoided by using an internal degree of freedom, like the spin of the electron, as then the transport of spatial superpositions is not required.

\section{Measurements as decoherence}

	\begin{figure}
		\centering
		\includegraphics[width=12cm]{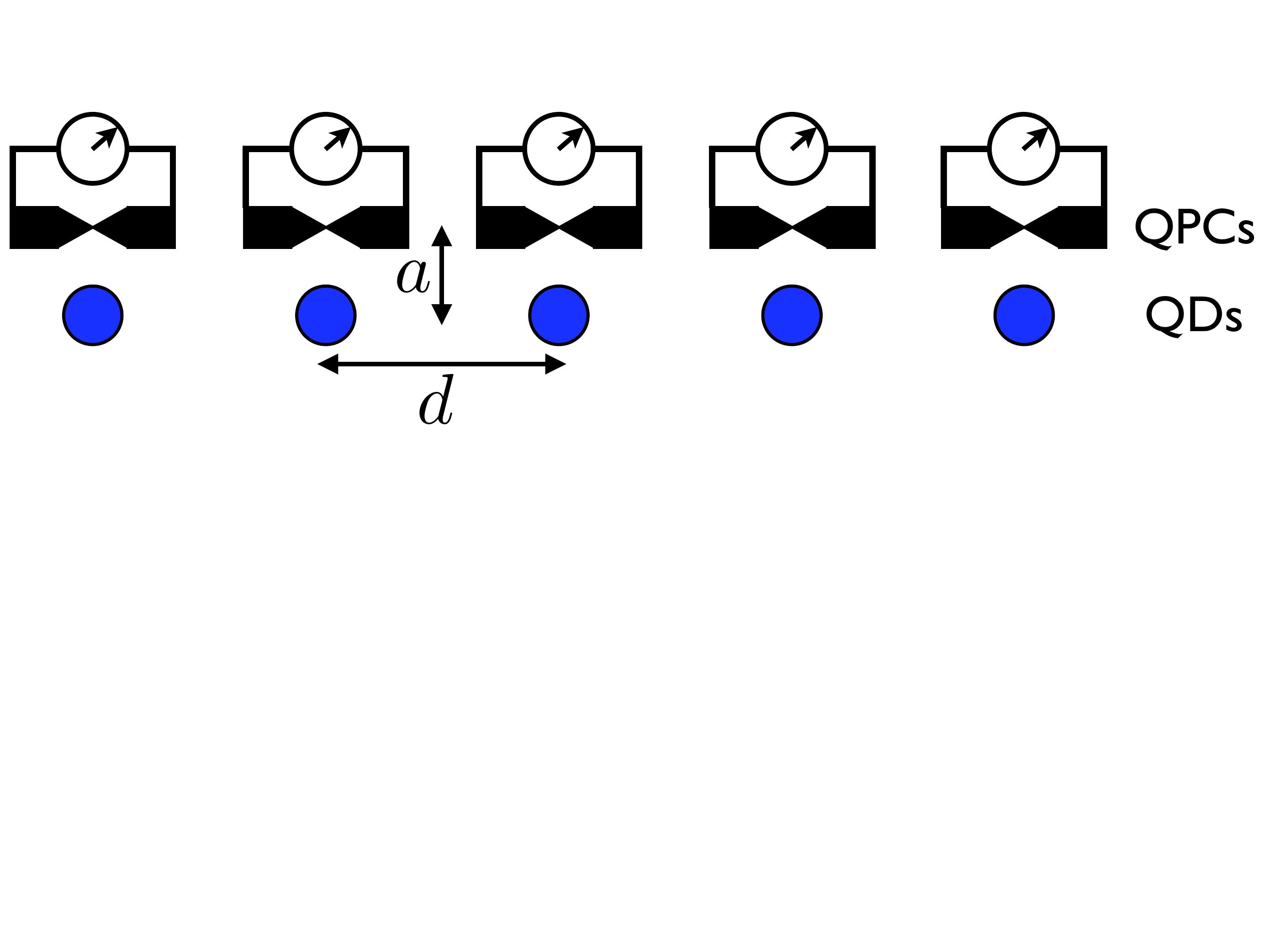}\vspace{-5cm}
		\caption{\small Realization of non-local measurement as proposed in \cite{QPC}. The electric current through a QPC is influenced by the appearance of an electronic charge in its near environment. Therefore, the electric current on the QPC measures the position of the electron. \label{figure3}}
	\end{figure}
As we mentioned above, we first consider each QD of the QD-chain to be continuously measured in a manner that gives rise to Markovian decoherence. We consider the measurements to be made by QPC situated close to the quantum dots as shown in Fig.~\ref{figure3}. In particular, we imagine that each time an electron travels through the QPC, this electron weakly measures whether there is an electron situated on a nearby quantum dot. The measurement executed by a QPC is caused by the modulation of the conductivity of the QPC due to the presence of a nearby electron (see section~\ref{things}).  The QPC conductivity is modulated by a factor $1-{\alpha}/{r}$, where $r$ is the distance between electron and QPC {\color{black} and $\alpha$ is a constant reflecting the properties of the QPCs}  (see~\cite{QPC}). This modulation results in an indirect position measurement of the electron's spatial position on the  rail of QDs.  However, it is a non-local measurement because even an electron on the neighboring QD influences the transmission through a QPC. The localness is parameterized by ${a}/{d}$, i.e. the distance between QPC rail and QD rail over the distance between two neighboring QDs, with small values representing more local measurements. Furthermore ${\alpha}/{a}$ parameterizes the sensitivity (signal over noise) of the measurements and is typically small and hence the measurements are weak ones. Such measurements are properly described in the language of positive operator valued measurements (POVM)~\cite{lectures,Busch}. 
	
The purpose of a measurement apparatus is the readout of quantum information for which one would like to use strong local measurements. This can be approximated by using a large number of weak, non-local measurements of the type described. A large number of measurements in a reasonably short time is achieved by using a high measurement rate which in turn can be realized, for instance, by applying a voltage to the QPCs. In this case there will be a small but macroscopic current through each QPC. By measuring this current, an observer learns the precise distance between the electron and the QPC, and essentially performs a perfect von Neumann-L\"uders measurement.  However, such measurements also act as a strong source of decoherence. During quantum unitary operations such as transportation by CTAP it is preferred that this decoherence is absent. However, switching off the measurements might not be completely achievable in practice and one may be left with a small current through the QPCs, either caused by non-zero voltage (possibly fluctuating), or by electrons which travel thermally across the QPC. Even if this current might be too small to be detected (no read out possible), each electron which travels across the QPC causes decoherence which is described by non-readout measurements\footnote{We are not interested in a possible readout during the transport anyway, as in a realistic setup we could not undo the decoherence effects of a measurement.}. Thus, non-readout measurements as a source of decoherence are included in the analysis presented here.
	
As in the previous chapter, we restrict our treatment to the case of having only one electron  in the rail of QDs. We also assume that the electron can only occupy the ground state of the QDs, and we take $\ket{i}_{\!A}$, to be the quantum state of the electron in the $i^{th}$ QD. Furthermore we neglect all interactions depending on the spin of the electron. Then, $\{ \ket{i}_{\!A},\; i=1,..,N\}$, form a basis for the Hilbert space $\mathcal{H}_A$ of this electron on the QD rail with $N$ dots. In the following we take the limit of a long rail, $N\to\infty$. We denote the distance between the $i$-th QPC and the $j$-th QD by 
	\begin{equation}
		r_{ij}=\sqrt{a^2+(|i-j|d)^2} \label{distance}
	\end{equation}
where $a$ and $d$ are defined in Fig~\ref{figure3}. 
	
The probability of the $j^{th}$ QPC detecting the presence of an electron on the QD rail  can be written as~\cite{lectures}
	\begin{equation}
		P_j(\rho^A) = \mbox{Tr}(\hat \pi_j\rho^A) \label{prob}
	\end{equation}
where $\rho^A$ is the state of the electron on the QD rail and $\hat \pi_j$ is the effect operator corresponding to the QPC measurement at site~$j$. If the electron is spatially localised to be only on the $i^{th}$ QD, i.e. in the state $\ket{i}_A$, \Eqref{prob} reduces to 
	\begin{equation}
		P_j(\ket{i}_A\bra{i}) =  \hspace{3mm}\bra{i}_{\hspace{-6mm}A}\hspace{4mm}\hat \pi_j\ket{i}_A. \label{egm}
	\end{equation}
As we noted above the measurement sensitivity of the QPC depends on the distance $r_{ij}$. The presence of an electron a distance $r_{ij}$ away from the QPC decreases the current flowing through the QPC by a factor  $1-{\alpha}/{r_{ij}}$  and this leads to a reduced detection probability,
	\begin{equation}
		P_j(\ket{i}_A\bra{i}) \propto 1-\frac{\alpha}{r_{ij}}\;\;. \label{relprob}
	\end{equation}
Fulfillment of \Eqref{relprob} is certainly achieved with the effect operators
	\begin{equation}
		\hat \pi_j = \frac{\bar{\kappa}}{N} \sum_{i=1}^N \left(1-\frac{\alpha}{r_{ij}}\right)\ket{i}_{\!A}\!\bra{i} \label{effect}
	\end{equation}
which can be checked by substitution into \Eqref{egm}. The constant
	\begin{eqnarray}
		\bar{\kappa}(N) &=& \frac{N}{ \sum_{i=1}^N \!\left(1-\frac{\alpha}{r_{ij}}\right)} \;=\; \frac1{1-\sum_{i=1}^N\frac{\alpha}{Nr_{ij}}}; \\
		 \kappa &=& \lim_{N\to\infty}\!\bar{\kappa}(N) \;=\; 1 + \lim_{N\to\infty}\sum_{j=1}^{N}\frac{\alpha}{Nr_{ij}}\label{gamma}
	\end{eqnarray}
is chosen to satisfy $\sum_{j=1}^{N}\hat \pi_j=\one$. Note that each effect operator is almost proportional to the unit operator (remember that $\alpha/r_{ij}$ is small) which reflects the weakness  of the measurements being performed.

We will assume that the measurements performed by the QPCs are efficient\footnote{This is quite realistic as we describe the measurement from a very microscopic viewpoint} (see section~\ref{genmeas}), that is they only introduce a minimum amount of decoherence. Measurement theory states that the transformation of the density operator due to such a measurement with result $j$ is described by \cite{lectures}
	\begin{equation}
		\rho^A \quad\parbox{1.2cm}{{\small $\;\;\hat \pi_j$}\\$\overrightarrow{\phantom{hihh}}$}  \hat A_{j} \rho^A \hat A_{j}^\dagger.
	\end{equation}
with $\hat A_j=\hat U_j\sqrt{\hat \pi_j}$ for some arbitrary unitary operators $\hat U_j$. This unitary depends on the interaction of the quantum system and measurement apparatus during the measurement process. In our case, if the electron is at site $i$, its position will not be changed by a detection event. For simplicity we also assume that a measurement does not introduce relative phases within the rail of QDs, leading to $\hat U_j\equiv \one$ and $\hat A=\hat A^\dagger$, which means the measurements influence the electron state as little as possible:
	 \begin{eqnarray}
		\hat A_j &=& \sqrt{\frac{\kappa}{N}} \sum_{i=1}^N \sqrt{1-\frac{\alpha}{r_{ij}}}\ket{i}_{\!A}\!\bra{i} \label{AAA}
	\end{eqnarray}
	
To derive the master equation we now assume that detection events in the QPCs occur uniformly at random and at a constant rate $R$. Furthermore, we ignore the detection result and average over all possibilities (non-readout measurements). Following \cite{MME}, we can write down the master equation describing the evolution of the density operator as
	\begin{equation}
		\frac{\dd \rho^A}{\dd t} = \frac1{\imath\hbar}[H_A,\rho^A] - R\rho^A + R\sum_{j=1}^N \hat A_j \rho^A \hat A_j .\label{4aa}
	\end{equation}
This equation is in Lindblad form with Lindblad operators $\hat A_j$ and thus the evolution is Markovian. The two incoherent terms can be interpreted in the following way: Each time a measurement occurs, the state $\rho^A$ is substituted by $\sum_j \hat A_j\rho^A\hat A_j$ corresponding to a non-readout measurement. This process happens with rate $R$.  We note that $R$ should scale proportional to $N$ as each QPC contributes equally to the overall measurement rate ($R/N$ is the average measurement rate of a single QPC). For the case when $H_A$ involves no electron transport along the QD rail, Eqn. \Eqref{4aa} possesses a stationary state which is diagonal in $\ket{i}_A$, and thus the evolution corresponds to a pure dephasing type of decoherence.  
	
For now we take $H_A=0$, but later we introduce a time dependent Hamiltonian to induce CTAP. Expressing \Eqref{4aa} in the basis $\ket{k}_A$ and using \Eqref{AAA},  we obtain $\dot{\rho}^A_{kk}=0$, for diagonal entries and $\dot{\rho}^A_{kl}=-D_{kl}\rho^A_{kl}$ for off-diagonal ones. The dephasing rate is
	\begin{equation}
		D_{kl} = R\left[1-\frac{\kappa}{N}\sum_{j=-\infty}^\infty \sqrt{1-\frac{\alpha}{r_{kj}}} \sqrt{1-\frac{\alpha}{r_{lj}}}\right] \label{5}
	\end{equation}
where $r_{ij}$ is defined in~\Eqref{distance}.

\subsection{Properties of the dephasing rate}

In this brief subsection we study the dephasing rate \Eqref{5} in more detail. First we show that it saturates for large distances \mbox{$|k-l|d\gg a$}. To this end we assume $l>k$ and split the sum into two parts 
	\begin{eqnarray}
		\frac{\kappa}{N}\sum_{j=-\infty}^\infty \sqrt{1-\frac{\alpha}{r_{kj}}} \sqrt{1-\frac{\alpha}{r_{lj}}} &=& \frac{\kappa}{N}\sum_{j=-\infty}^{(k+l)/2} \sqrt{1-\frac{\alpha}{r_{kj}}} + \frac{\kappa}{N}\sum_{j=(k+l)/2}^\infty  \sqrt{1-\frac{\alpha}{r_{lj}}} \qquad\quad
	\end{eqnarray}
where we also used that $r_{lj}$ is large in the first sum on the right hand side of the equation, and $r_{kj}$ is large in the second sum. With the same argument we can extend the sums to infinity 
	\begin{eqnarray}
		\frac{\kappa}{N}\sum_{j=-\infty}^{(k+l)/2} \sqrt{1-\frac{\alpha}{r_{kj}}} &=& \frac{\kappa}{N}\sum_{j=-\infty}^{\infty} \sqrt{1-\frac{\alpha}{r_{kj}}} - \frac{\kappa}{N}\sum_{j=(k+l)/2}^\infty 1
	\end{eqnarray}
to find
	\begin{eqnarray}
		\frac{\kappa}{N}\sum_{j=-\infty}^\infty \sqrt{1-\frac{\alpha}{r_{kj}}} \sqrt{1-\frac{\alpha}{r_{lj}}} &=& \frac{\kappa}{N}\sum_{j=-\infty}^{\infty} \left[ \sqrt{1-\frac{\alpha}{r_{kj}}} + \sqrt{1-\frac{\alpha}{r_{lj}}} -1 \right].
	\end{eqnarray}
Due to the periodic distribution of the QDs and QPCs, this is clearly independent of $k$ and $l$ and is therefore the limiting dephasing rate for large separations $|k-l|$.

Next we consider the limit of weak measurements $\frac{\alpha}{r_{ij}}\ll 1$ in \Eqref{5}, which should be well justified in experiments~\cite{QPC}. This weak measurement limit corresponds with the inability of the measurements to give detailed information on the position of the electron in the QD rail. Weak measurements of this nature feature in many models of continuous monitoring of a quantum particle's position \cite{Milburn-Caves}.  In the limit of weak measurements we can expand the square roots in \Eqref{5} to second order in $\alpha/R_{ij}$. Exploiting the periodicity of the setup we then use $\sum_j 1/r_{kj}=\sum_j 1/r_{lj}$ and applying \Eqref{gamma} we find
	\begin{equation}
		D_{kl} \approx \frac{R\alpha^2}{8N}\sum_{j=-\infty}^\infty \left( \frac{1}{r_{kj}}-\frac{1}{r_{lj}} \right)^2,
	\end{equation}
where it is again apparent that $R$ has to scale linear with $N$ as already discussed from a physical point of view.  For spatial separations larger than the threshold, \mbox{$|k-l|d\gg a$}, we can again split the sum into two parts, extend them to infinity and find 
	\begin{eqnarray}
		\sum_{j=-\infty}^\infty \left( \frac{1}{r_{kj}}-\frac{1}{r_{lj}} \right)^2 &\approx& 2\sum_{j=-\infty}^\infty \frac1{r_{kj}^2}\nn\\
		&=& \frac2{d^2}\sum_{j=-\infty}^\infty \frac1{(a/d)^2+(k-j)^2} \qquad\nn\\
		&=& \frac2{d^2}\sum_{j=-\infty}^\infty \frac1{(a/d)^2+j^2} \nn\\
		&=& \frac{2\pi}{d^2}\frac{\coth{(\pi a/d)}}{a/d} .
	\end{eqnarray}
The last equality is a standard formula which can be checked i.e. with Mathematica. Finally we get the large distance saturation value of the dephasing rate
	\begin{eqnarray}
		\lim_{|k-l|\to \infty} D_{kl}&=&\frac{\pi R\alpha^2}{4Nd^2}\frac{\coth{(\pi a/d)}}{a/d}, \label{coth}
	\end{eqnarray}
valid in the weak measurement limit $\alpha/a\ll1$.
	
We also note that we can model local measurements, where the measurement result of any QPC only depends on the charge of the nearest QD, in the limit $\frac{1}{r_{ij}}=\delta_{ij}$, while keeping $\alpha$ finite, e.g. in \Eqref{5}. In this case we find from \Eqref{5}
	\begin{equation}
		D=\frac{R}{N}\left( 2+\alpha-2\sqrt{1+\alpha}\right),\label{aaaa}
	\end{equation}
which is of course independent of $k$ and $l$.

The saturation of the decoherence rate in \Eqref{coth}, is somewhat surprising when one compares this with the similar situation for a free particle in a spatial superposition cat-state, experiencing continuous position measurements~\cite{Joos}.  In that case the decoherence rate suffered by the particle increases without bound according to the spatial separation of the cat, i.e. $D_{x_1,x_2}\propto (x_1-x_2)^2$. This however depends very much on the measurement model.

\section{Coupling to two level systems\label{coupling1}}
	\begin{figure}
		\centering
		\includegraphics[width=12cm]{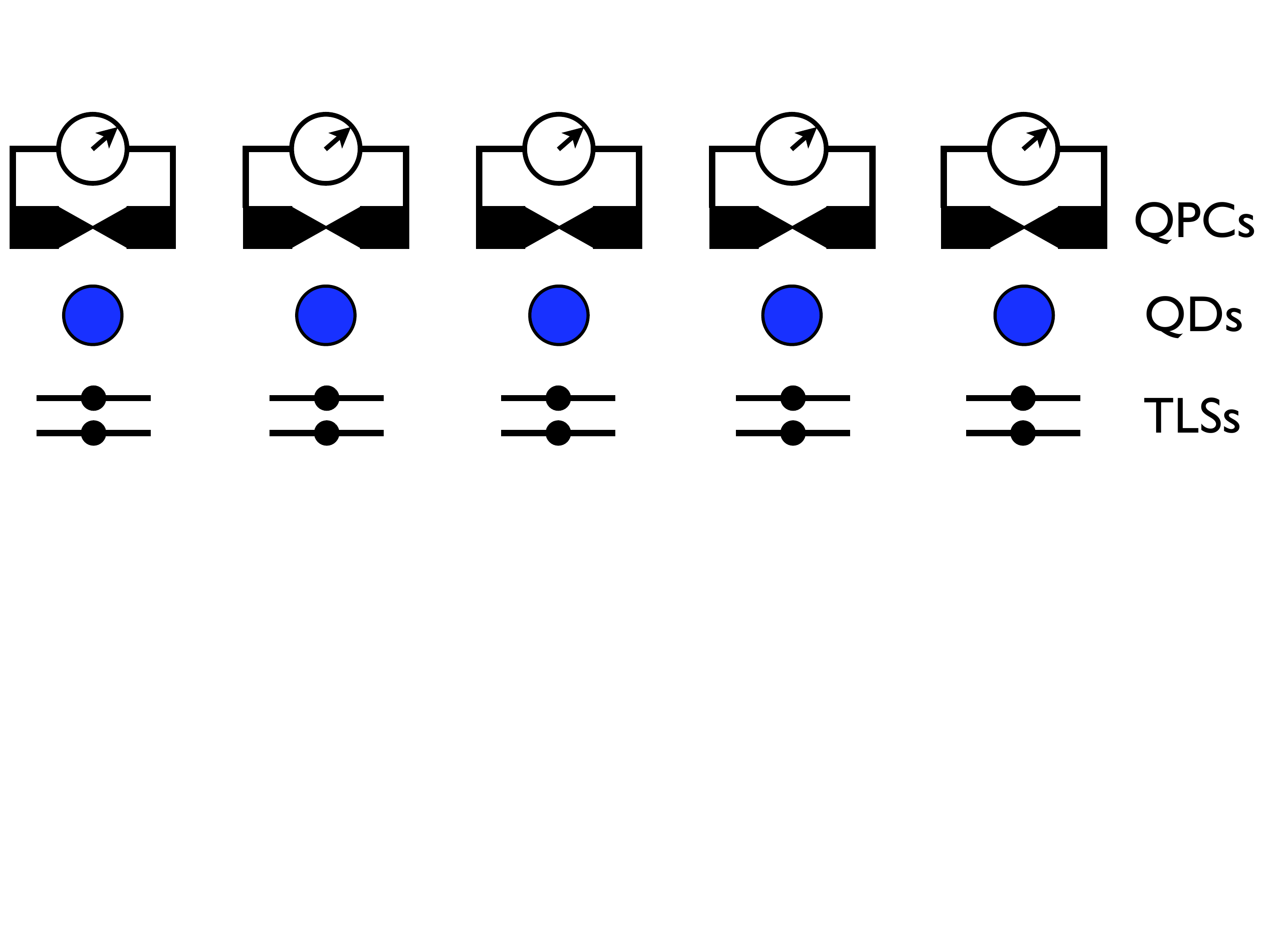}\vspace{-4.5cm}
		\caption{As in Fig~\ref{figure3} but with additional local coupling to two-level systems. \label{figure4}}
	\end{figure}

We now consider a further source of decoherence. It is highly likely that in any physical device there will be unknown accidental two level fluctuators nearby to the quantum dot rail (see Fig. \ref{figure4}). In fact, experiments in solid state physics \cite{TLF} (and references therein) often find decoherence due to coupling to two level fluctuators. Although not much is known about these two level fluctuators, they certainly have to be taken into account in many solid state devices. In fact, from the study of superconducting qubits~\cite{Clemens}, it seems likely that these TLF are within any layer of amorphous silicon oxide. Such layers are inevitable in most silicon devices, and would also be needed in the implementation of CTAP using phosphorus atoms embedded in a silicon matrix~\cite{Rahman}.

As a rail of QD's would most likely be embedded in a solid state matrix in any technical device, we include these mysterious systems in our studies. If these unknown two level systems (TLS) can couple to the electron on the rail, then these systems act as a source of decoherence which exhibits memory, i.e. is non-Markovian. That is, quantum coherences on the quantum dot rail can be transferred to the nearby TLSs, where they can remain for a period, before being transferred back. Typically the analysis of these types of non-Markovian effects are complex but in the following we are able to derive analytic solutions of the resulting reduced dynamics of the quantum dot rail. For simplicity we assume these fluctuators have no internal dynamics other than their coupling to the quantum dot rail\footnote{A possible internal Hamiltonian $H_{TLS}$ does not influence the QD rail if $[H_{TLS},H_{int}]=0$, because it can be removed by the unitary operator $e^{\imath H_{TLS}t/\hbar}$, without changing $H_{int}$.}, and that this coupling is local to the nearest QD only. Furthermore we assume that these TLSs do not experience significant decoherence on the time scales of the transport. We are aware that these assumptions may not be completely satisfied in all realistic situations. However this model will serve to highlight the striking difference between the Markovian and non-Markovian evolutions in, for instance, the transport of a spatial  superposition.
	
We use $\ket{1}_{\!j}$ and $\ket{0}_{\!j}$ as basis of the Hilbert space $\mathcal{H}_j$ of the $j-$th TLS such that the interaction Hamiltonian is diagonal. If the electron is on the $j$-th QD, it is assumed to induce a phase shift on the $j-$th TLS, so that the Hamiltonian acting in the {\color{black}product Hilbert space $\mathcal{H}=\mathcal{H}_A\bigotimes_{j=1}^N \mathcal{H}_j$ of electron and TLSs} reads
	\begin{equation}
		H_{int} = \sum_{n=-\infty}^{\infty}\,\left[ \chi_n\ket{n}_{\!A}\!\bra{n}\otimes\hat \sigma_{z,n}\bigotimes_{j\ne n} \one_j\right]\;\;. \label{inter}
	\end{equation}
{\color{black}The coupling constants $\chi_j$ are} considered to be constant in time, and $\hat \sigma_{z,j}$ and $\one_j$ are the Pauli $Z$-matrix and the identity operator, respectively, acting in $\mathcal{H}_j$. 
	
As typically assumed, we now take the initial state to be in product form $\varrho(t=0)=\rho_0^A\otimes\rho_{0}^{TLSs}$, where $\rho_{0}^{TLSs}$ does not have to be a product state of the different TLSs. We can now express the master equation describing the time evolution of the density matrix of the combined QD rail and TLSs under the effects of the above measurements  (see \Eqref{4aa}), to be 
	\begin{equation}
		\frac{\dd \varrho}{\dd t} = -\imath[H_{int},\varrho] - R\varrho + R\sum_{j=1}^N \hat A_j \varrho \hat A_j \label{8}
	\end{equation}
	with $\hat A_j=\sqrt{\hat \pi_j}\bigotimes \one^{\otimes\,N}$.
		
After some effort, one can trace out the TLSs to find the non-Markovian master equation for the reduced density matrix of the QD rail for an arbitrary initial product state of the TLSs, as
	\begin{eqnarray}
		\frac{\dd \rho^A}{\dd t} &=&  - R\rho^A+\sum_{n=1}^N \Big\{R \hat A_n\rho^A\hat A_n - \imath\Delta_n(t) \left[\ket{n}_{\!A}\!\bra{n},\rho^A\right]  \nn\\
		& & +  \gamma_n(t) \left[  \ket{n}_{\!A}\!\bra{n}, \left[ \rho^A,\ket{n}_{\!A}\!\bra{n} \right]\right]  \Big\} ,\label{stuff1}
	\end{eqnarray}
where the last two terms describe the effects of the TLSs on the QD rail. Here the definition 
	\begin{equation}
		\gamma_n(t)-\imath\Delta_n(t) = \chi_n\frac{\sin{\chi_nt}-\imath\omega_n\cos{\chi_nt}}{\cos{\chi_nt}+\imath\omega_n\sin{\chi_nt}} 
	\end{equation}
is used and $\omega_n=$Tr$[\rho_n\hat \sigma_{z,n}]$, is the inversion of the $n$-th TLS. Decoherence due to the TLSs is described by $\gamma_n$ whereas $\Delta_n$ represents the Lamb-shift. If $\omega_n=\pm 1$, then $\Delta_n=\pm\chi_n$ and $\gamma_n=0$. In this case the decoherence due to the coupling to TLSs vanishes and the coupling to the TLSs results only in a change of the energy of states $\ket{n}_A$ by $\pm\chi_n$. However, a more interesting case is when $\omega_n=0$, since it includes the case where the TLSs may initially be in a complete mixture $\rho_n(0)=\half\one$.  In  this case we find $\Delta_n=0$, and $\gamma_n=\chi_n\tan{\chi_nt}$. The presence of singularities in $\gamma_n(t)$, may cause difficulties in studying the time evolution of $\rho^A$, using normal methods. To avoid this we do not work directly with (\ref{stuff1}). Instead we return to solve the complete dynamics of the coupled QDs and TLSs and then trace out the latter to obtain $\rho^A(t)$.

To this end, we transform \Eqref{8} with
	\begin{equation}
		e^{\imath H_{int}t} = \sum_{n=1}^{N} \ket{n}_{\!A}\!\bra{n}\otimes \left(\cos{\!(\chi_nt)}\!\one_n +\imath \sin{\!(\chi_nt)} \hat \sigma_{z,n}\right) \bigotimes_{j\ne n} \!\one_j .
	\end{equation}
The combined density operator in this picture is given by  $\overline{\varrho}(t)=e^{\imath H_{int}t}\varrho(t)e^{-\imath H_{int}t}$, and is governed by the master equation
	\begin{equation}
		\frac{\dd \overline{\varrho}}{\dd t} =  - R\overline{\varrho} + R\sum_{j=1}^N \hat A_j \overline{\varrho} \hat A_j. \label{10aa}
	\end{equation}
Note that $e^{\imath H_{int}t}\hat A_je^{-\imath H_{int}t}=\hat A_j$. Taking the components of \Eqref{10aa} in the QD Hilbert space,  $\overline{\varrho}_{kl}(t)={}_A\!\bra{k}\overline{\varrho}\ket{l}_{\!A}$ (which is still an operator in the TLSs Hilbert space), we can repeat the calculation of the previous section to find 
	\begin{equation}
		\overline{\varrho}_{kl}(t) = e^{-D_{kl}t} \overline{\varrho}_{kl}(0)= e^{-D_{kl}t} {\varrho}_{kl}(0),
	\end{equation}
where  $D_{kl}$ is given in \Eqref{5}. After transforming back to the Schr\"odinger picture and tracing over the TLSs, we find for the components of the reduced density operator 
	\begin{eqnarray}
		\rho^A_{kk}(t) &=& \rho^A_{kk}(0) \\
		\rho^A_{kl}(t) &=& e^{-D_{kl}t} \cos{(\chi_kt)}\cos{(\chi_lt)}\rho^A_{kl}(0), \label{13}
	\end{eqnarray}
where we have assumed that the initial states of the TLSs is a completely mixed state
	\begin{equation}
		\rho_j(0)=\half\one . \label{initial}
	\end{equation}
Hence we see that information lost to the TLSs can return to the rail via the oscillatory terms in~\Eqref{13}, which is in contrast to Markovian decoherence induced by the measurements. Note that the non-Markovian behavior of~\Eqref{13} results from tracing out the environmental TLSs.

\chapter{Coherent Transport\label{full}}

Here we employ the technique of CTAP described in chapter~\ref{stirap} to the system of QDs coupled to QPCs as well as to TLSs as outlined in chapter~\ref{model}. A minor difference to chapter~\ref{stirap} is that the system here consists of an extended rail of QDs with an infinite dimensional Hilbert space $\mathcal H_A$. The aim is to transport an electron coherently from the $m$-th QD to the $n$-th one, where the number $(n-m+1)$ of involved QDs is required to be odd because of reasons discussed in section~\ref{transport}. Although we can restrict the system Hamiltonian $H_A$ as well as the local coupling to Hamiltonian $H_{TLS}$ to the subspace corresponding to these QDs, we have to take into account measurements performed by QPCs on sites other than $m,\cdots,n$. This is because of the non-local nature of these measurements.

	The  system Hamiltonian reads
	\begin{equation}
		\hspace{3mm}H_A = \left( \begin{array}{ccccccc}
			  0 & \Omega_m&0& \cdots& 0&0 &0 \\ \vspace{0mm}
			  \Omega_m & 0 &\!\!\!\Omega_{m+1} \!\!\! & &0& 0&0 \\ 
			  0 &\!\!\! \Omega_{m+1} \!\!\! & 0 & &0&0&0 \\
			  \vdots&  & &\ddots  &&&\vdots \\
			  0&0&0&&0&\!\!\!\Omega_{n-2}&0 \\
			  0 &0&0 && \!\!\!\Omega_{n-2} &0& \!\!\!\Omega_{n-1}  \!\!\! \\
			  0 & 0 &0&\cdots&0&  \!\!\!\Omega_{n-1}  \!\!\!& 0  \\
		\end{array} \right), \label{14aa}
	\end{equation}
where following section~\ref{transport} we call $\Omega_m=\Omega_P(t)$ the pump pulse and $\Omega_{n-1}=\Omega_S(t)$ the Stokes pulse, and set $\Omega_j=\Omega_{max}$ for all $j=m+1,\cdots,n-2$. Also, we will use the division of the process into three steps as introduced in section~\ref{transport}, and set $t=t_0$ at the beginning of step two, just before the population transfer starts, as well as $t=t_1$ at the end of step two, when the population transfer is finished.


	\section{Measurements\label{hello}}

To introduce our technique for solving a master equation in the adiabatic approximation, we first neglect the coupling to the TLSs. The master equation to be solved is \Eqref{4aa}
	\begin{equation}
		\frac{\dd \rho^A}{\dd t} = \frac1{\imath\hbar}[H_A,\rho^A] - R\rho^A + R\sum_{j=-\infty}^\infty \hat A_j \rho^A \hat A_j .\label{C5.2}
	\end{equation}	
with the system Hamiltonian \Eqref{14aa}.  At $t=t_0$ the $(n-m+1)$ eigenstates of $H_A$ which are superpositions of $\ket{m}_A,\cdots,\ket{n}_A$, are denoted by $\ket{\psi_j(t_0)}$ with $j=\frac{m-n}{2},\cdots,\frac{n-m}{2}$, ordered by their energy ${}E_j(t_0)$. These eigenstates evolve continuously to $\ket{\psi_j(t)}$ which depend on the choice of the pump and Stokes pulses $\Omega_P(t)$ and $\Omega_S(t)$, respectively. The un-normalized adiabatic state responsible for the transport is according to~\Eqref{C3.4}
	\begin{eqnarray}
		\ket{\psi_0(t)} &=& \cos{\Theta}\ket{m}_A+(-1)^{\frac{n-m}{2}}\sin{\Theta}\ket{n}_A  -X \sum_{j=2}^{\frac{n-m}{2}}(-1)^j\ket{2j-2+m}_A \label{psi0} \qquad
	\end{eqnarray}
with
	\begin{eqnarray*}
		\Theta = \arctan{\frac{\Omega_P}{\Omega_{S}}} ,&\;& X= \frac{\Omega_P\Omega_{S}}{\Omega_{max}\sqrt{\Omega_P^2+\Omega_{S}^2}}  .
	\end{eqnarray*}

Note that $\ket{\psi_0(t_0)}=\ket{m}_A$ and $\ket{\psi_0(t_{1})}=\ket{n}_A$, i.e. the states in which the electron is on the $m-$th and $n-$th QD, respectively. Furthermore during the entire process the energy  ${}E_0(t)=0$ of $\ket{\psi_0(t)}$ vanishes, which ensures that no dynamic phase appears for the state to be transported. We recall from section~\ref{adia} that if
	\begin{equation}
		|{}E_0(t)-{}E_j(t)| \gg\left|\scalar{\frac{\dd}{\dd t}\psi_0(t)}{\psi_j(t)}\right|\quad \label{15aa}
	\end{equation}
holds and if the system at $t=t_0$ is in $\ket{\psi_0(t_0)}\,$, then the adiabatic theorem states that the system will stay in $\ket{\psi_0(t)}\,$, provided it is a closed system. Therefore an electron starting in $\ket{n}_A$ will end up in $\ket{m}_A$. 
	 
To generalize this concept to the open system described here, we follow~\cite{adiabat} and transform \Eqref{4aa} with the unitary operators defined by
	\begin{equation}
		\hspace{-1mm}\begin{array}{rcll}
			\hat U^\dagger(t)\ket{\psi_j(t)} &\!\!=\!\!& \ket{\psi_j(t_0)}&\;\mbox{for }j=\frac{m-n}{2},\cdots,\frac{n-m}{2} \\
			\hat U^\dagger(t)\ket{j}_A &\!\!=\!\!& \ket{j}_A & \;\mbox{for } j< m \mbox{ and }j> n
		\end{array}\hspace{-1mm} \label{uni}
	\end{equation}
to get
	\begin{eqnarray}
		 \frac{\dd\widetilde{\rho}^A}{\dd t} &=& -\imath\left[ \sum_{j=\frac{m-n}{2}}^{\frac{n-m}{2}} {}E_j(t)\ket{\psi_j(t_0)}\bra{\psi_j(t_0)} - \imath \hat U^\dagger\frac{\dd \hat U}{\dd t} , \widetilde{\rho}^A \right]  - R\widetilde{\rho}^A + R\sum_{i=1}^N \widetilde{A}_i \widetilde{\rho}^A \widetilde{A}_i \qquad\quad \label{label}
	\end{eqnarray}
with $\widetilde{O}= \hat U^\dagger \hat O \hat U$ for any operator $\hat O$. If, in addition to adiabaticity \Eqref{15aa} we also assume weak coupling to the environment \footnote{This is usually justified in systems described by a Markovian master equation.}, one can neglect the term $\imath \hat U^\dagger\frac{\textrm{\footnotesize d} \hat U}{\textrm{\footnotesize d}t}$ in \Eqref{label} 
as is shown in~\cite{adiabat}. This is the generalization of the adiabatic theorem to systems described by a master equation of Lindblad form. Hence we have achieved the time independence of the eigenspaces of the transformed Hamiltonian $\widetilde{H}$:
	\begin{eqnarray}
		\frac{\dd\widetilde{\rho}^A}{\dd t} &=& -\imath\left[ \sum_{j=\frac{m-n}{2}}^{\frac{n-m}{2}} {}E_j(t)\ket{\psi_j(t_0)}\bra{\psi_j(t_0)} , \widetilde{\rho}^A \right]  - R\widetilde{\rho}^A + R\sum_{i=1}^N \widetilde{A}_i \widetilde{\rho}^A \widetilde{A}_i. \qquad \label{17}
	\end{eqnarray}

For $R=0$ we get the von Neumann equation
	\begin{equation}
		\frac{\dd\widetilde{\rho}^A}{\dd t} =-\imath\left[ \sum_{j=\frac{m-n}{2}}^{\frac{n-m}{2}} {}E_j\ket{\psi_j(t_0)}\bra{\psi_j(t_0)} , \widetilde{\rho}^A \right] 
	\end{equation}
and $\widetilde{\rho}^A_{00}(t):=\bra{\psi_0(t_0)}\widetilde{\rho}^A(t)\ket{\psi_0(t_0)}= {}_A\!\bra{m}\rho^A(t_0)\ket{m}_A$ is then a constant of motion, which ensures perfect transport of the electron.  At $t=t_{1}$ we use $U(t_{1})$ to transform back to the Schr\"odinger picture. From \Eqref{uni} we find again that an electron initially in state $\ket{m} _A =\ket{\psi_0(t_0)}  $ will be  in state $\ket{\psi_0(t_{1})}=\ket{n} _A $ at the end of step two.

For $R\ne 0$ we use \Eqref{17} to calculate the loss from perfect transport. To this end we note that the initial state is $\ket{\psi_0(t_0)}$, and in the transformed picture $\widetilde{\rho}^A$ we aim to stay in this state. Therefore the loss of transfer fidelity due to the measurements increases in time according to the projection of \Eqref{17} onto $\ket{\psi_0(t_0)}\,$:
	\begin{eqnarray}
		\frac{\dd}{\dd t}\widetilde{\rho}^A_{00}(t) &=& - R \bra{\psi_0(t_0)}\widetilde{\rho}^A(t)\ket{\psi_0(t_0)} + R\sum_{j=-\infty}^\infty \bra{\psi_0(t_0)} \widetilde{A}_j \widetilde{\rho}^A(t) \widetilde{A}_j \ket{\psi_0(t_0)}  \qquad \label{18}
	\end{eqnarray}
As we argued before, without measurements $\widetilde{\rho}^A(t)$ is a constant in time in the adiabatic approximation. To first order in the measurements \footnote{That is for small measurement rates $R/N$ and/or for weak measurements, i.e. small $\alpha/a$.} we can therefore substitute $\widetilde{\rho}^A(t)\to \widetilde{\rho}^A(t_0)=\ket{\psi_0(t_0)}\bra{\psi_0(t_0)}$ and $\rho^A(t)\to \ket{\psi_0(t)}\bra{\psi_0(t)}$ on the right side to find
	\begin{eqnarray}
		\frac{\dd}{\dd t}\widetilde{\rho}^A_{00}(t)	&=& - R + R\sum_{j=-\infty}^\infty \bra{\psi_0(t_0)} \widetilde{A}_j \ket{\psi_0(t_0)} ^2 \qquad \nn\\
		&=& - R + R\sum_{j=-\infty}^\infty \bra{\psi_0(t)} \hat A_j  \ket{\psi_0(t)}^2  \nn\\
		&=& -R\left[1 - \frac{\kappa}{N}\sum_{j=-\infty}^\infty \left(\sum_{i=m}^{n} \sqrt{1-\frac{\alpha}{r_{ij}}}  \left| \scalar{\psi_0(t)}{i}_{A}\right|^2 \!\right)^{2}  \right] \label{19}
	\end{eqnarray}
where we used the definition \Eqref{AAA} of the measurement operators $\hat A_j$. Now, $\ket{\psi_0(t)}$ can be substituted from \Eqref{psi0} to obtain numerical values and the error probability of the electron transport is obtained by integrating \Eqref{19}
	\begin{eqnarray}
		\mbox{Error Probability} &=& - \int_{t_0}^{t_{1}} \dd t\,\frac{\dd \widetilde{\rho}^A_{00}(t)}{\dd t}. \label{C5.11}
	\end{eqnarray}
	 
One might also ask what happens to coherences $\tilde{\rho}^A_{k0}(t):= {}_A\!\bra{k}\tilde{\rho}^A(t)\ket{\psi_0(t_0)}$ (for any $k<m$ or $k>n$)  during the transport. This question arises if one is interested in the transport of one part of a super position state, e.g. $ \left(c_1\ket{k}_A+c_2\ket{m} _A\right)\to  (c_1\ket{k}_A +c_2\ket{n} _A)$. In the same manner as \Eqref{18} and \Eqref{19} we arrive at
	\begin{eqnarray}
		\frac{\dd}{\dd t}\widetilde{\rho}^A_{k0}(t) &=& - R\bra{k}\widetilde{\rho}^A(t)\ket{\psi_0(t_0)}  + R \sum_{j=-\infty}^\infty  \bra{k} \widetilde{A}_j \widetilde{\rho}^A(t) \widetilde{A}_j \ket{\psi_0(t_0)}  \qquad \label{20} \\
		&\approx& - Rc_1c_2^*  + R \sum_{j=-\infty}^\infty c_1c_2^* \bra{k} \widetilde{A}_j  \ket{k}\bra{\psi_0(t_0)}\widetilde{A}_j \ket{\psi_0(t_0)} \nn\\
		&=& - Rc_1c_2^*  + R \sum_{j=-\infty}^\infty  c_1c_2^* \bra{k} \hat A_j  \ket{k}\bra{\psi_0(t)} \hat A_j \ket{\psi_0(t)} \nn\\
		&=& - Rc_1c_2^*\left[1- \frac{\kappa}{N}\sum_{j=-\infty}^\infty  \sqrt{1-\frac{\alpha}{r_{kj}}} \sum_{i=m}^{n}  \sqrt{1-\frac{\alpha}{r_{ij}}} \left|\scalar{\psi_0(t)}{i}_A\right|^2 \right] \qquad \qquad \label{21aa}
	\end{eqnarray}
where again \Eqref{psi0} can be substituted. In the line with the approximation sign, we substituted $\widetilde{\rho}^A(t)$ with $\widetilde{\rho}^A(t_0)$ on the right hand side of the equation which is again valid to first order in the measurements. Furthermore, we only kept $c_1c_2^*\ket{k}\bra{\psi_0(t_0)}$ as the other terms of $\widetilde{\rho}^A(t_0)$ vanish once the $\hat A_j$s are substituted in the last line.

\begin{figure}\begin{center}
		\includegraphics[width=0.6\linewidth]{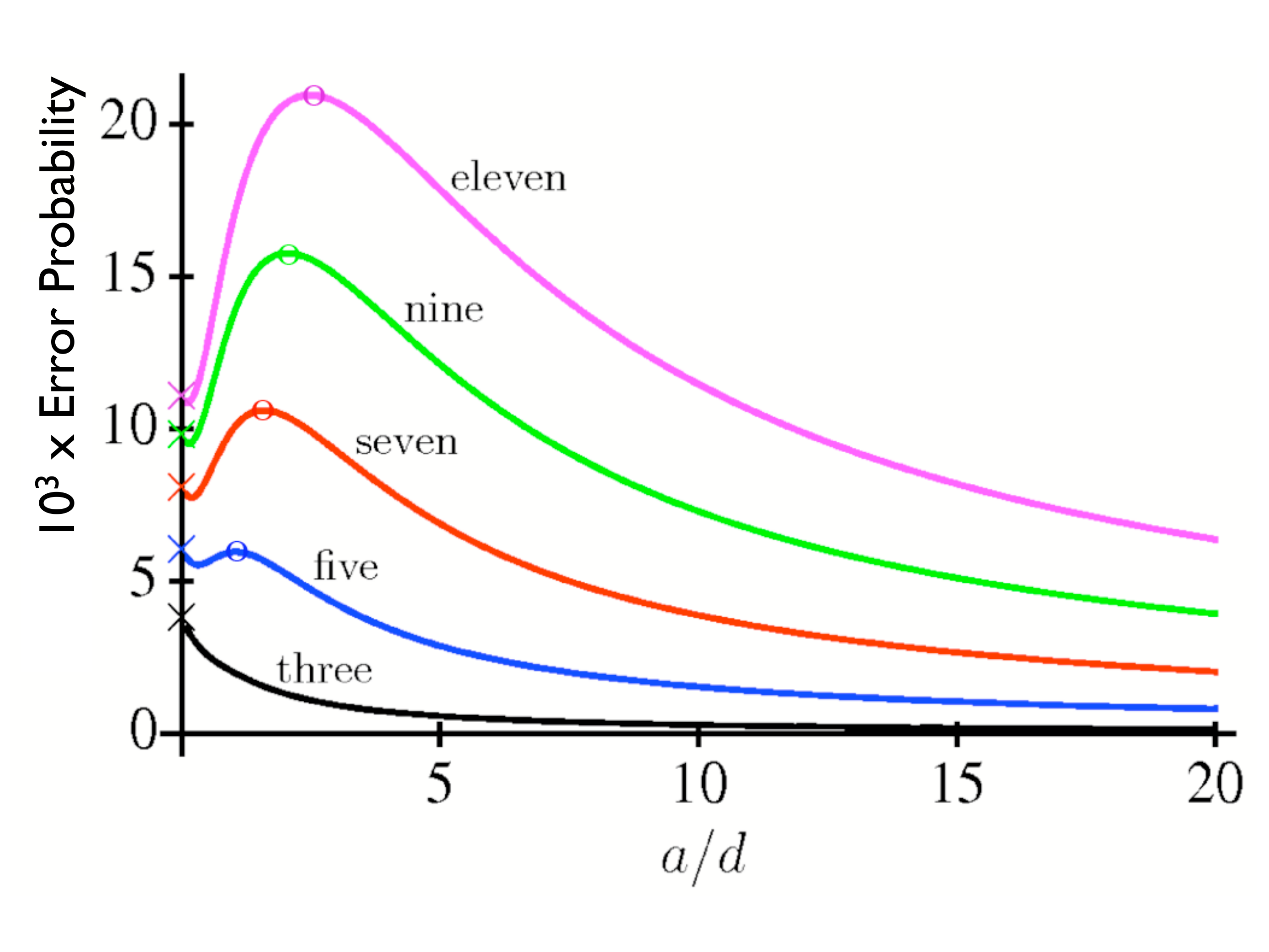} \vspace{-3mm}
		\caption{\small Transfer loss during the transport of an electron along three (black), five (blue), seven (red), nine (green), and eleven (purple) QDs as a function of $a/d$, i.e. the localness of the measurements. The crosses show the numerical solution of the exact equation \Eqref{C5.2} (i.e. before the adiabatic approximation is performed) for local measurments and the circles show the empirical cross-over between local and non-local measurements (see text). Pump and Stokes pulses are as in \Eqref{pulses} (see also Fig~\ref{figure5.2}(a)) with $T \Omega_{max}/\hbar$ chosen such that the additional loss due to non-adiabaticity is $24\times 10^{-6}$ for all curves, to allow reasonable comparison. This is $T=150,\;196,\;225,\;242,\;249$ for three, five, seven, nine, and eleven QDs, respectively. The intermediate couplings are equal one. System parameters are $R=N,\;\alpha=0.04a$. \label{figure5.1}}
	\end{center}\end{figure}

The terms in the square brackets of \Eqref{19} and \Eqref{21aa} become small when the measurements are sufficiently non-local, such that they can not distinguish well between the QDs involved in the transport (see also Fig~\ref{figure5.1}). Since the loss of information increases linearly in the time $t_{1}$ required for transport, it is crucial that the couplings $\Omega_j$ between the dots are as large as experimentally possible to give a big energy splitting which in turn allows a fast transport (see \Eqref{15aa}). 
	
Also note that in the case of local measurements \Eqref{21aa} reduces to $\frac{\rm{d}}{{\rm{d} }t}\widetilde{\rho}^A_{k0}=-D\widetilde{\rho}^A_{k0}$ where the decoherence rate $D=\frac{R}{N} \left(2+\alpha-2\sqrt{1+\alpha}\right)$ is the same as the one obtained in~\Eqref{aaaa} without transport. Therefore, the decoherence rate of a charge qubit on a quantum dot rail is the same during storage as during transport by CTAP if subject to local measurements. 	
	
The probability of not finding the electron in the desired state $\ket{n}_A$ after the transport, calculated from \Eqref{19} and \Eqref{C5.11}, is shown in Fig~\ref{figure5.1} as a function of the parameter $a/d$ for transport along 3, 5, 7, 9, and 11 dots. {\color{black}Remember that $a/d$ is the distance of the QPCs to their nearest QD compared to the distance between neighboring QDs, and therefore is a measure of the non-localness of the measurements performed by the QPCs. For any given $a$ this parameter can be changed with $d$ and we fix the sensitivity $\phantom{a}\bra{i}_{\hspace{-6mm}A}\hspace{4mm}\hat \pi_i\ket{i}_A$ of the measurements by setting $\alpha=0.04a$. Also the rail of dots and measurement apparatuses extends on both sides ($N\gg m-n$). In the special case of local measurements we also numerically integrated the exact master equation \Eqref{C5.2} (i.e. not applying the adiabatic approximation) and the crosses in Fig~\ref{figure5.1} are from this solution. The great agreement with the curves indicate a good performance of the extension of the adiabatic approximation to open systems.
	
One can distinguish two regimes in Fig~\ref{figure5.1}. First, for $\frac{a}{d}>\frac{n-m}{4}$ (right of the circles) the transport fidelity increases with the non-localness of the measurements. This is easily understood since sufficiently non-local measurements can not distinguish between the QDs involved in the transport, therefore not leading to decoherence.
	 
Second, for $\frac{a}{d}<\frac{n-m}{4}$ (left of the circles) we find the reverse and the transport fidelity increases with the localness of the measurements (except for transport along three QDs). This needs some explanation. In this regime the measurements are local enough to distinguish well between the states $\ket{m}_A$ and $\ket{n}_A$ (during CTAP most of the population is found on these two states) and hence one can not expect a further decrease of the fidelity with the localness of the measurements. To account for the apparent increase of the fidelity with localness we need another argument. For very local measurements, almost exclusively measurement results from QPCs at sites $n$ and $m$ contribute to decoherence, whereas if the measurements are slightly non-local, QPCs in the near neighborhood of $n$ and $m$ also contribute, therefore increasing the overall decoherence.
	 
The small decrease of the fidelity for extremely local measurements might be due to the small populations on QDs other than $m$ and $n$.


It is also evident from Fig~\ref{figure5.1}, that if local dephasing is the main source of decoherence, it is best to transport long distances in one step as is seen e.g. for the transfer $\ket{1}_A\rightarrow\ket{11}_A$. For the parameters given in Fig~\ref{figure5.1}, the chance of not finding the electron on the desired state after the transport is then about $11*10^{-3}$  (see left end of Fig~\ref{figure5.1}). This number increases to about $5*(4*10^{-3})$ when we first transfer to $\ket{3}_A$, then to $\ket{5}_A$, $\ket{7}_A$, $\ket{9}_A$, and finally to $\ket{11}_A$. On the contrary, if dephasing tends to be more global, a better transfer rate is achieved by breaking a long distance into several smaller ones (see right end of Fig~\ref{figure5.1}).

\section{Coupling to two level systems\label{last}}

We now apply CTAP to QDs coupled to TLSs as shown in Fig~\ref{figure4}. Hence the driving field $H_A$ from \Eqref{14aa} has to be added to $H_{int}$ from \Eqref{inter}, responsible for the QD-TLS-coupling. We first neglect the coupling to the measurement apparatus and point out its inclusion at the end of the subsection. Then, all couplings behave local and we can restrict ourself to study only the QDs of the rail which are involved in the electron transport, i.e. $N=n-m+1$.
	
To understand how CTAP works for this system, we note that the Hamiltonian $H=H_{int} +H_A$ can be written in a block diagonal form\footnote{A Hamiltonian $H=H_A+H_{TLS}+H_{int}$ with $[H_{TLS},H_{int}]=0$ always allows such a block diagonal form, because then $H_{TLS}$ does not induce transitions between different eigenstates of $H_{int}$.}, with $2^N$ blocks,  each block having dimension $N$. Each block gives the Hamiltonian of the electron on the rail of QDs, conditioned on the state of the TLSs. As an example we show part of the Hamiltonian for $N=3$ explicitly 
	\begin{eqnarray}
		\hspace{-0mm} H= \left(  \begin{array}{cccccccccc}
			\!-\chi_1\! & \Omega_1 & 0 \\
			\Omega_1 & -\chi_2 & \Omega_2&&0&&&0&&\cdots \\
			0 & \Omega_2 & \!-\chi_3 \!\\
			&&& \!\!-\chi_1\!\! & \Omega_1 & 0 \\
			&0&&\Omega_1 & -\chi_2 & \Omega_2 &&0&&\cdots\\
			&&&0 & \Omega_2 & \chi_3 \\
			&&&&&&\!-\chi_1 & \Omega_1 & 0 &\cdots\\
			&0&&&0&&\Omega_1 & \chi_2 & \Omega_2 &\cdots\\
			&&&&&&0 & \Omega_2 & -\chi_3\! &\cdots\\  
			&\vdots&&&\vdots&&\vdots&\vdots&\vdots&\ddots       \end{array} \right) \nn 
	\end{eqnarray}
 where the first block is for all TLSs in the $\ket{0}$ state, the second for the first and second TLSs in $\ket{0}$ and the third TLS in $\ket{1}\,$, and so on. 
		
We can now study each block individually because there is no coupling between them. If we want to transport the electron from the first QD to the third QD, regardless in which state or superposition the TLSs are, we have to make sure that CTAP works in each block. That is the state $\ket{1}_A\ket{0}\ket{0}\ket{0}$ should adiabatically evolve to $\ket{3}_A\ket{0}\ket{0}\ket{0}$ as well as $\ket{1}_A\ket{0}\ket{0}\ket{1}$ should evolve to $\ket{3}_A\ket{0}\ket{0}\ket{1}$ and so on. If this is achieved, then an electron initially found in $\ket{1}_A$ will evolve in the pure state $\ket{3}_A$. During the population transfer the reduced state $\rho^A=$Tr$_{TLSs}[\varrho]$ is not only in a superposition of $\ket{1}_A$, $\ket{2}_A$ and $\ket{3}_A$, but in a real mixture of these states (see Fig~\ref{mixture}(a)). This is because depending on the state of the TLSs, the electron will start moving earlier or later, which generally results in entaglement between the rail of QDs and the TLSs. However, once CTAP is completed the electron will be found in the desired state $\ket{3}_A$, and no entanglement will be left. This reduction of entanglement at certain times is a common feature of non-Markovian environments, where transfer of information between system and environment is possible in both directions~\cite{nonMarkov}.
	
Hence, for CTAP to be applicable despite coupling to TLSs, we only have to ensure that it works for a Hamiltonian of the form (where we return to an arbitrary long rail of dots)
	\begin{eqnarray}
		\hspace{-0mm}H = \left( \begin{array}{cccccc} \vspace{1mm}
						 \pm \chi_m & \Omega_m&0&\cdots & 0&0  \\ \vspace{-1mm}
			\Omega_P & \pm \chi_{m+1} &\Omega_{max} &  \cdots &0& 0 \\ 
			 0 & \Omega_{max} & \pm\chi_{m+2} &\ddots &0&0 \\
			\vdots &  \vdots &\ddots &\ddots & \ddots& \vdots  \\ \vspace{1mm}
			 0 &0&0 &\ddots& \pm\chi_{n-1} & \Omega_{S}  \! \\
			 0 & 0 &0&\cdots&  \Omega_{S}  & \!\pm\chi_{n}\!   			
		\end{array} \right) \label{12345}
	\end{eqnarray}
instead of the simpler form \Eqref{14aa}. CTAP should work for all possible permutations of + and $-$, since each permutation represents one block of the Hamiltonian corresponding to a certain combination of environmental TLS states. The crucial requirement for this is again relation~(\ref{15aa}), but the energy eigenstates $\ket{\psi_j(t)}$ as well as the energy eigenvalues ${}E_j(t)$ depend now on the couplings $\chi_m,\dots,\chi_n$ and the state of the TLSs. An important feature which is due to the non-zero values on the diagonals is that this relation can not always be fulfilled by just performing the transport sufficiently slowly. Additionally one has to ensure that no level crossings occur during step two, i.e. ${}E_0(t)-{}E_j(t)\ne0$. As before, ${}E_0(t)$ is the energy of the state $\ket{\psi_0(t)}$ which we want to evolve adiabatically from $\ket{\psi_0(t_0)}=\ket{m}_A$ to $\ket{\psi_0(t_{1})}=\ket{n}_A$. But unlike in the previous subsection, ${}E_0(t)\ne 0$ and depends on the state of the TLSs. The situation becomes even more complicated when anti-crossings are considered which appear quite naturally (see blue and red curves in Fig~\ref{figure5.2}). At an anti crossing two energy levels can get so close that adiabaticity can not be achieved with any reasonable transfer times.
	
Furthermore it might happen that the adiabatic state $\ket{\psi_0(t)}$ does not connect $\ket{m}_A$ to $\ket{n}_A$, but instead to another state $\ket{\psi_0(t_{1})}=\ket{j}_A$ with $j\neq n$. An example of this is shown in Fig~\ref{figure5.2} (green curves) where the population returns to the original QD.
	
To calculate the energies is generally only possible numerically and an exact treatment when level crossings or anti-crossings appear is highly non-trivial. Some work on this issue is done in~\cite{multilevel}, but the possibility of anti-crossings is not considered there (see also~\cite{detuning}). However, some qualitative statements can be made. If the energy level ${}E_0(t)$ of the adiabatic state is at all times during step two in the center of all energy levels involved in the transport, i.e.
	\begin{eqnarray}
		{}E_0(t)>{}E_j(t) &&\mbox{for}\quad j=\frac{m-n}{2},\dots ,-1\nn\\
		{}E_0(t)<{}E_j(t) &&\mbox{for}\quad j=1,\dots,\frac{n-m}{2},\label{cond}
	\end{eqnarray}
then we can be sure that no level crossings occur. In this case the process during  step two is qualitatively the same as without coupling to the TLSs. Since condition~(\ref{cond}) is valid with $\chi_j=0$ (previous subsection), it can always be achieved with sufficiently high driving fields $\Omega_j$, as then the comparatively small couplings $\chi_j$ do not influence the qualitative eigenvalue structure. 

Lets apply this condition to the block of the Hamiltonian in which we have minus for $\chi_m$ and pluses for all other coupling constants. Clearly we need sufficiently strong tunneling rates $\Omega_{max}$ and $\Omega_S(t_0)$ to ensure that ${}E_0(t_0)=-\chi_m$ is not the lowest energy at the beginning of the transport, but at the centre of all energies (see Fig~\ref{figure5.2}(c) where ${}E_0$ gets to the centre of the energy levels at $t\approx 3$ before the transport starts at $t_0\approx 20$). The key feature to realize is that level crossings or anti-crossings are actually needed to get ${}E_0$ to be at the centre of all energies. But it is crucial that this happens as long as $\Omega_P\approx 0$, i. e. before the population transfer starts.

For $n-m=2$ we find $\Omega_{j}^2(t_0)>(\chi_1-\chi_3)(\chi_1-\chi_2)$ for $j\neq P$. For transport along many QDs ($n-m\gg 2$) we get the approximate inequality $\Omega_j(t_0) \gtrapprox c\max(\chi_k)(n-m)$ for some constant $c$, which states that for given maximal $\Omega_j$ (due to experimental limitations) and given $\chi_j$, there exists a maximum number $n-m$ for which we can use CTAP. For transport along more QDs one has to break up the transport into several smaller sections and apply CTAP successively to each section. The shorter the rail gets (compared to the maximal length), the further we can stay away from any level crossing and the faster we can transport the electron without violating the adiabaticity condition~(\ref{15aa}).

As example we show in Fig~\ref{figure5.2} the eigenenergies and populations as a function of time for a  system with five QDs and constant $\chi_i=\chi$ for different values of $\chi$. We consider the block of the Hamiltonian acting in the subspace where the first TLS is in the state $\ket{0}$ and the other four TLSs are in the state $\ket{1}$, which reads
	 \begin{eqnarray}
		\hspace{-0mm}H_A = \left( \begin{array}{ccccc} 
			- \chi & \Omega_P&0 & 0&0  \\ 
			\Omega_P & + \chi &\Omega_{max}   &0& 0 \\ 
			 0 & \Omega_{max} & +\chi  &\Omega_{max}&0 \\
			 0 &0&\Omega_{max} & +\chi & \Omega_{S}   \\
			 0 & 0 &0&  \Omega_{S}  & +\chi  			
		\end{array} \right). \label{Hami}
	\end{eqnarray}
Of course, if we assume the TLSs to be in a statistical mixture one would have to calculate the populations for all blocks of  the Hamiltonian and average over them. Since all qualitative features can be seen with this particular block we restrict the discussion to this one for pedagogic reasons. We use Gaussian pump and Stokes pulses 
	 \begin{eqnarray}
	 	\Omega_P &=& \Omega_{max}\exp\left\{ -\frac{(t-3T/4)^2}{(T/4)^2}   \right\} \nn\\ 
		\Omega_S &=& \Omega_{max}\exp\left\{ -\frac{(t-T/4)^2}{(T/4)^2}   \right\} \label{pulses}
	\end{eqnarray}
with the parameter $T=150$. Because of the Gaussian pulse shapes the distinction of steps one, two and three is not sharp.  Looking at Fig~\ref{figure5.2}~(a) we can approximately say that step one is for $-50<t<20$, step two is for $20<t<130$ when all couplings are switched on, and step three  is for $130<t<200$. $T$ is  chosen long enough to achieve very good transfer fidelity (0.9996) for $\chi=0$, i.e. without coupling to TLSs. 
	
For $\chi=0.15$ there appears a level anti-crossing at which the energy gap (0.0002) is so small, that in can be treated as a crossing on the time scales set by $T$, the Gaussian pulse width. If one were to enlarge $T$ extremely, adiabaticity would be fulfilled even for this tiny energy gap which would result in no transport, similar to the case $\chi=0.45$ described below. However, as $\Omega_P\approx 0$ at the time of the anti-crossing, the state will remain in its initial state $\ket{m}$ with ${}E_m=-0.15$ during this anti-crossing for any reasonable pulse width $T$. That is, the state crosses the levels as indicated by the dashed line in the magnified part of Fig.~\ref{figure5.2}~(c). The smaller population transfer (fidelity 0.975) is because during step two, the energy level spacing is decreased compared to the case with $\chi=0$. Therefore, the transfer fidelity can be increased arbitrarily close to one by choosing larger $T$ \footnote{Non-adiabatic corrections scale with $\exp{(-cT|{}E_0-{}E_j|)}$ for some constant $c$ (see also section~\ref{adia}).}. 
		
	\begin{figure}{\begin{rotate}{270}
		\includegraphics[width=1.3\linewidth]{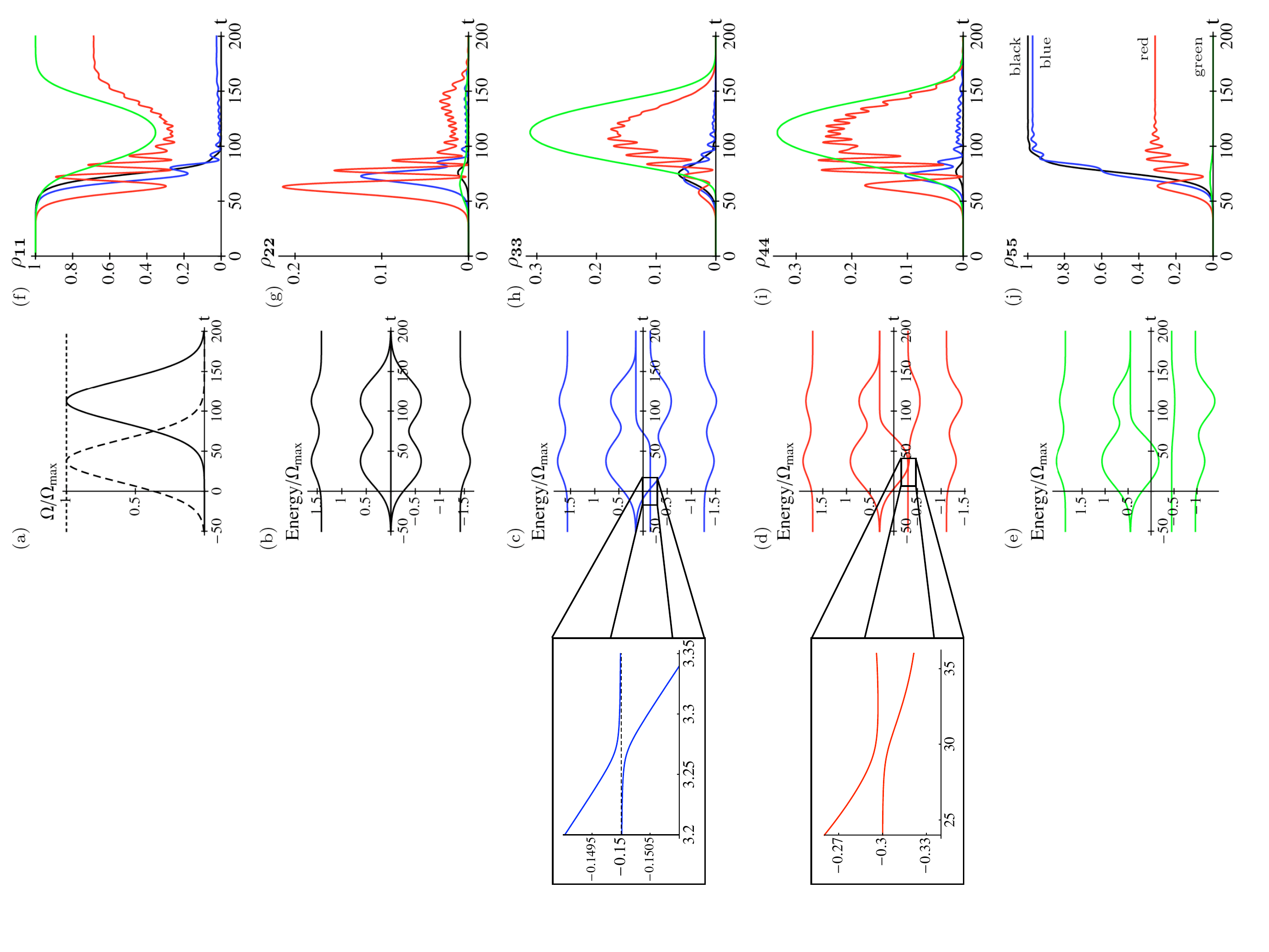}\end{rotate}}\vspace{18.7cm}
		\caption{\small (a): Tunneling rates (dashed:~$\Omega_1$, {\color{black}dotted:~$\Omega_{2,3}$, solid:~$\Omega_4$}), (b)-(e): energy levels of Hamiltonian~\Eqref{Hami}~and $\chi=0,\;0.15,\;0.3,\;0.45$ (see text), and (f)-(j): populations of $\ket{1}_A$ -- $\ket{5}_A$ obtained by the numerical integration of the Schr\"odinger equation. The colors (see (j) for grey scale) black, blue, red, and green correspond to $\chi=0,\;0.15,\;0.3,\;0.45$, respectively. Time is measured in units of $\Omega_{max}^{-1}$. What looks like level crossings in (c) and (d) are actually anti-crossings as can be seen in the magnified figures.  \label{figure5.2}}
	\end{figure}

Enlarging $\chi=0.3$ moves the anti-crossing into step two and adiabaticity can not be achieved any more. The result is wild oscillations of all populations and a low transfer fidelity. Finally at $\chi=0.45$ the anti-crossing disappears and adiabaticity is restored which is evident in the lack of oscillations in the populations. But because the eigenenergy -0.45 at $t=-\infty$ connects to the same energy at $t=\infty$, the population returns to the original state $\ket{m}_A$ and the population transfer failed.
		
This result should be quite surprising. Up to relatively large couplings to TLSs, $\frac{\chi_i}{\Omega_{j,max}}<0.2$ (for $m-n=4$) adiabatic transfer can be achieved and the transfer fidelity {\color{black}very quickly approaches one with sufficiently} large transfer times. This is contrary to the Markovian dephasing studied in the previous subsection, where the transfer loss increases with transfer time once the transfer time is long enough to ensure adiabaticity. 
	 	
There is however a disturbing effect due to the coupling to TLSs is that ${}E_0\ne 0$ and hence we get a dynamical phase 
	\begin{equation}
		\varphi=\int_0^{t_{1}} \dd t\, {}E_0(t) \label{dphase}
	\end{equation}
and worse even, this phase depends highly on the state of the TLSs. If the electron starts in a superposition $\frac{1}{\sqrt{2}}(\ket{k}_A +\ket{m}_A)$ it will not end up in $\frac{1}{\sqrt{2}}\left(\ket{k}_A +e^{\imath\varphi}\ket{n}_A\right)$, but in a real mixture of $\ket{k} _A $ and $\ket{n} _A $. That is, during the transport, the electron gets entangeled with the TLSs. This should not be surprising since we saw a similar behavior already in subsection~\ref{coupling1} without transport (see \Eqref{13}). However, transport prevents the lost information from returning back to the electron periodically. This can be seen in Fig~\ref{mixture}(b) for transport along three QDs, $\chi_i\equiv\chi$, and $\Omega_P ,\;\Omega_S$ as in \Eqref{pulses} with $T=150$. The initial states  $\rho_j(t_0)=\half\one$ of the TLSs are taken to be a complete mixture. As a measure of purity we use Tr$\left[\left(\rho^A\right)^2\right]$.
	
	\begin{figure}[htb]\begin{center}
		\includegraphics[width=\linewidth]{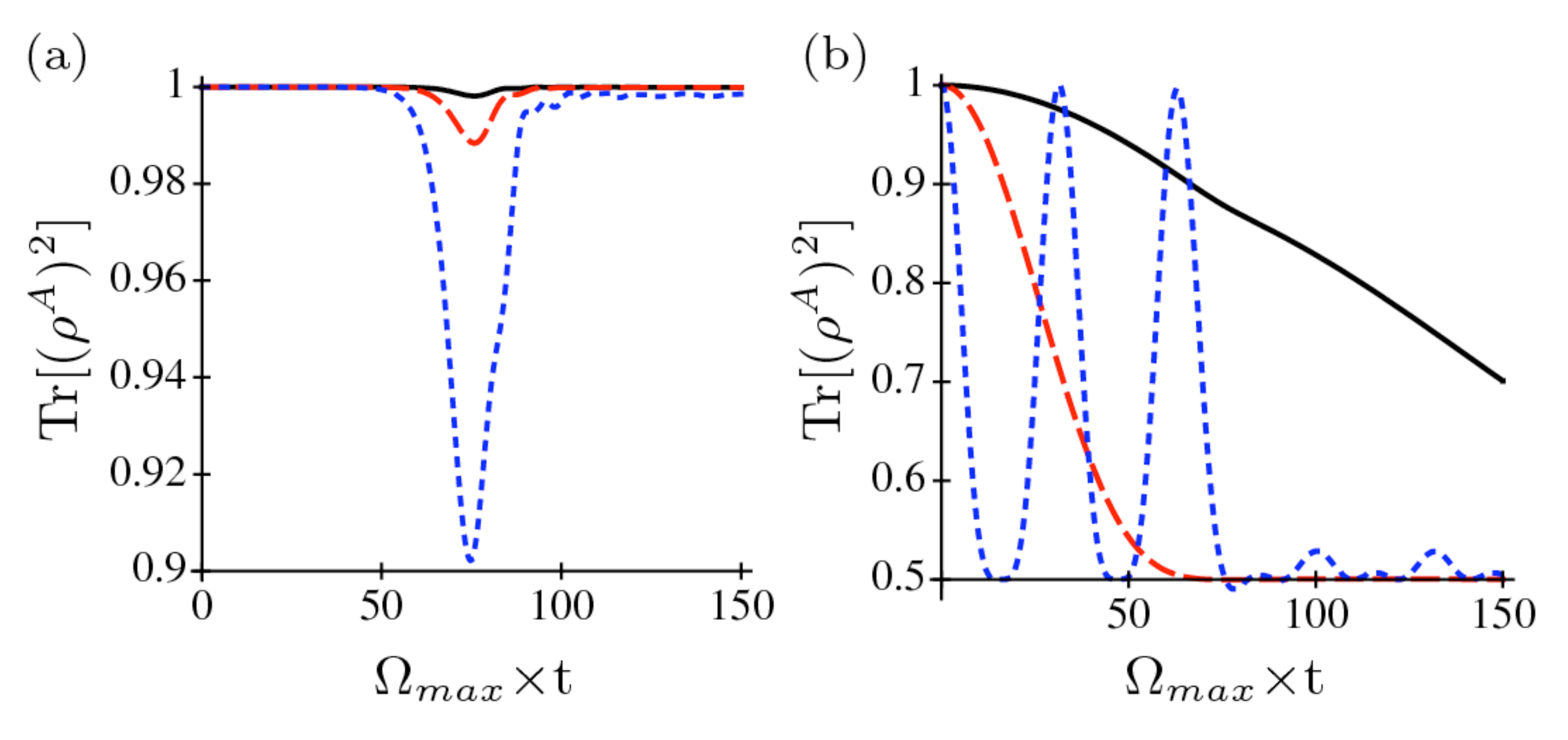}
		\caption{\small The purity of the reduced density matrix $\rho^A(t)$ during the transfer $\ket{1}_A\to\ket{3}_A$. The couplings to environmental TLS are homogeneous, i.e. $\chi_1=\chi_2=\chi_3=\chi$, and their states are in a complete mixture.\newline
		(a): Transport of a position eigenstate $\ket{\psi(t_0)}=\ket{1}_A$ to $\ket{\psi(t_1)}=\ket{3}_A$ with couplings to environmental TLS $\chi=0.02$~(solid), 0.05~(dashed), 0.15~(dotted).  Because of the dependance of the adiabatic state $\ket{\psi_0(t)}$ on the states of the TLSs, the electron gets entangled with the TLSs during transport. However, this entanglement disappears with the completion of CTAP, and the electron on the quantum dot rail finishes the transport in the desired pure state.\newline
		(b): Transport of half of a superposition state. $\ket{\psi(t=0)}=\frac{1}{\sqrt{2}}\left(\ket{0}_A+\ket{1}_A\right)$ and $\chi=0.005$~(solid), 0.02~(dashed), 0.1~(dotted). The transported half of the superposition state picks up a dynamical phase which depends on the state of the TLSs. Hence the electron gets entangled with the TLSs and finishes CTAP in a statistical mixture of $\ket0_A$ and $\ket{3}_A$, rather than in a superposition.  \label{mixture}}
	\end{center}\end{figure}
	
On the other hand, as described earlier this section, if the electron starts in $\ket{1}_A$ it will get transfered to the pure state $\ket{3}_A$ despite being in a real mixture during the transfer, as is shown in Fig~\ref{mixture}(a). In this case, $\varphi$ is an undetectable global phase.
		
Now we briefly discuss the inclusion of measurements once again. In the previous subsection the adiabatic theorem was assumed to hold and therefore we first have to make sure that it holds also with coupling to TLSs. This is best done by solving the Schr\"odinger equation without measurements as was done to get Fig~\ref{figure5.2}. The solution shows whether the transport is adiabatic (almost no fast oscillations) or not (much oscillations). If not, the driving fields $\Omega_j$ as well as the transfer time can be increased, until the solution shows adiabatic behavior. Then equations (\ref{18})-(\ref{21aa}) derived in the previous section can be applied here as well, if one uses the appropriate eigenstate $\ket{\psi_0(t)}$, which now depends on the state of the TLSs. However, $\ket{\psi_0(t)}$ can only be calculated numerically which makes an exact treatment very difficult. Help comes from the fact that the effects of dephasing do not depend much on the exact form of this state. Therefore one might prefer to use the unperturbed eigenstate from \Eqref{psi0} as in section~\ref{hello} to calculate transport fidelities. This leads essentially to two independent examinations of transport losses, one for measurements and one for TLSs. This separate treatment of the two decoherence effects is justified if they do not influence each other, which is generally the case in applications where decoherence is weak.

\chapter{Discussion and Conclusion\label{conclusion}}

We first analyzed decoherence effects on a charge qubit without transportation. The main result is that while storing information as a charge qubit on a quantum dot rail, some of it will leak because of decoherence due to measurements performed by the QPCs. Information loss from decoherence arising from coupling to TLSs, on the contrary, returns periodically. However, as the period is usually not known,  the information is essentially lost as well.
	
The differences between Markovian and non-Markovian noise becomes more striking in the electron transport by CTAP. The transport fidelity decrease due to the measurements is essentially linear in the transport time, therefore the time should be kept as short as the adiabatic theorem allows. But quite to the contrary, the reason for less fidelity in the presence of TLSs is because the spacing between energy levels decreases. This however can be offset by a longer transfer time, as long as the coupling to the TLS is not too strong compared to the driving field $\Omega_j$, and the QDs involved are few enough. This is good news for non-Markovian decoherence during adiabatic processes.
	
One negative effect of the TLSs is that the transported state will acquire a generally unknown dynamic phase during the transport. This limits applications were one wants to transport a charge qubit, which might be in a superposition of position eigenstates. But as the couplings to the TLSs are unknown, any diabatic transport will face similar problems.
		
Summarizing we can say, if we can reduce Markovian and non-Markovian dephasing sufficiently to be able to store a charge qubit on a rail of QDs, then we can also transport it using CTAP. Furthermore, adiabatic transport protocols also have the natural advantage that they are robust against experimental parameter variations. 
	
It is often tried to get only a small population in the intermediate QDs, by using weak pulses $\Omega_S$ and $\Omega_P$ compared to the coupling $\Omega_n$ between intermediate QDs~\cite{main}. This is possibly due to the belief, that decoherence is suppressed if only two QDs are populated. This is in contrast to our results, because we find that for both, Markovian and non-Markovian decoherence, the transport fidelity mainly depends on the energy spacing. A large spacing, in turn, is achieved by large couplings between all QD, including the pump pulse and Stokes pulse, which results in some population of intermediate QDs. The above statement might stem from analyses of STIRAP, where the intermediate atomic states are exposed to spontaneous decay, whereas the initial and final states are not. This is not the case for CTAP and there is no advantage of not populating the intermediate states. 
	
We should note, that often the use of the spin degree of freedom as qubit is considered to be more promising than the position degree of freedom, because of the generally larger decoherence time. Then the transport (as well as storage) of position superpositions is not necessary, therefore avoiding the biggest challenge for CTAP. As we saw in subsection~\ref{last}, the transport fidelity for position eigenstates remains very large despite appreciable coupling to TLSs. Therefore, this study provides further evidence, that the spin should be the preferred degree of freedom.

\part{\textbf{Quantum Brownian Motion}}

\chapter{Introduction\label{introII}}

Well over a hundred years ago, Boltzmann derived the linear Boltzmann equation \cite{Boltzmann} to describe a tracer particle affected by molecules of a dilute gas. The special case of a heavy tracer particle, or Brownian particle, was studied in more detail by Einstein~\cite{Einstein} and Smoluchowski~\cite{Smoluchowski}, to describe the phenomenon of Brownian motion, first observed by botanist Brown~\cite{Brown}. 

In more recent times, physicists became interested in quantum generalizations, which lead to the well established fields of collisional decoherence~\cite{Joos,Gallis,Hornberger}, as well as the Caldeira-Leggett model of quantum Brownian motion~\cite{Caldeira,Hu}, where the Brownian particle experiences a random ``force" due to the coupling to a bath of harmonic oscillators. Later, quantum versions of the linear Boltzmann equation (QLBE) \cite{Diosi,Vacchini,Hornberger2} were developed, to describe the interplay between collisional decoherence, friction and diffusion. Not only is such a quantum description of collisional Brownian motion desirable, but it also sheds light on the non-classical process of decoherence, which is believed to be of importance in the quantum classical transition. Furthermore, experimental observations of the quantum nature \cite{Exp2} and decoherence \cite{Exp} of large molecules improve rapidly, and are reviewed in~\cite{Expreview}.

In the study of collisional Brownian motion (classical and quantum), one usually assumes that each collision is independent of earlier collisions, which requires that the collision time is small compared to the collision frequency. This requirement is typically satisfied in the high temperature (small collision time) and low density (low collision frequency) limit.  Also some coarse grained time scale (large compared to collision time) is used, such that collisions appear to be instantaneous and the dynamics become Markovian.

To fully appreciate the quantum discussion in this field, we state some well known results of the classical theory~\cite{Fokker}. On short time scales, the momentum of a particle initially at rest diffuses according to $\Delta p^2\propto t$, whereas the position variance behaves like $\Delta x^2 \propto t^3$. The reason is that an almost instantaneous collision changes the momentum, but not the position of a particle. Only on a much larger time scale, when the momentum distribution is close to thermal equilibrium, the position diffuses according to $\Delta x^2\propto t$. Although for \emph{large} classical particles one is more interested on time scales on which many collisions occur, in the study of decoherence of a quantum particle the short time scales are of importance, because no quantum behavior is observable after a large number of collisions.

Interestingly, microscopic derivations of QLBEs (which are all Markovian, i.e. do not have memory) predict an additional quantum contribution to position diffusion (QPD), acting already on the short time scale of momentum diffusion. Although such a process is impossible in classical dynamics as it derives from finite position jumps, the very same process is needed for a QLBE to be of Lindblad form, necessary for completely positive Markovian dynamics. Therefore, QPD is currently believed to arise as a quantum effect accompanying collisional quantum friction~\cite{Diosi,review,Hornberger3}. In this work, one of the main aims is to show the contrary: We find that QPD is not a real physical process, but results from using a coarse grained time scale on which collisions appear instantaneous. Unproblematic in classical dynamics where collision times of hard-core particles are indeed short, the assumption of instantaneous collisions has to be used with care in quantum dynamics, where collision times depend on the widths of the colliding wave packets.

Although position diffusion is generally needed to get a positive and Markovian master equation, one has to keep in mind that any quantum Brownian motion (QBM) Markovian master equation can only be valid in the high temperature and low density limit, and on a coarse grained time scale. It can hence be expected, that every derivation of a QBM master equation in Lindblad form will find position diffusion in some form, even if it is not an actual physical process. Indeed, in the limit of high temperature and low density, the predicted QPD vanishes in all proposed master equations.

To clarify our point, it is convenient to draw a quick comparison to classical BM. If no coarse grained time scale is used at all, the collision of a gas particle and the Brownian particle will not be instantaneous, but instead, the momentum transfer will take some finite time, which depends on the steepness of the interaction potential. However, to a very good approximation one can use a hard core potential which is infinitely steep, resulting in instantaneous collisions. The effect are random momentum jumps and momentum diffusion of the Brownian particle. In quantum mechanics on the other hand, even a hard core potential does not justify instantaneous collisions, as the colliding wave packets have a nonzero width. Therefore, there is a finite collision time depending on the width of the wave packets divided by their relative velocity, and every Markovian master equation can only be valid on a coarse grained time scale large compared to the collision time.

Important contributions to QLBEs include \cite{Diosi,Vacchini,Hornberger2,mainII,Diosi2}, and are reviewed in~\cite{review}. Mostly, the analysis is based on using scattering theory to describe the effects of a single collision with a gas particle, which is assumed to be in an momentum eigenstate. Unfortunately, such collisions with momentum eigenstates impose a major unphysical feature regarding decoherence. For simplicity, we use one spatial dimension to point out our concern, but the three dimensional case is along the same lines. Assume that before the collision, the tracer particle is in a superposition of two momentum eigenstates $(\ket p+\ket{p'})/\sqrt 2$, whereas the colliding gas particle is in the state $\ket{p_g}$. A collision will lead to
	\begin{eqnarray}
		\frac1{\sqrt 2}\left( \ket{\bar p(p,p_g)}\otimes \ket{\bar p_g(p,p_g)} + e^{\imath\phi}\ket{\bar p(p',p_g)}\otimes \ket{\bar p_g(p',p_g)} \right)\!, \nn
	\end{eqnarray}
where $\phi$ depends on the interaction potential, and the momenta after the collision are dictated by momentum and energy conservation
	\begin{eqnarray}
		\bar p = \frac{2mp_g+(m-m_g)p}{m+m_g}, && \bar p_g=\frac{2m_gp-(m-m_g)p_g}{m+m_g}.\nn
	\end{eqnarray}
Because the gas particle's momentum after the collision depends on the tracer particle's momentum before the collision, it follows that the two particles become entangled during the collision process. In fact, $\scalar{\bar p_g(p,p_g)}{\bar p_g(p',p_g)}=0$ for $p\ne p'$, and tracing out the gas particle one then finds that all coherences between $p$ and $p'$ are lost in a single collision, no matter how small $|p-p'|$. Furthermore, it follows from linearity that for a general incoming state of the tracer particle, the collision leads to a density matrix which is perfectly diagonal in the momentum representation. Therefore, a well localized state will be spread out infinitely after a single collision. In a three dimensional collision, one finds that the tracers particles density matrix will be perfectly diagonal only in the momentum component parallel to the momentum transfer, which was also discovered in \cite{Diosi2} independently from the present author.

This seemingly unphysical situation can be understood, if the collision time is considered. As the collision time of two colliding wave packets depends on their width, it must be assumed to be infinity if the gas particle is in a non-localized momentum eigenstate. After an infinite time, it is acceptable that all momentum coherences disappear, even if the interaction is only with a single gas particle.

One might wonder how \cite{Diosi,Vacchini,Hornberger2,mainII} were able to derive a QLBE to describe the tracer particle at finite times, without encountering the problem of \emph{infinite} momentum decoherence in a single collision. For this purpose, we briefly review Hornberger's contribution~\cite{Hornberger2}, because it seems to be the most complete one. Due to momentum conservation, a two particle collision reduces to a one particle problem, where the transition operator $\hat T$ with momentum matrix elements related to the scattering amplitude $\langle\mathbf p_f|\hat T|\mathbf p_i\rangle = \delta(\mathbf p_f^{2}-\mathbf p_i^2)f(\mathbf p_f,\mathbf p_i)/(\pi\hbar)$ is of importance. In particular, an expression of the form 
	\begin{eqnarray}
		X=\frac{(2\pi\hbar)^3}V \langle\mathbf p_f + \mathbf p_s |\hat T|\mathbf p_i+\mathbf p_s\rangle  \langle\mathbf p_i-\mathbf p_s|\hat T^\dag|\mathbf p_f-\mathbf p_s\rangle, \quad \label{problem}
	\end{eqnarray}
appears behind an integral over the momentum transfer $(\mathbf p_f-\mathbf p_i)$. Here $V\to\infty$ is a box-normalization volume which is taken to infinity. The diagonals of the tracer particle's density operator after the collision correspond to $\mathbf p_s=0$, whereas $\mathbf p_s\neq 0$ describes coherences. For $\mathbf p_s=0$ this term contains an ill-defined square of the Dirac delta function and a physically motivated replacement rule
	\begin{eqnarray}
		\frac{(2\pi\hbar)^3}{V}|\langle\mathbf p_f|\hat T|\mathbf p_i\rangle |^2 &\to& \delta\!\left(\frac{\mathbf p_f^{2}-\mathbf p_i^2}2\right) \frac{|f(\mathbf p_f,\mathbf p_i)|^2}{\sigma(\mathbf p_i)|\mathbf p_i|}.\qquad \label{replacement}
	\end{eqnarray}
is used. One should mention that for any representation of the delta function in terms of a series of functions, the square of the Dirac delta function goes to infinity. Hence, for $\mathbf p_s=0$, $X$ contains an infinite term divided by an infinite volume $V$, justifying the use of a replacement rule to assign a finite value to the expression.  On the contrary, for $\mathbf p_s \neq 0$ the product of delta functions in \Eqref{problem} is well defined. In fact, using standard relations for Dirac delta functions,
	\begin{eqnarray}
		X &=& \frac{\pi\hbar}{V p_i q} \delta(p_i-p_f)\delta\!\left(\mathbf p_{s,\|\mathbf q}\right) f(\mathbf p_f+\mathbf p_s,\mathbf p_i+\mathbf p_s) f^*(\mathbf p_f-\mathbf p_s,\mathbf p_i-\mathbf p_s)\qquad \label{correct}
	\end{eqnarray}
is found, where $\mathbf q=\mathbf p_f-\mathbf p_i$ is the momentum transfer,  $\mathbf q_{\|\mathbf p_s}$ is $\mathbf q$ projected onto $\mathbf p_s$, and $q=|\mathbf q|$. The product of delta functions encountered here is well defined, and upon integration will lead to terms of finite value. Therefore, $X$ is now a term of finite value divided by an infinite volume $V$, which can only result in zero. This confirms our earlier argument, that momentum coherences vanish completely upon a single collision with a gas particle in a momentum eigenstate.

Di\'osi~\cite{Diosi} and Vacchini~\cite{Vacchini} did not encounter this decoherence problem because they assumed  ``quasi diagonality" of the density operator in momentum basis and only studied $\mathbf p_s=0$. Although Hornberger did consider $\mathbf q\neq 0$, he substituted the square root of the replacement rule \Eqref{replacement} for $\langle\mathbf p_f\pm \mathbf{p_s}|\hat T|\mathbf p_i\pm\mathbf{p_s}\rangle$ in \Eqref{problem}. This assigns a non-zero value to $X$, effectively bringing back the otherwise lost coherences of momentum states. The use of the replacement rule for a term which would otherwise vanish is not easily justified. Whether decoherence in momentum bases, and the closely related QPD are correctly described by approaches of this kind, is surely not certain.

Also physically, the use of non-localized momentum eigenstates for the gas particle seems problematic, because of the resulting infinite collision time. First, any Markovian QLBE relies on the assumption of short collision times. Second, successive collisions are not independent of each other if the collision time is long compared to the inverse collision frequency, leading to the necessity of studying multi particle collisions. Considering that diffusion processes usually depend on the considered time scales, an approach where a collision time diverges might not be appropriate to discuss QPD.

To resolve this matter of momentum decoherence as well as of QPD, we study a single collision in terms of localized gas states, which avoids the problems of a squared delta function. Then a collision time can be precisely defined, and the low-density and high-temperature limit will be quantified by `\emph{collision time}' $\times$ `\emph{collision rate}' $\ll 1$. To examine a collision event time resolved, and to investigate under which conditions a complete collision of two wave packets occurs, we analytically solve the two particle Schr\"odinger equation. This is in contrast to scattering calculations, which postulate complete collisions of wave packets. Although solving Schr\"odinger's equation requires more effort, the result will lead us to conclude that quantum collisions can not contribute to QPD. 

As our main aim is to clarify fundamental issues of QBM, and not specific details involving the interaction potential between the particles, we use the simplest possible model: QBM in one dimension and a delta type interaction between Brownian particle and gas particle. Furthermore, the gas particles do not interact with each other and are assumed to be in a thermal state using Boltzmann statistics. These are both valid approximations in the high temperature and low density limit.

Our approach is based on the theory of generalized measurements, and was introduced by Barnett and Cresser \cite{mainII} in a heuristic way. Although the observation that a colliding gas particle carries away information about the position and momentum of the Brownian particle is certainly correct, they had to guess the measurement operators and needed some unspecified parameters. Here we derive a measurement approach to QBM from first principles. In addition to rigorously justifying the approach, this also determines various parameters and corrects some flaws of the earlier treatment, such as the collision rate.

This work is outlined as follows. We start by deriving the most important results of classical Brownian motion in chapter~\ref{classmo}, and introducing some mathematical concepts used in this part of the thesis in chapter~\ref{tools}. This should prepare us to draw sensible conclusion when studying the quantum counterpart in the following chapters. In chapter \ref{lit} the existing literature is reviewed. Our examination of QBM starts in chapter \ref{col}, with the study of a single collision of two particles with Gaussian wave packets. Here we find, that, if the widths $W_g$ and $W $ of the wave packets of the colliding particles are related according to $m_gW_g^2=m W ^2$, where $m_g$ and $m $ are the masses of the respective particles, then the two particles are not entangled after the collision. This presents us with a preferred basis, which will significantly simplify the analysis of QBM. Furthermore, the centers of the wave packets in position as well as in momentum representation behave as in a classical collision.

In chapter \ref{meas}, we work out the measurement, which the gas particle performs on the Brownian one. For this we imagine a simultaneous position-momentum measurement on the gas particle, which serves as an indirect measurement of the position and momentum of the Brownian particle. We use the theory of generalized measurements \cite{Breuer,Busch} to derive the appropriate effect and Kraus operators, describing the measurement in the Brownian particle's Hilbert space. We can then use the Kraus  operators to describe the collision induced change of a general Brownian particle density operator.

In chapter \ref{rate}, we study the statistics of the collisions, assuming a gas in a thermal equilibrium. To this end, we decompose the thermal gas density operator in Gaussian wave packets. It turns out, that there is a considerable freedom in choosing the width of these wave packets, which we will do as to satisfy the conditions introduced in chapter~\ref{col}. Then we use a projection type operator to regard only the wave packets which collide with the Brownian particle in some time interval $\delta$. 
Putting things together, we finally arrive at our master equation in chapter \ref{master}. We will explain why this is a valid approach to QBM, even though the measurements on the gas particle are not actually carried out. Furthermore, we will review all approximations used in the previous chapters and discuss under which circumstances they can be fulfilled simultaneously, which will naturally lead us to the high temperature and low density limit.

In chapter \ref{expectation}, we use the master equation to derive equations of motion for the two lowest order moments of position and momentum. We will find classical behavior for all but the square displacement in position, where we recover QPD as in earlier approaches to QBM. However, we are able to track down the origin of this unphysical feature and find its connection to a coarse grained time scale and instantaneous collisions. We will establish that position jumps do not really happen, but instead are the result of approximations. These approximations in turn are necessary to derive a memory free QBM master equation.

Chapter \ref{final} is devoted to the important study of collisional decoherence, which is carried out using the Wigner function of the Brownian particle. We will see that decoherence is caused by phase averaging, as a collision will change the relative phase of superposition states. This phase change depends on the state of the colliding gas particle, and is therefore random. The often cited information transfer between the Brownian particle and the colliding gas particles, which is due to the measurements performed by the colliding gas particles, turns out to be small in the range of validity of our work. Furthermore, the decoherence rate depends on the spatial separation of the two wave functions making up the superposition state, and does not directly depend on the momentum separation. These results will lead us to the conjecture, that classical Fokker-Planck type equations might be applicable to the Wigner function of a quantum particle in a dilute gas.

\chapter{Classical Brownian Motion\label{classmo}}

We devote this chapter to the classical counterpart to our work. This discussion will later enable us to compare quantum Brownian motion with classical Brownian motion, and to debate whether the effects discovered in quantum Brownian motion are of true quantum nature.

All of the calculations and results below can most likely be found in the literature. However, it turned out to be extremely difficult to find this sort of work. First, most work in standard textbooks is in three spatial dimensions. Second, usually the Brownian particle is assumed to be much heavier and much slower than the gas particles. 

To draw a comparison between quantum and classical dynamics which is as complete as possible, we also need classical results for our situation, i.e.\ one spatial dimension and non vanishing mass ratio between gas and Brownian particle. Although such work was most certainly done, possible around a century ago, it is apparently not readily available in current standard textbooks. Consequently, we will present a detailed, classical derivation of the dynamics of a classical Brownian particle, based on statistical mechanical concepts introduced in \cite{Risken}.



\section{Classical collision\label{classcol}}

The study of a single collision in one dimension is a very simple one if a hard core interaction potential between the two colliding particles is assumed. Then a collision is instantaneous and its effect is an exchange of momentum of the two colliding particles. If $p_g$ and $p $ are the respective momenta of the gas and the Brownian particle before the collision, and $\bar p_g$ and $\bar p $ the momenta after the collision, then momentum conservation
	\begin{eqnarray}
		p_g+p  &=& \bar p_g+\bar p 
	\end{eqnarray}
and energy conservation
	\begin{eqnarray}
		\frac{p_g^2}{2m_g}+\frac{p ^2}{2m } &=& \frac{\bar p_g^2}{m_g}+\frac{\bar p ^2}{m }
	\end{eqnarray}
lead to
	\begin{eqnarray}
		\bar{p}_g &=& \frac{2m_g p-(m -m_g)p_g}{m +m_g} \\
		\bar{p} &=& \frac{2mp_g+(m -m_g)p}{m +m_g}. \label{qlql}
	\end{eqnarray}
Therefore the Brownian particle will gain the momentum $q=\bar p -p $
	\begin{eqnarray}
		q &=& \frac{2mp_g-2m_gp }{m +m_g}\label{mkmk}
	\end{eqnarray}
which depends on both, the Brownian particle's and the gas particle's momentum.

\section{Collision probability}

The probability of finding a gas particle with momentum between $p_g$ and $p_g+\dd p_g$ and position between $x_g$ and $x_g+\dd x_g$ is
	\begin{eqnarray}
		\rho_g(x_g,p_g) \,\dd x_g\,\dd p_g &=& n_g\mu_T(p_g)\,\dd x_g\,\dd p_g,\label{gaspp}
	\end{eqnarray}
where $n_g$ is the gas particle density, and 
	\begin{eqnarray}
		\mu_{T}(p_g) &=& \frac{1}{\sqrt{2\pi m_g k_B{T}}} e^{-p_g^2/(2m_gk_B{T})} \label{mbdis}
	\end{eqnarray}
is the Maxwell-Boltzmann momentum distribution for a thermal ideal gas at temperature $T$ in one dimension.

During a time interval $\dd t$, a gas particle with a momentum $p_g$ will move a distance $(v_g-v)\,\dd t$ relative to the Brownian particle. Therefore, it will collide with the Brownian particle, if its position is within the interval $\big(x-(v_g-v)\dd t,x)$. The probability of the Brownian particle colliding with a gas particle with momentum between $p_g$ and $p_g+\dd p_g$ is then
	\begin{eqnarray}
		P_p(p_g) \,\dd p_g\,\dd t &=& \int_{x-(v_g-v) d t}^x \dd x_g\, \rho_g(x_g,p_g) \dd p_g \nn\\
		&=& n_g\mu_T(p_g)|v_g-v|\dd t\,\dd p_g \nn\\
		&=& \frac{n_g}{\sqrt{2\pi m_gk_BT}}\exp\!\left(-\frac{p_g^2}{2m_gk_BT}\right) \left|\frac{p_g}{m_g}-\frac pm\right| \dd p_g\,\dd t, \label{psq}
	\end{eqnarray}
which, of course, depends on the momentum of the Brownian particle itself.

A number of articles~\cite{Diosi,mainII,Diosi2,Halliwell,Dodd} regarding QBM use the Maxwell-Boltzmann distribution for the probability distribution for a collision with a gas particle with momentum $p_g$. This happens to give the correct thermal state\footnote{The reason is, that using \Eqref{mbdis} instead of \Eqref{psq} is equivalent to dividing the scattering cross section by the relative velocity of the colliding particles, which, of course, also results in the correct thermal state.}, but fails to give the correct friction and diffusion constants. Therefore we want to emphasize, that for classical as well as quantum Brownian motion, one strictly has to use \Eqref{psq} for the collisional probability distribution, if one aims for quantitative results. 

To get the probability for the Brownian particle to experience a momentum ``kick" $q$, we substitute $p_g=\frac{m+m_g}{2m}q+\frac{m_g}{m}p$, which is according to \Eqref{mkmk}, into \Eqref{psq}, and find
	\begin{eqnarray}
		P_p(q)\,\dd q\,\dd t &=&  \frac{n_g}{\sqrt{2\pi m_gk_BT}} \frac{(m+m_g)^2}{4m^2m_g}|q| \exp\!\left\{ -\frac{\left[q(m+m_g)+2m_gp\right]^2}{8m^2m_gk_BT}\right\} \dd q \,\dd t. \qquad\quad\label{gngn}
	\end{eqnarray}
That is, a momentum transfer with $q\to0$ is impossible. The reason is, that zero momentum transfer means that the colliding particles have the same velocity, therefore not being able to approach each other. \\
	\begin{center}\begin{minipage}{6.5cm} \small
	Dotted: The exponential in \Eqref{gngn} is shifted in the opposite direction of the Brownian particle momentum, because according to \Eqref{qlql} it looses some of its momentum during a collision.\vspace{2mm}
	
	Dashed: Because of $|q|$ in \Eqref{gngn}, more gas particles with momentum directed in the opposite direction of $p$ collide with the gas particle.\vspace{2mm}
	
	Solid: Both effects contribute to the momentum transfer \Eqref{gngn} being in average directed opposite to the Brownian particle momentum, therefore resulting in friction. 
	\end{minipage}\hspace{3mm}
	\begin{minipage}{7.5cm}\vspace{5mm}
	\includegraphics[width=7.5cm]{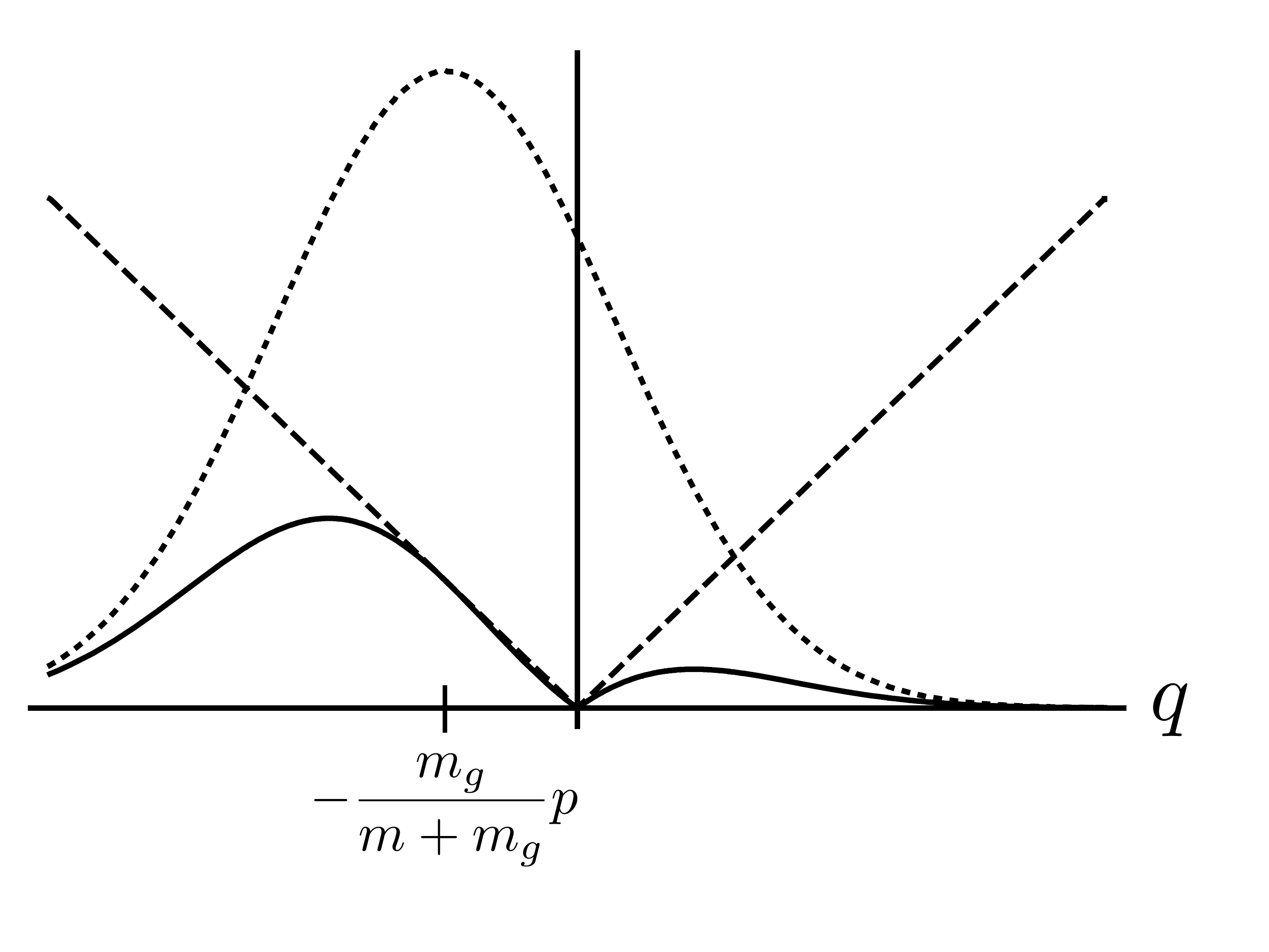}
	\end{minipage}\end{center}

\section{Master equation\label{rfko}}

We denote the phase space probability density for the Brownian particle by $\rho(x,p,t)$. The aim of this section is do derive a differential equation, which is called a master equation, according to which $\rho(x,p,t)$ evolves. If the particle evolves freely, then its position moves according to its velocity, resulting in $\rho(x,p,t+\dd t)=\rho(x-p/m\,\dd t,p,t)$. If, on the other hand, the particle experiences a momentum kick $q$, then we find $\rho(x,p,t+\dd t)=\rho(x,p-q,t)$. Strictly speaking, the position will also change. But because the probability of a momentum kick is of the order $\dd t$, we do not have to take into account the infinitesimal position shift. Adding both possibilities together, we find
	\begin{eqnarray}
		\rho(x,p,t+\dd t) &=& \rho\!\left(x-\frac pm \dd t,p,t\right)\left[ 1-\int\dd q\, P_p(q) \dd t\right]  + \int \dd q\, \rho(x,p-q,t)P_{p-q}(q) \dd t, \nn\\ \label{ksdj}
	\end{eqnarray}
where $\left[ 1-\int\dd q\, P_p(q) \dd t\right]$ is the probability that no collision occurred, and $P_{p-q}(q)\dd t$ is the probability of a momentum kick $q$, conditioned on the Brownian particle momentum $p-q$ at time $t$.

Next we substitute
	\begin{eqnarray}
		\rho\!\left(x-\frac pm \dd t,p,t\right) &=& \rho(x,p,t)-\frac pm \dd t\frac{\partial\rho(x,p,t)}{\partial x} \label{zpzp}
	\end{eqnarray}
into the right hand side of \Eqref{ksdj}, to arrive at the master equation
	\begin{eqnarray}
		\frac{\partial\rho(x,p,t)}{\partial t} &=& -\frac pm \frac{\partial\rho(x,p,t)}{\partial x} -\rho(x,p,t)\int\dd q \, P_p(q)  +\int\dd q\,\rho(x,p-q,t)P_{p-q}(q). \nn\\ \label{clasmas}
	\end{eqnarray}
The first term represents the free evolution, the second one corresponds to momentum jumps away from $p$, and the third to momentum jumps towards $p$.

This master equation is correct for finite mass ratios $m_g/m$, and not limited to a slow Brownian particle. The only approximations used are the hard core interaction potential, and a low density and high temperature gas such that we can assume independent collisions. In particular, we did not use a time coarse graining approximation.

\section{Expectation values}

We can use the master equation \Eqref{clasmas} to derive equations of motion for the expectation values of the lowest order moments of position and momentum. Using the definition of an expectation value
	\begin{eqnarray}
		\langle f(x,p)\rangle &=& \int\!\!\!\!\int \dd x\,\dd p\, f(x,p)\rho(x,p,t)
	\end{eqnarray}
we find for the position
	\begin{eqnarray}
		\frac{\dd}{\dd t} \langle x \rangle &=& \int\!\!\!\!\int \dd x\,\dd p\, x\left\{ -\frac pm \frac{\partial\rho(x,p,t)}{\partial x} + \int\dd q\,[\rho(x,p-q,t)P_{p-q}(q) - \rho(x,p,t) P_p(q)] \right\}\!. \nn\\ 
	\end{eqnarray}
In the second term we change the order of the $p$ and $q$ integrations, and substitute $u=p-q$. It is then seen that the second and third terms cancel each other. The first term is integrated by parts and we find
	\begin{eqnarray}
		\frac{\dd}{\dd t} \langle x \rangle &=&  \int\!\!\!\!\int \dd x\,\dd p\, \frac pm\rho(x,p,t) \nn\\
		&=& \frac{\langle p\rangle}{m}. \label{classx}
	\end{eqnarray}
While integrating by parts, we assumed $\lim_{x\to\pm\infty}[x\rho(x,p,t)] =0$, which is certainly true for any normalized probability distribution.

For the momentum we find
	\begin{eqnarray}
		\frac{\dd}{\dd t} \langle p \rangle &=& \int\!\!\!\!\int \dd x\,\dd p\, p\left\{ -\frac pm \frac{\partial\rho(x,p,t)}{\partial x} + \int\dd q\,[\rho(x,p-q,t)P_{p-q}(q) - \rho(x,p,t) P_p(q)] \right\}\!. \nn\\ 
	\end{eqnarray}
This time the first part vanishes because of $\lim_{x\to\pm\infty}[\rho(x,p,t)] =0$, but the second term does not cancel the third upon substituting $u=p-q$, and we find 
	\begin{eqnarray}
		\frac{\dd}{\dd t} \langle p \rangle &=&  \int\!\!\!\!\int \dd x\,\dd p\left[\rho(x,p,t) \int \dd q\, q P_p(q) \right]\nn\\
		&=& \left\langle  \int \dd q\, q P_p(q)\right\rangle. \label{classp}
	\end{eqnarray}

In much the same way we derive
	\begin{eqnarray}
		\frac{\dd}{\dd t} \left\langle x^2 \right\rangle &=& 2\frac{\langle xp\rangle}{m}, \label{classxx} \\
		\frac{\dd}{\dd t} \left\langle p^2 \right\rangle &=&  \left\langle  \int \dd q\, \left(q^2+2pq\right) P_p(q)\right\rangle, \label{classpp} \\
		\frac{\dd}{\dd t} \langle xp \rangle &=&   \frac{\left\langle p^2\right\rangle}{m} + \left\langle x \int \dd q\, q P_p(q)\right\rangle . \label{classxp}
	\end{eqnarray}
The integrations over $q$ can not be carried out analytically in general, unless one assumes the limit that the Brownian particle is either much slower or much faster than a thermal gas particle. We will discuss both limits in chapter~\ref{expectation}.

\section{Fokker-Planck equation}

In order to derive a Fokker-Planck equation, we have to assume that the collisional momentum transfer $q$ is small compared to the momentum uncertainty of $\rho(x,p,t)$.\footnote{That is typically satisfied if the gas particles are very light compared to the Brownian particle.} Then we can approximate 
	\begin{eqnarray}
		\rho(x,p-q,t)P_{p-q}(q) &\approx& \rho(x,p,t)P_{p}(q) -q\frac\partial{\partial p}[\rho(x,p,t)P_{p}(q)]+\frac{q^2}2 \frac{\partial^2}{\partial p^2}[\rho(x,p,t)P_{p}(q)], \nn\\ \label{xpxp}
	\end{eqnarray}
which leads us to the Fokker-Planck equation
	\begin{eqnarray}
		\frac{\partial\rho(x,p,t)}{\partial t} &\!\!=\!\!& -\frac pm \frac{\partial\rho(x,p,t)}{\partial x} -\frac\partial{\partial p}\!\left[\rho(x,p,t)\!\int\! \dd q\, qP_{p}(q)\right]\! + \frac{\partial^2}{\partial p^2}\!\left[\rho(x,p,t)\!\int\! \dd q \frac{q^2}2 P_{p}(q)\right] \!\!\:\!. \nn\\ \label{fockerplanck}
	\end{eqnarray}
The first term on the right hand side shifts the position according to the Brownian particle velocity. The second term is responsible for momentum damping, and the third one for momentum diffusion.

The integrations over the momentum transfer $q$ can not be carried out analytically without further approximations. Therefore, we now assume that the Brownian particle is slow compared to the thermal velocity of the gas particles, i.e.\ $\rho(x,p,t)$ is only non-vanishing for $p/m\ll\sqrt{k_BT/m_g}$. We substitute \Eqref{mkmk} for $q$ and then evaluate the integrals according to the appendix of this chapter. We then find for the friction constant $\gamma$, and momentum diffusion constant $mk_BT\gamma$
	\begin{eqnarray}
		-\gamma p &=& \int \dd q\, qP_{p}(q) \;\approx\; -\frac{4n_g\sqrt{2m_gk_BT}}{\sqrt\pi}\frac pm  , \\
		mk_BT\gamma &=& \int \dd q\, \frac{q^2}2 P_{p}(q) \;\approx\; 4n_gk_BT\frac{\sqrt{2m_gk_BT}}{\sqrt\pi} ,
	\end{eqnarray}
respectively. Substitution into \Eqref{fockerplanck} leads us to the Fokker-Planck equation for a slow and heavy particle
 	\begin{eqnarray}
		\frac{\partial\rho(x,p,t)}{\partial t} &\!=\!& -\frac pm \frac{\partial\rho(x,p,t)}{\partial x} + \gamma \frac{\partial[p\rho(x,p,t)]}{\partial p} + \gamma mk_BT\frac{\partial^2\rho(x,p,t)}{\partial p^2}, \qquad \label{fockerplanck2}
	\end{eqnarray}
which is often referred to as Kramer's equation. It satisfies the equipartition theorem.

\section{What is a diffusion process?\label{what}}

It is interesting to note that while we expanded \Eqref{zpzp} only to first order, we needed a second order expansion in \Eqref{xpxp} to derive a sensible Fokker-Planck equation. The reason is that the first expansion is exact because $\dd t$ is an infinitesimal, whereas $q$ is finite and therefore the second expansion is an approximation. In fact, if we had expanded \Eqref{xpxp} only to first order, then the Focker-Planck equation would not incorporate momentum diffusion. This in turn would lead to a steady state phase space distribution $\rho(x,p,t\to\infty)$ with exact momentum $p=0$, and therefore violate the very condition for the justification of the expansion \Eqref{xpxp} (see beginning of this section).\footnote{If the momentum of the Brownian particle is much larger than $mk_BT$, then the diffusion of the momentum distribution will be small compared to the damping, and important physical results about friction can indeed be obtained without considering the second order of the expansion in \Eqref{xpxp}.}

Although the friction term is due to random momentum jumps of finite size, which are in average directed opposite to the momentum of the particle, one would derive a similar term from a continuously applied force which depends on the particles momentum. But then the momentum change in an infinitesimal time would also be infinitesimal, and the expansion of $\rho(x,p+\dd p,t)$ to second order in $p$ would be an infinitesimal of second order and therefore vanish. For this reason, for a momentum diffusion process to be possible, a single collision has to change the momentum by a finite value.\footnote{Of course, this value can be small and even the limit to zero can be taken after the expansion is performed, but that has to be accompanied with an infinite collision rate.}

Because an instantaneous collision does not change the position of the Brownian particle, the Fokker-Planck equation can not exhibit position diffusion. In fact, solving the equation for a Brownian particle with initial position $x'$ and initial momentum $p'$, i.e.\ $\rho(x,p,0)=\delta(x-x')\delta(p-p')$, we find for short times $t\ll\gamma^{-1}$, that the position variance increases according to $\Delta x^2\propto t^3$, while the momentum variance diffuses with $\Delta p^2\propto t$.

The well known position diffusion $\Delta x^2\propto t$ of a Brownian particle is only valid on a time scale large compared to the momentum damping time $\gamma^{-1}$, that is, once the momentum distribution is a thermal one. On this large time scale, the first term on the right hand side of the Fokker-Planck equation is responsible for position diffusion, because of the randomness of the particle momentum.

During the thesis, we use the terminology generally used in literature about quantum linear Boltzmann equations (see~\cite{Diosi} and \cite{review}). That is, a diffusion process has to act on a short time scale of single collisions. Therefore, classical dynamics include momentum diffusion, but position diffusion is an impossible process as it would correspond to finite position jumps of the Brownian particle. A further reasoning for this terminology is that in quantum dynamics, the short time scale is of interest because after a large number of collisions, any quantum behavior is lost.

\section*{Appendix}

We will need several integrals of the form
	\begin{eqnarray}
		&& \int\dd p_g \left|\frac{p_g}{m_g}-\frac pm\right| \mbox{pol}(p_g) \exp\!\left(\frac{-p_g^2}{2m_gk_BT}\right) \nn\\
		&=& \left[ \int_{-\infty}^{\alpha p}\dd p_g-\int_{\alpha p}^\infty \dd p_g\right] \left[\left(\frac{p_g}{m_g}-\frac pm\right) \mbox{pol}(p_g) \exp\!\left(\frac{-p_g^2}{2m_gk_BT}\right)\right],
	\end{eqnarray}
where pol$(p_g)$ is a low order polynomial. The integral operators in the left square brackets act on the  term to its right in the obvious way. We now split the term in the right square brackets in a symmetric term $S$ and an asymmetric one $A$, and find
	\begin{eqnarray}
		\left[ \int_{-\infty}^{\alpha p}\dd p_g-\int_{\alpha p}^\infty \dd p_g\right]S &=& \left[ \int_{-\infty}^{\alpha p}\dd p_g-\int_{-\infty}^{-\alpha p} \dd p_g\right]S \;=\; 2\int_0^{\alpha p} \dd p_g \,S \qquad\quad \\
		\left[ \int_{-\infty}^{\alpha p}\dd p_g-\int_{\alpha p}^\infty \dd p_g\right]A &=& \left[ \int_{-\infty}^{\alpha p}\dd p_g+\int_{-\infty}^{-\alpha p} \dd p_g\right]A \;=\; 2\int_{-\infty}^{\alpha p} \dd p_g\, A.
	\end{eqnarray}
The integration of the asymmetric part can be carried out analytically, in particular
	\begin{eqnarray}
		\int_{-\infty}^{\alpha p} \dd p_g\, \frac{p_g}{m_g}\exp\!\left(\frac{-p_g^2}{2m_gk_BT}\right) &=& -k_BT \exp\!\left(\frac{-\alpha p^2}{2mk_BT}\right) \\
		&\approx& -k_BT+\frac{\alpha p^2}{2m}-\frac{\alpha^2p^4}{8m^2k_BT} \label{pg1} \\
		\int_{-\infty}^{\alpha p} \dd p_g\, \frac{p_g^3}{m_g^3} \exp\!\left(\frac{-p_g^2}{2m_gk_BT}\right) &=& -\left(\frac{2 k_B^2T^2}{\alpha m} + k_BT\frac{p^2}{m^2}\right) \exp\!\left(\frac{-\alpha p^2}{2mk_BT}\right) \qquad\quad \\
		&\approx& -\frac{2 k_B^2T^2}{\alpha m}+\frac{\alpha p^4}{4m^3}. \label{pg3}
	\end{eqnarray}
Here, we defined $\alpha=m_g/m$. The symmetric part can not be integrated analytically. Therefore, we first expand the exponential and then integrate to find
	\begin{eqnarray}
		\int_0^{\alpha p} \dd p_g\, \exp\!\left(\frac{-p_g^2}{2m_gk_BT}\right) &\approx&  \alpha p - \frac{\alpha^2p^3}{6mk_BT} + \frac{\alpha^3p^5}{40 (mk_BT)^2} \qquad \label{pg0} \\
		\int_0^{\alpha p} \dd p_g\, \frac{p_g^2}{m_g^2}\exp\!\left(\frac{-p_g^2}{2m_gk_BT}\right) &\approx& \frac{\alpha p^3}{3m^2} - \frac{\alpha^2p^5}{10m^3k_BT} . \label{pg2}
	\end{eqnarray}

\chapter{Tools and Definitions\label{tools}}

We have already discussed general aspects of quantum mechanics in chapter~\ref{quantmech}, but there are more specific results associated with simultaneous position-momentum measurements, as well as quasi-probability distributions, that find frequent use in later chapters. Here, we are concerned with certain mathematical results, grouped into one self-contained chapter, from the point-of-view of their use as tools in later calculations, rather than providing a detailed discussion of their properties.

\section{Simultaneous position-momentum measurements}

We will see in the following chapters that in some sense, a gas particle which collides with the Brownian particle performs a simultaneous measurement of position and momentum of the latter particle. If the position and momentum uncertainties of the measurement\footnote{That are the uncertainties related to the measurement, which are on top of the intrinsic uncertainties of the measured state.} is at least as large as given by the Heisenberg uncertainty limit $\Delta \tilde p \,\Delta \tilde x \geq \hbar /2$, then these are perfectly valid quantum measurements which can be described in the language of POVM which was introduced in section~\ref{genmeas}.\footnote{Although Heisenberg said that it is not possible to measure both, momentum and position with arbitrary precision, what he really had in mind was that the momentum and position of a state can only be simultaneously well defined up to some uncertainty. In his time, the simultaneous measurement of non-commuting observables was not yet formally developed, and therefore the difference of both statement was not clear. Today, some people feel that uncertainty limit related to the measurement itself should not be called Heisenberg limit.} To be consistent with following chapters, we denote measurement outcomes with tildes.

\subsection{Quantum-limited measurements}

We start with the special case of a quantum-limited measurement, in which the Heisenberg uncertainty principle is satisfied with an equal sign. With such a measurement we associate effect operators which are proportional to projectors onto Gaussian minimum uncertainty states. If the system to be measured is in a state $\rho$, then the probability that a measurement of position and momentum yields results in the respective intervals $(\tilde x,\tilde x+\dd \tilde x)$ and $(\tilde p,\tilde p+\dd \tilde p)$  is
	\begin{eqnarray}
		P_0(\tilde x,\tilde p)\,\dd \tilde x\,\dd \tilde p &=& \mbox{Tr}[\rho \hat\pi_0(\tilde x,\tilde p)]\, \dd \tilde x\,\dd \tilde p, \label{kkkk}
	\end{eqnarray}
where 
	\begin{eqnarray}
		\hat\pi_0(\tilde x,\tilde p) &=& \frac1{2\pi\hbar}\ket{\tilde x,\tilde p}_W\!\bra{\tilde x,\tilde p},
	\end{eqnarray}
and 
	\begin{eqnarray}
		\ket{x,p}_W &=& \int \dd x' \frac{e^{-ix p /2\hbar}}{\sqrt{\sqrt\pi W }} e^{ix 'p /\hbar}e^{-(x -x ')^2/2W ^2} \ket{x'}.
	\end{eqnarray}
is a minimum uncertainty state with mean position $x$, mean momentum $p$, and position variance $W$. The uncertainties introduced by these measurements on top of the uncertainties $\Delta x(\rho)$ and $\Delta p(\rho)$ of the state itself are
	\begin{eqnarray}
		\Delta \tilde x &=& \frac{W}{\sqrt 2} \\
		\Delta \tilde p &=& \frac{\hbar}{\sqrt 2 W}.
	\end{eqnarray}
Therefore, the parameter $W$ specifies whether the measurement is more precise in position (small $W$) or in momentum (large $W$). 

\subsection{Imperfect measurements\label{dwog}}

What we will need in the following chapters are imperfect measurements, which means the uncertainties  in position and momentum are greater than required by the Heisenberg uncertainty principle. These can be obtained from the quantum-limited ones by convolving the ideal distribution $P_0(x,p)$ with a weighting factor $w(x,p)$
	\begin{eqnarray}
		P(\tilde x,\tilde p) &=& \int\!\!\!\!\int \dd x\,\dd p\, w(x,p)P_0(\tilde x+x,\tilde p+p). 
	\end{eqnarray}
By using \Eqref{kkkk} it follows that the effect operators for such a ``smeared-out" measurement are
	\begin{eqnarray}
		\hat\pi_{}(\tilde x_{},\tilde p_{}) &=&\frac1{2\pi\hbar} \int\!\!\!\!\int\dd x\,\dd p\, w(x,p) \ket{\tilde x_{}+x,\tilde p_{}+p}\!\bra{\tilde x_{}+x,\tilde p_{}+p}.
	\end{eqnarray}
In particular, we will be interested in a Gaussian convolution factor
	\begin{eqnarray}
		w(x,p) &=&  \frac{1}{2\pi\hbar \bar n} \exp\!\left[ -\frac{1}{2 \bar n} \left(\frac{x^2}{ W_{}^2}+\frac{ W_{}^2p^2}{\hbar^2}\right) \right], \label{pqy}
	\end{eqnarray}
where $ \bar n$ specifies the uncertainty due to imperfectness of the measurement. Now we find 
	\begin{eqnarray}
		\Delta \tilde x &=& W\sqrt{ \bar n+\half} \\
		\Delta \tilde p &=& \frac{\hbar}{W}\sqrt{ \bar n+\half}.
	\end{eqnarray}
For $ \bar n\to 0$ the convolution factor becomes a Dirac delta function and we regain the quantum-limited case.

We can relate the effect operators according to
	\begin{eqnarray}
		\hat\pi_{}(\tilde x_{},\tilde p_{}) &=& \hat D(\tilde x,\tilde p) \hat \sigma \hat D^\dag(\tilde x,\tilde p) \label{jak}
	\end{eqnarray}
where $\hat \sigma=\hat \pi(0,0)$, and $\hat D(\tilde x, \tilde p)=e^{i(\tilde p\hat x-\tilde x \hat p)/\hbar}$ is the unitary displacement operator. We will also need the square root of the effect operators
	\begin{eqnarray}
		\sqrt{\hat\pi_{}(\tilde x_{},\tilde p_{})} &=& \hat D(\tilde x,\tilde p) \sqrt{\hat \sigma} \hat D^\dag(\tilde x,\tilde p). \label{ekd}
	\end{eqnarray}

\section{Phase space distribution functions}

In classical dynamics, the phase space probability density $\rho(x,p)$ is of great importance. It gives the probability density of finding the particle at position $x$ with momentum $p$. Because the Heisenberg uncertainty principle forbids that a particle has definite position and momentum at the same time, such a probability density can not exist in quantum mechanics. Nevertheless, there are several possibilities to assign a phase space distribution function to any given quantum mechanical density operator, but each of them will lack at least one property of a phase space probability density. For this reason, they are also called \emph{quasiprobability distributions}.

\subsection{Wigner function\label{Wfunc}}

The Wigner function $W_\rho(x,p)$ is defined by~\cite{Busch,Leonhardt}
	\begin{eqnarray}
		W_\rho(x,p) &=& \frac1{\pi\hbar}\int \dd x'\, e^{-2\imath px'}\rho(x+x',x-x'), \label{wfunc}
	\end{eqnarray}
where $\rho(x,x')=\bra{x}\rho\ket{x'}$ are the position matrix elements of the density operator. Its marginals give the position and momentum densities
	\begin{eqnarray}
		\int \dd x \,W_\rho(x,p) &=& \rho(p,p),\\
		\int \dd p \,W_\rho(x,p) &=& \rho(x,x),
	\end{eqnarray}
where $\rho(p,p')=\bra{p}\rho\ket{p'}$ are the momentum matrix elements of the density operator. In this sense, it is the closest possible to a phase space probability density, but it is lacking the property of positivity. It is well known that only the Gaussian wave functions and mixtures thereof have positive Wigner functions.

\begin{figure}[htbp]
\begin{center}
	\includegraphics[width=\linewidth]{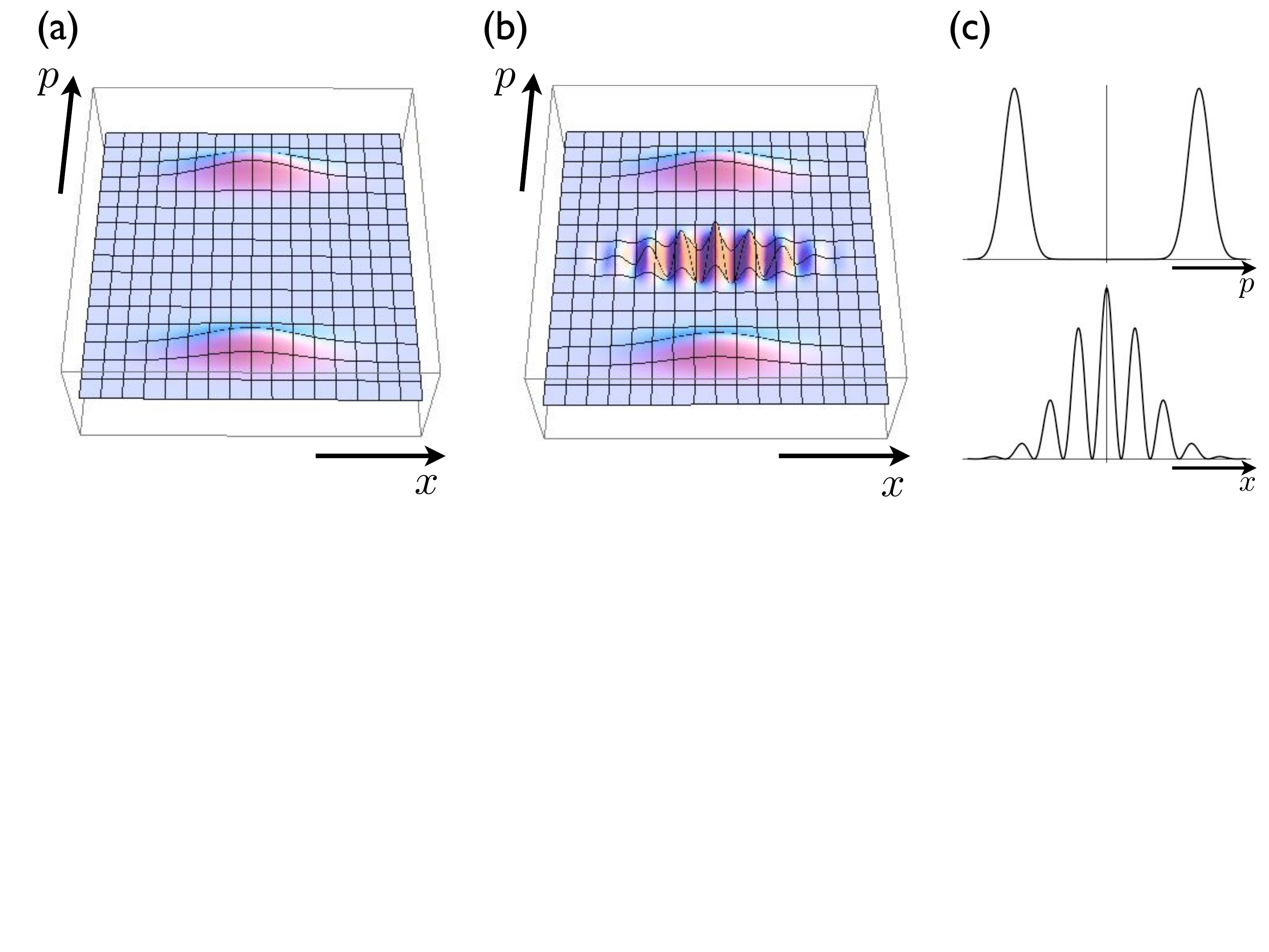}\vspace{-50mm}
\caption{\small (a): The Wigner function of the incoherent mixture $1/2(\ket{x,p_1}\!\bra{x,p_1}+\ket{x,p_2}\!\bra{x,p_2})$ reminds of a classical phase space density. \newline (b): The Wigner function of the coherent superposition $1/2(\ket{x,p_1}+\ket{x,p_2})(\bra{x,p_1}+\bra{x,p_2})$ oscillates from negative to positive values in the phase space region between the two ``classical" humps. \newline (c): Position and momentum marginals of the same superposition state as in (b). Although the momentum distribution is very similar to an incoherent mixture, the position distribution shows an interference pattern which is absent in the mixed state.}
\label{fig22}
\end{center}
\end{figure}

The Wigner function can be generalized to operators other than density operators. The Wigner function is real for all Hermitean operators $\rho$, and it is normalized $\int\!\!\int \dd x\,\dd p\, W_\rho(x,p)=1$ for operators $\rho$ with trace one. Another remarkable property of the Wigner functions is the overlap formula
	\begin{eqnarray}
		\mbox{Tr}(\rho_1\rho_2) &=& 2\pi\hbar \int\!\!\!\!\int\dd x\,\dd p\,{W_{\rho_1}(x,p)}W_{\rho_2}(x,p),
	\end{eqnarray}
which asserts that similar density operators also have similar Wigner functions and vice versa, and is not true for other popular phase space probability distributions.

Negative and/or oscillating regions in the Wigner function can be seen as signatures of nonclassical behavior (see figure~\ref{fig22}). Therefore, decoherence is associated with the vanishing of these negative regions. 

A further usefull property of the Wigner function of a free particle\footnote{This property is also true for a particle in an Harmonic potential.} is that it evolves according to the classical equation for the phase space probability distribution
	\begin{eqnarray}
		\frac{\partial W_{\rho(t)}(x,p)}{\partial t} &=& -\frac pm \frac{\partial W_{\rho(t)}(x,p)}{\partial x},
	\end{eqnarray}
which means that the Wigner function $W_{\rho(t)}(x,p)$ ``flows" according to its velocity $v=p/m$. An example is shown in figure~\ref{fig23}
\begin{figure}[htbp]
\begin{center}
	\includegraphics[width=\linewidth]{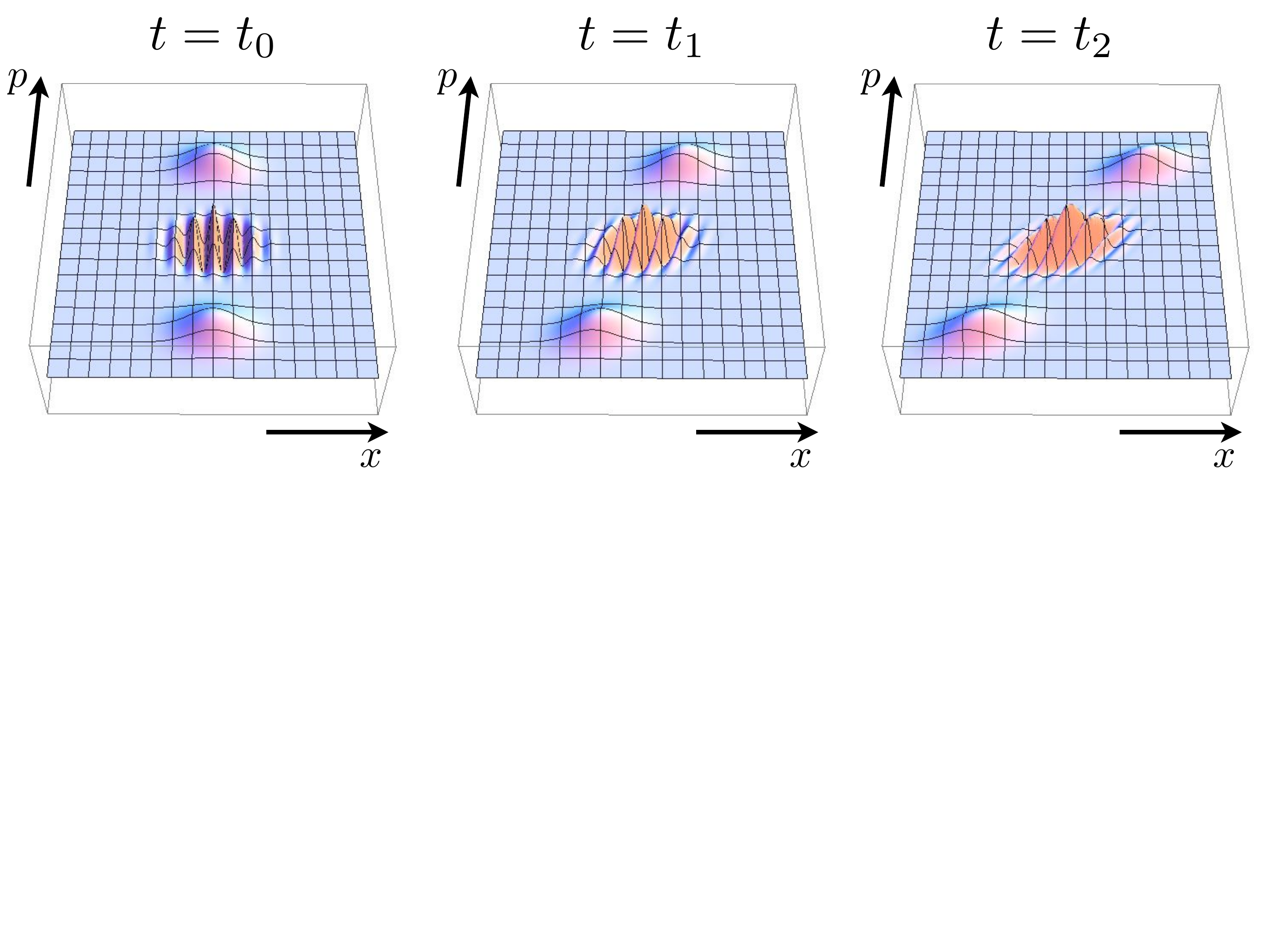}\vspace{-55mm}
\caption{\small The Wigner function of the initial coherent superposition $1/2(\ket{x,p_1}+\ket{x,p_2}\:\!)(\bra{x,p_1}+\bra{x,p_2})$ during free evolution at three different times. Each point in the distribution moves according to the velocity $v=p/m$.}
\label{fig23}
\end{center}
\end{figure}

\subsection{Husimi $Q$ function\label{Qfunc}}

One may obtain a new phase space distribution function by convolving the Wigner function with a Gaussian phase space distribution having the marginals as a minimum uncertainty wave packet which is located at the origin
	\begin{eqnarray}
		Q_\rho(x, p) &=& \frac1{\pi\hbar}\int\!\!\!\!\int \dd x'\dd p' W_\rho(x',p')\exp\!\left[ -\frac{(x-x')^2}{W^2} -\frac{(p-p')^2W^2}{\hbar^2} \right].
	\end{eqnarray}
As it turns out, the $Q$~function is precisely the phase space distribution obtained from a quantum-limited position-momentum measurement \Eqref{kkkk}
	\begin{eqnarray}
		Q_\rho(\tilde x,\tilde p) &=& P_0(\tilde x,\tilde p) \nn\\
		&=& \bra{\tilde x,\tilde p}{\rho}\ket{\tilde x,\tilde p}.
	\end{eqnarray}
The convolution accounts for a smearing of the marginal distributions, establishing the positivity which was missing in the Wigner function.

Similar to the Wigner function, there is a one to one correspondence between state operators and $Q$~functions. Therefore, knowing the precise measurement statistics $P_0(\tilde x,\tilde p)$ is equivalent to knowing the state $\rho$. This property will turn out to be useful at times, because if we need to show the equivalence of two operators, it is sufficient to compare their diagonals with respect to Gaussian wave packets. However, contrary to the Wigner function, very similar $Q$ functions can represent very different states. For example, it can not distinguish well between superpositions and mixtures, which makes the $Q$ function not suited to the study of decoherence.

\subsection{Glauber-Sudarshan $P$~function\label{Pfunc}}

The $P$ function $P_\rho(x,p)$ is a phase space distribution, which diagonalizes a density operator in terms of minimum uncertainty wave packets
	\begin{eqnarray}
		\rho &=& \int\!\!\!\!\int \dd x\,\dd p\, P_\rho(x,p)\ket{x,p}\!\bra{x,p}.
	\end{eqnarray}
Although this formula is valid for general states $\rho$, the above formula seems like a statistical mixture of Gaussian wave functions. For states which are not such statistical mixtures, the $P$~function will become negative, and might include delta functions and derivatives thereof. In fact, the Wigner function, which itself can become negative in small regions, is obtained from the $P$~function by a smearing process
	\begin{eqnarray}
		W_\rho(x,p) &=& \frac1{\pi\hbar}\int\!\!\!\!\int \dd x'\dd p' P_\rho(x',p')\exp\!\left[ -\frac{(x-x')^2}{W^2} -\frac{(p-p')^2W^2}{\hbar^2} \right].
	\end{eqnarray}

The thermal state of a free particle is an important example where the $P$ function is positive, as we will see in section~\ref{convex decomp}. For this reason, we can assume that a thermal gas particle is in a Gaussian state $\ket{x,p}$, with a probability given by $P_\rho(x,p)$.

\section{Useful formulas\label{formulas}}

\subsubsection{Representations of position and momentum operators}

It will turn out to be useful to write position and momentum operators in a representation diagonal in the Gaussian pure states, resembling the Glauber $P$ representation of density operators.
	\begin{eqnarray}
		\hat x &=& \int\!\!\!\!\int\frac{\dd x\,\dd p}{2\pi\hbar}x\ket{x,p}\!\bra{x,p} \label{x} \\
		\hat p &=& \int\!\!\!\!\int\frac{\dd x\,\dd p}{2\pi\hbar}p\ket{x,p}\!\bra{x,p} \label{p} \\
		\hat x^2 +\frac{W^2}2 &=& \int\!\!\!\!\int\frac{\dd x\,\dd p}{2\pi\hbar}x^2\ket{x,p}\!\bra{x,p} \label{xx} \\
		\hat p^2 +\frac{\hbar^2}{2W^2} &=& \int\!\!\!\!\int\frac{\dd x\,\dd p}{2\pi\hbar}p^2\ket{x,p}\!\bra{x,p} \label{pp} \\
		\half\left\{\hat x,\hat p\right\} &=& \int\!\!\!\!\int\frac{\dd x\,\dd p}{2\pi\hbar}xp\ket{x,p}\!\bra{x,p} \label{xp} \\
	\mbox{for }W\to\infty : \hspace{18.8mm}	f(\hat p) &=& \int\!\!\!\!\int\frac{\dd x\,\dd p}{2\pi\hbar}f(p)\ket{x,p}\!\bra{x,p} \qquad\qquad \label{f(p)}  \\
	\mbox{for }W\to\infty : \qquad	\half\left\{\hat x,f(\hat p)\right\} &=& \int\!\!\!\!\int\frac{\dd x\,\dd p}{2\pi\hbar}xf(p)\ket{x,p}\!\bra{x,p} \qquad\qquad \label{xf(p)}
	\end{eqnarray}
The correctness of these formulas is easily shown by taking their expectation values for states described by Gaussian wave functions, i.e.\
	\begin{eqnarray}
		\langle x,p|\hat x |x,p\rangle &=& \int\!\!\!\!\int\frac{\dd x'\dd p'}{2\pi\hbar}x'\scalar{x,p}{x',p'}\!\scalar{x',p'}{x,p} \nn\\
		&=&  \int\!\!\!\!\int\frac{\dd x'\dd p'}{2\pi\hbar}x' \exp\!\left[ -\frac{(x-x')^2}{2W^2}-\frac{W^2(p-p')^2}{2\hbar^2} \right]\nn\\
		&=& x.
	\end{eqnarray}
Because Gaussian states form an overcomplete set of states, an operator is uniquely defined by its diagonal values (see subsection~\ref{Qfunc}), and therefore the operator \Eqref{x} is indeed the position operator. The other operators are shown along the same lines.

\subsubsection{Commutators of position and momentum operators}

We will also frequently need the commutator of arbitrary analytic functions of $\hat x$ and $\hat p$
	\begin{eqnarray}
		[f(\hat x),g(\hat p)] &=& -\sum_{n=1}^\infty \frac{(-\imath \hbar)^n}{n!} f^{(n)}(\hat x) g^{(n)}(\hat p),  \label{momposcom}
	\end{eqnarray}
where $f^{(n)}$ denotes the $n$-th derivative of $f$. This can be derived from 
	\begin{eqnarray}
		[\hat x^k,\hat p^l] &=& -\sum_{n=1}^{\min(k,l)} \frac{(\imath \hbar)^n k!l!}{n!(k-n)!(l-n)!}\hat x^{k-n}\hat p^{l-n},
	\end{eqnarray}
which in turn is shown by induction.

\chapter{Literature Review\label{lit}}

In this chapter, we review some of the existing literature on QBM. We can roughly divide the topic into three categories, each of which is reviewed in one section. The first assumes, that the Brownian particle in infinitely heavy. This simplification leads to the study of collisional decoherence, where any change of the Brownian particle position is neglected. The Caldeira-Leggett model in comparison, incorporates friction and diffusion of the Brownian particle, but achieves this using an environment of harmonic oscillators rather than gas particles. Furthermore, the interaction potential between Brownian and gas particle is linearized. Finally, there are quantum versions of the linear Boltzmann equation, which were developed to study the full interplay of friction, diffusion and decoherence of a Brownian particle in a gaseous environment. This approach can be considered the genuine quantum counterpart of classical Brownian motion, as studied, e.g.\ by Einstein~\cite{Einstein} and Smoluchowski~\cite{Smoluchowski}. Because our own work belongs to the latter category, we will review publication of this type in more detail. We close this chapter with a brief section, to draw some critical conclusion which are of relevance to our work.

There are yet other approaches to QBM. Some do not specify the Brownian environment of the particle, but rather postulates some properties of QBM. These properties, like spatial invariance or momentum damping, are related to classical Brownian motion. The master equation is then constructed such that the dynamics of the Brownian particle fulfills these properties. The articles of Gallis~\cite{Gallis2} and Holevo~\cite{Holevo} use such a postulation ansatz.

Our work questions some of the popular results of quantum linear Boltzmann equations. Therefore, we will review some of the papers in more detail, to point out where difficulties in each approach might occur. However, the literature is too large to account for every contribution in such detail. We therefore restrict a comprehensive review to only a view of these, and mention only some of the others rather briefly, referring to the well written review article of Vacchini and Hornberger~\cite{review}. 

In order to avoid difficulties arising from the variety of notations used in these papers, we changed some of the notations of the original articles. Variables and values with index $g$ correspond to gas particles whereas no index corresponds to the Brownian particle.

\section{Collisional decoherence}

The paper of \textbf{Joos and Zeh}~\cite{Joos} initiated the quantitative study of environmental decoherence due to scattering of photons and molecules. They discuss this phenomenon on various macroscopic systems, like the position of a macroscopic particle and the chirality of macro molecules. Due to our interest in QBM, we restrict this review on the first of these examples, and refer the reader to the very well written original article if interested in the non-observability of coherences in our macroscopic world. We should mention that the authors were mainly interested in a qualitative understanding of the decoherence process, and therefore made use of several qualitative formulas.

Their discussion about collisional localization of a macroscopic particle is restricted to recoil-free collisions, which effectively means that the Brownian particle is infinitely heavy. The scattering of a gas particle by the Brownian one is described by 
	\begin{eqnarray}
		\ket{\mathbf r }\ket{\psi_g} &\rightarrow& \ket{\mathbf r }\ket{\psi_g(\mathbf r )} \;=\; \ket{\mathbf r }\hat S_{\mathbf r }\ket{\psi_g}
	\end{eqnarray}
where $\hat S_{\mathbf r }$ is the gas particle scattering matrix for a Brownian particle at position $\mathbf r $, and $\ket{\psi_g}$ is the state of the gas particle before the scattering event. Therefore, the Brownian particle's density matrix after the scattering event is
	\begin{eqnarray}
		\rho(\mathbf r ,\mathbf r ') &=& \rho_0(\mathbf r ,\mathbf r ') \bra{\psi_g} \hat S_{\mathbf r '}^\dag \hat S_{\mathbf r }\ket{\psi_g},
	\end{eqnarray}
where $\rho_0(\mathbf r ,\mathbf r ')$ is the Brownian particle density matrix prior to a collision. If the scattering interaction is invariant under translation in space, $S_{\mathbf r }(\mathbf k_g,\mathbf k_g') = S(\mathbf k_g,\mathbf k_g')e^{-\imath(\mathbf k_g-\mathbf k_g')\mathbf r }$ can be used. They continue with
	\begin{eqnarray}
		\bra{\psi_g} \hat S_{\mathbf r '}^\dag\hat  S_{\mathbf r }\ket{\psi_g} &\!=\!& \int \dd^3 k_g\, \dd^3 k_g' \,\dd^3 k_g''\, S(\mathbf k_g,\mathbf k_g')S^*(\mathbf k_g,\mathbf k_g'') e^{\imath \mathbf k_g(\mathbf r' -\mathbf r)} e^{\imath\mathbf k_g'\mathbf r -\imath\mathbf k_g''\mathbf r '} c(\mathbf k_g')c^*(\mathbf k_g''),\! \nn\\ \label{JoosScatter}
	\end{eqnarray}
where $c(\mathbf k_g)$ is the momentum representation of $\ket{\psi_g}$. Now they approximate the gas particle state by a momentum eigenstate $c(\mathbf k_g) = L^{-3/2}\delta^3(\mathbf k_g-\mathbf p_g)$, where $\mathbf p_g$ is the momentum of the incident particle and $L^3$ is the box normalization volume. Next the scattering matrix is expressed in terms of the scattering amplitude
	\begin{eqnarray}
		S(\mathbf k_g,\mathbf k_g') &=& \delta^3(\mathbf k_g-\mathbf k_g') + \frac\imath{2\pi k_g}f(\mathbf k_g,\mathbf k_g')\delta(k_g-k_g').
	\end{eqnarray}
Inserting this relation into \Eqref{JoosScatter} leads to a problem shared by all approaches which use momentum eigenstates of a gas particle to study a collision: \textbf{The occurrence of an ill defined squared Dirac delta function.} Joos and Zeh therefore simply substitute\footnote{They do not mention this procedure, but it is clearly carried out.} $\left[\delta^3(\mathbf k_g-\mathbf k_g')\right]/L^3 \to \delta^3(\mathbf k_g-\mathbf k_g')$, as well as $\left[\delta(k_g-k_g')\right]^2/L \to \delta(k_g-k_g')$.

Expanding the exponential in \Eqref{JoosScatter} to second order and performing the integral over the angles of $\mathbf k_g$ eventually yields 
	\begin{eqnarray}
		\rho(\mathbf r ,\mathbf r ') &=& \rho_0(\mathbf r ,\mathbf r ') \left( 1-\frac{(p_g|\mathbf r -\mathbf r '|)^2}{8\pi^2L^2}  \sigma_{eff} \right) \nn\\
		&\approx& \rho_0(\mathbf r ,\mathbf r ')\exp\!\left( \frac{(p_g|\mathbf r -\mathbf r '|)^2}{8\pi^2L^2}  \sigma_{eff} \right)
	\end{eqnarray}
where
	\begin{eqnarray}
		\sigma_{eff} &=& \frac\pi2\int\dd\cos\Theta\, |f(\cos\Theta)|^2 \left[ (2-\cos\Theta)^2-1\right].
	\end{eqnarray}
The effect of $N$ subsequent collisions is given by multiplication of the exponent by $N$. The number of collisions in a time interval $\delta$ can be given in terms of the flux $n _g/v_g$
	\begin{eqnarray}
		N &=& L^2n _gv_g\delta, \label{numberd}
	\end{eqnarray}
although the appearance of a surface related to the box normalization volume instead of the total scattering cross section seems questionable. Taking the limit $\delta \to0$, they finally find
	\begin{eqnarray}
		\frac{\partial\rho(\mathbf r ,\mathbf r ')}{\partial t} &=& -\Lambda (\mathbf r -\mathbf r ')^2\rho(\mathbf r ,\mathbf r '),\label{JoosMain}
	\end{eqnarray}
with the localization rate
	\begin{eqnarray}
		\Lambda &=& \frac{p_g^2\sigma_{eff}n _gv_g}{8\pi^2}. 
	\end{eqnarray}

It is important to note that because of the expansion of the exponential in \Eqref{JoosScatter}, the result \Eqref{JoosMain} is only valid for small distances $(\mathbf r -\mathbf r ')\cdot \mathbf p_g\ll 0$. For large distances the localization rate in \Eqref{JoosMain} increases to infinity, which is clearly not physical as the localization rate should be bound by the collision rate. 

They continue to calculate $\Lambda$ for various special cases by averaging over the momentum distribution $p_g$, including not only scattering of gas particles, but also thermal photons. Coherence length after $t=1$s are given. Next Joos and Zeh include the internal system Hamiltonian $H=\hat{\mathbf p} ^2/2m $. As this part is quite long and does not qualitatively change any results, we skip its discussion an refer to the original paper~\cite{Joos}.

A problem in \Eqref{JoosMain} is that it leads to localization without bound, which due to Heisenberg's uncertainty principle corresponds to an increase of mean energy towards infinity. This problem is discussed in~\cite{Ballentine}, and occurs because any movement of an infinitely heavy Brownian particle is neglected from the beginning. The inclusion of the unitary free particle dynamics later in their paper can not avoid this ``heating" of the Brownian particle.

Joos and Zeh's results were refined by the article of \textbf{Gallis and Fleming}~\cite{Gallis}, where they discuss the problem of the transition from the quantum world to the classical world. In one part of their work, environmentally induced decoherence is discussed. In particular, Gallis and Fleming assume an environment which scatters off the system of interest, much as in QBM.

The explicit calculation is very similar to the one of Joos and Zeh, and not repeated here. The main difference is that they do not expand the exponential in \Eqref{JoosScatter}, and therefore their result 
	\begin{eqnarray}
		\frac{\partial\rho (\mathbf{r,r'})}{\partial t} &=& -F(\mathbf{r-r'})\rho (\mathbf{r,r'}) \nn\\
		F(\mathbf{r}) &=& \int\dd p_g\, \mu_T(p_g)\frac{p_g}{m_g}\int \frac{\dd \Omega\,\dd\Omega'}{2}  \left( 1- e^{\imath(\mathbf p_g-\mathbf p_g')\mathbf r} \right)  | f(\mathbf p_g,\mathbf p_g') |^2,\quad\label{GallisMain}
	\end{eqnarray}
where $\mu_T(p_g)$ is the Maxwell-Boltzmann distribution, should be valid for general spatial separations of superposition states. Indeed, for large $\mathbf r$, the exponential oscillates quickly and does not contribute to the decoherence function $F(\mathbf r)$, guaranteeing that the latter is bound for large separations.  For small separations $\mathbf r\to0$, they obtain the familiar result \Eqref{JoosMain}.

\textbf{Hornberger and Sipe}~\cite{Hornberger} finally managed to avoid the ill defined square of the Dirac delta function, by applying scattering theory to normalized states. In doing so, they corrected Gallis and Fleming's result \Eqref{GallisMain} by a numerical factor. 

Performing a careful scattering calculation, and assuming the limit of an heavy gas particle, they find that the scattering operator $\hat S$ transforms the two particle state $\ket{\mathbf r }\ket{\psi_g}$ according to
	\begin{eqnarray}
		\hat S\left(\ket{\mathbf r }\ket{\psi_g}\right) &=&  \ket{\mathbf r } \left( e^{-\imath\hat{\mathbf p}_g \mathbf r } \hat S_g e^{\imath\hat{\mathbf p}_g \mathbf r }\right) \ket{\psi_g} \nn\\
		&=&  \ket{\mathbf r } \ket{\psi_g^{\mathbf r }} .
	\end{eqnarray}
Here, $\ket{\psi_g}$ is a general, but normalized gas particle state, $\hat S_g$ is the one particle scattering operator acting on the gas particle, and $\hat{\mathbf p}_g$ is its momentum operator. Therefore, if the two particle state before the scattering event is $\varrho=\rho_0 \otimes \ket{\psi_g}\!\bra{\psi_g}$ with
	\begin{eqnarray}
		\rho  &=& \int \dd^3 r\,  \dd^3 r ' \, \rho (\mathbf r ,\mathbf r ') \ket{\mathbf r {\phantom '}\!}\!\bra{\mathbf r '},
	\end{eqnarray}
then the scattering of a gas particle state $\ket{\psi_g}$ results in
	\begin{eqnarray}
		\rho (\mathbf r ,\mathbf r ') &=& \scalar{\psi_g^{\mathbf r '}}{\psi_g^{\mathbf r }} \rho_0 (\mathbf r ,\mathbf r ').
	\end{eqnarray}

Now they proceed by introducing the transition operator $\hat S_g=\one+\imath \hat T_g$. The scalar product is then
	\begin{eqnarray}
		\scalar{\psi_g^{\mathbf r '}}{\psi_g^{\mathbf r ^{\phantom 1}}} &=& 1+ \langle\psi_g | \hat {\mathcal A} |\psi_g \rangle,
	\end{eqnarray}
where the operator $\hat {\mathcal A}$ can be expressed in terms of the transition operator $\hat T_g$. To calculate the change of the Brownian particles density matrix due to one collision, they insert two complete sets of momentum eigenstates to find
	\begin{eqnarray}
		\langle\psi_g | \hat {\mathcal A} |\psi_g \rangle &=& \int\dd^3 p_g \, \dd^3 p_g'\, \langle\psi_g | \mathbf p_g' \rangle\langle \mathbf p_g'| \hat {\mathcal A} | \mathbf p_g \rangle\langle \mathbf p_g | \psi_g \rangle.
	\end{eqnarray}
In this equation, the advantage of using momentum eigenstates instead of $\ket{\psi_g}$ is apparent, as only diagonals of $\hat {\mathcal A}$ would be needed. But $\hat {\mathcal A}$ involves terms like $\hat T_g\hat T_g^\dagger$, and if a complete set of momentum eigenstates is inserted between these transition operators (because the transition operator involves an energy conserving Dirac delta function), one would again encounter the square of a Dirac delta function.

To avoid this problem, they chose Gaussian minimum uncertainty wave packets $\ket{\mathbf r_g,\mathbf p_g}$ for the gas particle states. The consequence is a very long calculation, which we drop here, but rather present the result
	\begin{eqnarray}
		\langle\psi_g | \hat {\mathcal A} |\psi_g \rangle &=& \int \dd^3 q \frac{e^{-|\mathbf q-\mathbf p_g|^2/b^2}}{\left(\pi b^2\right)^{3/2}} A^{\mathbf r_g}(\mathbf q),\label{hoho}
	\end{eqnarray}
where $b$ is the momentum variance of the state $\ket{\mathbf r_g,\mathbf p_g}$. Furthermore they defined
	\begin{eqnarray}
		A^{\mathbf r_g}(\mathbf q) &=& \Gamma_{\mathbf q}\!\left(\mathbf r_g-\overline{\mathbf R}\right) \int \dd{\mathbf n}\, e^{\imath(\mathbf q-q{\mathbf n})\cdot(\mathbf r -\mathbf r ')/\hbar} |f(q{\mathbf n},\mathbf q)|^2 \nn\\
		&& -\half \left[ \Gamma_{\mathbf q}(\mathbf r_g-\mathbf r ) + \Gamma_{\mathbf q}(\mathbf r_g-\mathbf r ') \right]  \int \dd{\mathbf n}\,  |f(q{\mathbf n},\mathbf q)|^2 \nn\\
		&& +\frac{2\pi\imath\hbar}{q} \left[ \Gamma_{\mathbf q}(\mathbf r_g-\mathbf r ) + \Gamma_{\mathbf q}(\mathbf r_g-\mathbf r ') \right] \mbox{Re} [f(\mathbf q,\mathbf q)],
	\end{eqnarray}
where ${\mathbf n}$ is a unit vector, $\overline{\mathbf R}=(\mathbf r +\mathbf r ')/2$, 
	\begin{eqnarray}
		\Gamma_{\mathbf q}(\mathbf R) &=& \frac{\exp\!\left\{-\left[R^2-(\mathbf R \cdot{\mathbf q}/q)^2\right]/a^2\right\}}{\pi a^2},
	\end{eqnarray}
and $a$ is the position variance of the state $\ket{\mathbf r_g,\mathbf p_g}$.

Now they turn to a collision with a thermal Boltzmann gas. Most important for the application of the previous part of their paper is the convex decomposition of a thermal gas particle state into minimum uncertainty wave packets $\ket{\psi_g}=\ket{\mathbf r_g,\mathbf p_g}$, with mean position $\mathbf r_g$ and mean momentum $\mathbf r_g$. This is reviewed in subsection \ref{convex decomp}, and the result is
	\begin{equation}
		\rho_g = \int \frac{\dd^3 r_g}{V} \int \dd^3 p_g\, \mu_{\widetilde{T}}(\mathbf p_g) \ket{\mathbf x_g,\mathbf p_g}\!\bra{\mathbf x_g,\mathbf p_g}, \label{thermalgas2}
	\end{equation}
where $V$ is some normalization volume. The momentum distribution $\mu_{\widetilde{T}}(\mathbf p_g)$ is similar to the Maxwell-Boltzmann distribution $\mu_{T}(\mathbf p_g) = (2\pi m_g k_B{T})^{-3/2} e^{-p_g^2/(2m_gk_B{T})}$, but with the temperature $T$ substituted by a slightly smaller value, as some of the particles energy is related to the momentum uncertainty of the states $\ket{\mathbf x_g,\mathbf p_g}$. In the following these states are chosen such that this momentum uncertainty is small, and the approximation $\mu_{\widetilde{T}}(p_g)\approx\mu_{T}(p_g)$ can be used.

\begin{figure}[htbp]
\begin{center}
	\includegraphics[width=0.6\linewidth]{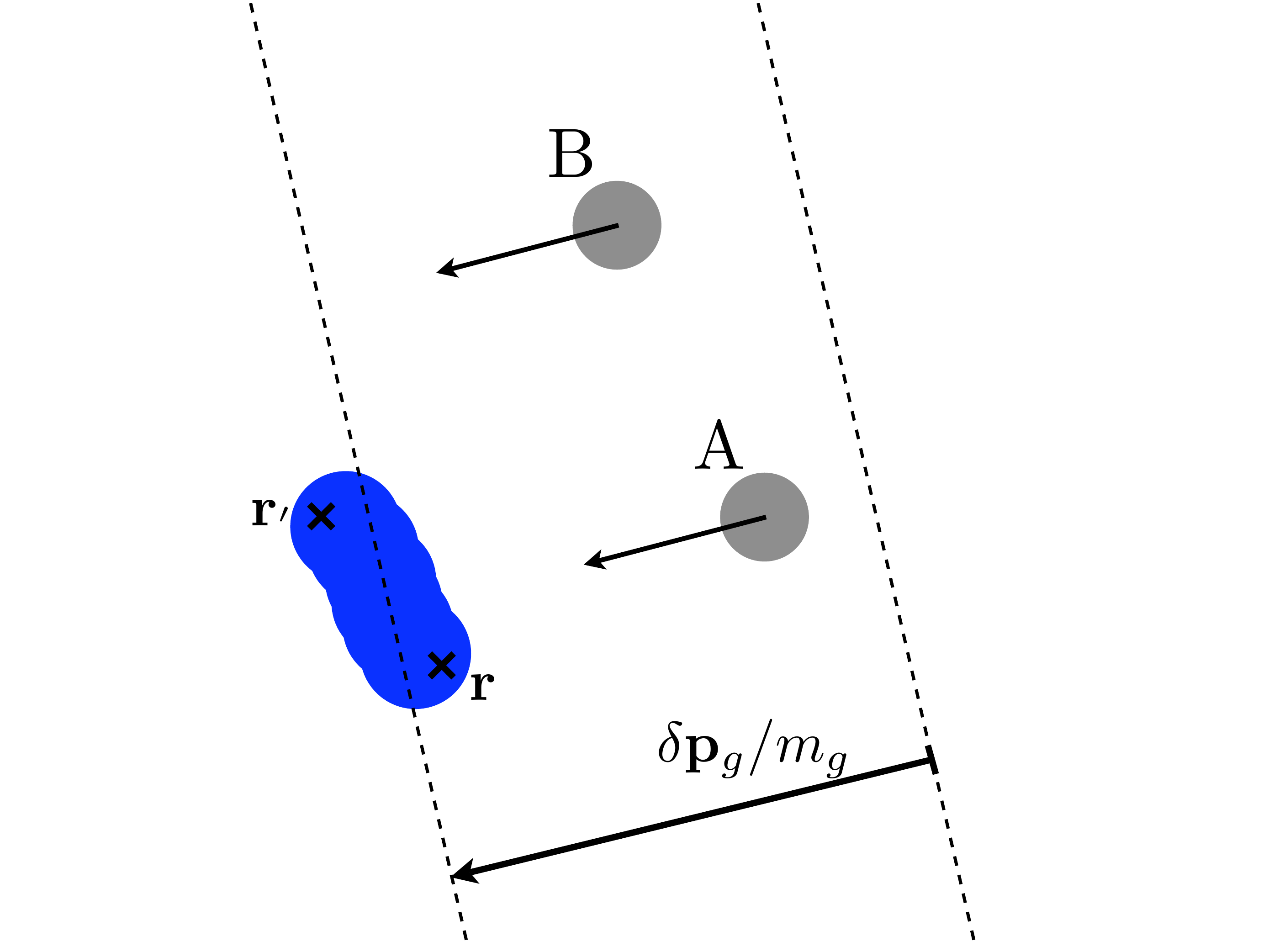}
\caption{\small The volume between the dashed lines is $R(\mathbf p_g)$. During a time interval $(0,\delta)$, a collision with the Brownian particle (blue) can occur, if a gas particle (grey) with momentum $\mathbf p_g$ is in $R(\mathbf p_g)$.\label{figVolume}}
\end{center}
\end{figure}
Now a collision can be pictured in an almost classical way. Assuming that the Brownian particle is located within a small area, as are the gas particle states $\ket{\psi_g}=\ket{\mathbf r_g,\mathbf p_g}$, then we can think of a volume $R(\mathbf p_g)$ depicted in Fig.~\ref{figVolume}. A gas particle will collide with the Brownian one during a time interval $\delta$, exactly if it is in this volume. 

Although particles like the particle B in Fig.~\ref{figVolume} will miss the Brownian particle, here it is considered to be a scattering particle. That is acceptable, as the authors show that the scattering of a gas particle which ``misses" the Brownian particle does not contribute much to the decoherence of the Brownian particle's state.

Therefore, the change of the density matrix during a time intervall $\delta$ is
	\begin{eqnarray}
		\Delta \rho (\mathbf r ,\mathbf r ') &=& N\rho (\mathbf r ,\mathbf r ') \int\dd^3 p_g\, \mu_{T}(\mathbf p_g) \int_{R(\mathbf p_g)} \frac{\dd^3 r_g}{V} \bra{\mathbf r_g,\mathbf p_g} \hat{\mathcal A} \ket{\mathbf r_g,\mathbf p_g}, \quad
	\end{eqnarray}
where $N$ is the number of gas particles in the volume $V$. After substitution of \Eqref{hoho} and some calculation, they arrive at Gallis and Fleming's result \Eqref{GallisMain}, with a small correction in that the right hand side is multiplied by $2\pi$.

Finally, they repeat the calculation of Gallis and Fleming~\cite{Gallis} with a modified and physically motivated replacement rule for the squared Dirac delta function
	\begin{eqnarray}
		\delta(p_g-p'_g) &\to& \frac{V}{2\pi\hbar\sigma(\mathbf{p}_g)},
	\end{eqnarray}
where $\sigma(\mathbf{p}_g)$ is the total scattering cross section for an incoming gas particle momenta $\mathbf{p}_g$. The recovery of their earlier results shows that the much simpler approach, using delocalized momentum eigen states can also be successful. It further gives some confidence in the application of the replacement rule, at least for an infinitely heavy Brownian particle.

The work of Hornberber and Sipe is, up to now, the only one found in the literature to avoid squares of Dirac delta functions. Their results should therefore correctly describe collisional decoherence. However, their calculations are only valid for an infinitely heavy Brownian particle. As such an particle does not move in position space, issues around friction and position diffusion can not be resolved.

A different replacement rule is motivated by \textbf{Adler}~\cite{Adler}. He uses a particular representation of the Dirac delta function
	\begin{eqnarray}
		\delta(E) &=& \frac 1{2\pi\hbar}\int \dd t\, e^{\imath Et/\hbar} \label{deltaE},
	\end{eqnarray}
from which he follows
	\begin{eqnarray}
		\delta(0) &=& \frac{\delta}{2\pi\hbar},
	\end{eqnarray}
where $\Delta$ is an elapsed time interval large compared to the collision time. He continues along the lines of Hornberger and Sipe's replacement rule calculation, and obtains their results. 

It is however interesting to note, that, by using a different, but equally valid representation $\delta(E) = \frac A{2\pi\hbar}\int \dd t\, \exp(\imath AEt/\hbar)$ of the delta function, Adler's argumentation leads to a different replacement rule $\delta(0) = \frac {A \delta}{2\pi\hbar}$, where $A$ is an arbitrary real number.

\textbf{Dodd and Halliwell}~\cite{Dodd} used non-relativistic many body quantum field theory, to essentially derive the result found by Hornberger and Sipe. They also need to rely on the replacement rule. At the end of their article, extension to a finite Brownian particle mass are considered, but these are even in need of a replacement rule for the derivation of a Dirac delta function.

\section{Caldeira-Leggett model}

The Caldeira-Leggett model of QBM, which is readily available in textbooks ~\cite{Breuer}, consists of a Brownian particle in a potential $V(x)$, coupled to a bath of harmonic oscillators via a linear interaction. The Hamiltonian of this system is
	\begin{eqnarray}
		H &=& H_S + H_E + H_I + H_c \\
		H_S &=& \frac{\hat p^2}{2m}+V(\hat x), \\
		H_E &=& \sum_i \frac{\hat p_i^2}{2m}+\frac{2m_i\omega_i^2\hat x_i^2}{2} ,\\
		H_I &=& -\hat x \sum_i \kappa_i\hat x_i ,\\
		H_c &=&  \hat x^2 \sum_i \frac{\kappa_i^2}{2m_i\omega_i},
	\end{eqnarray}
where $\omega_i$ is the frequencies of the $i$-th harmonic oscillator, $m_i$ is its mass, and $\kappa_i$ is its coupling constant to the Brownian particle. The interaction term leads to a renormalization of the Brownian particle potential, and the counter-term $H_c$ is included to counteract this potential. 

In the derivation of the master equation, the spectral density
	\begin{eqnarray}
		J(\omega) &=& \sum_i \frac{\kappa_i^2}{2m_i \omega_i} \delta(\omega-\omega_i) \label{spectraldensity}
	\end{eqnarray}
is of importance. The spectral density can be a smooth function, if the limit of a continuous distribution of oscillator frequencies is used. In particular, an Ohmic spectral density
	\begin{eqnarray}
		J(\omega) &=& \frac{2m\gamma}{\pi}\omega
	\end{eqnarray}
is often used, where $\gamma$ turns out to be the damping rate. In the limit of high temperature, it is possible to derive a simple master equation
	\begin{eqnarray}
		\frac{\dd \rho_{}(t)}{\dd t} &=& -\frac\imath\hbar \left[ H_{},\rho_{}(t)\right] - \imath \frac{\gamma}{\hbar}[\hat x,\{\hat p,\rho\}] - \frac{2m\gamma k_BT}{\hbar^2}[\hat x,[\hat x,\rho]], \label{grdk}
	\end{eqnarray}
which was first achieved in the seminal paper of \textbf{Caldeira and Leggett}~\cite{Caldeira}. Although they used the complicated Feynman-Vernon theory of the influence functional, there are now simpler derivations of these results available~\cite{Breuer}.

Because this master equation is not of Lindblad form, it can not be completely positive. However, it can be cast into Lindblad form by adding a position diffusion term
	\begin{eqnarray}
		-\frac{\gamma}{8mk_BT}[\hat p,[\hat p,\rho]],
	\end{eqnarray}
which is small for high temperatures. It is interesting to see, how approximations used to derive Markovian dynamics can lead to unphysical results, as the lacking positivity of \Eqref{grdk} shows. To establish positivity, a position diffusion term has to be added, even if there is no physical reason to do so. However, as long as any unphysical property is small under the conditions of validity of the master equation, the master equation can perfectly be used to calculate physical quantities, which can be tested against experiments.

The exact master equation for the Caldeira-Leggett model has also been derived~\cite{Hu}, and is known as the Hu-Paz-Zhang master equation. As an exact equation valid at any temperatures, it is non-Markovian and does not suffer any problems such as the violation of positivity.

\section{Quantum linear Boltzmann equation}

The contribution of \textbf{Di\'osi} \cite{Diosi} was the first one to address the interplay of friction, diffusion and decoherence in collisional QBM. He also started by considering a single collision in the interaction picture, and uses an initially uncorrelated state $\varrho=\rho \otimes\rho_g$, where $\rho_g$ is a thermal state with a Maxwell-Boltzmann momentum distribution $\mu_T(p_g)$ from \Eqref{mbdis}. He introduces the transition operator by $\hat S=\one+\imath \hat T$, where $\hat S$ is the unitary scattering operator. He finds for the change of the two particle density matrix due to one collision
	\begin{eqnarray}
		\Delta \varrho &=& \frac\imath 2 \left[ \hat T+\hat T^\dagger,\varrho \right] +\hat T\varrho \hat T^\dagger - \half \left\{\hat T^\dagger \hat T,\varrho \right\}. \label{gda}
	\end{eqnarray}
For the evolution of the Brownian particle, he approximates $\dd\rho /\dd t$ by
	\begin{eqnarray}
		\frac{\Delta \rho }{\Delta t} &=& \frac 1{\Delta t}\mbox{Tr}_g \Delta\varrho,
	\end{eqnarray}
where $\Delta t$ is a time long compared to the scattering time.

Now he substitutes the standard form of the transition operator
	\begin{eqnarray}
		\hat T &=& \frac 1{2\pi m^*}\int \dd^3 p \, \dd^3 p_g\, \dd^3 p_g'\, f\!\left(\mathbf p_g^*,{\mathbf p_g'\!}^*\right) \delta\!\left(E_{\mathbf p_g^*}-E_{{\mathbf p_g'\!}^*}\right) \ket{\mathbf p -\mathbf p_g'+\mathbf p_g,\mathbf p_g'}\!\bra{\mathbf p ,\mathbf p_g}, \nn\\
	\end{eqnarray}
where quantities marked by stars are centre of mass quantities (see \cite{Diosi} for precise definition). After performing the partial trace and changing the integration variables, he finds for the second term in \Eqref{gda}
	\begin{eqnarray}
		\mbox{Tr}_g\left(\hat T\varrho \hat T^\dagger\right) &=& \frac{2\pi n_g}{m^{*2}}\left(\frac{m +m_g}{m }\right)^{\!3}  \int  \dd^3 p \, \dd^3 p '\, \dd^3 p_g^*\, \dd^3 {p_g'}^*\, \mu\!\left( \mathbf p_g^*+\frac{m_g}{m }(\hat{\mathbf p} +{\mathbf p_g'\!}^*)   \right) \nn\\ 
		& \times&\left|f\!\left(\mathbf p_g^*,{\mathbf p_g'\!}^*\right)\right|^2 \left[\delta\!\left(E_{\mathbf p_g^*}-E_{{\mathbf p_g'\!}^*}\right)\right]^2  \ket{\mathbf p -\mathbf p_g'+\mathbf p_g}\!\bra{\mathbf p }\!\rho \!\ket{\mathbf p '}\!\bra{\mathbf p '-\mathbf p_g'+\mathbf p_g} \nn\\
	\end{eqnarray}
where $n_g$ is the density of the gas particles. In the derivation of this formula, he assumed that the density operator of the Brownian particle is \emph{almost} diagonal in momentum representation. 

In Di\'osi's approach, there also appears the square of the Dirac delta function. He removes this ill-defined term by applying $\delta(E=0)=\Delta t/2\pi$. By rewriting the previous equation in operator form he finds
	\begin{eqnarray}
		\mbox{Tr}_g\!\left(\hat T\varrho \hat T^\dagger\right) &\!=\!& \frac{n_g}{m^{*2}}\!\left(\frac{m +m_g}{m }\right)^{\!3}  \!\int\! \dd^3 p_g^*\, \dd^3{p_g'\!}^*\, \delta\!\left(E_{\mathbf p_g^*}-E_{{\mathbf p_g'\!}^*}\right) \! \left|f\!\left(\mathbf p_g^*,{\mathbf p_g'\!}^*\right)\!\right|^2 \hat V_{\mathbf p_g^*{\mathbf p_g'\!}^*}\rho \hat V^\dag_{\mathbf p_g^*{\mathbf p_g'\!}^*} \nn\\
		 \hat V_{\mathbf p_g^*{\mathbf p_g'\!}^*} &\!=\!& \sqrt{\mu\left( \mathbf p_g^*+\frac{m_g}{m }(\hat{\mathbf p} +{\mathbf p_g'\!}^*)   \right)} e^{-\imath ({\mathbf p_g'\!}^*-\mathbf p_g^*)\hat{\mathbf x} }.
	\end{eqnarray}

The third term on the right hand side of \Eqref{gda} can be calculated in a similar way, whereas the first term can be neglected in rarified gases. Performing an integration over the Dirac delta function, he finds the master equation for the Brownian particle to be
	\begin{eqnarray}
		\frac{\dd\rho }{\dd t} &=& n_g \left(\frac{m +m_g}{m }\right)^{\!3} \int\dd E^*\, \dd\Omega^*\,\dd{\Omega'}^* \, k^{*2} \frac{\dd\sigma(\theta^*,E^*)}{\dd{\Omega'}^*} \nn\\
		&\times& \left( \hat V_{\mathbf p_g^*{\mathbf p_g'\!}^*}\rho \hat V^\dagger_{\mathbf p_g^*{\mathbf p_g'\!}^*} - \half\left\{ \hat V^\dagger_{\mathbf p_g^*{\mathbf p_g'\!}^*} \hat V_{\mathbf p_g^*{\mathbf p_g'\!}^*},\rho  \right\} \right) ,\label{diodio}
	\end{eqnarray}
where $\dd\sigma/\dd\Omega=|f|^2$ is the differential cross-section. This equation is Di\'osi's central result. From the condition $\dd\rho/\dd t=0$, he obtains the correct thermal equilibrium state 
	\begin{eqnarray}
		\rho (t\to\infty) &\propto& \exp\!\left(-\frac{\hat p ^2}{2k_BTm }\right)
	\end{eqnarray}
at temperature $T$, and therefore, Di\'osi was the first to avoid the unbound heating of infinite mass approaches.

Di\'osi also derives the familiar interaction picture master equation for QBM,
	\begin{eqnarray}
		\frac{\dd \rho(t)}{\dd t} &=&  - \imath \frac{\gamma}{2\hbar}[\hat x ,\{\hat p ,\rho \}] - \frac{D_{pp}}{\hbar^2}[\hat x ,[\hat x ,\rho ]] - \frac{D_{xx}}{\hbar^2}[\hat p ,[\hat p ,\rho ]], \qquad\label{dididi}
	\end{eqnarray}
by using the limit of a heavy Brownian particle, with a small coherent extension.

Both, \Eqref{diodio} and \Eqref{dididi} show position diffusion which led Di\'osi to the conclusion that quantum friction always has to be accompanied by finite, random position jumps. We will show in our approach that this is in fact not true.

Despite rather nice results, there are some shortcomings in Di\'osi's calculations. First, he assumes that the probability of a collision of a gas particle with momentum $p_g$ with the Brownian particle is independent of its momentum. This can certainly not be quantitatively correct, as e.g.\ two particles with the same velocity can never collide. Gallis and Fleming accounted for this difficulty by introducing a particle flux, which is proportional to the particle's velocity. For a Brownian particle with finite mass this issue becomes less trivial.

Second, the approximation of the Brownian particle density operator being almost diagonal in the momentum basis is not trivial, as in the calculation, it is rather assumed that the density operator is in fact diagonal. Although this assumptions seemed appropriate at the time (it was also used by Vacchini~\cite{Vacchini}), we~\cite{our} as well as Di\'osi~\cite{Diosi2} himself, showed later that a collision with a gas particle in a momentum eigenstate removes all coherences parallel to the momentum transfer (see also chapter~\ref{introII}). To account for this effect, one has to consider off-diagonals of the Brownian particle density operator in momentum basis.

\textbf{Vacchini}~\cite{Vacchini} uses the formalism of non-relativistic multi particle quantum field theory. Similar to Di\'osi, he also assumes that the Brownian particle's density matrix is almost diagonal in momentum basis to derive a master equation. By taking the limit of light gas particles, he obtains the familiar form \Eqref{dididi}. However, his coefficients $\gamma$, $D_{pp}$, and $D_{xx}$ are quite different to the ones found by Di\'osi~\cite{Diosi}.

In the derivation of the master equation, there appears no square of an energy conserving Dirac delta function, because  $E_{initial}-E_{final}=0$ is avoided by adding $\imath \varepsilon$ to the energy change. There are no reasons given for this procedure, and its validity is difficult to judge.

The letter of \textbf{Hornberger}~\cite{Hornberger2} seems to be the most complete approach to collisional QBM to date, in that he is the first to explicitly consider off-diagonals in the momentum representation of the density operator. To do so, he introduces an extension to the replacement rule, as to be applicable to off-diagonals. 

As Di\'osi, he starts from \Eqref{gda}. The incoherent part can be expressed in terms of the kernel
	\begin{eqnarray}
		\bra{\mathbf p \phantom {\!\!'}}\mbox{Tr}_g\! \left( \hat T\left[\ket{\mathbf p ''}\bra{\mathbf p '''}\otimes\rho_g\right] \hat T^\dag \right)\ket{\mathbf p '}. \label{uyt}
	\end{eqnarray}
Using the standard representation of the transition matrix, he finds, that the momentum transfer $\textbf q$ has to be the same for both cohering momenta, i.e.\ $\mathbf p-\mathbf p''=\mathbf p'-\mathbf p'''$. Inserting a momentum diagonal representation of the gas particle density operator, Hornberger arrives, for $\mathbf p =\mathbf p '$, at the familiar squared Dirac delta function. He uses the replacement rule
	\begin{eqnarray}
		\frac{(2\pi\hbar)^3}{V}|\langle\mathbf p_f|\hat T_0|\mathbf p_i\rangle |^2 &\to& \delta\!\left(\frac{\mathbf p_f \phantom{\!\!l}^{\!2}-\mathbf p_i^2}2\right) \frac{|f(\mathbf p_f,\mathbf p_i)|^2}{\sigma(\mathbf p_i)|\mathbf p_i|},\label{ah}
	\end{eqnarray}
where $V$ is a box normalization volume, $\hat T_0$ is the single particle transition operator, $f(\mathbf p_f,\mathbf p_i)$ is the scattering amplitude and $\sigma(\mathbf p_g)$ is the total scattering cross section. Next, Hornberger turns to the the off-diagonals $\mathbf p \neq\mathbf p '$, which were avoided by previous studies. He finds terms like
	\begin{eqnarray}
		\langle\mathbf p_f+\mathbf p_s|\hat T_0|\mathbf p_i+\mathbf p_s\rangle\langle\mathbf p_i-\mathbf p_s|\hat T_0^\dag|\mathbf p_f-\mathbf p_s\rangle, \label{poi}
	\end{eqnarray}
where $\mathbf p_s=m_g(\mathbf p'-\mathbf p)/[2(m+m_g)]$ corresponds to the momentum separation of the cohering momenta. He then substitutes twice the square root of the replacement rule~(\ref{ah}), which yields the square root of a product of two energy conserving $\delta$-functions, with arguments $\frac{\mathbf p_f^2-\mathbf p_i^2}2\pm(\mathbf p_f-\mathbf p_i)\cdot\mathbf p_s$. He argues therefore, that $\mathbf p_s$ should be replaced by its projections $\mathbf p_{s,\bot \mathbf q}$ onto the plane vertical to the momentum transfer $\mathbf q=\mathbf p_f-\mathbf p_i$ in all expressions\footnote{To give a one-dimensional analog, instead of the rule $\int \dd x\,f(x)\delta(x-y)=f(y)$, one uses $\int \dd x\,f(x)\delta(x-y)=\int \dd x\, f(y)=f(y)\int \dd x$, and cancels the now infinite integral, with an infinite normalization length.}, and is left with a single, proper Dirac delta function $\delta\Big( \frac{\mathbf p_f^2-\mathbf p_i^2}2 \Big)$.

To account for the collisional statistics, he introduces a two particle rate operator 
	\begin{eqnarray}
		 \hat R &=& \int \dd\mathbf p  \,\dd\mathbf p_g\, n _gv(\mathbf p ,\mathbf p_g)\sigma(\mbox{rel}(\mathbf p , \mathbf p_g)) \ket{\mathbf p }\bra{\mathbf p }\otimes \ket{\mathbf p_g}\bra{\mathbf p_g},
	\end{eqnarray}
to determine the collision rate with the gas. Here, $n _g$ is the particle density of the gas, $v(\mathbf p ,\mathbf p_g)$ is the relative velocity of gas particle and Brownian particle, and $\mbox{rel}(\mathbf p , \mathbf p_g)=\frac{m }{m +m_g}\mathbf p_g - \frac{m_g}{m +m_g}\mathbf p $ is their relative momentum. He then returns to \Eqref{gda} and (\ref{uyt}), and substitutes $\hat T$ with $\hat T\sqrt  {\hat R}$. Eventually, he finds the master equation
	\begin{eqnarray}
		\frac{\dd\rho }{\dd t} &=& \frac1{\imath\hbar}\left[\frac{\hat {\mathbf p}^2}{2m },\rho \right]  +\int\!\dd\mathbf q\int_{\mathbf q^\bot}\!\frac{\dd\mathbf k}{q}\left( \hat L_{\mathbf q,\mathbf k}\rho   \hat L_{\mathbf q,\mathbf k}^\dag -\half\rho  \hat L_{\mathbf q,\mathbf k}^\dag  \hat L_{\mathbf q,\mathbf k}-\half  \hat L_{\mathbf q,\mathbf k}^\dag  \hat L_{\mathbf q,\mathbf k}\rho  \right)\!, \qquad\qquad
	\end{eqnarray}
where $\hat L_{\mathbf q,\mathbf k} = e^{\imath\hat{\mathbf r} \cdot\mathbf q/\hbar} F(\mathbf k,\hat{\mathbf p} ;\mathbf q)$ and
	\begin{eqnarray}
		 F(\mathbf k,{\mathbf p};\mathbf q) &=& \frac{\sqrt{n _g}(m+m_g)}{\sqrt{m_g}m}f\!\left[  \mbox{rel}\!\left(\mathbf k_{\bot\mathbf q},\mathbf p_{\bot\mathbf q}\right)-\frac{\mathbf q}2,\mbox{rel}\!\left(\mathbf k_{\bot\mathbf q},\mathbf p_{\bot\mathbf q}\right)+\frac{\mathbf q}2  \right]  \nn\\
		 && \times \sqrt{ \mu_{T}\!\left( \mathbf k_{\bot\mathbf q}+\frac{m_g+m }{m }\frac{\mathbf q}2+\frac{m_g}{m }\mathbf p_{\|\mathbf q} \right) }.\nn
	\end{eqnarray}
The second integration is over the plane perpendicular to the momentum transfer $\mathbf q$, and a vector with index $\|\mathbf q$ denotes the part of the vector parallel to $\mathbf q$. We wish to note, although it is not pointed out in the article, that this QLBE also displays QPD.

We showed in the introduction (chapter~\ref{introII}), that the use of the modified replacement rule for momentum coherences is problematic, because the term~(\ref{poi}) is perfectly well defined without any replacement rule. However, if Hornberger had not used the replacement rule, he would have found that a collision with a momentum eigenstate destroys all momentum coherences, which is also explained in the introduction. These observations indicate, that one might have to use normalized gas particle states, to describe decoherence aspects of QBM.

There is a further questionable element in the use of the modified replacement rule. To outline this objection, we write for the density matrix before a collision
	\begin{eqnarray}
		\rho(\mathbf p'',\mathbf p''')&=&\rho\!\left(\mathbf p'',\mathbf p''+2(m+m_g)\mathbf p_s/m_g\right).
	\end{eqnarray}
A strict application of Hornberger's argumentation, that the momentum separation $\mathbf p_s$ should be replaced by $\mathbf p_{s\bot \mathbf q}$ in arguments of all functions, would require it to be used also in the argument of the density matrix. However, replacing the off-diagonals of the density operator by entries which are on the diagonal in the direction $\mathbf q$, clearly does not make sense, and was not done in the article. It seems unsatisfactory, that one has to choose when to use a projected vector, such as not to obtain unphysical results. 

On the other hand, applying the delta functions \Eqref{correct} without a replacement rule, one finds that a collision with momentum transfer $\mathbf q$ leads to
	\begin{eqnarray}
		\rho(\mathbf p,\mathbf p') &\propto& \delta(\mathbf p_{s}\cdot\mathbf q)   \rho\!\left(\mathbf p''-\mathbf q,\mathbf p''-\mathbf q+2(m+m_g)\mathbf p_s/m_g\right),
	\end{eqnarray}
which should be valid for $\mathbf p_s\neq 0$. That is, a mathematically strict treatment using the delta functions shows, that only coherences with momentum separation orthogonal to the momentum transfer survive a collision, which is exactly what  we derived in the introduction by a different method.

Despite this criticism, we would like to emphasize, that Hornberger was the first to address the very complicated problem posed by the off-diagonals of the density operator. As such, his valuable article is the starting point for a number of further investigations~\cite{Breuer2}, and heavily influenced the work presented in this thesis. A somewhat more detailed derivation of his master equation, including an alternative, although not much more stringent motivation of his replacement rule, is found in~\cite{Hornberger4}.

A somewhat alternative approach to QBM was offered by \textbf{Barnett and Cresser}~\cite{mainII}. Motivated by the difficulty of collisional approaches, they formulated a measurement approach to QBM, recognizing that a colliding gas particle performs a measurement on the Brownian particle. They could then use the machinery of generalized measurements to derive their Lindblad operators, and did not encounter the dubious squared Dirac delta function. Most reasonable seemed simultaneous position-momentum measurements, and as a gas particle can not be considered as a perfect measurement apparatus, imperfect measurements of the form described in subsection~\ref{dwog} where postulated.

These measurements are specified by two parameters. First, $W$ determines whether the measurement is more precise in position or momentum, and second, $\bar n$ specifies the imperfectness of the measurements. In fact, $\bar n=0$ corresponds to a quantum-limited phase space measurement.

The Kraus operators for the collisional transformation of the Brownian particle density operator are then given by
	\begin{eqnarray}
		\hat A_{p_g}(x ,p ) &=&\sqrt{\frac{\mu_{T}(p_g)}{2\pi\hbar}} \hat{D}\!\left(x ,\frac{m -m_g}{m +m_g}p +\frac{2m }{m +m_g}p_g\right)\hat \sigma^{1/2}\hat{D}^\dag(x ,p ),\qquad\; \label{lss}
	\end{eqnarray}
and include the back action of the measurement, which is the classical momentum transfer due to a collision. Furthermore, they included the statistics of the collision due to the thermal momentum distribution $\mu_{T}(p_g)$ of the gas particle.

Assuming that the measurements (or collisions) are instantaneous, the master equation 
	\begin{eqnarray}
		\frac{\dd \rho }{\dd t} &=& -\frac\imath\hbar\left[\frac{\hat p ^2}{2m },\rho \right] - R\rho +R\int\!\!\!\!\int\!\!\!\!\int \dd p_g\,\dd x \,\dd p \, \hat A_{p_g}(x ,p )\rho \hat A_{p_g}^\dag(x ,p )\qquad\quad \label{ghh}
	\end{eqnarray}
is derived, where $R$ denotes the total collision rate. They applied the limits $R\to\infty$, $m_g\to 0$, and $\bar{n}\to\infty$, such that $\gamma={2m_g}R /({m +m_g})$ and $R/\bar n=$ are kept finite, to find a master equation of the well known form \Eqref{dididi}. Their diffusion coefficients are
	\begin{eqnarray}
		D_{pp} &=& \gamma \left(m k_BT+\frac{m_g}{m }\Delta_\sigma p^2\right) + \frac{R\hbar^2}{8\Delta_\sigma x^2}, \\
		D_{xx} &=& \frac{R\hbar^2}{8\Delta_\sigma p^2} ,
	\end{eqnarray}
where the variances $\Delta_\sigma x^2$ and $\Delta_\sigma p^2$ correspond to the uncertainties associated with the simultaneous measurement of position and momentum, over and above the particle's intrinsic variances. 

It is interesting to note, that this approach gives some physical insight to the mysterious position diffusion, as it can be traced back to the state reduction postulate of measurements in quantum mechanics. The better the momentum resolution of the measurements, the larger the position diffusion.

There are some inconsistencies in their derivation. As in Di\'osi's article, the collision statistics do not account for the relative velocity of the two colliding particles. Better collision statistics would be obtained by replacing $\mu_{T}(p_g)$ with $|p_g/m_g-\hat p /m |\mu_{T}(p_g)$. Next, the collision rate $R$ is not given in terms of physical properties of the gas. In fact, by using correct collision statistics, we will show in chapter~\ref{rate}, that the total collision rate is $R=n _g\sqrt{2k_BT}/\sqrt{\pi m_g}$. Furthermore, there are several more parameters, which can not be obtained from first principles.

More important is that the Brownian particle, if described by the master equation \Eqref{ghh} (or the limiting one for $m_g\to 0$), does not approach the correct thermal state. The authors claim that the difference compared to the expected thermal state is due to quantum corrections. However, by considering the special case of a Brownian particle, which is of the same type as the gas particle, i.e.\ $m_g=m $, it is apparent that this can not be true, as then the Brownian momentum distribution must approach the very same momentum distribution of the surrounding gas.

Nevertheless, many of their ideas, especially in regard to the measurement interpretation of a collision, are of importance throughout our work.


\textbf{Breuer and Vacchini}~\cite{Breuer2} performed a quantum trajectory analysis of Vacchini's~\cite{Vacchini} QLBE. The quantum trajectory method is a popular technique of simulating general master equations, and is well described in the book of Breuer and Petruccione~\cite{Breuer}.

They define the scaled Brownian particle momentum 
	\begin{eqnarray}
		\mathbf u = \mathbf p/(mv_{\mbox{\scriptsize mp}}), \label{breue}
	\end{eqnarray}
where $v_{\mbox{\scriptsize mp}}=\sqrt{2k_BT/m_g}$ is the most probable velocity of a thermal gas particle, and $m$ is the mass of the Brownian particle. Therefore, $\mathbf u=1$ means that the Brownian particle has the same velocity as an average gas particle. Although the results about  dissipation seem very nice, the physical parameters chosen in the decoherence chapter are not of much relevance for QBM. For the most part, the masses of the gas particle and the Brownian particles are the same. When they finally take the limit of a light Brownian particle, they also assume $u\gg1$, which upon definition \Eqref{breue} means that the Brownian particle is fast compared to the gas particles. For a typical gas, even at cold temperatures, one would have to prepare a Brownian particle in a superposition of velocities $\pm \mathbf v$, where $v$ is of the order of several hundred meter per second.

In particular, superposition states of the form $\ket{\mathbf u_0}+\ket{-\mathbf u_0}$ are considered. Although one would expect, that a single collision with a particle of the same mass destroys all coherences, they find (their figure 10) that coherences vanish only once the scaled momentum approaches one. We further wish to point out, that the `decoherence per collision' to lowest order in $\mathbf u_0$ is $2\mathbf u_0^2/3$, with the only temperature dependence being due to the scaled momentum. This is obtained by dividing their decoherence rate from their Eq.~(67) by their collision rate from their Eq.~(31). For a given Brownian momentum superposition state $\ket{\mathbf p_0}+\ket{-\mathbf p_0}$, that means that from \Eqref{breue}, the `decoherence per collision' decreases as the temperature of the gas increases, which certainly can not be true.

We believe, that the reason for these counter intuitive results could be found in their definition of coherence in their Eq.~(62), which does not seem to take into account dephasing effects. As we will see in chapter~\ref{final}, dephasing is the main contribution to collisional decoherence.

\section{Difficulties with current theories}

There are two main observations from this literature review: First, it seems well accepted, that quantum dissipation goes along with position diffusion, and hence finite position jumps.  This opinion originated even before detailed studies on QLBE emerged (see e.g.~\cite{Dekker}), and is obtained throughout later articles by different authors, including the recent review article~\cite{review} by Vacchini and Hornberger. 

The other observation is about the ambiguity of current derivations. Despite the fact that QPD is clearly related to momentum decoherence, so far there is no satisfactory treatment of the change of momentum coherences due to a collision. Furthermore, all derivations of QLBEs assume instantaneous collisions, which are at odds with the diverging collision time associated with colliding gas particles in delocalized momentum eigenstates. This is especially problematic, because diffusion processes often depend on the considered time scales.

It is interesting to note, that, the master equation of the Caldeira-Leggett model in the Markov limit predicts slightly negative probabilities for certain initial states~\cite{Barnett2}. Yet, it is clearly the case that these negative probabilities are unphysical, i.e.\ that they simply arise as a consequence of approximations used to derive the master equation. Similar, the master equation for collisional decoherence predicts slow, but unbound heating of the Brownian particle. Also this effect is clearly not of physical nature, but a result of mathematical approximations. Nevertheless, to our knowledge, the possibility that the same could be true for QPD never occurred in the literature, despite such a position diffusion process demands non-continuous evolution of the density probability (in the sense described in detail in Chapter~\ref{classmo}), and therefore also for the wave function. In this sense, the discovery in the following chapters, that QPD can \emph{not} arise from random collisions, is not really that surprising.


\chapter{Single Collision\label{col}}

The obvious starting point of the study of a particle undergoing random collisions with gas particles, is the discussion of a single collision. Therefore, this chapter is devoted to the examination of a one dimensional collision of two point particles. For reasons discussed in chapter~\ref{introII}, we assume that both particles are initially described by Gaussian wave functions
	\begin{eqnarray}
		\scalar{x'}{x,p}_W &=& \frac{e^{-ix p /(2\hbar)}}{\sqrt{\sqrt\pi W }} e^{ix 'p /\hbar}e^{-(x -x ')^2/2W ^2} . \label{abel}
	\end{eqnarray}
The relative phases of these states are chosen such that $\ket{x,p}=\hat D(x,p)\ket{0,0}$ is satisfied, where 
	\begin{equation}
		\hat D(x,p) = e^{i(p\hat{x}-x\hat{p})/\hbar} 
	\end{equation}
is the Glauber displacement operator. The index $W$ indicates the position variance, and will be dropped at times for shorter notation.

We assume a hardcore interaction potential such that the two particles cannot tunnel through each other, i.e.\ $V(x_g, x )=a\delta(x -x_g)$, $a\to\infty$. This potential is a good approximation to real interaction potentials if the wave length of the particles is large compared to their interaction range. In regards to the topic of this part of the thesis, we call the two colliding particles the Brownian particle and the gas particle, respectively. In the following, the index $g$ will be used for the gas particle and no index corresponds to the Brownian particle. The two particle Hamiltonian is then given by
\begin{equation}
	H=\frac{\hat{p}_g^2}{2m_g}+\frac{\hat{p} ^2}{2m }+a\delta(\hat{x} -\hat{x}_g),\quad a\to\infty, \label{1}
\end{equation}
where $\hat{x}$ and $\hat{p}$ are the position and momentum operators for the particles.

We will find that if the width of the two Gaussian wave packets relate in a certain way, then the collision does not produce any entanglement. In this case, the mean positions and momenta of both particles behave like in a classical collision, and the only non-classical feature will turn out to be the usual spreading of the wave packets. These features will enable us to think almost classically of QBM, which in turn will guide us in setting up the QBM master equation.

There are two approaches to tackle the described collision. The first and easier one is along the lines of Hornberger and Sipe \cite{Hornberger}, but generalized to finite mass ratios of the colliding particles. One decomposes the Gaussian wave functions into momentum eigenstates $\ket{p_g}$ and $\ket{p }$, and applies scattering theory to their collision. For particles moving only in one dimension, the only possible scattering operator is of the form
	\begin{eqnarray}
		\hat S(\ket{p_g}\otimes\ket{p }) &=& e^{\imath \varphi(p,p_g)}\ket{\frac{2\alpha p -(1-\alpha)p_g}{1+\alpha}}\otimes\ket{\frac{2p_g+(1-\alpha)p }{1+\alpha}}, \label{ssmatrix}
	\end{eqnarray}
which is easily seen from conservation of energy and momentum. The parameter $\alpha=m/m_g$ denotes the mass ratio of the colliding particles. For the hard core interaction, one finds that the phase $\varphi(p,p_g)=\pi$ is independent of the momenta. For completeness and for a first taste of what follows, we start with such a scattering calculation in the first section of this chapter. 

In scattering theory, a complete collision of the two Gaussian wave packets is postulated. But there are situations in which the wave packets collide only partially, in particular when the velocity uncertainty of the colliding particles is comparable to their relative mean velocity. Therefore we will employ the more complicated approach of solving the time dependent Schr\"odinger equation. This will enable us to precisely link the situation in which complete collisions occur with the high temperature and low denisty limit. Furthermore and contrary to the scattering theory calculation, this approach enables us to look at the two particle state not only before and after the collision takes place, but also during the collision event. In position representation this will serve as proof that no position jumps take place in collisional Brownian motion, whereas in momentum representation we will find the expected momentum jumps.

\section{Scattering theory calculation}

\subsection{Some remarks about scattering theory}

Scattering theory is a very powerful tool for the following type of question. Assume at time $t=0$ we have two sufficiently far separated particles whose initial state can be approximated by a product state (asymptotic-in state) $\varrho(t=0)\approx\varrho_{in}=\rho_1\otimes\rho_2$. Because of their mutual separation, one can neglect their initial interaction which justifies the approximation of an initial product state, but if they travel towards each other they will interact at some time. What is their state sufficiently long after their interaction?

Quantum scattering theory provides an answer to this question without looking at the actual interaction time. For this one introduces the two particle scattering operator $\hat S$ which maps asymptotic-in states to asymptotic-out states\footnote{Of course, the scattering operator depends on the interaction potential. But once the scattering operator is known, either from experiments or from theory, conclusions can be drawn without considering the microscopic details of the interaction potential.}. 
	\begin{eqnarray}
		\varrho_{out} &=& \hat S \varrho_{in} \hat S^\dag
	\end{eqnarray}
At no time is $\varrho_{out}$ the actual two particle state, but if the time $t=\tau$ is such that the two particles do not interact any more, the actual two particle state can be obtained approximately by using the free particle evolution operator for each particle
	\begin{eqnarray}
		\varrho(\tau) &\approx& U_1(\tau)\otimes U_2(\tau)\varrho_{out}U_1^\dag(\tau)\otimes U_2^\dag(\tau) \nn\\
		&\approx&  U_1(\tau)\otimes U_2(\tau)\hat S \varrho(0) \hat S^\dag U_1^\dag(\tau)\otimes U_2^\dag(\tau) .
	\end{eqnarray}
The approximation is good if the interaction of the particles at times $t=0,\tau$ can be neglected.

The scattering operator preserves the energy of the two particle state, and as such it commutes with the two particle free evolution operator $[U_1(t)\otimes U_2(t),\hat S]=0$. Therefore, we get the same result if we first apply the free particle evolution operators and then the scattering operator as is shown in Fig.~\ref{smatrix}. The calculation in this section is along Fig.~\ref{smatrix}~(b).
\begin{figure}[htbp]
\begin{center}
	\includegraphics[width=\linewidth]{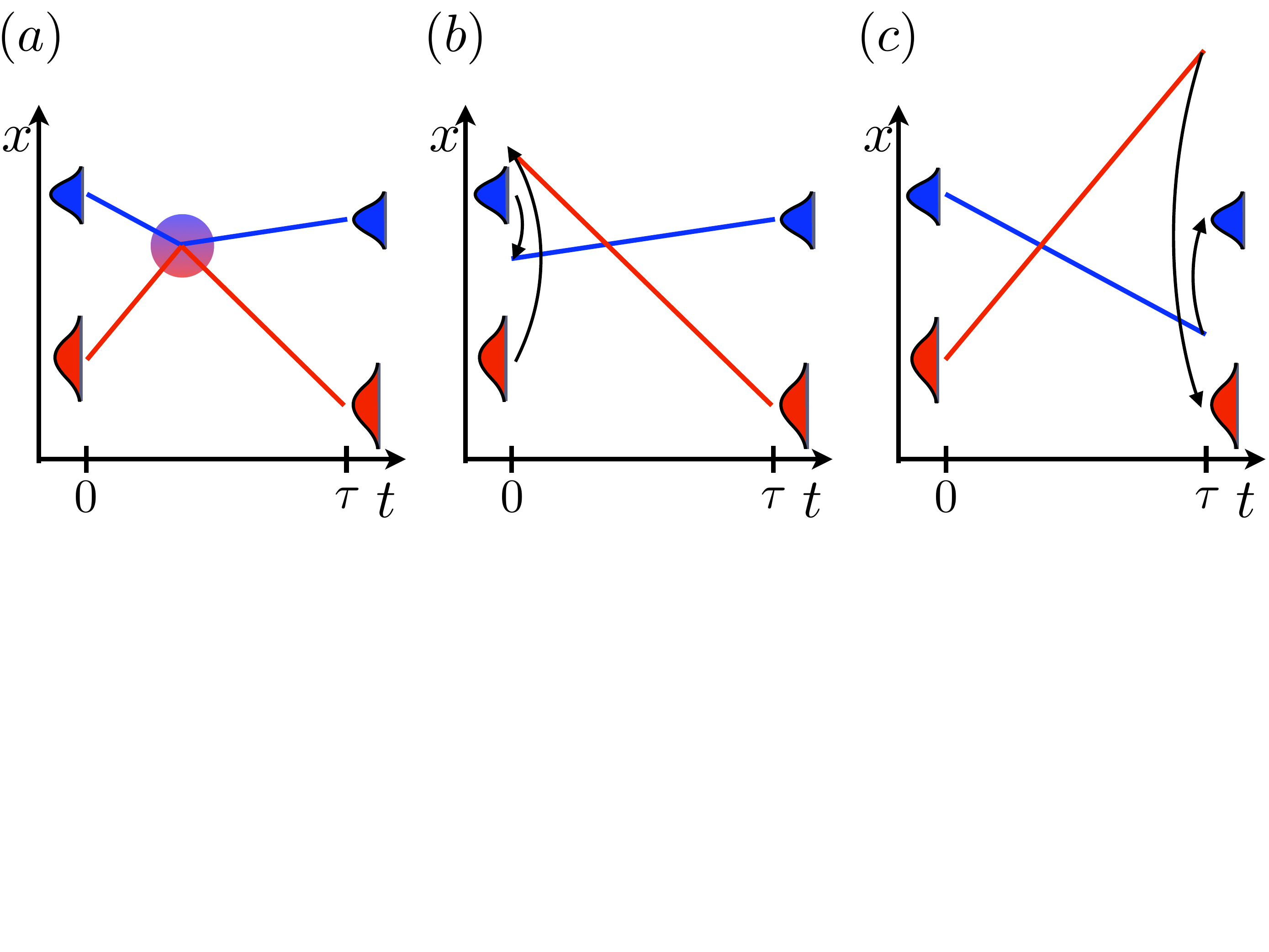}\vspace{-50mm}
\caption{\small Different representations of the same physical process. Each time the same initial two particle state ($t=0$) evolves into the same final state ($t=\tau$). Blue and red lines represent free evolution of the respective particle and arrows represent the action of the scattering operator. (a) The initial states evolve freely up to the interaction time, and afterwards continue their free evolution towards the final state. (b) The scattering operator acts on the initial state and the transformed state undergoes free evolution. (c) The initial state evolves freely until $t=\tau$, and then the scattering operator is applied to yield the final state.}
\label{smatrix}
\end{center}
\end{figure}

The popularity of scattering theory is due to its simplicity compared to calculations using the Schr\"odinger equation. On the other hand, scattering theory calculations have several drawbacks in that they do not tell us 
	\begin{itemize}
		\item what happens during a collision? \vspace{-5mm}\\
		\item what is the collision time $t_c$ (sometimes called interaction time)?\vspace{-5mm} \\
		\item if the collision is complete ($t_c<\infty$)?\vspace{-5mm} \\
		\item if a collision occurs at all? 		
	\end{itemize}
In typical applications of scattering theory, one prepares an incoming beam of particles which scatters off a target\footnote{Often two beams of particles are prepared, which scatter off each other.}, and one is interested in the momentum distribution of the particles after they collided with the target. The beam is prepared long before it hits the target, and the momentum distribution is measured long after the target was hit, often by using a screen. In such a setup one is not interested in any of the above questions, which explains the huge success of scattering theory.

The situation is different in QBM, where we aim to derive the time resolved dynamics of a Brownian particle. In principle, once the approximation of instantaneous collisions is applied, it should be possible to derive a master equation for the Brownian particle only using scattering theory. However, such approaches lead to some questionable results in the past. Therefore, we choose to instead apply Schr\"odinger's equation to gain a deeper insight in the scattering event, eventually leading to a better understanding of QBM as well as the conditions under which we can expect Markovian behavior. 

Nevertheless, we present a scattering theory calculation in this section. On the one hand this serves as confirmation of the later Schr\"odinger type calculation, on the other hand it indicates where problems might occur in such an approach. We will assume an initial product state of gas and Brownian particle, both being described by a Gaussian wave packet. 

Although the incoming gas particle is assumed to be reasonably localized in momentum, we will see that the outgoing Brownian particle state depends on the precise nature of the incoming gas particle state. In fact, from linearity of the scattering operator as well as the partial trace, we find\footnote{This is correct for any decent distance measure of density matrices.}
	\begin{eqnarray}
		\mbox{Tr}_g\!\left[\hat S (\rho\otimes\ket{\psi_g}\bra{\psi_g})\hat S^\dag\right]\,\approx\;\mbox{Tr}_g\!\left[\hat S \left(\rho\otimes\ket{\psi'_g}\bra{\psi'_g}\right)\hat S^\dag\right] &\mathbf{if}& \scalar{\psi_g}{\psi'_g}\approx1 \qquad
	\end{eqnarray}
Therefore we can approximate a gas state $\ket{\psi_g}$ by $\ket{\psi'_g}$ if their overlap is close to one, without changing results for the Brownian particle. But because the overlap ${}_{W_g}\!\scalar{x_g,p_g}{x_g,p_g}_{W'_g}$ of two Gaussian wave packets [\Eqref{abel}] approaches zero if $W'_g$ goes to infinity (for any fixed value of $W_g$), it can not be guaranteed to find correct results by approximating the incoming gas particle state $\ket{x_g,p_g}_{W_g}$ with a momentum eigenstates (which is $\lim_{W_g'\to\infty}\ket{x_g,p_g}_{W_g'}$ up to normalization). 

If one is only interested in the momentum distribution after a collision, it might be sufficient to consider incoming gas particles in momentum eigenstates. But as we are interested in the entire Brownian particle state, including the coherences of different momenta as well as the position, this can no longer be assured and one has to consider the actual incoming gas particle state.\footnote{Of course, one can write $\int dp_g\,\tilde\psi(p)\ket {p_g}$ for the incoming gas particle state and perform the calculation in the momentum basis, as is done in the next subsection.}

\subsection{The calculation}

We use an asymptotic-in state in the following momentum representation
	\begin{eqnarray}
		\ket{\Psi_{in}} &=& \ket{x,p}_W\otimes\ket{x_g,p_g}_{W_g} \nn\\
		&=& \frac{\sqrt{WW_g}}{\sqrt\pi\hbar} e^{i(xp+x_gp_g)/2\hbar} \nn\\
		&& \times \int\!\!\!\!\int\! \dd p'\,\dd p'_g\, e^{-i(xp'+x_gp'_g)/\hbar} e^{-\left[(p'-p)^2W^2+(p'_g-p_g)^2W_g^2\right]/2\hbar^2} \ket{p'}\ket{p'_g}. \qquad
	\end{eqnarray}
Applying the transformation \Eqref{ssmatrix} and substituting
	\begin{eqnarray}
		u=\frac{(1-\alpha)p'+2p'_g}{1+\alpha}, && u_g=\frac{(\alpha-1)p'_g+2\alpha p'}{1+\alpha}
	\end{eqnarray}
we find
	\begin{eqnarray}
		\ket{\Psi_{out}} &=& \hat S \ket{\Psi_{in}} \nn\\
		&=&  \frac{\sqrt{WW_g}}{\sqrt\pi\hbar} e^{i(xp+x_gp_g)/2\hbar} \int\!\!\!\!\int\! \dd u\,\dd u_g\, \ket u \ket{u_g}\nn\\
		&&\!\!\times \exp\!\left[ -i\frac{x[(1\!-\!\alpha)u+2u_g]+x_g[(\alpha\!-\!1)u_g+2\alpha u]}{(1+\alpha)\hbar}\right]\nn\\
		&&\!\!\times \exp\!\left[-\frac{W^2[(1\!-\!\alpha)u+2u_g-(1\!-\!\alpha)]p^2}{2(1+\alpha)^2\hbar^2} - \frac{W_g^2[(\alpha\!-\!1)u_g+2\alpha u-(1\!-\!\alpha)p_g]^2}{2(1+\alpha)^2\hbar^2} \right]\!\! \nn\\
	\end{eqnarray}
To see whether this is a product state, we now separate $u$ and $u_g$ in the last exponential. To this end we find a term
	\begin{eqnarray}
		\exp\!\left[-\frac{4(1\!-\!\alpha)uu_g\left(W^2-\alpha W_g^2\right)}{2(1+\alpha)^2\hbar^2}\right]\label{entang}
	\end{eqnarray}
which shows that $u$ and $u_g$ can in general not be separated, and therefore, $\varrho_{out}$ is not a product state. But it is also apparent from \Eqref{entang} that the entanglement of the two outgoing particles is lifted, if the width of the colliding wave packets relate according to their masses 
	\begin{eqnarray}
		mW^2&=&m_gW_g^2. \label{relation}
	\end{eqnarray}
This relation will hugely simplify the calculation as well as the physical interpretation, and therefore the (over)complete basis $\ket{x,p}_W$ with $W$ satisfying \Eqref{relation} is the basis of choice for studying a single collision with a gas particle state $\ket{x_g,p_g}_{W_g}$. If the initial state of the Brownian particle is not of the form $\ket{x,p}_W$, then is can be written as a superposition of such states and one can still use the results from the simpler calculation with $\ket{x,p}_W$. For such an initial state, of course, the two particle state after a collision will show entanglement. An important example is a so called cat-state $\ket{x_a,p_a}_W+\ket{x_b,p_b}_W$, which is presented in section~\ref{cat}.

Next we introduce 
	\begin{eqnarray}
		\bar{x}_{} = \frac{2\alpha x_g+(1-\alpha)x_{}}{1+\alpha}, && \bar{x}_g = \frac{2x_{}-(1-\alpha)x_g}{1+\alpha}, \nn\\
		\bar{p}_{} = \frac{2p_g+(1-\alpha)p_{}}{1+\alpha}, && \bar{p}_g = \frac{2\alpha p_{}-(1-\alpha)p_g}{1+\alpha} ,\label{apa}
	\end{eqnarray}
and after replacing $u$ by $p'$ we find
	\begin{eqnarray}
		\ket{\Psi_{out}} &=& -\frac{\sqrt{WW_g}}{\sqrt\pi\hbar} e^{i(\bar x\bar p+\bar x_g\bar p_g)/2\hbar} \nn\\
		&& \times \int\!\!\!\!\int\! \dd p'\,\dd p'_g\, e^{-i(\bar xp'+\bar x_gp'_g)/\hbar} e^{-\left[(p'-\bar p)^2W^2+(p'_g-\bar p_g)^2W_g^2\right]/2\hbar^2} \ket{p'}\ket{p'_g}\nn\\
		&=& -\ket{\bar x,\bar p}_W\otimes\ket{\bar x_g,\bar p_g}_{W_g}
	\end{eqnarray}
This surprisingly simple result shows that all the collision does, is to change the position and momenta of the particles according to \Eqref{apa}. We want to point out that this does not mean that the position necessarily changes during the collision process, because as discussed in the previous subsection, the state $\ket{\Psi_{out}}$ is not the actual state at any time. To find the physical state at time $t=\tau$ sufficiently long after the collision, we need to apply the free particle evolution operators as in
	\begin{eqnarray}
		\ket{\Psi(\tau)} &=& U(\tau)\ket{\bar x,\bar p}_W\otimes U_g(\tau)\ket{\bar x_g,\bar p_g}_{W_g}.
	\end{eqnarray}
Taking into account the free particle evolution operators, which shift the position of the wave packets according to their momentum, we will find in the following sections that the position distribution follows a continuous flow.

\section{Eigenfunctions}

The following sections will be concerned with solving the two particle Schr\"odinger equation \Eqref{1}. The first task will be to find the eigenstates of the Hamiltonian, which will be done in position representation. The time independent Schr\"odinger equation reads
\begin{eqnarray}
	\left(-\frac{\hbar^2}{2m_g}\frac{\partial^2}{\partial x_g'^2}-\frac{\hbar^2}{2m }\frac{\partial^2}{\partial x '^2}+a\delta(x '-x_g')-E_{tot} \right)\Psi_{E_{tot}}(x_g',x ')&=& 0. \quad\label{2}
\end{eqnarray}
Two orthogonal sets of solutions are found with the ansatz
\begin{eqnarray}
	\Psi_{E_{tot}}(x_g',x ') \;=\; e^{i\bar{k}(x '+\alpha x_g')}\sin(\tilde{k}(x_g'-x ')),&\quad& \bar{k}\in \mathbb{R},\;\tilde{k}\in\mathbb{R}_+ \quad\label{3}\\
	\Psi_{E_{tot}}(x_g',x ') \;=\; e^{i\bar{k}(x '+\alpha x_g')}\sin(\tilde{k}|x_g'-x '|),\:&\quad& \bar{k}\in \mathbb{R},\;\tilde{k}\in\mathbb{R}_+. \quad\label{4}
\end{eqnarray}
Note that in the second set the relative coordinate is replaced by its absolute value. Substitution in \Eqref{2} gives 
\begin{eqnarray}
	\alpha&=&\frac{m_g}{m }\\
	E_{tot}&=&E +E_g=\widetilde{E}+\overline{E} \label{6}\\
	\widetilde{E}&=&(1+\alpha)\frac{\tilde{k}^2\hbar^2}{2m_g}\\
	\overline{E}&=&(1+\alpha)\frac{\bar{k}^2\hbar^2}{2m }.
\end{eqnarray}
We will refer to the eigenfunctions \Eqref{3} and \Eqref{4} as $\Psi_{\tilde{k}\bar{k}}^a$ and $\Psi_{\tilde{k}\bar{k}}^s$, where the superscripts stand for antisymmetric and symmetric. With 
\begin{equation}
	-i\hbar\left( \frac{\partial}{\partial x_g'} + \frac{\partial}{\partial x '} \right) \Psi_{\tilde{k}\bar{k}}^{a/s} = \hbar\bar{k}(1+\alpha)\Psi_{\tilde{k}\bar{k}}^{a/s}
\end{equation}
these states are eigenstates of the total momentum operator $\hat{p}_{tot}=\hat{p}_g+\hat{p} $  and have therefore definite total momentum
\begin{equation}
	p_{tot} = \hbar\bar{k}(1+\alpha).\label{ptot}
\end{equation}
Similar, $\Psi_{\tilde{k}\bar{k}}^{a/s}$ are also eigenstates of the square of the relative velocity operator $\hat{v}_{rel}=(\alpha \hat{p} -\hat{p}_g)/m_g$ with eigenvalues
\begin{equation}
	 v_{rel}^2 = \left(\frac{1}{m_g}+\frac{1}{m }\right)^2\hbar^2\tilde{k}^2.\label{vrel}
\end{equation}

\section{Initial state}

In this section, we construct the appropriate initial state for the wave packet scattering. The problem is that the most obvious candidate for this state,
\begin{equation}
	\ket{\Psi(t=0)}\;=\;\ket{x_g,p_g}\otimes\ket{x ,p }, \label{init}
\end{equation}
does not satisfy constraints implied by the interaction between the two particles, so a more careful derivation is required, and is detailed below.

The state \Eqref{init} can be written in position representation
\begin{equation}
	\Psi(t=0, x_g',x ') \;=\; \frac{e^{-i(x_gp_g+x p )/(2\hbar)}}{\sqrt{\pi W_gW }}e^{ix_g'p_g/\hbar}e^{-(x_g-x_g')^2/2W_g^2}\; e^{ix 'p /\hbar}e^{-(x -x ')^2/2W ^2}. \label{9}
\end{equation}
Note that this state is only a good approximation to the real two particle state if the Gaussian wave functions have little overlap. To make this statement precise, we define the overlap as
	\begin{equation}
		\int \dd x_g' \left| \Psi(t=0,x_g',x '=x_g')\right|^2 \;=\; \frac{1}{\sqrt{\pi\left(W_g^2+W ^2\right)}}\exp\!\left(-\frac{(x_g-x )^2}{W_g^2+W ^2}\right)
	\end{equation}
from which we find the condition for small overlap
	\begin{eqnarray}
		|x_g-x |&\gg&\sqrt{W_g^2+W ^2}. \label{overlap}
	\end{eqnarray}

To simplify the calculations, we will use the centre-of-mass reference frame, and assume the gas particle approaching the Brownian one from the left side, i.e.\
\begin{eqnarray}
	p_g&=&-p  \,\;>\,\;0 \label{10}\\
	\alpha x_g&=&-x  \,\:<\;\,0. \label{11}
\end{eqnarray}

The task at hand is to expand \Eqref{9} in the eigenfunctions $\Psi_{\tilde{k}\bar{k}}^a$ and $\Psi_{\tilde{k}\bar{k}}^s$. Here we encounter a problem as the two particle state \Eqref{9} does not satisfy the boundary condition $\Psi(0, x_g'=x ')$. Hence it is not possible to expand this state into a linear combination of $\Psi_{\tilde{k}\bar{k}}^a$ and $\Psi_{\tilde{k}\bar{k}}^s$, and we have choose a more elaborate approach to find an initial state $\ket{\Psi_i}$ which is close to \Eqref{9} and has the property $\Psi_i(0,x_g'=x ')=0$. This will prove quite tedious, but the very nice result of an almost classical collision will make this effort worthwhile. 

To do so, we first expand \Eqref{9} into a set of functions which is related to the set energy eigenfunctions by substituting the sine function with an exponential function. In a second step we will remove the overlap of the two Gaussian wave packets and by doing so we will recover the energy eigenfunctions. 

We start by comparing two functional expansions, one being the two-coordinate Fourier transform, the other one being closely related to our eigenfunctions by replacing the sine function with an exponential function:
\begin{eqnarray}
	\Psi(0,x_g',x ') &=& \frac{1}{2\pi}\int\!\!\!\!\int \dd k_g\,\dd k \,e^{ik_gx_g'}e^{ik x '}\,\widetilde{\Psi}(0,k_g,k )\label{14} \\
	&=& \frac{1}{2\pi}\int\!\!\!\!\int \dd \tilde{k}\,\dd \bar{k} \,e^{i\tilde{k}\tilde{x}'}e^{i\bar{k}\bar{x}'}\,\widetilde{\overline{\Psi}}(0,\tilde{k},\bar{k}). \label{15}
\end{eqnarray}
Here we introduced $\bar{x}'=x '+\alpha x_g'$ and $\tilde{x}'=x_g'-x '$. The Fourier transform of \Eqref{9} is easily found to be
\begin{equation}
	\widetilde{\Psi}(0,k_g,k ) = \sqrt{\frac{W_gW }{\pi}} e^{i(x_gp_g+x p )/2\hbar} e^{-i(x_gk_g+x k )} e^{-W_g^2(k_g-\frac{p_g}{\hbar})^2/2} e^{-W ^2(k -\frac{p }{\hbar})^2/2} \label{16}
\end{equation}
whereas \Eqref{15} can be written as
\begin{equation}
	\Psi(0,x_g',x ') = \frac{1}{2\pi}\int\!\!\!\!\int \dd \tilde{k}\,\dd \bar{k}\,e^{i(\tilde{k}+\alpha \bar{k})x_g'}e^{i(\bar{k}-\tilde{k})x '}\,\widetilde{\overline{\Psi}}(0,\tilde{k},\bar{k}).
\end{equation}
Now we substitute 
\begin{equation}
	\left( \begin{array}{c}k_g\\k \end{array} \right) = \left( \begin{array}{c}\tilde{k}+\alpha \bar{k}\\\bar{k}-\tilde{k}\end{array} \right) \quad \Rightarrow \quad \left( \begin{array}{c}\tilde{k}\\\bar{k}\end{array} \right) = \left( \begin{array}{c} \frac{k_g-\alpha k }{1+\alpha} \\ \frac{k_g+k }{1+\alpha} \end{array} \right).
\end{equation}
With the determinant of the Jacobi matrix
\begin{equation}
	J=\left( \begin{array}{cc} \frac{1}{1+\alpha} & \frac{-\alpha}{1+\alpha} \\ 
		\frac{1}{1+\alpha} & \frac{1}{1+\alpha} \end{array} \right)
	 \quad \Rightarrow \quad \det(J)=\frac{1}{1+\alpha}
\end{equation}
we find 
\begin{equation}
	\Psi(0,x_g',x ') = \frac{1}{2\pi}\int\!\!\!\!\int \dd k_g\,\dd k\, e^{ik_gx_g'}e^{ik x '}\frac{1}{1+\alpha}\widetilde{\overline{\Psi}}\left(0,\frac{k_g-\alpha k }{1+\alpha},\frac{k_g+k }{1+\alpha}\right).
\end{equation}
Comparison with \Eqref{14} then results in
\begin{equation}
	\widetilde{\overline{\Psi}}(0,\tilde{k},\bar{k}) = (1+\alpha)\widetilde{\Psi}(0,\tilde{k}+\alpha \bar{k}, -\tilde{k}+\bar{k})\label{21}
\end{equation}
where we can substitute \Eqref{16} to finally find
\begin{eqnarray}
	\widetilde{\overline{\Psi}}(0,\tilde{k},\bar{k}) &=& (1+\alpha) \sqrt{\frac{W_gW }{\pi}} e^{i(x_gp_g+x p )/2\hbar} e^{-i\tilde{k}(x_g-x )} e^{-i\bar{k}(\alpha x_g+x )}\nn\\
	& \times & \exp\left\{ -\half \bar{k}^2(\alpha^2W_g^2 +W ^2) + \bar{k}\left(\frac{\alpha p_gW_g^2}{\hbar}+\frac{p W ^2}{\hbar}\right) - \frac{W_g^2p_g^2}{2\hbar^2} - \frac{W ^2p ^2}{2\hbar^2} \right\} \nn\\
	& \times & \exp\left\{ -\half \tilde{k}^2(W_g^2+W ^2) + \tilde{k}\left( \frac{p_gW_g^2}{\hbar}-\frac{p W ^2}{\hbar} \right) - \tilde{k}\bar{k}(\alpha W_g^2-W ^2) \right\}.\nn\\
\end{eqnarray}
In the centre of mass reference frame (\Eqref{10} and \Eqref{11}) this can be simplified to
\begin{eqnarray}
	\widetilde{\overline{\Psi}}(0,\tilde{k},\bar{k}) &=& (1+\alpha) \sqrt{\frac{W_gW }{\pi}} e^{ix p (1+\alpha)/2\alpha\hbar} e^{i\tilde{k}x (1+\alpha)/\alpha} \nn\\
	&\times& \exp\left\{ -\half  \bar{k}^2(\alpha^2W_g^2 +W ^2) + \bar{k}\left(\tilde{k}+\frac{p }{\hbar}\right)(W ^2-\alpha W_g^2) \right\} \nn\\
	&\times& \exp\left\{ -\half\left(\tilde{k}+\frac{p }{\hbar}\right)^2(W_g^2+W ^2) \right\}.
\end{eqnarray}
Let us have a closer look at this equation. We see in the first line, that the phase is independent of $\bar{k}$ which tells us that the expectation value of the position of the centre of mass is zero, as should be expected in the reference frame of the centre of mass. In the second line (in combination with the third line) we find that the expectation value of the total momentum is also zero, as it should be in the reference frame of the centre of mass. Furthermore, the second term in the second line describes an entanglement of the values $\tilde{k}$ and $\bar{k}$ (there is no entanglement between gas and Brownian particle, but entanglement of the total momentum and the relative velocity), which is lifted if and only if
\begin{equation}
	W ^2 = \alpha W_g^2. \label{33}
\end{equation}
This is the situation we are interested in, as it will turn out to be the condition for Brownian and gas particles not being entangled after the collision. Using \Eqref{33} we can now write 
\begin{eqnarray}
	\widetilde{\overline{\Psi}}(0,\tilde{k},\bar{k}) &=& (1+\alpha) \frac{W }{\sqrt{\pi\sqrt{\alpha}}} e^{ix p (1+\alpha)/2\alpha\hbar} e^{i\tilde{k}x (1+\alpha)/\alpha} \nn\\
	&\times& \exp\left\{ -\half  \bar{k}^2(1+\alpha)W ^2  -\half\left(\tilde{k}+\frac{p }{\hbar}\right)^2 \frac{1+\alpha}{\alpha} W ^2 \right\}. \label{28}
\end{eqnarray}

\begin{figure}[htb]
\begin{center}
	\includegraphics[width=0.8\linewidth]{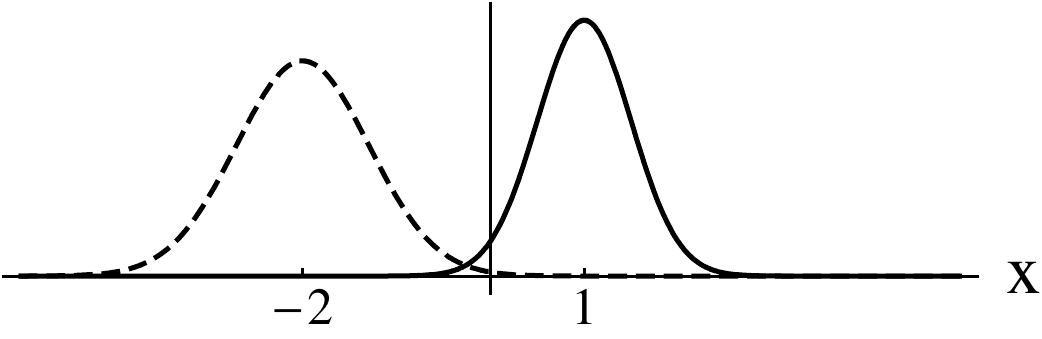}
\caption{\small Position representation  $|\scalar{x_g}{x_gp_g}|^2$ (dashed) and $|\scalar{x}{x p }|^2$ (solid). The overlap is not physical and has to be removed before the initial state can be written as a superposition of energy eigenstates.}
\label{fig1}
\end{center}
\end{figure}

With \Eqref{15} and \Eqref{28} we have constructed an initial state \Eqref{9} which is of the form seen in Fig.~\ref{fig1}. As pointed out earlier, there is some probability of finding the gas and Brownian particle at the same position which is not physical for the hard core interaction. Although, if relation~(\ref{overlap}) is satisfied this overlap is small, it is non-zero and hence it is not possible to decompose this state into energy eigenstates \Eqref{3}-(\ref{4}). Therefore we have to construct a state $\ket{\Psi_i(t=0)}$ which is close to $\ket{x_gp_g}\otimes\ket{x p }$ for $|x_g-x |\gg W_g+W $, and has the property 
\begin{equation}
	\Psi(0,x_g',x ')=0 \quad\mbox{for all}\quad x '=x_g'. \label{29}
\end{equation}
\begin{figure}[htb]
\begin{center}
	\includegraphics[width=0.8\linewidth]{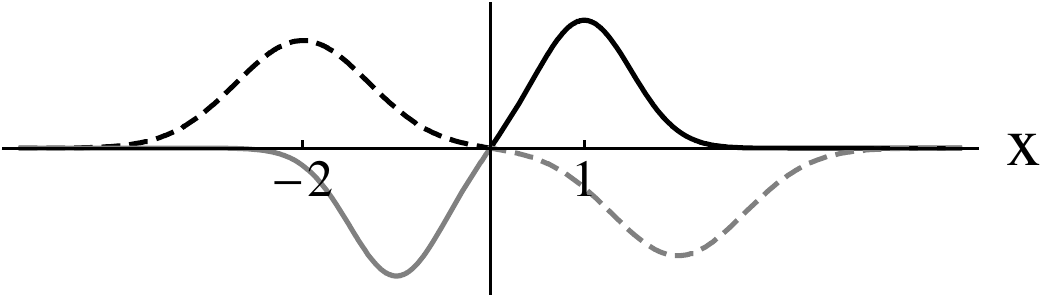}
\caption{\small Position representation of  $\ket{\Psi_{sup}}$ (gas: dashed. Brownian: solid). There is no overlap and therefore this represents a physical state. The black parts represent the gas particle coming from the left, whereas the gray part represents the gas particle coming from the right.}
\label{fig2}
\end{center}
\end{figure}
As a first step we use the antisymmetric superposition 
\begin{equation}
	\ket{\Psi_{sup}}=C(\ket{x_gp_g}\otimes\ket{x p }-\ket{-x_g,-p_g}\otimes\ket{-x ,-p })\label{30}
\end{equation}
which automatically fulfills \Eqref{29}. The normalization constant 
\begin{equation}
	C=\sqrt{2-2\exp\left[ -\frac{1+\alpha}{\alpha}\left( \frac{W ^2p ^2}{\hbar^2}+\frac{x ^2}{W ^2} \right) \right] }^{-1}
\end{equation}
approaches $\sqrt{1/2}$ if initially Brownian and gas particle are sufficiently separated. With \Eqref{28} it is an easy task to find
\begin{eqnarray}
	\widetilde{\overline{\Psi}}_{sup}(0,\tilde{k},\bar{k}) &=& C(1+\alpha) \frac{W }{\sqrt{\pi\sqrt{\alpha}}} e^{ix p (1+\alpha)/2\alpha\hbar} \exp\left[ -\half  \bar{k}^2(1+\alpha)W ^2 \right] \nn\\
	&\times&  \left\{e^{i\tilde{k}x (1+\alpha)/\alpha} \exp\!\left[ -\left(\tilde{k}+\frac{p }{\hbar}\right)^2 \frac{1+\alpha}{2\alpha} W ^2 \right] \right. \nn\\
	&& \left. - e^{-i\tilde{k}x (1+\alpha)/\alpha} \exp\!\left[ -\left(\tilde{k}-\frac{p }{\hbar}\right)^2 \frac{1+\alpha}{2\alpha} W ^2 \right]\right\} . 
\end{eqnarray}
Substituting in \Eqref{15}, we find that the $\cos(\tilde{k}\tilde{x}')$ cancels and therefore $\widetilde{\overline{\Psi}}_{sup}(0,\tilde{k},\bar{k})$ expands $\ket{\Psi_{sup}}$ into energy eigenstates
\begin{equation}
	\Psi_{sup}(0,x_g',x ') = \frac{i}{\pi}\int_\infty^\infty \dd\bar{k} \int_0^\infty \dd\tilde{k}\, e^{i\bar{k}\bar{x}'}\sin(\tilde{k}\tilde{x}')\widetilde{\overline{\Psi}}_{sup}(0,\tilde{k},\bar{k}).
\end{equation}
The possibility to decompose $\Psi_{sup}(0,x ',x_g')$ into energy eigenstates is a direct consequence of \Eqref{29}. The state $\ket{\Psi_{sup}}$ can be pictured as in Fig.~\ref{fig2} (this is not completely correct since in this state Brownian and gas particle are entangled). To achieve \Eqref{29} we paid the price that the gas particle does not approach the Brownian particle from one definite direction. In order to return to the original picture where the gas particle approaches the Brownian one from the left, we want to get rid of the parts in Fig.~\ref{fig2} indicated in grey. This is simply done by adding the symmetric set of eigenfunctions \Eqref{4} to the antisymmetric one:
\begin{eqnarray}
	\hspace{-5mm}\Psi_i(0,x_g',x ') &\!\!=\!\!& \frac{i}{\sqrt{2}\pi}\int_\infty^\infty \!\! \dd\bar{k}  \int_0^\infty \!\! \dd\tilde{k}\, e^{i\bar{k}\bar{x}'}\left(\sin(\tilde{k}\tilde{x}')-\sin(\tilde{k}|\tilde{x}'|)\right) \widetilde{\overline{\Psi}}_{sup}(0,\tilde{k},\bar{k})\qquad\\
	&\!\!=\!\!& \frac{i}{\sqrt{2}\pi}\int_\infty^\infty \!\! \dd\bar{k} \int_0^\infty \!\! \dd\tilde{k} \left(\Psi_{\tilde{k}\bar{k}}^a(x_g',x ')-\Psi_{\tilde{k}\bar{k}}^s(x_g',x ') \right) \widetilde{\overline{\Psi}}_{sup}(0,\tilde{k},\bar{k}) . \qquad
\end{eqnarray}
This state has all the required properties and is the initial state we will use in the following.

There are of course other ways to construct initial states which fulfill \Eqref{29} and are approximately the tensor product of two Gaussian states. But the choice here has the appealing property that $\ket{\Psi_{sup}(t=-\infty)}$ is an exact product state as can be seen in the next section. This should be required since the two incoming particles should not be entangled at $t=-\infty$. This property is of course lost for all finite times, but if $|x_g-x |\gg W_g+W $ then the two particle state is still approximately a product state at $t=0$.

\section{Time evolution\label{timeevo}}

Since $\Psi_{\tilde{k}\bar{k}}^{a/s}$ are eigenfunctions of the Hamiltonian  with energy \Eqref{6} the time evolution is an easy matter:
\begin{equation}
	\hspace{0mm}\Psi_i(t,x_g',x ') \;=\; \frac{i}{\sqrt{2}\pi}\int_\infty^\infty \!\! \dd\bar{k}\! \int_0^\infty \!\! \dd\tilde{k} \,e^{-i{E}_{tot}(\tilde{k},\bar{k})t/\hbar} \left[\Psi_{\tilde{k}\bar{k}}^a(x_g',x ')-\Psi_{\tilde{k}\bar{k}}^s(x_g',x ') \right] \widetilde{\overline{\Psi}}_{sup}(0,\tilde{k},\bar{k}),
\end{equation}
where ${E}_{tot}(\tilde{k},\bar{k})$ is given in \Eqref{6}. To perform the integrals for $x '>x_g'$, we use the exponential instead of the sine (remember that the cosine cancels upon multiplication of the exponential with $\widetilde{\overline{\Psi}}_{sup}(0,\tilde{k},\bar{k})$)
\begin{equation}
	\Psi_i(t,x_g',x ') \;=\; \left\{ \begin{array}{lc}
		\!\!\frac{1}{\sqrt{2}\pi}\int_{-\infty}^\infty \int_{-\infty}^\infty \dd \tilde{k}\,\dd \bar{k}\, e^{-i{E}_{tot}(\tilde{k},\bar{k})t/\hbar}e^{i\tilde{k}\tilde{x}'}e^{i\bar{k}\bar{x}'}\,\widetilde{\overline{\Psi}}_{sup}(0,\tilde{k},\bar{k}) & \mbox{for } x_g'<x' \\
		\!\!0 & \mbox{for } x_g'>x' \end{array} \right.
\end{equation}
and find after a straightforward integration
\begin{eqnarray}
	\Psi_i(t,x_g',x ') &=&  \frac{CW \sqrt{2\sqrt{\alpha}}}{\sqrt{\pi}\left(W ^2+\frac{i\hbar t}{m }\right)}     \exp\!\left[ i\frac{x_gp_g+x p }{2\hbar}\right] \nn\\
	&\times& \exp\!\left[   - \frac{ \left(\frac{1+\alpha}{\alpha}\right)^2W ^2\left(x +\frac{p t}{m }\right)^2  +  \left(\tilde x'^2+\frac{\bar{x}'^2}{\alpha}\right)\!\left(W ^2-\frac{i\hbar t}{m }\right)}    {2\left(\frac{1+\alpha}{\alpha}\right)\left(W ^4+\frac{\hbar^2t^2}{m ^2}\right)}    \right] \nn\\
	&\times& \left\{  \exp\!\left[  -\tilde x'\frac{W ^2\left(x +\frac{p t}{m }\right)+i\left(\frac{p W ^4}{\hbar}-\frac{x \hbar t}{m }\right)}    {\left(W ^4+\frac{\hbar^2t^2}{m ^2}\right)}  \right] \right.  \nn\\  
	&& - \left.  \exp\!\left[  \tilde x'\frac{W ^2\left(x +\frac{p t}{m }\right)+i\left(\frac{p W ^4}{\hbar}-\frac{x \hbar t}{m }\right)}    {\left(W ^4+\frac{\hbar^2t^2}{m ^2}\right)}  \right]   \right\} \label{38}
\end{eqnarray}
for $\tilde x'<0$. Let us have a closer look at the two terms in curly brackets. The first term dominates if $\left(x +\frac{p t}{m }\right)>0$, which is the case for $t=0$ because we chose $x >0$ at the beginning of the calculation. The second term dominates for $t\to\infty$ as $p <0$. That the smaller term vanishes for $t\to\pm\infty$ requires a more detailed comparison which can be found in the appendix of this chapter. It turns out that the smaller term only vanishes if 
\begin{equation}
	|p |\gg \sqrt{\frac\alpha{1+\alpha}}\frac\hbar{W }.\label{39}
\end{equation}
That is, the absolute value of the mean momentum of each particle has to be greater than its momentum uncertainty. This is actually quite intuitive, as otherwise the wave packets would spread faster than they move towards each other, and parts of the wave packets would not collide at all, which we will refer to as ``incomplete collisions". If this condition holds we are also able to define a collision time (see appendix)
	\begin{equation}
		t_c=\sqrt{\frac{8}{1+\alpha}}W_g\frac{m_g}{|p_g|}\label{collisiontime}
	\end{equation}
as the time it takes from the first term to dominate to the second term to dominate.

Let us assume that  condition~(\ref{39}) is fulfilled. The first term in the curly brackets can then be neglected sufficiently long after the collision and the two particle state can be brought into the form
\begin{eqnarray}
	\Psi_i(t,x_g',x ') &\!=\!& \frac{e^{-\frac{i}{2\hbar}\left(x_g+\frac{p_gt}{m_g}\right)p_g}}   {\sqrt{\sqrt{\pi}\left(W_g+\frac{i\hbar t}{W_gm_g}\right)}}   e^{-ip_gx_g'/\hbar}    \exp\!\left[  \!-\!\left(1-\frac{i\hbar t}{m_gW_g^2}\right)   \frac{\left(x_g'+x_g+\frac{p_gt}{m_g}\right)^2}   {2\left(W_g^2+\frac{\hbar^2t^2}{m_g^2W_g^2}\right)}    \right] \nn\\
	&\!\times\!& \frac{e^{-\frac{i}{2\hbar}\left(x +\frac{p t}{m }\right)p }}   {\sqrt{\sqrt{\pi}\left(W +\frac{i\hbar t}{W m }\right)}}   e^{-ip x '/\hbar}    \exp\!\left[  \!-\!\left(1-\frac{i\hbar t}{m W ^2}\right)   \frac{\left(x '+x +\frac{p t}{m }\right)^2}   {2\left(W ^2+\frac{\hbar^2t^2}{m ^2W ^2}\right)}     \right] \nn\\ \label{41}
\end{eqnarray}
where $C=\frac{1}{\sqrt{2}}$ as well as $\tilde x'=x_g'-x '$ and $\bar{x}'=\alpha x_g'+x '$ were used.

This is a surprising simple result since it is a product of a function of $x_g'$ and a function of $x '$ and therefore the position representation of a product state. Even more, these are the Gaussian states
\begin{equation}
	\framebox{\phantom{\LARGE ]}$\ket{\Psi(t)} = U_g(t)\ket{-x_g,-p_g}\otimes U (t)\ket{-x ,-p } $\phantom{\LARGE ]}} \label{42}
\end{equation}
where $U_g(t)$ and $U (t)$ are the unitary evolution operator for a free gas and Brownian particle, respectively. Of course, this equation also applies if the gas particle approaches the Brownian one from the right. Therefore, in the centre-of-mass reference frame, all the collision does is mirroring the initial values $x_g$, $p_g$, $x $, and $p $ which is exactly what happens in a classical collision. The only extra requirement in a quantum collision is that the widths of the colliding wave packets fulfill \Eqref{33}. This result agrees with Schm\"user and Janzing \cite{Schmuser} who are looking at entanglement creation during the collision of two hard core particles in one dimension. They find that the collision does not produce any entanglement if the width of the Gaussian wave packets relate to their masses like $m W ^2=m_g W_g^2$.

Now we briefly return to the remark at the end of the last section. If \Eqref{39} holds, then the second term in the curly brackets in \Eqref{38} vanishes for $t\to -\infty$. We then find
\begin{equation}
	\ket{\Psi(t)} = U_g(t)\ket{x_g,p_g}\otimes U (t)\ket{x ,p },
\end{equation}
which is a Gaussian product state as promised.

\section{General reference frame}

So far we used the centre of mass reference frame. In this section we turn to a collision of the two particles with an initial state
\begin{equation}
	\ket{ {\Psi}(t=0)}=\ket{ {x}_g, {p}_g}\otimes\ket{ {x} , {p} } \label{44}
\end{equation}
with general $ {x}_g$, $ {p}_g$, $ {x} ,$ and $ {p} $. It would seem that we could just change the reference frame as in classical mechanics, but that would not necessarily give us the phase of the final state. Therefore we will perform the change of reference frame in a more formal way by using the Glauber displacement operators $\hat D_g(\Delta{x}_g,\,\Delta{p}_g)$ and $\hat D (\Delta{x} ,\,\Delta{p} )$, such that the state \Eqref{init} transforms to the state \Eqref{44}
\begin{equation}
	\ket{ {\Psi}(t=0)}=\hat D_g(\Delta{x}_g,\,\Delta{p}_g)\hat D (\Delta{x} ,\,\Delta{p} )\ket{\widetilde x_g,\widetilde p_g}\ket{\widetilde x ,\widetilde p }.\label{45}
\end{equation}
To avoid confusion, in this section we use a wide tilde to denote quantities in the centre-of-mass reference frame.

Since we have a number of `$x$'-symbols, it is worth summarizing them at this point to avoid confusion. The `$p$'-symbols are defined in an analogous way.
\begin{center}
\begin{tabular}{|l|c|c|}
	& \parbox{20mm}{gas particle\vspace{2mm}} & \parbox{30mm}{Brownian particle\vspace{2mm}} \\ \hline
	\parbox{50mm}{\vspace{2mm} position operator\vspace{1.5mm}} & $\hat{x}_g$ & $\hat{x} $  \\ 
	\parbox{57mm}{\vspace{1mm} initial position of wave packet\vspace{1.5mm}} & $ {x}_g$ & $ {x} $ \\
	\parbox{57mm}{\vspace{1mm} initial position of wave packet \vspace{-1mm}\\in centre  of mass reference frame\vspace{1.5mm}} & $\widetilde x_g$ & $\widetilde x $ \\
	\parbox{50mm}{\vspace{2mm} $ {x} -\widetilde x $\vspace{1mm}} & $\,\Delta{x}_g$ & $\,\Delta{x} $  \\
	\parbox{57mm}{\vspace{1mm} variable in position representation\vspace{1.5mm}} &$x_g'$ & $x '$   \\
	\parbox{57mm}{\vspace{1mm} $x '+\alpha x_g'$  \vspace{1.5mm}} & \multicolumn{2}{|c|}{$\bar{x}'$\phantom{pppp}} \\
	\parbox{57mm}{\vspace{1mm} $x_g'-x '$ \vspace{1.5mm}} & \multicolumn{2}{|c|}{$\tilde x'$\phantom{pppp}}\\
	\parbox{57mm}{\vspace{1mm} position of wave package induced\vspace{-1mm}\\ by collision\vspace{1.5mm}} &  ${ \bar{x}}_g$ & ${ \bar{x}} $
\end{tabular}
\end{center}

The following relations will be useful:
\begin{equation}
	\begin{array}{rcl}
		x_g+\Delta{x}_g&=& \widetilde{x}_g\\
		x +\Delta{x} &=& \widetilde{x} \\
		p_g+\Delta{p}_g&=& \widetilde{p}_g\\
		p +\Delta{p} &=& \widetilde{p} \\
		-\alpha\widetilde x_g&=&\widetilde x \\
		-\widetilde p_g&=&\widetilde p \\
		\Delta{x}_g&=&\Delta{x} \\
		\Delta{p}_g&=&\alpha\,\Delta{p} 
	\end{array} 
	\quad\Longrightarrow\quad
	\begin{array}{rcl}
		\widetilde x_g&=&\frac{ {x}_g- {x} }{1+\alpha}\vspace{1mm}\\
		\widetilde x &=&\alpha\frac{ {x} - {x}_g}{1+\alpha}\vspace{1mm}\\
		\,\Delta{x}_g=\,\Delta{x} &=&\frac{ {x} +\alpha {x}_g}{1+\alpha}\vspace{1mm}\\
		\widetilde p_g&=&\frac{ {p}_g-\alpha {p} }{1+\alpha}\vspace{1mm}\\
		\widetilde p &=&\frac{\alpha {p} - {p}_g}{1+\alpha}\vspace{1mm}\\
		\,\Delta{p}_g=\alpha\,\Delta{p} &=&\alpha\frac{ {p}_g+ {p} }{1+\alpha}
	\end{array}\label{46}
\end{equation}
The first four relations on the left hand side follow directly from \Eqref{45}, the next two are the properties of the centre-of-mass reference frame, and the last two are because $\widehat D=\hat D_g(\Delta{x}_g,\,\Delta{p}_g)\hat D (\Delta{x} ,\,\Delta{p} )$ represents a change of reference frame. Actually, it is because of the last four relations that there is no additional phase in \Eqref{45}. For given initial values $ {x}_g$, $ {p}_g$, $ {x} ,$ and $ {p} $, we can use the right hand side of \Eqref{46} to find the appropriate values $\,\Delta{x}_g$, $\,\Delta{p}_g$, $\,\Delta{x} ,$ and $\,\Delta{p} $ as well as $\widetilde {x}_g$, $\widetilde {p}_g$, $\widetilde {x} ,$ and $\widetilde {p} $ to use in \Eqref{45}.

Let us return to the time evolution
\begin{equation}
	\ket{ {\Psi}(t)} = e^{-iHt/\hbar}\ket{ {\Psi}(0)} = \widehat D\widehat D^{-1}e^{iHt/\hbar}\widehat D\ket{{\widetilde \Psi}(0)}, \label{47}
\end{equation}
where we also used a wide tilde to denote the two particle state in the centre-of-mass reference frame. We therefore need to calculate the operator 
\begin{eqnarray}
	\widehat D^{-1}e^{-iHt/\hbar}\widehat D &\!\!=\!\!& \exp\!\left[i\frac{\hat{p}_g\,\Delta{x}_g-\Delta{p}_g\,\hat{x}_g}{\hbar} + i\frac{\hat{p} \,\Delta{x} -\Delta{p}\, \hat{x} }{\hbar}  \right]   \nn\\
	&& \times   \exp\!\left[\frac{it}{\hbar}\!\left(\frac{\hat{p}_g^2}{2m_g}+\frac{\hat{p} ^2}{2m}+a\delta(\hat{x}_g\!-\!\hat{x} )\!\right)\right]  \nn\\
	&& \times   \exp\!\left[i\frac{\Delta{p}_g\,\hat{x}_g-\hat{p}_g\,\Delta{x}_g}{\hbar} + i\frac{\Delta{p} \,\hat{x} -\hat{p} \,\Delta{x} }{\hbar}  \right] \\
	&=& e^Ae^Be^{-A} \nn\\
	&=& \exp\!\left( B+[A,B]+\frac12[A,[A,B]] + \cdots \right). \label{qmf}
\end{eqnarray}
In the last line we used the Baker-Campbell-Hausdorff formula
	\begin{equation}
		\exp(X)\exp(Y)=\exp\!\left(X+Y+\half[X,Y]+\frac{1}{12}[X,[X,Y]]-\frac{1}{12}[Y,[X,Y]]+\cdots\right).\label{baker}
	\end{equation}
With the last two relations of the left hand side of \Eqref{46} the following commutators are found
	\begin{eqnarray}
		\frac{i\hbar}{t}[A,B] &=& \frac{\hat{p}_g\,\Delta{p}_g}{m_g}+\frac{\hat{p} \,\Delta{p} }{m } + (\Delta{x} -\Delta{x}_g)a\delta'(\hat{x} -\hat{x}_g) \nn\\
		&=& \frac{\hat{p}_g\,\Delta{p}_g}{m_g}+\frac{\hat{p} \,\Delta{p} }{m }\\
		\frac{i\hbar}{t}[A,[A,B]] &=& \frac{\Delta{p}_g^2}{m_g}+ \frac{\Delta{p} ^2}{m } 
	\end{eqnarray}
where $\delta'$ denotes the derivative of the delta function in its argument. Substituting the commutators into \Eqref{qmf}, we arrive at
	\begin{eqnarray}
		\widehat D^{-1}e^{-iHt/\hbar}\widehat D &=& \exp\!\left[\frac{-it}{\hbar} \left(\frac{(\hat{p}_g+\Delta{p}_g)^2}{2m_g} + \frac{(\hat{p} +\Delta{p} )^2}{2m } + a\delta(\hat{x}_g-\hat{x} ) \right) \right] \nn\\
		&=& \exp\!\left[  \frac{-it}{\hbar} \left(\frac{\Delta{p}_g^2}{2m_g} + \frac{\Delta{p} ^2}{2m }\right)  \right]   \exp\!\left[  \frac{-it}{\hbar} \left( \frac{\hat{p}_g\,\Delta{p}_g}{m_g}+\frac{\hat{p} \,\Delta{p} }{m } \right) \right] \exp\!\left[ \frac{-it}{\hbar}H\right], \nn\\
	\end{eqnarray}
where $(\alpha\,\Delta{p} -\Delta{p}_g)\delta'(\hat{x} -\hat{x}_g)=0$ was used in the second line. Since we know from the previous subsection that $e^{-iHt/\hbar}\ket{{\widetilde\Psi}(0)}=\ket{{\widetilde\Psi}(t)}$ with $\ket{{\widetilde\Psi}(t)}$ given in \Eqref{42}, this operator can now be substituted in \Eqref{47} to give
	\begin{eqnarray}
		\ket{ {\Psi}(t)} &=& \exp\!\left[  \frac{-it}{\hbar} \left(\frac{\Delta{p}_g^2}{2m_g} + \frac{\Delta{p} ^2}{2m }\right)  \right] \nn\\
		&& \times\; \hat D_g(\Delta{x}_g,\Delta{p}_g)\hat D_g\!\left(\frac{t\,\Delta{p}_g}{m_g},0\right)   \hat D (\Delta{x} ,\Delta{p} )\hat D \!\left(\frac{t\,\Delta{p} }{m },0\right)   \ket{{\widetilde \Psi}(t)}  \nn\\
		&=& \hat D_g\left(\Delta{x}_g+\frac{t\,\Delta{p}_g}{m_g},\Delta{p}_g\right)\hat D \left(\Delta{x} +\frac{t\,\Delta{p} }{m },\Delta{p}\right)  \ket{{\widetilde \Psi}(t)}  . \label{59}
	\end{eqnarray}

Next we need to determine the effect of the displacement operator on a single particle wave function. For that we apply the operator to the position eigenstates
	\begin{eqnarray}
		\hat D(x,p)\ket{x'} &=& e^{ixp/2\hbar}e^{-ix\hat{p}/\hbar}e^{i\hat{x}p/\hbar}\ket{x'} \nn\\
		&=& e^{ixp/2\hbar}e^{ix'p/\hbar}\ket{x'+x}.
	\end{eqnarray}
Therefore the transformation of the position representation of a wave function can easily be shown to be\vspace{2mm}
\begin{equation}
	\psi(x') \quad\parbox{13mm}{\small$\hat D(x,p)$\vspace{-2mm}\\ $-\!\!\!-\!\!\!-\!\!\!-\!\!\!\longrightarrow\vspace{-2mm}$\\\phantom{d}}\quad e^{-ixp/2\hbar}e^{ix'p/\hbar}\psi(x'-x). \label{61}
\end{equation}

By substituting the position representation \Eqref{41} into \Eqref{59} and by using \Eqref{61}, $\ket{ {\Psi}(0)}=\ket{ {x}_g, {p}_g}\otimes\ket{ {x} , {p} }$ evolves to
	\begin{equation}
		\framebox{\phantom{\LARGE ]}$\ket{ {\Psi}(t)} = U_g(t)\ket{{ \bar{x}}_g,{ \bar{p}}_g} \otimes U (t)\ket{{ \bar{x}} ,{ \bar{p}} }$\phantom{\LARGE ]}}   \label{main}
	\end{equation}
where ${ \bar{x}}$ and ${ \bar{p}}$ are given in terms of the initial values $ {x}$ and $ {p}$ (using \Eqref{46}) as follows

	\hspace{40mm}\framebox{\begin{minipage}{5.1cm}\vspace{-2mm}
		\begin{eqnarray}
			\quad{ \bar{x}}_g &=& \frac{2 {x} -(1-\alpha) {x}_g}{1+\alpha} \hspace{55mm} \label{noname2}\\
			{ \bar{p}}_g &=& \frac{2\alpha {p} -(1-\alpha) {p}_g}{1+\alpha} \label{noname3}\\
			{ \bar{x}}  &=& \frac{2\alpha {x}_g+(1-\alpha) {x} }{1+\alpha} \label{noname4}\\
			{ \bar{p}}  &=& \frac{2 {p}_g+(1-\alpha) {p} }{1+\alpha}.\label{noname} 
		\end{eqnarray} \vspace{-4mm}
	\end{minipage}}	\vspace{3mm}
	
As the evolution operator $U(t)$ shifts the position of a wave packet according to $ \bar{x}(t) = { \bar{x}}+ { \bar{p}}t/m$ (in addition to spreading the wave packet), these equations show that the centres of the wave packets in position as well as momentum space behave precisely the same as in a classical collision (see chapter~\ref{classcol}). This remarkable result will allow us to think almost classically when setting up the QBM master equation in chapters~\ref{rate} and~\ref{master}. The only difference will be that the interaction of the colliding particles is not instantaneous.

\section{Momentum and position jumps\label{jumps}}

After we have examined a single collision, we are in the position to discuss the appearance of position and momentum jumps (recall the definition of a jump from section~\ref{what}) during the collision process. The two particle wave packet during the collision in position representation is given in \Eqref{38} (without loss of generality we use the center-of-mass reference frame in this section).  In the first term the Brownian momentum is localized around $p $ and in the second term around $-p $. During the collision the first term decreases continuously while the second term increases. As the momentum distribution is effectively zero between $p$ and $-p$, this process accounts for momentum ``jumps" (although the jump is not instantaneous), and if these occur at random we are lead to momentum diffusion. In fact, already \Eqref{vrel} shows that the absolute relative velocity of gas and Brownian particle is a constant of motion (this is a consequence of the delta type interaction potential). Therefore, during the collision the eigenvalues for the relative velocity can swap the sign, but not change continuously. 

The situation is different in position space, as during the collision ($x +\frac{p t}{m }\approx0$) both terms  of \Eqref{38} are localized around the same position. To be more precise, we have a look at the position probability distribution $p(t,x ')$ of the Brownian particle which we obtain from \Eqref{38}
	\begin{eqnarray}
		p(t,x ') &=& \int \dd x_g'\, \left|\Psi(t,x_g',x ')\right|^2 \nn\\
		&=& \frac1{\sqrt{\pi\left(W ^2+\frac{\hbar^2t^2}{m ^2W ^2}\right)}}
		\left\{  Er\!\left[  \frac{\alpha x '-x -\frac{p t}{m }}{\sqrt{\alpha\left(W ^2+\frac{\hbar^2t^2}{m ^2W ^2}\right)}}  \right]      \exp\!\left[  \frac{\left(x '+x +\frac{p t}{m }\right)^2}{W ^2+\frac{\hbar^2t^2}{m ^2W ^2}}  \right] \right. \nn\\
		&& \left. +\;   Er\!\left[  \frac{\alpha x '+x +\frac{p t}{m }}{\sqrt{\alpha\left(W ^2+\frac{\hbar^2t^2}{m ^2W ^2}\right)}}  \right]      \exp\!\left[  \frac{\left(x '-x -\frac{p t}{m }\right)^2}{W ^2+\frac{\hbar^2t^2}{m ^2W ^2}}  \right]    \right\} .\label{posdens}
	\end{eqnarray}
Here we defined $Er(x)=\frac1{\sqrt\pi}\int_{-\infty}^x \dd t\, e^{-t^2}$ which is closely related to the error function and appears because \Eqref{38} is restricted to $x_g'<x '$. We further used $p_g\gg\frac{W_g}{\hbar}$ to avoid the interference of the two terms in \Eqref{38}, but this is only done to simplify the expression and not necessary to avoid position jumps. A plot of the position probability distribution over time during a collision can be seen in Fig.~\ref{figdis}~(a), and reminds of a classical collision. The probability distribution ``flows" continuously in position space and no jumps can be associated with such an behavior. As such, random collisions can not result in position diffusion according to $\left\langle\Delta x^2\right\rangle\propto t$ on the short time scale of few collisions. 
\begin{figure}[htbp]
	\includegraphics[width=\linewidth]{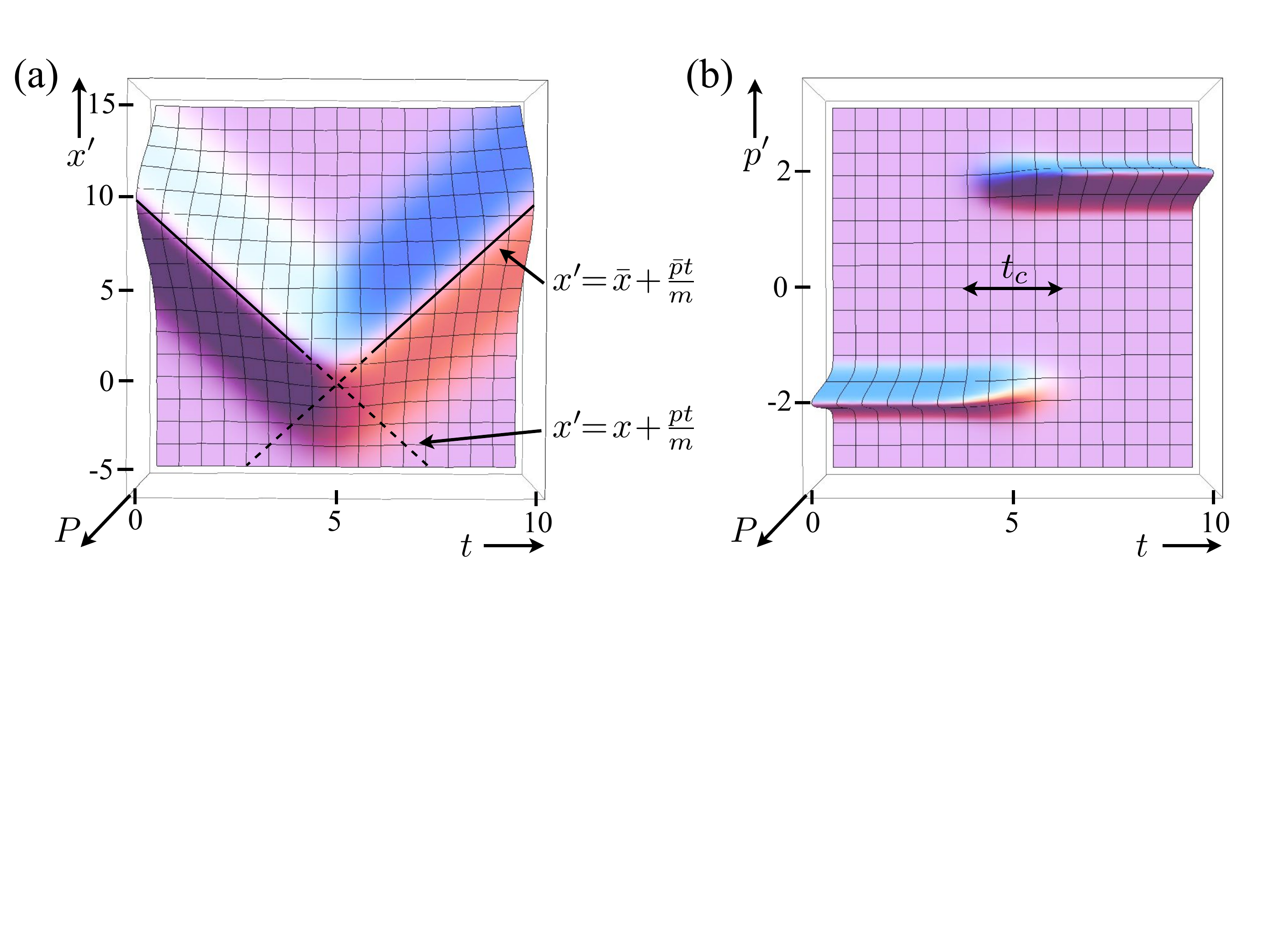}\vspace{-46mm}
\caption{\small Position (a) and momentum (b) probability distributions for the tracer particle during a collision in the reference frame of centre of mass. While the position probability distribution ``flows", the momentum distribution ``jumps" from the initial value to the final one. This disproves QPD. The solid lines in (a) are the corresponding classical trajectories. The collision time indicated in (b) is taken from \Eqref{collisiontime}.  Parameters in the centre of mass reference frame are: $x = 10$, $p = -2$, $m = 1$, $\alpha = 0.3$, $\hbar = 1$, and $W = 4$.}
\label{figdis}
\end{figure}

This has to be compared with the momentum probability distribution $p(t,p ')$ for which we again use \Eqref{38}
	\begin{eqnarray}
		p(t,p ') &=&  \int \dd x_g'\, \left|\Psi(t,x_g',p ')\right|^2 \nn\\
		\Psi(t,x_g',p ') &=& \frac1{\sqrt{2\pi\hbar}} \int_{x_g}^\infty \dd x '\, e^{-\imath p 'x '/\hbar} \Psi(t,x_g',x ').
	\end{eqnarray}
The lower bound of the integration is because \Eqref{38} is only valid for $x '>x_g'$ and $\Psi(t,x_g',x ')$ is zero for $x '<x_g'$. The integrations are rather lengthly and involves non-analytic functions. We therefore only show the resulting plot in Fig.~\ref{figdis}~(b) obtained numerically. We see that the momentum distribution does not ``flow" from the initial value towards the final value, but rather continuously decreases at the initial value and increases at the final value. It is this sort of behavior, which, if occurring at random, results in a diffusion process according to $\left\langle\Delta p^2\right\rangle\propto t$.

\subsection{Discussion of position diffusion\label{subb}}

In light of the conclusions of the previous subsection, one might wonder why all proposed QLBE exhibit QPD. In particular the necessity of QPD to achieve positivity of any Markovian QLBE seems to challenge our results at first glance. This apparent contradiction can however be resolved by taking into account that any Markovian master equation is only valid on a coarse grained time scale $\delta$, which has to be large compared to the collision time $t_c$, such that a collision can be considered as an instantaneous event. During a time interval $\delta$, a collision does not only change the momentum distribution, but also significantly influences the position distribution as is indicated by the arrow in Fig.~\ref{figjump}. That is, looking only at a coarse grained time grid, a collision results in apparent position jumps, therefore leading to fictitious position diffusion.
\begin{figure}[htbp]\begin{center}
	\includegraphics[width=0.7\linewidth]{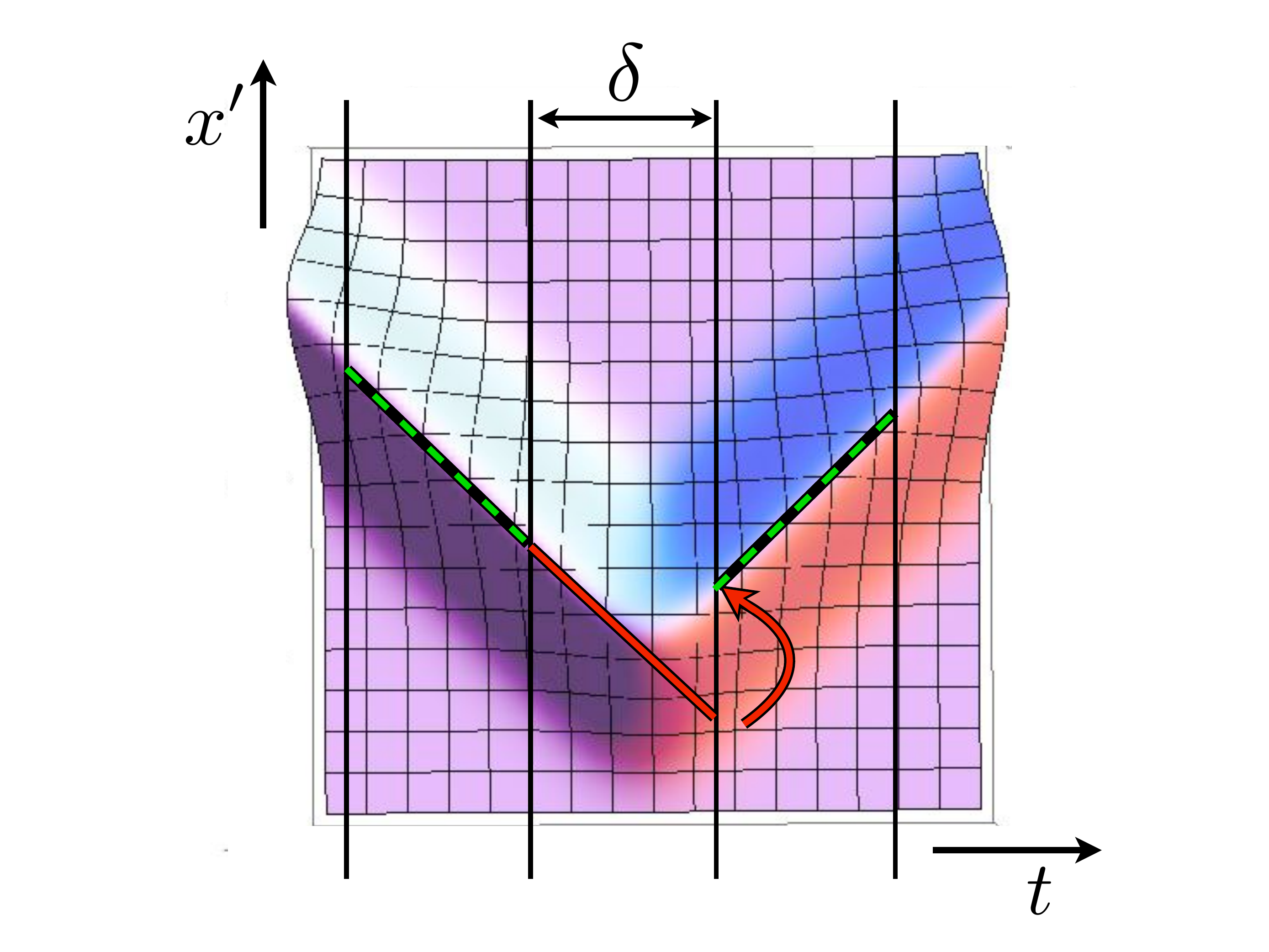}
\caption{\small During a time interval $\delta$, there can either be a collision (red) or no collision (green). On a coarse grained time grid a collision results in a discrete change of the position distribution (indicated by the arrow) compared to a no collision trajectory. As this discrete change occurs at random, it gives raise to position diffusion. It is also clear from the picture, that the same is true for classical trajectories.}
\label{figjump}
\end{center}\end{figure}

This is by no means a quantum feature, as also classical dynamics lead to position diffusion if they are derived using a coarse grained time scale. In particular Smoluchowski \cite{Einstein} used the damping time as coarse grained time scale to derive Brownian motion, one of the most fundamental diffusion processes in nature. However, in classical dynamics it is very much accepted that position diffusion results from a coarse graining time approximation and is not apparent on the short time scale of momentum diffusion. In fact, if the interaction between gas and tracer particle is of the hard core type, it is possible to derive classical Markovian dynamics without using a coarse grained time approximation~\cite{Fokker} (see also section~\ref{classmo}). The resulting Fokker-Planck equation for the phase space probability distribution (also called Kramers equation) then exhibits momentum diffusion, but not position diffusion.

The wave nature of quantum particles forbids instantaneous collisions even for hard core interaction potentials, and therefore the introduction of a time coarse graining approximation is  necessary to derive Markovian quantum dynamics. For the same reason as for classical calculations, the coarse grained time scale then leads to fictitious position diffusion on the short time scales. But again, it has to be realized that this contribution to position diffusion is not a physical process, even if any possible QLBE would indicate so.

\section{Scattering off a cat-state\label{cat}}

To get a first qualitative feeling about the process of collisional decoherence, we consider a gas particle in the state $\ket{\psi_g(0)}=\ket{{x}_g,{p}_g}$ to be scattered off the Brownian particle, which is initially in a superposition state $\ket{\psi (0)}=\ket{x_a ,p_a}+\ket{x_b ,p_b}$ (not normalized), known as a \emph{cat state} in the literature. Because the gas particle state after the collision depends on the initial position and momentum of the Brownian particle, we expect that the collision creates entanglement if the Brownian particle is initially in a cat state. Using \Eqref{main}~-~(\ref{noname}) we find for the two particle state after the collision
	\begin{eqnarray}
		\ket{\Psi(t)} &=& U_g(t)\ket{\bar{x}_g(x_g,x_a),\bar{p}_g(p_g,p_a)}\otimes U (t)\ket{\bar{x} (x_g,x_a),\bar{p} (p_g,p_a)} \nn\\
		&+& U_g(t)\ket{\bar{x}_g(x_g,x_b),\bar{p}_g(p_g,p_b)}\otimes U (t)\ket{\bar{x} (x_g,x_b),\bar{p} (p_g,p_b)},
	\end{eqnarray}
where $\bar x(x_g,x_a),\dots$ are given in \Eqref{noname2}~-~(\ref{noname}). Building the two particle density operator and tracing out the gas particle we find the Brownian particles density operator
	\begin{eqnarray}
		\rho (t) &=& U (t)\ket{\bar{x} (x_g,x_a),\bar{p} (p_g,p_a)}  \!\bra{\bar{x} (x_g,x_a),\bar{p} (p_g,p_a)}U ^\dagger \nn\\
		&+& U (t)\ket{\bar{x} (x_g,x_b),\bar{p} (p_g,p_b)}  \!\bra{\bar{x} (x_g,x_b),\bar{p} (p_g,p_b)}U ^\dagger \nn\\
		&+&  U (t)\ket{\bar{x} (x_g,x_a),\bar{p} (p_g,p_a)}  \!\bra{\bar{x} (x_g,x_b),\bar{p} (p_g,p_b)}U ^\dagger \nn\\
		&&\times \scalar{\bar{x}_g(x_g,x_b),\bar{p}_g(p_g,p_b)}{\bar{x}_g(x_g,x_a),\bar{p}_g(p_g,p_a)} \nn\\
		&+& \mbox{h.c.}, \label{472}
	\end{eqnarray}
where h.c. denotes the Hermitian conjugate. We see that due to the collision, the diagonals get shifted according to \Eqref{noname2}~-~(\ref{noname}). That also means, that the two humps in the position probability distribution get closer to each other, as is shown in figure~\ref{picsup}. One might think that after a large number of collision, both humps will eventually be on top of each other, which would contradict uniformity in space. However, this apparent problem is resolved by taking into account gas particles which are initially located between the two humps, therefore only colliding with one of them. When constructing up our master equation, we will therefore have to be careful with the collision statistics, which will be determined by a rate operator.\vspace{2mm}
\begin{figure}[htbp]
	\includegraphics[width=\linewidth]{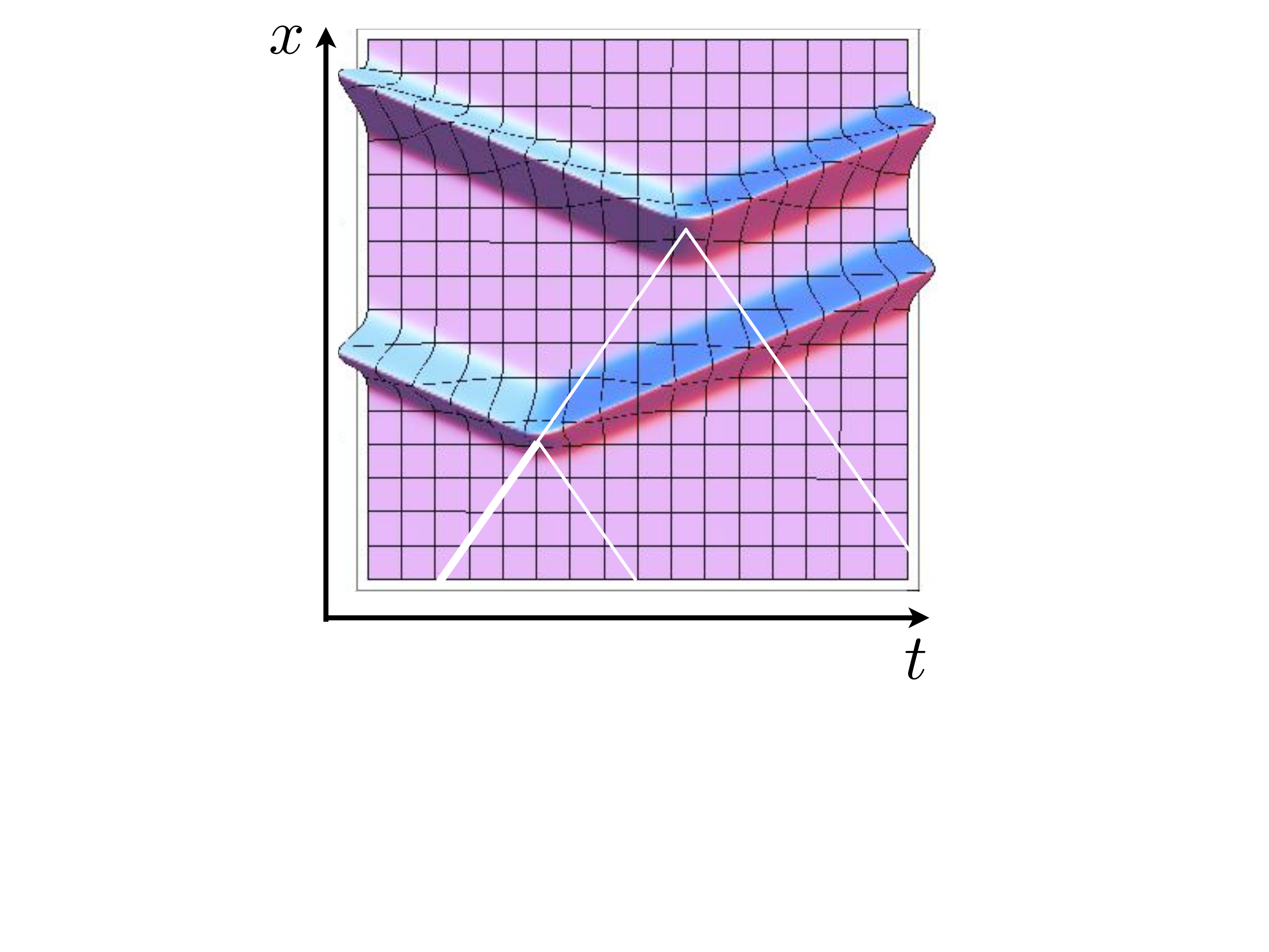}\vspace{-30mm}
\caption{\small Position distribution of an initial superposition of two Gaussian wave functions, both having the same mean momentum, but different mean positions. The lower hump experiences the collision first and then moves towards the upper hump, until this one also experiences the collision. The white lines correspond to the mean positions of the colliding gas particle.}
\label{picsup}
\end{figure}

In addition to a change of the distributions of momentum and position, a collision also results in a multiplication of the coherences by
	\begin{eqnarray}
		\scalar{\bar{x}_g(x_g,x_b),\bar{p}_g(p_g,p_b)}{\bar{x}_g(x_g,x_a),\bar{p}_g(p_g,p_a)} &=&\nn\\
		&& \hspace{-54mm} \exp\!\left[  \imath\frac{2\alpha(x_b p_a-x_a p_b) + (1-\alpha)p_g(x_a-x_b) - \alpha(1-\alpha)x_g(p_a-p_b)}{(1+\alpha)^2\hbar}  \right] \nn\\
		&\hspace{-104mm}\times& \hspace{-54mm} \exp\!\left[ -\frac\alpha{(1+\alpha)^2} \left(\frac{(x_a-x_b)^2}{W ^2} + \frac{W ^2(p_a-p_b)^2}{\hbar^2} \right)  \right].
	\end{eqnarray}
The collision has two effects on the off-diagonals. In the first line of the right hand side, there is a relative phase of the two wave packets, which depends on the state of the gas particle. In the second line, decoherence occurs in both, position and momentum, which reflects the post collisional entanglement of the two particles. Later, when we discuss a measurement interpretation of Brownian motion, the second line corresponds to decoherence due to a measurement, which the gas particle performs on the Brownian particle. 

It is important to realize that, if the state of the gas particle is taken randomly from a thermal distribution, also the first part will contribute to decoherence due to phase averaging and, in fact, in chapter~\ref{final} it will turn out to be by far the dominant contribution.

\section*{Appendix}

To compare the two contributions of \Eqref{38}, we take their $L^2$-norm defined by
	\begin{eqnarray}
		\| f \|^2 &=& \int\!\!\!\!\int \dd x_g'\,\dd x '\, | f(x_g',x ')|^2.
	\end{eqnarray}
 Rewriting the absolute value of each term of \Eqref{38} we find
 	\begin{eqnarray}
		\|\Psi_\pm\| &=&   \frac{2C^2{\sqrt{\alpha}}}{{\pi}\left(W ^2+\frac{\hbar^2 t^2}{m ^2W ^2}\right)}    \int\!\!\!\!\int_{x '>x_g'} \!\dd x_g'\,\dd x '\,  \exp\!\left\{  -\frac{\frac{\bar{x}'^2}{\alpha}+\left[\tilde{x}'\pm \frac{1+\alpha}{\alpha}\left(x +\frac{p t}{m }\right)\right]^2}    {\frac{1+\alpha}{\alpha}\left(W ^2+\frac{\hbar^2t^2}{W ^2m ^2}\right)}  \right\}\! \nn\\
		&=&    \frac{2C^2{\sqrt{\alpha}}}{{\pi}\left(W ^2+\frac{\hbar^2 t^2}{m ^2W ^2}\right)(1+\alpha)}    \int_{-\infty}^0 \!\!\dd\tilde x'  \!\int_{-\infty}^\infty\!\! \dd\bar x'  \exp\!\left\{  -\frac{\frac{\bar{x}'^2}{\alpha}+\left[\tilde{x}'\pm \frac{1+\alpha}{\alpha}\left(x +\frac{p t}{m }\right)\right]^2}    {\frac{1+\alpha}{\alpha}\left(W ^2+\frac{\hbar^2t^2}{W ^2m ^2}\right)}  \right\}\! \nn\\
		&=&  2C^2 \int_{-\infty}^{\pm\sqrt{\frac{1+\alpha}\alpha }\left(x +\frac{p t}{m }\right) \mbox{\Large/} \!\sqrt{W ^2+\frac{\hbar^2t^2}{m ^2W ^2}}} \dd\tilde x' \,e^{-\tilde x'^2}  \qquad \label{75}
	\end{eqnarray}
where the plus corresponds to the first term in the second line of \Eqref{38} and the minus to the second one. We used the transformation of the integration variables according to
	\begin{eqnarray}
		\left( \begin{array}{c} x_g'\\x '
		\end{array} \right)
		=
		\left( \begin{array}{c} \frac{\bar x'+\tilde x'}{1+\alpha}\\ \frac{\bar x' -\alpha \tilde x'}{1+\alpha} 
		\end{array} \right)
		&\Longrightarrow&
		\det J = \frac1{1+\alpha}
	\end{eqnarray}
 where $J$ is the Jacobi matrix. A term is small if the upper integration boundary is significantly negative. As $x >0$, the second term in \Eqref{38} is small at $t=0$ exactly if \Eqref{overlap} is satisfied. For $t\to\infty$ we find that the first term vanishes exactly if
 	\begin{eqnarray}
		p &\gg& \sqrt{\frac\alpha{1+\alpha}}\frac\hbar{W } \label{71}
	\end{eqnarray}
If inequality~(\ref{71}) does not hold, then the relative velocity of the wave packets is not large compared to their spreading, and therefore, there is a non-vanishing probability of having no collision at all. For later purposes it is useful to put \Eqref{71} into a form which is invariant in respect to Galilean transformations. Relation~(\ref{71}) was derived in the center-of-mass reference frame in which $(1+ \alpha)|p |=| \alpha  p -p_g|$ is valid. Therefore we find
	\begin{eqnarray}
		| \alpha p -p_g|&\gg& \sqrt{1+\alpha}\frac\hbar{W_g} \label{71b}
	\end{eqnarray}
which is valid for all $p $ and $p_g$ (not only in the centre-of-mass reference frame).

We can also deduce the collision time $t_c$ from \Eqref{75} as the time it takes for the upper integration boundary to change from $\pm 1$ to $\mp 1$. Using also relation~(\ref{71}) we find
	\begin{eqnarray}
		t_c &=& 2\sqrt{\frac{\alpha}{1+\alpha}}W \frac{m }{|p |} \label{77}
	\end{eqnarray}
or in a general reference frame
	\begin{eqnarray}
		t_c &=& 2\sqrt{{\alpha}(1+\alpha)}W \frac{m }{| \alpha p -p_g|} \nn\\
		&=&  \frac{2\sqrt{W_g^2+W^2}}{|v -v_g|}. \label{77b}
	\end{eqnarray}
 This is a very intuitive and expected result in that it is the time non-interacting wave packets need to cross each other.

\chapter[Measurement Interpretation]{State transformation from a measurement interpretation\label{meas}}

In this chapter we will derive a Kraus operator representation of the collisional transformation of the Brownian particle state. In doing so, we will work out measurements which the gas particle performs on the Brownian one. For that, we take a well prepared gas particle state $\ket{x_g,p_g}$ at $t=0$, and let the gas particle collide with the Brownian one which is assumed to be in an unknown state. Then we ask ourselves what information we can one obtain about the initial Brownian particle state, by performing a measurement on the gas particle after the collision. The answer depends, of course, on the type of measurement performed on the gas particle, and we have to find some kind of optimal measurement. 

The outline is much the same as usual in indirect quantum measurements (see section~\ref{genmeas}): One wants to get information about the system's state at time $t=0$. For that one first prepares some known state for the meter (here the gas particle). Then, the meter is made to interact with the system (the Brownian particle), and finally, at some time $t_f$ after the interaction one performs a measurement on the meter. If the initial state of the meter and the interaction are chosen appropriately, then the result of the meter measurement gives the sought information about the system.

Several standard textbooks discuss measurements of this outline (see e.g.\ \cite{Breuer} and \cite{Busch}). However, the examples are usually limited to two simplifications: First, the interaction of meter and system is taken to be instantaneous, meaning that it is switched on and off quickly enough such that the free evolution of meter and system can be neglected during the interaction time. This can certainly not be true in our case, as actually the free evolution of gas and Brownian particle ``switches" on the interaction by bringing the particles together. Second, the measurement on the meter is usually of projection type (or a smeared out version), whereas the measurement we will apply on the gas particle is not of this type, but a quantum-limited position-momentum measurement.

We are not aware of any examples of the most general type (none of the above simplifications) of indirect measurements. Because of this we feel that the study of collisional measurements is a very interesting one by itself. But most importantly, we will employ results of this chapter later when setting up a master equation for the Brownian particle.

\section{Effect operators}

Although the state of the Brownian particle prior to the collision is not known, it is sufficient to consider Brownian particle states of the form $\ket{x_{},p_{}}$ to uniquely determine the effect operators, because $\{\ket{x,p}\}$ is a (over-)complete set of states. We recall from the preceding chapter that if $m_{}W_{}^2 = m_g W_g^2$, the scattering process can be described by
	\begin{eqnarray}
		\ket{x_g,p_g}\otimes\ket{x_{},p_{}} \qquad\parbox{2cm}{{\small scattering}\\$\overrightarrow{\phantom{scattering}}$} \quad		U_g(t_f)\ket{\bar{x}_g,\bar{p}_g}\otimes U_{}(t_f) \ket{\bar{x}_{},\bar{p}_{}} \qquad\label{one}
	\end{eqnarray}
where $U_g(t)$ and $U_{}(t)$ are the evolution operator for the free gas particle and the free Brownian particle, respectively, and $t_f$ is any time after the collision took place. The momentums and positions after the collision relate with the initial values as in \Eqref{noname2}~-~(\ref{noname}). Since at time $t_f$ the two particles are in a product state, a measurement on the gas particle can not change the state of the Brownian one. 

We want to perform a measurement on the gas particle, which is \emph{optimal} (see efficient measurements in section~\ref{genmeas}) in obtaining information about the Brownian particle. The right hand side of \Eqref{one} suggests to use $U_g(t_f)\ket{x_{g},p_{g}}$ as measurement basis on the gas particle. The corresponding effect operators are
\begin{equation}
	\hat{\pi}_g(\tilde{x}_g,\tilde{p}_g)\;\;=\;\;\frac{1}{2\pi\hbar}U_g(t_f)\ket{\tilde{x}_g,\tilde{p}_g}\bra{\tilde{x}_g,\tilde{p}_g}U_g^\dagger (t_f), \label{effectgas}
\end{equation}
where measurement outcomes are denoted with a tilde. Keep in mind that the measurement on the gas particle is done after the interaction took place, and therefore the gas particle effect operators \Eqref{effectgas} are applied at time $t_f$. As the process should give information about the Brownian particle's state before the collision, the Brownian particle's effect operators are applied at time $t=0$.

After we have chosen the effect operators \Eqref{effectgas} acting on the gas particle's Hilbert space, we can determine the corresponding effect operators $\hat{\pi}_{}(\tilde{x}_{},\tilde p_{})$ for the Brownian particle by comparing the respective probabilities (see Eq.~(3.16) of~\cite{Busch})
	\begin{eqnarray}
		&&\bra{x_{},p_{}}\hat{\pi}_{}(\tilde{x}_{},\tilde p_{})\ket{x_{},p_{}} \dd\tilde x_{}\,\dd\tilde p_{}  \nn\\
		& \!\!\equiv\!& \!\left[ \bra{\bar x_g,\bar p_g}U^\dagger_g(t_f)\otimes  \bra{\bar x_{},\bar p_{}}U^\dagger_{}(t_f)\right] \!\left[ \hat{\pi}_g^{\phantom{2}}(\tilde x_g,\tilde p_g) \!\otimes \!\one \right]\! \left[ U_{}(t_f)\ket{\bar x_{},\bar p_{}}\otimes U_g^{\phantom{\dagger}}(t_f) \ket{\bar x_g,\bar p_g}\right]\! \dd\tilde x_g\,\dd\tilde p_g \nn\\
		& \!\!=\!& \frac 1{2\pi\hbar} |\!\!\scalar{\bar x_g,\bar p_g}{\tilde x_g,\tilde p_g}\!|^2 \,\dd\tilde x_g\,\dd\tilde p_g \nn\\
		& \!\!=\!&   \frac {\dd\tilde x_g\,\dd\tilde p_g}{2\pi\hbar}   \exp\!\left\{  -\frac{[2x_{}-(1\!-\!\alpha)x_g-(1\!+\!\alpha)\tilde x_g]^2}{2(1\!+\!\alpha)^2W_g^2} -\frac{[2\alpha p_{}-(1\!-\!\alpha)p_g-(1\!+\!\alpha)\tilde p_g]^2W_g^2}{2(1\!+\!\alpha)^2\hbar^2}  \right\}\!, \nn\\  \label{5.7}
	\end{eqnarray}
where \Eqref{noname2}~-~(\ref{noname}) were used in the last equality. This equation can be solved for all $(x_{},p_{})$ with the ansatz
	\begin{eqnarray}
		\hat\pi_{}(\tilde x_{},\tilde p_{}) &=&\frac1{2\pi\hbar} \int\!\!\!\!\int\dd x'\,\dd p'\, w(x',p') \ket{\tilde x_{}+x',\tilde p_{}+p'}\!\bra{\tilde x_{}+x',\tilde p_{}+p'}\nn\\
		w(x,p) &=&  \frac{2\alpha}{\pi\hbar(1-\alpha)^2} \exp\!\left[ -\frac{2\alpha}{(1-\alpha)^2} \left(\frac{x^2}{ W_{}^2}+\frac{ W_{}^2p^2}{\hbar^2}\right) \right] \label{5.8}
	\end{eqnarray}
with $W_{}^2=\alpha W_g^2$. Substituting \Eqref{5.8} into the left hand side of \Eqref{5.7}, we find after performing two Gaussian integrals, that
	\begin{eqnarray}
		\bra{x_{},p_{}}\hat{\pi}_{}(\tilde{x}_{},\tilde p_{})\ket{x_{},p_{}} \dd\tilde x_{}\dd\tilde p_{} &\!\!=\!\!& \frac {\dd\tilde x_{}\,\dd\tilde p_{}}{2\pi\hbar}\frac{4\alpha}{(1\!+\!\alpha)^2} \exp\!\left[  -
		\frac{2(x_{}-\tilde x_{})^2}{W_g^2(1\!+\!\alpha)^2} - \frac{2\alpha^2W_g^2(p_{}-\tilde p_{})^2}{\hbar^2(1\!+\!\alpha)^2} \right]\!. \qquad\;\; \label{doe}
	\end{eqnarray}
This has to be compared to the right hand side of \Eqref{5.7} to find
	\begin{eqnarray}
		\tilde x_{} &=& \frac{1-\alpha}2 x_g + \frac{1+\alpha}2 \tilde x_g \label{5.10}\\
		\tilde p_{} &=& \frac{1-\alpha}{2\alpha} p_g + \frac{1+\alpha}{2\alpha} \tilde p_g \label{5.11}\\
		\dd\tilde x_{}\,\dd\tilde p_{} &=& \frac{(1+\alpha)^2}{4\alpha}\dd\tilde x_g\,\dd\tilde p_g.
	\end{eqnarray}
That is, from the measured values $(\tilde x_g,\tilde p_g)$ we can use \Eqref{5.10} and (\ref{5.11}) to infer the approximate position $\tilde x$ and momentum $\tilde p$ of the Brownian particle. The measurement on the Brownian particle described by the effect operators \Eqref{5.8} is a smeared out version of a quantum-limited simultaneous position-momentum measurement. Whether the measurement is more precise in position or momentum depends on the choice of $W_g^2$ of the initial gas state. The smearing out results in additional uncertainty on top of the Heisenberg uncertainty limit, and can be reduced by choosing a gas particle with a similar mass as the Brownian particle. 

In the limit of a light gas particle we see from \Eqref{doe} that the position precision of the indirect measurement is half the position variance $W_g$ of the initial meter state $\ket{x_g,p_g}_{W_g}$. On the other hand, the momentum precision is reduced by a factor $\alpha$ compared to half the momentum variance $\hbar/W_g$. The reason is that in this limit the gas particle measures twice the velocity of the Brownian particle\footnote{This is because in the limit $\alpha\to 0$ the gas particle velocity after the collision is $\bar v_g=-v_g+2v$. A measurement of $\bar v_g$ with some uncertainty $\Delta \bar v_g$ then results in half of this uncertainty for the velocity $v$.}, and the precision is its own velocity variance. As momentum is velocity times mass, the momentum precision is reduced by the ratio $m_g/m$.

For $\alpha=1$ we find $w(x,p)=\delta(x,p)$ and the effect operators take the form of a quantum-limited position-momentum measurement. Of course, if $\alpha=1$ then a collision just swaps the states of the two colliding particles, and therefore a measurement on the gas particle after the collision is equivalent to a measurement on the Brownian one before the collision. Hence, a quantum-limited indirect measurement is no surprise if both particles have the same mass. Furthermore, because both particles swap their state, the Brownian particle carries no information about its pre collisional state and a subsequent measurement can not reveal any more information, therefore assuring that the Heisenberg uncertainty principle is not violated.

Note that the effect operators \Eqref{5.8} were already used (but not derived from first principals) in \cite{mainII} to phenomenologically describe collisional measurements. But the authors used free parameters to specify the precision of momentum and position measurements, and were not able to give a relation to the mass and the initial state of the gas particle.

It is often stated in the literature (e.g.\ \cite{review}), that a colliding particle measures the position of a tracer particle, but not the momentum. In this section we showed clearly that this statement is not generally true. Instead, we find a beautiful symmetry between measurements of momentum and position, despite an interaction potential which solely depends on position variables.

\section{Kraus operators}

In this section we are concerned about the change of the Brownian particle's state vector due to a measurement. Without loss of generality we assume that the Kraus operators acting on the gas particle's Hilbert space are the square root of the effect operators (using more general Kraus operators give the same result for the Kraus operators acting on the Brownian partcle)
	\begin{eqnarray}
		\hat A_g(\tilde x_g,\tilde p_g) &=& \frac1{\sqrt{2\pi\hbar}}U_g(t_f)\ket{\tilde x_g,\tilde p_g}\!\bra{\tilde x_g,\tilde p_g}U_g^\dagger(t_f).
	\end{eqnarray}
Therefore, a measurement with result $(\tilde{x}_g,\tilde{p}_g)$ changes the two particle state \Eqref{one} to (omitting normalization)
\begin{eqnarray}
	\frac{1}{\sqrt{2\pi\hbar}}\scalar{\tilde{x}_g,\tilde{p}_g}{\bar{x}_g,\bar{p}_g}U_g(t_f)\ket{\tilde{x}_g,\tilde{p}_g}U_{}(t_f)\ket{\bar{x}_{},\bar{p}_{}}. \label{seven}
\end{eqnarray}

The Kraus operators acting on the Brownian particle depend on the measurement result $(\tilde{x}_{},\tilde{p}_{})$ as well as on the state of the gas particle which was used for the measurement, i.e.\ $(x_g,p_g)$. Measurement theory~\cite{Busch} states that the Kraus operators can be written as
\begin{equation}
	\hat A_{}(x_g,p_g;\tilde{x}_{},\tilde{p}_{})\;\;=\;\;\hat V_{}(x_g,p_g;\tilde{x}_{},\tilde{p}_{})\sqrt{\hat{\pi}_{}(\tilde{x}_{},\tilde{p}_{})}, \label{5.15}
\end{equation}
 with some unitary operator $\hat V_{}(x_g,p_g;\tilde{x}_{},\tilde{p}_{})$ which is yet to be specified. Together with \Eqref{seven} we see that this operator has to satisfy
 \begin{equation}
	U_{}^\dagger(t_f)\hat V_{}(x_g,p_g;\tilde{x}_{},\tilde{p}_{})\sqrt{\hat{\pi}_{}(\tilde{x}_{},\tilde{p}_{})}\ket{x_{},p_{}}\;\;=\;\;\frac{1}{\sqrt{2\pi\hbar}}\scalar{\tilde{x}_g,\tilde{p}_g}{\bar{x}_g,\bar{p}_g}\ket{\bar{x}_{},\bar{p}_{}}. \label{thirteen}
\end{equation}
The phase of the scalar product depends on the state of the Brownian particle and because the Kraus operators will act on general states (i.e.\, on superposition states) we have to keep track of it. We find for this phase
\begin{eqnarray}
	\arg\scalar{\tilde{x}_g,\tilde{p}_g}{\bar{x}_g,\bar{p}_g} &=& \frac{2\alpha(p_{}\tilde{x}_{}-\tilde{p}_{}x_{}) + \alpha(1-\alpha)x_g(\tilde{p}_{}-p_{})+(1-\alpha)p_g(x_{}-\tilde{x}_{})}{\hbar(1+\alpha)^2},\qquad\qquad \label{fourteen}
\end{eqnarray}
where the relations \Eqref{noname2}, (\ref{noname3}), (\ref{5.10}), and (\ref{5.11}) were used. We further need
\begin{eqnarray}
	\sqrt{\hat{\pi}_{}(\tilde{x}_{},\tilde{p}_{})}\ket{x_{},p_{}} &\!=\!& c\exp{\!\left( i\frac{\alpha(p_{}\tilde{x}_{}-\tilde{p}_{}x_{})}{\hbar(1+\alpha)} \right)}\ket{\frac{1-\alpha}{1+\alpha}x_{}+\frac{2\alpha}{1+\alpha}\tilde{x}_{},\frac{1-\alpha}{1+\alpha}p_{}+\frac{2\alpha}{1+\alpha}\tilde{p}_{}}, \qquad\quad\; \label{fifteen}
\end{eqnarray}
where $c$ is a real number. The derivation of \Eqref{fifteen} is a bit longer and can be found in the appendix of this chapter. Next we substitute \Eqref{noname4} and (\ref{noname}) into the right hand side of \Eqref{thirteen}, and \Eqref{fifteen} into the left hand side, to find
\begin{eqnarray}
	U_{}^\dagger(t_f)\hat V_{}(x_g,p_g;\tilde{x}_{},\tilde{p}_{}) &=& \exp\!\left( i\frac{\alpha x_g\tilde{p}_{}-\tilde{x}_{}p_g}{\hbar(1+\alpha)} \right) \hat{D}\!\left( \frac{2\alpha(x_g-\tilde{x}_{})}{1+\alpha}, \frac{2(p_g-\alpha\tilde{p}_{})}{1+\alpha} \right)\!,\qquad\quad\; \label{sixteen}
\end{eqnarray}
where \Eqref{fourteen} was used, as well as the Glauber displacement operator $\hat{D}(x,p)$ for the Brownian particle. Substituting into \Eqref{5.15} we finally get the Kraus operators
	\begin{eqnarray}
		\hat A_{}(x_g,p_g;\tilde{x}_{},\tilde{p}_{}) &=& \exp\!\left( i\frac{\alpha x_g\tilde{p}_{}-\tilde{x}_{}p_g}{\hbar(1+\alpha)} \right) U_{}(t_f) \hat{D}\!\left( \frac{2\alpha(x_g-\tilde{x}_{})}{1+\alpha}, \frac{2(p_g-\alpha\tilde{p}_{})}{1+\alpha} \right)\!\sqrt{\hat{\pi}_{}(\tilde{x}_{},\tilde{p}_{})}. \nn\\ \label{5.20}
	\end{eqnarray}

As required, the Kraus operators do depend on the measurement outcome $(\tilde{x}_{},\tilde{p}_{})$ and the state of the measurement apparatus, i.e.\ $(x_g,p_g)$, but not on the state of the Brownian particle before the measurement.

The displacement in position seems to be odd, but that can be resolved by considering the finite time $t_f$ the measurement (or the collision of the wave packets) takes, as well as the free Brownian particle evolution operator $U_{}(t_f)$. In fact, if one assumes the collision takes place instantaneously at the time when the centers of the wave functions of the gas and the Brownian particle are equal, i.e.\ $t_f=0$ and $x_{}=x_g$, then the displacement operator \Eqref{sixteen} just shifts the state \Eqref{fifteen} back to the original position $x_{}$. Therefore the non-zero position value in the displacement operator is needed to counteract the position movement due to the effect operator $\sqrt{\hat{\pi}_{}(\tilde{x}_{},\tilde{p}_{})}$.

One might wonder why the effect operator decomposes with only one Kraus operator rather than the more general decomposition $\hat\pi=\sum_i \hat A_i^\dagger \hat A_i$. The reason for this is twofold. First, we used a gas state which is initially pure, thus not introducing classical uncertainties. Second, we used an optimal measurement on the gas particle in the sense that we measured the basis in which correlations between the two particles are most pronounced.

Now we are in the position to write down the state of the Brownian particle at time $t_f$ after a collision with a gas particle, which initially is in the state $\ket{x_g,p_g}$. Because in a real collision we do not actually perform any measurement on the gas particle, we need to use the general formula for a non-readout measurement (not measuring the meter accounts for not reading out the measurement result)
	\begin{equation}
		\rho_{}(t_f) \;\;=\;\; \int\!\!\!\!\int\dd \tilde x_{}\,\dd\tilde p_{}\, \hat A_{}(x_g,p_g;\tilde x_{},\tilde p_{})\rho_{}(0)\hat A_{}^\dagger(x_g,p_g;\tilde x_{},\tilde p_{}). \label{5.21}
	\end{equation}
A nice and important property of this transformation is that it is valid for general states of the Brownian particle, and not just for Gaussian ones considered so far.

Later on we will need the Kraus operators without the free evolution part. Furthermore, as \Eqref{5.21} shows, we do not have to take care of phase factors which only result in a global phase. Therefore we define
	\begin{eqnarray}
		\hat B(x_g,p_g;\tilde{x}_{},\tilde{p}_{}) &\dot =& \hat{D}_{}\!\left( \frac{2\alpha}{1+\alpha}(x_g-\tilde{x}_{}), \frac{2}{1+\alpha}(p_g-\alpha\tilde{p}_{}) \right)\sqrt{\hat{\pi}_{}(\tilde{x}_{},\tilde{p}_{})}  \nn\\
		&\dot =& \frac1{\sqrt{2\pi\hbar}} \hat{D}\!\left( \frac{(1-\alpha)\tilde x +2\alpha x_g}{1+\alpha}, \frac{(1-\alpha)\tilde  p+2p_g}{1+\alpha} \right) \sqrt{\hat \sigma}\, \hat{D}(\tilde x,\tilde p) .\qquad\qquad \label{5.22}
	\end{eqnarray}
where $\dot =$ denotes equal up to a phase, and
	\begin{eqnarray}
		\hat \sigma &=& 2\pi\hbar\,\hat\pi(0,0) \nn\\ 
		&=&  \int\!\!\!\!\int\dd x'\,\dd p'\,w(x',p')\ket{x',p'}\bra{x',p'}  \label{5.23}
	\end{eqnarray}
is a trace one operator with vanishing first moments of position and momentum. Then, \Eqref{5.21} reads
	\begin{equation}
		\rho_{}(t_f) \;\;=\;\; U(t_f)\left[ \int\!\!\!\!\int\dd \tilde x_{}\,\dd\tilde p_{}\, \hat B(x_g,p_g;\tilde x_{},\tilde p_{})\rho_{}(0)\hat B^\dagger(x_g,p_g;\tilde x_{},\tilde p_{}) \right] U(t_f)^\dag . \label{5.24}
	\end{equation}

\section*{Appendix}

Here we will derive the action of the square root of the effect operators on the Gaussian states. This can be done either in position representation, resulting in a very messy calculation, or, by a much neater algebraic calculation, which we choose here. To do so, we introduce some harmonic oscillator algebra along the lines of Barnett and Cresser~\cite{mainII}. They defined annihilation, creation, and number operators via 
	\begin{eqnarray}
		\hat a &=& \frac{\hat x}{\sqrt2 W}+i\frac{W\hat p}{\sqrt2\hbar} \\
		\hat a^\dag &=& \frac{\hat x}{\sqrt2 W}-i\frac{W\hat p}{\sqrt2\hbar} \\
		\hat n \;=\; \hat a^\dag\hat a &=& \frac{\hat x^2}{2W^2}+\frac{W^2\hat p^2}{2\hbar^2}-\half ,
	\end{eqnarray}
and the eigenstates $\ket n$, $n=0,\,1,\,2,\,\dots$ of $\hat n$,
	\begin{eqnarray}
		\hat n\ket n&=& n\ket n.
	\end{eqnarray}

They then found that (see \Eqref{jak} for the relation of $\hat \sigma$ to the effect operator)
	\begin{eqnarray}
		\hat \sigma &=& \left(1-e^{-\lambda}\right)e^{-\lambda\hat n} \\
		&=& \frac{1}{1+\bar n} \sum_{n=0}^\infty  \left(\frac{\bar n}{1+\bar n}\right)^n\ket n\!\bra n
	\end{eqnarray}
with $\lambda = \ln(1+1/\bar n)$. From the first line we deduce
	\begin{eqnarray}
		\sqrt{\hat \sigma} &=& \sqrt{1-e^{-\lambda}}e^{-\lambda \hat n/2}. \label{ijb}
	\end{eqnarray}

Also useful is that the Gaussian pure states are eigenstates of the creation operator
	\begin{eqnarray}
		\hat a \ket{x,p}_W &=& \left( \frac{x}{\sqrt2 W}+i\frac{W p}{\sqrt2\hbar}\right) \ket{x,p}_W \nn\\
		&\equiv& \alpha(x,p) \ket{x,p}_W ,
	\end{eqnarray}
which is easily shown, e.g.\ in position representation. From this, we can derive the following formulas
	\begin{eqnarray}
		\scalar{n}{x,p} &=& \frac1{\sqrt n}\langle n-1|\hat a|x,p\rangle \;=\; \frac{\alpha(x,p)}{\sqrt n}\scalar{n-1}{x,p} \nn\\
		&=&  \frac{\alpha^n(x,p)}{\sqrt{n!}} \scalar{n=0}{x,p} \;=\; e^{-|\alpha(x,p)|^2/2} \frac{\alpha^n(x,p)}{\sqrt{n!}}, \qquad \nn\\
		\left\langle e^{-\lambda/2}x,e^{-\lambda/2}p \left| e^{-\lambda\hat n/2} \right| x,p \right\rangle 
		&=& \sum_{n=0}^\infty \left\langle e^{-\lambda/2}x,e^{-\lambda/2}p \left| e^{-\lambda\hat n/2} \right| n\right\rangle\scalar{n}{x,p} \nn\\
		&=& \exp\!\left[-\frac{|\alpha(x,p)|^2}{2} \left(1-e^{-\lambda}\right) \right] \label{qnda}
	\end{eqnarray}

We can now use these formulas to derive the action of the square root of the effect operators on the Gaussian states, which will turn out to be necessary to determine the Kraus operators corresponding to each effect operator. To this end we use $\hat a e^{-\lambda \hat n/2}=e^{-\lambda/2} e^{-\lambda \hat n/2} \hat a$ to show that $e^{-\lambda \hat n/2} \ket{x,p}$ is an eigenstate of $\hat a$ with eigenvalue $e^{-\lambda/2}\alpha(x,p)$, and therefore $e^{-\lambda \hat n/2} \ket{x,p}\propto \ket{e^{-\lambda/2}x,e^{-\lambda/2}p}$. The proportionality constant is  obtained from \Eqref{qnda} and with \Eqref{ijb} we arrive at
	\begin{eqnarray}
		\sqrt{\hat \sigma}\ket{x,p} &=& \sqrt{1-e^{-\lambda}} \exp\!\left[-\frac{|\alpha(x,p)|^2}{2} \left(1-e^{-\lambda}\right) \right] \ket{e^{-\lambda/2}x,e^{-\lambda/2}p} \qquad
	\end{eqnarray}
By using \Eqref{ekd} it is now straightforward to find the action of the square root of the effect operator on a Gaussian state
	\begin{eqnarray}
		\sqrt{\hat\pi(\tilde x,\tilde p)} \ket{x,p} &=& \frac{\sqrt{1-e^{-\lambda}}}{\sqrt{2\pi\hbar}}\exp\!\left\{  \frac{-i}{2\hbar} \left[ \tilde px-p\tilde x-e^{-\lambda/2}[\tilde p(x-\tilde x)-\tilde x(p-\tilde p)]\right]\right\} \quad\nn\\
		&&\times\; \exp\left\{ -\frac14\left(1-e^{-\lambda}\right) \left[\left(\frac{x-\tilde x}{W}\right)^{\!2} + \left(\frac{W(p-\tilde p)}{\hbar}\right)^{\!2} \right]  \right\} \nn\\
		&&\times\; \ket{e^{-\lambda/2}(x-\tilde x)+\tilde x, e^{-\lambda/2}(p-\tilde p)+\tilde p}.
	\end{eqnarray}
To use this result for the measurement a gas particle performs on the Brownian particle, we compare \Eqref{5.8} in chapter~\ref{meas} with \Eqref{pqy} to relate
	\begin{equation}
		e^{-\lambda} \;=\; \frac{\bar n}{\bar n+1} \;=\; \left(\frac{1-\alpha}{1+\alpha}\right)^{\!2},
	\end{equation}
where $\alpha=m_g/m$ is the ratio of the masses of the two particles. Then we find 
	\begin{eqnarray}
		\sqrt{\hat\pi(\tilde x,\tilde p)} \ket{x,p} &=& \frac{2\sqrt\alpha}{\sqrt{2\pi\hbar}(1+\alpha)} \exp\left\{ \frac{-\alpha}{(1+\alpha)^{2}} \left[\left(\frac{x-\tilde x}{W}\right)^{\!2} + \left(\frac{W(p-\tilde p)}{\hbar}\right)^{\!2} \right]  \right\} \nn\\
		&&\times\;  \exp\!\left[ \frac{i\alpha(\tilde xp-x\tilde p)}{(1-\alpha)\hbar}\right] \ket{\frac{2\alpha\tilde x +(1-\alpha)x}{1+\alpha},\frac{2\alpha\tilde p +(1-\alpha)p}{1+\alpha}} .\qquad\qquad\;
	\end{eqnarray}

\chapter{Rate Operator\label{rate}}

Since we know the effect of a collision if the gas particle is in a Gaussian state $\ket{x_g,p_g}$, we now have to ask what is the rate of such collisions. As in classical collisions, the rate will depend on the state of the Brownian particle. For example, the Brownian particle will never collide with a gas particle moving with the same velocity. Because of this state dependence, the quantum mechanical collision rate will have to be operator valued. 

In the first section we will discuss the probability of finding a gas particle in a state $\ket{x_g,p_g}$. In the next section we show, for a given gas particle state, under what conditions a collision with the Brownian particle will occur.

\section{Gas density operator\label{convex decomp}}

Before we derive the probability of finding a gas particle in a state $\ket{x_g,p_g}$, we start with some important remarks about the density operator of a gas particle of an ideal Boltzmann gas in a thermal state. In particular we wish to point out that there are many different convex decompositions
	\begin{eqnarray}
		\rho_g &=& \int\dd\alpha\, P(\alpha) \ket{\psi_g(\alpha)}\!\bra{\psi_g(\alpha)} \label{convex}
	\end{eqnarray}
of such a thermal density operator. Here $\alpha$ is a finite set of parameters, $\ket{\psi_g(\alpha)}$ is a set of normalized gas particle states (not necessarily orthogonal) spanning the Hilbert space of a gas particle, and $P(\alpha)\geq0$ is the probability density of the gas particle to be in the state $\ket{\psi_g(\alpha)}$. This freedom has some important consequences, one of which is that the notion of a collision time $t_c$, which is roughly the time required for a wave packet of the tracer particle to cross  the width of the wave packet of a gas particle (a more precise definition is given in section~\ref{timeevo}), is not to be viewed as a physical parameter whose value is intrinsic to the gas, but rather a parameter whose value can be chosen to suit the needs of the analysis. But already here it is clear that momentum eigenstates which are often used in \Eqref{convex} are not appropriate to define a collision time.

The thermal state of an ideal gas particle decomposed in momentum eigenstates is usually given in terms of the Maxwell-Boltzmann distribution $\mu_{T}(p)$
	\begin{eqnarray}
		\rho_g &=& \frac{2\pi\hbar}{L} \mu_{T}(\hat p_g) \nn\\
		&=& \frac{\hbar}{L}\sqrt{\frac{2\pi}{m_gk_BT}} \exp\!\left( \frac{-\hat p_g^2}{2m_gk_B T} \right) \\
		&=& \frac{\hbar}{L}\sqrt{\frac{2\pi}{m_gk_BT}} \int \dd p_g\, \exp\!\left( \frac{-p_g^2}{2m_gk_B T} \right) \ket{p_g}\!\bra{p_g}\nn
	\end{eqnarray}
where $T$ is the temperature of the environment and $L$ is a normalization length. Hornberger and Sipe~\cite{Hornberger} showed how this density operator can also be decomposed in projectors to minimum uncertainty wave packets $\ket{x_g,p_g}$. For this purpose we need use the equality 
	\begin{equation}
		\exp\!\left(\frac{-\hat p_g^2}{2m_gk_BT}\right) = \sqrt{\frac{T}{2\pi m_gk_B\widetilde{T}\overline T}}\int\dd p_g\, \exp\!\left( \frac{-p_g^2}{2m_gk_B\widetilde{T}} \right)  \exp\!\left( \frac{-(\hat p_g-p_g)^2}{2m_gk_B\overline T} \right)
	\end{equation}
which is true if $\overline{T}+\widetilde{T}=T$ as is easily shown by performing the integration. Then we split the  second term and include the identity operator $\one_g=\int_L\dd x_g\,\ket{x_g}\!\bra{x_g}$ in position representation to find
	\begin{equation}
		\rho_g =\sqrt{\frac{2\pi\hbar^2}{m_gk_B\overline{T}}} \int_L\frac{\dd x_g}{L}\int\dd p_g\, \mu_{\widetilde{T}}(p_g) \exp\!\left( \frac{-(\hat p_g-p_g)^2}{4m_gk_B\overline T} \right) \ket{x_g}\!\bra{x_g} \exp\!\left( \frac{-(\hat p_g-p_g)^2}{4m_gk_B\overline T}\right)
	\end{equation}
where 
	\begin{equation}
		\mu_{\widetilde{T}}(p_g) = \frac{1}{\sqrt{2\pi m_g k_B\widetilde{T}}} e^{-p_g^2/2m_gk_B\widetilde{T}} \label{12}
	\end{equation}
is the Maxwell-Boltzmann distribution with modified temperature $\widetilde{T}$. Finally we have to show that $\sqrt{\frac{2\pi\hbar^2}{m_gk_B\overline{T}}}\exp\!\left( \frac{-(\hat p_g-p_g)^2}{4m_gk_B\overline T} \right) \ket{x_g}$ are the desired Gaussian states by going to the position representation
	\begin{eqnarray}
		 \bra{x_g'} \left(\frac{2\pi\hbar^2}{m_gk_B\overline{T}}\right)^{\!\frac14} &&\!\!\!\!\!\!\!\!\!\!\!\!\!  \exp\!\left( \frac{-(\hat p_g-p_g)^2}{4m_gk_B\overline T} \right) \ket{x_g}\nn\\
		&\!\!=& \left(\frac{2\pi\hbar^2}{m_gk_B\overline{T}}\right)^{\!\frac14} \int \!\dd p\,\exp\!\left( \frac{-( p-p_g)^2}{4m_gk_B\overline T} \right) \scalar{x_g'}{p}  \!\scalar{p}{x_g} \nn\\
		&\!\!=&  \left(\frac{1}{(2\pi)^3\hbar^2m_gk_B\overline{T}}\right)^{\!\frac14} \int \!\dd p\,\exp\!\left( \frac{-( p-p_g)^2}{4m_gk_B\overline T} +\imath\frac{p(x_g'-x_g)}{\hbar} \right)\nn\\
		&\!\!=&\left(\frac{1}{(2\pi)^3\hbar^2m_gk_B\overline{T}}\right)^{\!\frac14} \int \!\dd p\,\exp\!\left[ -\!\left(\frac{p\!-\!p_g}{4m_gk_B\overline T}-\imath\frac{x_g'-x_g}{2\hbar}\right)^{\!2} \right. \nn\\
		&& \phantom{hallohallohallohallohallohh}  \left.+\imath\frac{p_g(x_g'\!-\!x_g)}{\hbar} - \frac{m_gk_B\overline T(x_g'\!-\!x_g)^2}{\hbar^2}\right] \quad\nn\\
		&\!\!=& \left(\frac{2m_gk_B\overline{T}}{\pi\hbar^2}\right)^{\!\frac14} \exp\!\left( \imath\frac{p_g(x_g'-x_g)}{\hbar} - \frac{m_gk_B\overline T(x_g'-x_g)^2}{\hbar^2}\right) \nn\\
		&\!\!=& \frac{e^{\imath p_g(x_g'-x_g)/\hbar}}{\sqrt{\sqrt{\pi}W_g}} \exp\!\left(-\frac{(x_g'-x_g)^2}{2W_g^2}\right)
	\end{eqnarray}
where $W_g=\frac\hbar{\sqrt{2m_gk_B\overline T}}$ is defined. Up to a global phase, that is exactly the position representation of $\ket{x_g,p_g}$, and we arrive at the desired decomposition of the thermal density operator into Gaussian states
	\begin{equation}
		\rho_g = \int \frac{\dd x_g}{L} \int \dd p_g\, \mu_{\widetilde{T}}(p_g) \ket{x_g,p_g}\bra{x_g,p_g}. \label{thermalgas}
	\end{equation}
We see that part of the thermal energy is associated with the wave packets themselves, while the rest is in the motion of the centres of the wave packets via $\mu_{\widetilde{T}}(p_g)$. As our collisional calculation is only valid if the mean momentum of the gas particle is large compared to its momentum uncertainty, we will use \Eqref{thermalgas} with 
	\begin{equation}
		T\gg\overline T,\quad\widetilde{T}\approx T\label{6.7}
	\end{equation}
in the following.

\section{Definition of the rate operator \label{ratedef}}

The probability for the Brownian particle to collide with a gas particle state $\ket{x_g,p_g}$, during a time interval $\delta{}$, should be determined by the expectation value of a rate operator $ \hat R_{\delta{}}(x_g,p_g)$ times the time interval $\delta{}$. Because we know the state of the thermal gas particles, and we are looking for an operator acting on the Brownian particle, it is advantageous to rephrase the question: \emph{If at time $t$ we are given a gas particle in the state $\ket{x_g,p_g}$, what is the probability that it collides with the Brownian particle in the interval $(0,\delta{})$?}

In classical physics, for a given gas particle with position $x_g$ and momentum $p_g$, there will be a collision with the Brownian particle exactly if its position and momentum are within the phase space region
\begin{eqnarray}
	S_\delta(x_g,p_g) &=& \left\{ (x,p) \left|  0<\frac{x_{}-x_g}{p_g/m_g-p_{}/m}<\delta{} \right\}\right. . \label{6.8}
\end{eqnarray}
That is, the particles either collide or do not collide. There are no partial collisions because classical hard core collisions are instantaneous. This phase space region is shown in figure~\ref{lct}~(a).

\begin{figure}[htbp]\begin{center}
	\includegraphics[width=0.8 \linewidth]{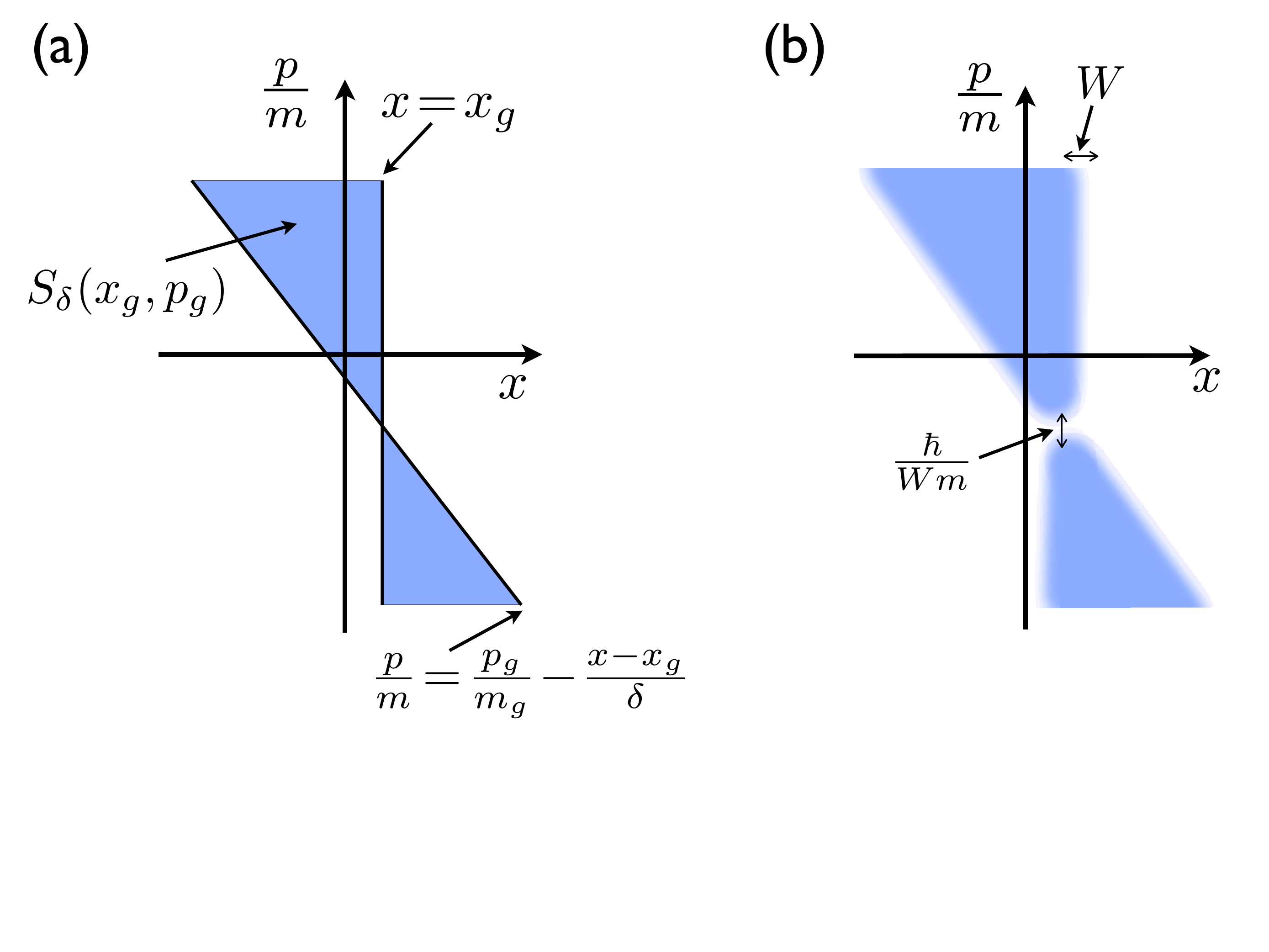} \vspace{-20mm}
\caption{\small(a): If a classical particle is in the phase space region $S_\delta(x_g,p_g)$, then it will collide with a classical gas particle with position $x_g$ and velocity $p_g$ during a time interval $\delta$. (b): $\hat \Gamma_{\delta}(x_g,p_g)$ acts like a projection operator well within $S_\delta(x_g,p_g)$, but not near the edges of this phase space region which is indicated by the blurry area. If $\delta$ is sufficiently large, the edges can be neglected in comparison to the interior.}
\label{lct}
\end{center}\end{figure}

In a quantum collision we have to be more careful. Because the collision of wave packets is not instantaneous, we have to account for the possibility that at time $\delta{}$ a collision is not complete. However, if we choose $\delta{}$ sufficiently large
\begin{eqnarray}
	\delta{} &\gg& t_c \;=\; \frac{2\sqrt{(1+\alpha)}W_g}{|v-v_g|}, \label{6.10}
\end{eqnarray}
where $t_c$ is the collision time \Eqref{77b}, then the probability of a complete collision will outweight the probability of an incomplete one, and we can neglect the latter. Note that relation~(\ref{6.10}) can not be fulfilled for $v_{}$ close to $v_g$, but for sufficiently large $\delta{}$ (and/or for high temperatures) relation (\ref{6.10}) can be achieved for most gas particle velocities. We can then use guidance from the above classical statements also for a quantum collision, at least in an approximate manner.

The aim now is to restrict the collision to Brownian particle states which are located within the phase space region $S_\delta(x_g,p_g)$. To this end we define the quantum mechanical ``phase space projection operator"
\begin{equation}
	 \hat \Gamma_{\delta}(x_g,p_g) = \frac{1}{2\pi\hbar} \int\!\!\!\!\int_{S_\delta(x_g,p_g)}\dd x_{}\,\dd p_{}\, \ket{x_{},p_{}}\bra{x_{},p_{}}. \label{6.11}
\end{equation}
Strictly speaking, phase space projections can not exist in quantum mechanics because of the Heisenberg uncertainty principle. But if the Brownian particle is in a Gaussian state $\ket{x_{},p_{}}$, with values of $x_{}$ and $p_{}$ well within $S_\delta(x_g,p_g)$, then
\begin{equation}
	 \hat \Gamma_{\delta}(x_g,p_g)\ket{x_{},p_{}} \approx \ket{x_{},p_{}} \label{6.12}
\end{equation}
and there will certainly be a collision. On the other hand, if $(x,p)$ is well outside of $S_\delta(x_g,p_g)$, then
\begin{equation}
	 \hat \Gamma_{\delta}(x_g,p_g)\ket{x_{},p_{}} \approx 0  \label{6.13}
\end{equation}
and there will be no collision. If $\delta{}$ obeys relation~(\ref{6.10}), then the weight of the values $x_{}$ and $p_{}$ which are near the boundary of $S_\delta(x_g,p_g)$ becomes small, and 
	\begin{equation}
		 \hat \Gamma_{\delta}^2(x_g,p_g)\;\approx\;  \hat \Gamma_{\delta}(x_g,p_g) \label{6.14}
	\end{equation}
is indeed an approximate projector operator. The action of this operator can be visualized in Fig.~\ref{lct}~(b).

To get the rate operator we also have to consider the probability of having a gas particle in the state $\ket{x_g,p_g}$. If the gas is an ideal one and in a thermal state, we saw in the previous section that this probability is
\begin{equation}
	n_g\mu_{\widetilde{T}}(p_g) = \frac{n _g}{\sqrt{2\pi m_gk_B\widetilde{T}}} \exp\!\left[\frac{-p_g^2}{2m_gk_B\widetilde{T}}\right],  \label{6.15}
\end{equation}
where $n_g$ is the particle density of the gas and $\widetilde{T}=T-\frac{\hbar^2}{2m_gk_BW_g^2}$. Furthermore, we divide the operator by the time interval $\delta{}$ to get a rate, and arrive at
\begin{eqnarray}
	 \hat R_{\delta{}}(x_g,p_g) &\!=\!& n_g\mu_{\widetilde{T}}(p_g) \frac{ \hat \Gamma_{\delta}(x_g,p_g)}{\delta{}} \nn\\
	&\!=\!& \frac{n _g}{\delta{}\sqrt{2\pi m_gk_B\widetilde{T}}} \exp\!\left[\frac{-p_g^2}{2m_gk_B\widetilde{T}}\right] \int\!\!\!\!\int_{S_\delta(x_g,p_g)} \frac{\dd x_{}\,\dd p_{}}{2\pi\hbar}  \ket{x_{},p_{}}\bra{x_{},p_{}} . \qquad\quad \label{6.16}
\end{eqnarray}
Note that the rate operator depends on the time interval $\delta{}$, which is not uniquely fixed by physical parameters, but can be chosen freely as long as it satisfies \Eqref{6.10}. However, the trace norm Tr$[ \hat R_{\delta{}}(x_g,p_g)]$ of the rate operator does not depend on $\delta{}$. Also we will see that on a coarse grained time scale the resulting master equation does not depend on this parameter.

\subsection{Integrated rate operators}


We also define a rate operator corresponding to any collision with a Gaussian gas state with mean momentum $p_g$
\begin{eqnarray}
	 \hat R(p_g) &=& \int \dd x_g\,  \hat R_{\delta{}}(x_g,p_g) \label{6.17}\\
	&=& n_g \frac{\mu_{\widetilde{T}}(p_g)}{\delta{}}  \int \!\!\!\!  \int\!\!\!\!\int_{0<\frac{x_{}-x_g}{\frac{p_g}{m_g}-\frac{p_{}}{m_{}}}<\delta{} }\dd x_g\frac{\dd x_{}\,\dd p_{}}{2\pi\hbar}  \ket{x_{},p_{}}\bra{x_{},p_{}}   \nn\\
	&=& \frac{n _g}{\sqrt{2\pi m_gk_B\widetilde{T}}} \exp\!\left(\frac{-p_g^2}{2m_gk_B\widetilde{T}}\right) \int\!\!\!\!\int  \frac{\dd x_{}\,\dd p_{}}{2\pi\hbar}  \left| \frac{p_g}{m_g}-\frac{p_{}}{m_{}} \right| \ket{x_{},p_{}}\bra{x_{},p_{}} \quad \nn\\
	&\approx& \frac{n _g}{\sqrt{2\pi m_gk_BT}} \exp\!\left(\frac{-p_g^2}{2m_gk_BT}\right)  \left| \frac{p_g}{m_g}\!-\!\frac{\hat p_{}}{m_{}} \right|  \label{6.18}
\end{eqnarray}
where \Eqref{6.7} as well as \Eqref{f(p)} was used. The integration over $x_g$ is easily performed because the integrand does not depend on this value (see \Eqref{8.19} for more detail). Furthermore we define the total collision rate
\begin{eqnarray}
	 \hat R &\!=\!& \int \dd p_g\,  \hat R(p_g) \label{6.19b}\\
	&\!=\!& \frac{n _g\sqrt{2k_B\widetilde{T}}}{\sqrt{\pi m_g}} \! \int\!\!\!\!\int \frac{\dd x_{}\,\dd p_{}}{2\pi\hbar}  \!\left\{\! \exp\!\left[ -\frac{m_gp_{}^2}{2k_B\widetilde{T}m_{}^2}\right]\!  +  \frac{\sqrt{\pi m_g}p_{}}{\sqrt{2k_B\widetilde{T}}m_{}} \mbox{erf}\!\left[\frac{\sqrt{ m_g}p_{}}{\sqrt{2k_B\widetilde{T}}m_{}}\right] \! \right\}\! \ket{x_{},p_{}}\bra{x_{},p_{}}\! \nn\\
	&\!\approx\!& \frac{n _g\sqrt{2k_BT}}{\sqrt{\pi m_g}}  \left\{ \exp\!\left[ -\frac{m_g\hat p_{}^2}{2k_B{T}m_{}^2}\right]  +  \frac{\sqrt{\pi m_g }\hat p_{}}{\sqrt{2k_B{T}}m_{}} \mbox{erf}\!\left[\frac{\sqrt{ m_g}\hat p_{}}{\sqrt{2k_B{T}}m_{}}\right]  \right\}\!, \label{6.19}
\end{eqnarray}
where we used again \Eqref{6.7} and \Eqref{f(p)}. The integration over $p_g$ is also straightforward if the cases $ \frac{p_g}{m_g}\!-\!\frac{p_{}}{m_{}} \gtrless 0$ are considered separately. Note that neither $ \hat R(p_g)$ nor $ \hat R$ depend on $\delta{}$. In fact, these collision rates are precisely the ones found in the classical calculations in section~\ref{classmo}.

Also useful will be the total collision rate in the limit of slow Brownian velocities $|v_{}|\ll |v_g|\approx\sqrt{k_BT/m_g}$
\begin{equation}
	 \hat R\;\,\approx\;\, \frac{n _g\sqrt{2k_BT}}{\sqrt{\pi m_g}} \left( \one+\frac{m_g}{2k_BT}\frac{\hat p_{}^2}{m^2_{}}\right)\!, \label{6.20}
\end{equation}
as well as 
	\begin{eqnarray}
		U_{}^\dagger(t) \hat R(p_g)U_{}(t)&=&  \hat R(p_g) \label{6.22} \\
		U_{}^\dagger(t) \hat R U_{}(t)&=& \hat R. \label{6.23}
	\end{eqnarray}

\chapter{Master Equation\label{master}}

Now, as we have developed formulas for the change of the state due to a collision  \Eqref{5.21}, as well as for the rate of a collision \Eqref{6.16}, we are in the position to derive a master equation describing the Brownian particle experiencing random collisions with gas particles of an ideal gas in a thermal state.

\section{Discrete master equation}

Let us first consider the change of the Brownian particle state during some finite time $\delta{}$. We assume that $\delta{}$ is small compared to the collision frequency
	\begin{equation}
		\delta{}\, \mbox{Tr} [ \hat R \rho_{}(0)]\ll1,
	\end{equation}
such that we can neglect the possibility of two collisions in $(0,\delta{})$. If there is no collision in this time period, then the density operator undergoes the unitary transformation
	\begin{eqnarray}
		\rho_{}(\delta{}) &=& U_{}(\delta{})\rho_{}(0)U_{}^\dagger(\delta{}),
	\end{eqnarray}
whereas if there is a collision with a gas particle in the state $\ket{x_g,p_g}$ the transformation is according to \Eqref{5.24}
	\begin{eqnarray}
		\rho_{}(\delta{}) &=& U(\delta) \left[ \int\!\!\!\!\int \dd \tilde x_{}\,\dd\tilde p_{} \, \hat B(x_g,p_g;\tilde x_{},\tilde p_{})\rho_{}(0)\hat B^\dagger(x_g,p_g;\tilde x_{},\tilde p_{}) \right] U^\dag(\delta).\label{7.3}
	\end{eqnarray}
To find the actual density operator at time $t=\delta{}$, we have to account for collisions with all possible gas particle states with the correct probabilities. Quantum trajectory theory \cite{Gardiner} suggests to multiply each Kraus operator by the square root of the probability operator
	\begin{eqnarray}
		\rho_{}(\delta{}) &=& U_{}(\delta{}) \bigg[ \sqrt{1-\delta \hat R}\rho_{}(0)\sqrt{1-\delta \hat R} + \int\!\!\!\!\int\!\!\!\!\int\!\!\!\!\int\dd x_g\,\dd p_g\,\dd\tilde x_{}\,\dd\tilde p_{}\,    \nn\\
		&& \times\; \hat B(x_g,p_g;\tilde x_{},\tilde p_{})\sqrt{\delta \hat R_\delta{}(x_g,p_g)}\rho_{}(0)\sqrt{\delta \hat R_\delta{}(x_g,p_g)}\hat B^\dag(x_g,p_g;\tilde x_{},\tilde p_{})\bigg] U_{}^\dag(\delta{}), \qquad \;\;\; \label{7.4}
	\end{eqnarray}
where $1-\delta \hat R$ gives the probability of having no collision. The probability operator $\delta \hat R_\delta{}(x_g,p_g)$ ensures that only the part of $\rho(0)$ which is within the ``range'' of $\ket{x_g,p_g}$ undergoes the transition \Eqref{7.3}, weighted by the thermal distribution $\mu_{T}(p_g)$. The ``range'' is defined by \Eqref{6.8}. The integration in \Eqref{7.4} is over all possible gas particle states $(x_g,p_g)$, as well as over all possible ``measurement outcomes" $(\tilde x, \tilde p)$. 

It seems advantageous to use the interaction picture $\rho_{int}(t)=U_{}^\dagger(t)\rho_{}(t)U_{}(t)$ for the following
	\begin{eqnarray}
		\rho_{int}(\delta{}) &=& \sqrt{1-\delta \hat R}\rho_{int}(0)\sqrt{1-\delta \hat R} + \delta{} \int\!\!\!\!\int\!\!\!\!\int\!\!\!\!\int\dd x_g\,\dd p_g\,\dd\tilde x_{}\,\dd\tilde p_{}\,    \nn\\
		&& \times \; \hat B(x_g,p_g;\tilde x_{},\tilde p_{})\sqrt{  \hat R_\delta{}(x_g,p_g)}\rho_{int}(0)\sqrt{  \hat R_\delta{}(x_g,p_g)}\hat B^\dag(x_g,p_g;\tilde x_{},\tilde p_{}). \qquad\;\;  \label{7.5}
	\end{eqnarray}
Because $\delta$ is small compared to the inverse collision rate, we can expand $\sqrt{1- \delta \hat R}\approx 1- \delta \hat R/2$ to find
	\begin{eqnarray}
		\frac{\rho_{int}(\delta{})-\rho_{int}(0)}{\delta{}} &=& -\half [ \hat R\rho_{int}(0)+\rho_{int}(0) \hat R] + \int\!\!\!\!\int\!\!\!\!\int\!\!\!\!\int \dd x_g\,\dd p_g\,\dd\tilde x_{}\,\dd\tilde p_{}\, \nn\\
		&&  \times\; \hat B(x_g,p_g;\tilde x_{},\tilde p_{})\sqrt{  \hat R_\delta{}(x_g,p_g)}\rho_{int}(0)\sqrt{  \hat R_\delta{}(x_g,p_g)}\hat B^\dag(x_g,p_g;\tilde x_{},\tilde p_{}) \quad \nn\\ 
		&=& \mathcal L_\delta{}[\rho_{int}(0)] \label{7.7}
	\end{eqnarray}
which is the discrete master equation in the interaction picture. Let us have a quick look at this equation. The operators $\hat B$ and $\hat B^\dagger$, integrated over $\tilde x$ and $\tilde p$, transform $\ket{x,p}\to\ket{\bar x,\bar p}$. The rate operator ensures that only states undergo this transformation which are in the ``range'' of a gas particle within the state $\ket{x_g,p_g}$ during the time interval $(0,\delta{})$. Furthermore, this operator includes the probability of finding a gas particle in the respective state. Finally, the integration over $x_g$ and $p_g$ averages over all possible gas states. The anti commutator arises from the non-collision part, and ensures a constant trace of the density operator.

\section{Continuous master equation \label{7.2}}

To derive a continuous master equation from \Eqref{7.7}, one would wish to take the limit $\delta\to 0$. Here we encounter a problem. From the discussion in section~\ref{ratedef}, we know that we can not take this limit in the rate operator $ \hat R_\delta{}(x_g,p_g)$, because no complete collisions can occur in an infinitesimal time interval, and in the derivation of the rate operator we required the ability to neglect incomplete collision. In fact, as can be seen from \Eqref{6.11}, taking this limit in the rate operator would account for performing quantum-limited position-momentum measurements on the Brownian particle, and these are by no means performed by a colliding gas particle. Therefore $\delta$ strictly has to satisfy relation~(\ref{6.10}).\footnote{One could say that the rate operator in the master equation is supposed to measure the collision rate without disturbing the state much. This can only be achieved approximately if the rate operator performs a very imprecise measurement, which in turn is achieved by a large $\delta$.}

If we now perform the continuity transition 
	\begin{eqnarray}
		\frac{\rho_{int}(\dd t)-\rho_{int}(t)}{\dd t} &\approx & \frac{\rho_{int}(\delta{})-\rho_{int}(0)}{\delta{}}  ,\label{7.8}
	\end{eqnarray}
we arrive at a master equation, which accounts for the free evolution in an exact manner (because we take the continuity transition in the interaction picture), and for the collisions in first order in $\delta{}$. This transition essentially takes all the possible collisions which might occur in $(0,\delta{})$ and performs the corresponding state transformations (with probability reduced by $\dd t/\delta{}$) at time $t=0$ rather than spread out during $(0,\delta{})$. \textbf{It is this step where position diffusion is introduced.} (See also figure~\ref{figjump}.) If we look at the real evolution of a Brownian particle (initial state $\ket{x_{},p_{}}$) colliding with a gas particle (initial state $\ket{x_g,p_g}$), we find that the position of the Brownian wave packet evolves continuously from $x_{}$ to $\bar x_{} + \delta{} \bar p_{}/m_{}$  during the time $\delta{}$, with $\bar x_{}$ and $\bar p_{}$ given by \Eqref{noname4} and~(\ref{noname}), as is clearly seen in Fig~\ref{figdis}. However, in the continuous master equation, which is derived from the discrete master equation, there is a jump from $x_{}$ to $\bar x_{}$ occurring at time $t=0$, and therefore position jumps are introduced. In fact, it is well known that any attempt to derive a Markovian master equation (and with our set up we can only derive this type of master equation) describing QBM without introducing position jumps must be ill-fated~\cite{Diosi}, as it can not be brought into the required Lindblad form.


If we were to use the continuity transition~(\ref{7.8}), the position jumps would not be uniform. In fact we would find for the position expectation value (for simplification we take $|v_{}|\ll\sqrt{k_BT/m_g}$, more about the derivation of expectation values in chapter~\ref{expectation})
	\begin{equation}
		\frac{\dd \langle \hat x_{}\rangle}{\dd t} \;\,=\,\; \frac{\langle\hat p_{}\rangle}{m_{}} \left( 1- \frac{4\alpha\sqrt{k_BT}n _g}{(1+\alpha)\sqrt{2\pi m_g}}\delta{} \right), \label{7.9}
	\end{equation}
where the latter part is due to the non-uniform position jumps. This part is proportional to the number of collisions and therefore small and hence \Eqref{7.9} would be an unorthodox, but acceptable result of a master equation. However, it is straightforward to avoid such unphysical contributions linear in $\delta{}$ by noting that for any smooth function $\frac{f(\delta)-f(0)}{\delta}=\frac{f(\delta/2+d t)-f(\delta/2)}{d t} + O(\delta^2)$ is valid. Therefore we will perform the continuity transition to second order in $\delta{}$
	\begin{eqnarray}
		 \frac{\rho_{int}\left(\frac{\delta{}}2+\dd t\right)-\rho_{int}\left(\frac{\delta{}}2\right)\! }{\dd t} &\approx & \frac{\rho_{int}(\delta{})-\rho_{int}(0)}{\delta{}} \nn\\
		 &=& \mathcal L_\delta{}[\rho_{int}(0)] .\label{7.10}
	\end{eqnarray}
This transition essentially takes all the possible collisions which might occur in $(0,\delta{})$ and performs the corresponding state transformations (with probability reduced by $\dd t/\delta{}$) at time $t=\delta/2$. As this time lies symmetrically in the above time interval, the (non-physical) position jumps will be uniform and not show up in the position expectation value. But they will show up in second order in $\delta{}$ in the position variance, which is unavoidable if one is seeking for a Markovian description of QBM.

After we have done the continuity transition, we can return Schr\"odinger picture by using
	\begin{eqnarray}
		\frac{\rho_{int}\!\left(\frac{\delta{}}2\!+\!\dd t\right)-\rho_{int}\!\left(\frac{\delta{}}2\right)}{\dd t}  
		&=& U_{}^\dagger\!\left(\tfrac{\delta{}}2\right)\!   \frac{U_{}^\dagger(\dd t)\rho\!\left(\frac{\delta{}}2\!+\!\dd t\right)U_{}(\dd t) - \rho\!\left(\frac{\delta{}}2\right)\! }{\dd t} U_{}\!\left(\tfrac{\delta{}}2\right)\!  \nn\\
		&=& U_{}^\dagger\!\left(\tfrac{\delta{}}2\right)\!   \left[  \frac{\dd \rho_{}\!\left(\frac{\delta{}}2\right)\! }{\dd t} + \frac\imath\hbar \left[ H_{},\rho_{}\!\left(\frac{\delta{}}2\right)\! \right]   \right]  U_{}\!\left(\tfrac{\delta{}}2\right)\!  .\qquad\quad
	\end{eqnarray}
In the second step we expanded the free evolution operator $U_{}(\dd t) = \one - \frac\imath\hbar H_{}\dd t$. Finally we substitute the discrete master equation \Eqref{7.7} in the right hand side of \Eqref{7.10}, noting that $\rho_{int}(0)=\rho(0)$, and multiply with the free particle evolution operator from the left and right to find
	\begin{eqnarray}
		\frac{\dd \rho_{}\!\left(\frac{\delta{}}2\right)\! }{\dd t} &\!=\!& -\frac\imath\hbar \left[ H_{},\rho_{}\!\left(\tfrac{\delta{}}2\right)\! \right] - U_{}\!\left(\tfrac{\delta{}}2\right)\!  \bigg\{ \half [ \hat R\rho_{}(t)+\rho_{}(t) \hat R]  \int\!\!\!\!\int\!\!\!\!\int\!\!\!\!\int\dd x_g\,\dd p_g\,\dd\tilde x_{}\,\dd\tilde p_{}\, \bigg. \nn\\
		&& -    \hat B(x_g,p_g;\tilde x_{},\tilde p_{})\sqrt{  \hat R_\delta{}(x_g,p_g)}\rho_{}(0)\sqrt{  \hat R_\delta{}(x_g,p_g)}\hat B^\dag(x_g,p_g;\tilde x_{},\tilde p_{}) \bigg\} U_{}^\dagger\!\left(\tfrac{\delta{}}2\right)\!.\qquad  \quad\;\; \label{7.12}
	\end{eqnarray}
 As last step we substitute $\rho_{}(0)\approx U_{}^\dagger\!\left(\tfrac{\delta{}}2\right)\!  \rho_{}\!\left(\frac{\delta{}}2\right)\!  U_{}^\dagger\!\left(\tfrac{\delta{}}2\right)\! $ into the incoherent part of \Eqref{7.12}, which is a justified as long as two collisions during $\delta$ are negligible. Finally we use homogeneity in time to replace $\left(\frac{\delta{}}{2}\right)$ by $t$ to arrive at our final master equation (also using \Eqref{6.23}): \vspace{2mm}\\
  	\framebox{\begin{minipage}{14.4cm}
	\begin{eqnarray}
		\frac{\dd \rho_{}(t)}{\dd t} &=& -\frac i\hbar \left[ H_{},\rho_{}(t)\right] -\half [ \hat R\rho_{}(t)+\rho_{}(t) \hat R] + \int\!\!\!\!\int\!\!\!\!\int\!\!\!\!\int\dd x_g\dd p_g\dd\tilde x_{}\dd\tilde p_{} \nn\\
		& \hspace{-22mm}\times&\hspace{-12mm}  \!U_{}\!\left(\tfrac{\delta{}}2\right)\! \hat B(x_g,p_g;\tilde x_{},\tilde p_{})\sqrt{  \hat R_\delta{}(x_g,p_g)}U_{}^\dagger\!\left(\tfrac{\delta{}}2\right)\! \rho_{}(t)U_{}\!\left(\tfrac{\delta{}}2\right)\! \sqrt{  \hat R_\delta{}(x_g,p_g)}\hat B^\dag(x_g,p_g;\tilde x_{},\tilde p_{}) U_{}^\dagger\!\left(\tfrac{\delta{}}2\right)\!. \! \nn\\ \label{7.13}
	\end{eqnarray}
	\end{minipage}}\vspace{2mm}\\
This equation is one of the main results of our work. 

Note that the rate operator still depends on $\delta{}$. However, as long as $\delta{}$ is chosen appropriately and all other limits and approximations are satisfied (see next section), \Eqref{7.13} should be a good description of the actual physical situation. The influence of the precise value of $\delta{}$ on the evolution of the density operator is of second order in $\delta{}$ and negligible on a coarse grained time scale. A demonstration of this point can be found in the evolution of the expectation values of position and momentum as well as their variances in chapter~\ref{expectation}.

If we had used (\ref{7.8}) instead of the more correct (\ref{7.10}), we would have found the same master equation except without free particle evolution operators. In this case the influence of the precise value of $\delta{}$ would be of first order. As discussed earlier, that is not strictly forbidden, but using this slightly more complex master equation (\ref{7.13}) results in a neater form of equations of motion for the lowest order moments.

\section{Review of approximations\label{refapprox}}


In the preceding chapters we used several approximations and limits. At this point we want to summarize them, and to discuss under what physical conditions they can simultaneously be fulfilled.

The collisional calculation in chapter~\ref{col} is valid exactly if
	\begin{eqnarray}
		|x_g-x|&\gg&\sqrt{1+\alpha}W_g \label{dddd}\\
		| v_g-v|&\gg& \sqrt{1+\alpha}\frac\hbar{m_gW_g}. \label{cccc}
	\end{eqnarray}
The first relation states that there is negligible initial overlap of the two wave functions, and the second one states that their mean relative velocity has to be large compared to the uncertainty of their relative velocity (small overlap of the initial wave functions in momentum space).

In the definition of the rate operator we required
	\begin{eqnarray}
		\delta{} &\gg& \sqrt{1+\alpha}\frac{W_g}{|v_g-v_{}|}.  \label{bbbb}
	\end{eqnarray}
to ensure that most collisions in $\delta{}$ are complete.

Finally, the derivation of the master equation required the low-density limit
	\begin{eqnarray}
		\delta{}\, \mbox{Tr} [ \hat R \rho_{}(t)]  \;\;\approx\;\; \delta{}\frac{n _g\sqrt{2k_BT}}{\sqrt{\pi m_g}} &\ll&1\label{eeee}
	\end{eqnarray}
to ensure that we have to consider at most one collision during $\delta{}$.

Of course the above relations cannot be fulfilled for all $x_g$ and $p_g$ occurring in a thermal gas. For our description of QBM to be valid, they should be fulfilled for most gas momenta $p_g$, and for a given gas momentum they should be fulfilled for most gas positions $x_g$ which can reach the Brownian particle's wave function in the time interval $\delta{}$. 

Using that a gas particle can reach the Brownian one only if $|x_g-x|\leq \delta{}|v_g-v|$, relation~(\ref{bbbb}) follows directly from relation~(\ref{dddd}). For a typical collision we can use $|v_g-v|\approx \sqrt\frac{k_BT}{m_g}$, and relation~(\ref{cccc}) becomes 
	\begin{eqnarray}
		W_g&\gg&\frac{\sqrt{1+\alpha}\hbar}{\sqrt{m_gk_BT}}, \label{7.18}
	\end{eqnarray}
which gives a lower bound to the parameter $W_g$. Substituting this result into relation~(\ref{bbbb}) gives a lower bound for the coarse grained time scale in terms of the temperature
	\begin{eqnarray}
		\delta{} &\gg& \frac{(1+\alpha)\hbar}{k_BT}, \label{7.19}
	\end{eqnarray}
which can be interpreted as the high-temperature limit. Substituting into \Eqref{bbbb}, we further find
	\begin{eqnarray}
		W_g &\ll& \sqrt{\frac{\pi}{2(1+\alpha)}}\frac1{n_g} \label{7.100}
	\end{eqnarray}

Putting relations \Eqref{eeee} and \Eqref{7.19}, as well as \Eqref{7.18} and \Eqref{7.100} together, we find see that the parameters $\delta$ and $W_g$ are bound from above and below by the respective relations (disregarding constants)
	\begin{eqnarray}
		\frac{\hbar}{k_BT} \;\,\ll &\delta& \ll \; \, \frac{\sqrt{m_g}}{n_g\sqrt{k_BT}} \label{deltalim} \\
		 \frac{\hbar}{\sqrt{m_gk_BT}} \,\;\ll &W_g& \ll\;\, \frac1{n_g}. \label{Wglim}
	\end{eqnarray}

Both of these relations can be simultaneously satisfied in the high-temperature and low-density limit
	\begin{eqnarray}
		\frac{n _g\hbar}{\sqrt{m_gk_BT}} &\ll&1. \label{7.20}
	\end{eqnarray}
This is the crucial relation for the validity of our QBM approach. As such, it does not contain any freely chosen parameters like $W_g$ or $\delta{}$, but only depends on physical parameters like the temperature and density of the gas.  As expected, high temperature can be traded off with low density, but as we approach lower temperatures, the coarse graining time scale \Eqref{7.19} on which our description is valid becomes larger.

We close this section with a comparison to classical Brownian motion (see chapter~\ref{classmo}). There, the limit of a small momentum transfer per collision is often used together with the limit of frequent collisions. It is apparent from \Eqref{7.20}, that this limiting procedure is not possible in a strict treatment of QBM. The underlying reason is that the wave functions of the individual gas particles would overlap in this limit. Then, the ideal gas can not be described by Boltzmann statistics any more, and has to be considered as a quantum gas using the appropriate quantum statistics (Fermi-Dirac for fermions, and Bose-Einstein for Bosons).

\chapter{Expectation Values\label{expectation}}

In this chapter, we will use \Eqref{7.13} to calculate equations of motions for the lowest order moments of position and momentum. The general approach to find an equations of motion for an expectation value of an arbitrary time-independant observable $\widehat O$ is as follows. Recalling the formula of an expectation value of an observable from section~\ref{lueder}, we find
	\begin{eqnarray}
		\langle \widehat O \rangle &=& \mbox{Tr}\left[\widehat O\rho(t)\right]\nn\\
		\Longrightarrow\quad \frac{\dd \langle \widehat O \rangle}{\dd t} &=& \mbox{Tr}\left[\widehat O\frac{\dd \rho(t)}{\dd t}\right]. \label{8.1}
	\end{eqnarray}
Now we can substitute the master equation $\frac{\mbox{\scriptsize d} \rho(t)}{\mbox{\scriptsize d} t}=\mathcal L \rho(t)$ into the right hand side. As the master equation is linear in $\rho(t)$ and local in time, it is always possible to use the invariance of the trace under cyclic permutations to bring \Eqref{8.1} in the from
	\begin{eqnarray}
		\frac{\dd \langle \widehat O \rangle}{\dd t} &=& \mbox{Tr}\left[  \left(\mathcal L^\dagger \widehat O\right) \rho(t)  \right]  \nn\\
		&=& \left\langle \mathcal L^\dagger \widehat O \right\rangle  \label{8.2}
	\end{eqnarray}
where $\mathcal L^\dagger$ is called the adjoint Liouville operator. 

In the following we will discuss the observables $\hat x,\;\hat x^2,\;\hat p,\;\hat p^2$, and $\{\hat x,\hat p\}$. The challenge will be to express, e.g.\ $\mathcal L^\dagger \hat x^2$, in terms of these observables to find a closed set of equations for the lowest order moments.

\section{Mean position\label{sec8.1}}

In this section we will explain each step in detail, whereas in the following sections we will limit ourself to outline the calculation and to refer to this section for the details. Substituting \Eqref{7.13} into \Eqref{8.1} and using the invariance of the trace under cyclic permutations, we find for the expectation value of the position operator 
	\begin{eqnarray}
		\frac{\dd \langle \hat x \rangle}{\dd t} &=& \mbox{Tr}\left[ \frac\imath{2\hbar m}\left[\hat p^2,\hat x\right]\rho \right]  -  \mbox{Tr}\left[  \half(\hat x \hat R +  \hat R \hat x)\rho  \right]  +  \mbox{Tr}\left[  \int\!\!\!\!\int\!\!\!\!\int\!\!\!\!\int\dd x_g\,\dd p_g\,\dd\tilde x_{}\,\dd\tilde p_{}  \right.  \nn\\
		& \hspace{-10mm}\times&\hspace{-7mm}  U_{}\!\left(\tfrac{\delta{}}2\right)\! \sqrt{  \hat R_\delta{}(x_g,p_g)} \hat B^\dag(x_g,p_g;\tilde x_{},\tilde p_{}) U_{}^\dagger\!\left(\tfrac{\delta{}}2\right)\! \hat x U_{}\!\left(\tfrac{\delta{}}2\right)\! \hat B(x_g,p_g;\tilde x_{},\tilde p_{}) \sqrt{  \hat R_\delta{}(x_g,p_g)} U_{}^\dagger\!\left(\tfrac{\delta{}}2\right)\! \rho \bigg] .\nn\\ \label{8.3}
	\end{eqnarray}
Therefore we have 
	\begin{eqnarray}
		 \mathcal L^\dagger \hat x   &\!=\!&  \frac\imath{2\hbar m}\left[\hat p^2,\hat x\right]  -  \half(\hat x \hat R +  \hat R \hat x)  +  \int\!\!\!\!\int\!\!\!\!\int\!\!\!\!\int\dd x_g\,\dd p_g\,\dd\tilde x_{}\,\dd\tilde p_{}  \nn\\
		& \!\times\!& U_{}\!\left(\tfrac{\delta{}}2\right)\! \sqrt{  \hat R_\delta{}(x_g,p_g)} \hat B^\dag(x_g,p_g;\tilde x_{},\tilde p_{}) U_{}^\dagger\!\left(\tfrac{\delta{}}2\right)\! \hat x U_{}\!\left(\tfrac{\delta{}}2\right)\! \hat B(x_g,p_g;\tilde x_{},\tilde p_{}) \sqrt{  \hat R_\delta{}(x_g,p_g)} U_{}^\dagger\!\left(\tfrac{\delta{}}2\right)\!.  \nn\\ \label{8.4}
	\end{eqnarray}
The first term results from the unitary evolution of the Brownian particle. Evaluating the commutator, this term is the expected Brownian velocity operator $\hat p/m$. On the second line we start in the centre at the position operator and work outwards. Using \Eqref{momposcom}, we find $[U(t),\hat x]=-\frac{\hat p t}{m}U(t)$ from which we deduce 
	\begin{equation}
		U_{}^\dagger\!\left(\tfrac{\delta{}}2\right)\! \hat x U_{}\!\left(\tfrac{\delta{}}2\right)\!=\hat x +\frac{\hat p\delta{}}{2m}. \label{8.5}
	\end{equation}

Next, we substitute \Eqref{5.22} for $\hat B(x_g,p_g;\tilde x_{},\tilde p_{})$. Furthermore, we apply 
	\begin{eqnarray}
		\hat{D}^\dagger(x,p) \hat x \hat{D}(x,p) &=& \int\dd x' \,x' \hat{D}^\dagger(x,p) \ket{x'}\!\bra{x'} \hat{D}(xp) \nn\\
		&=& \int\dd x' \,x'  \ket{x'-x}\!\bra{x'-x}  \nn\\
		&=& \int\dd x'\,(x'+x)   \ket{x'}\!\bra{x'} \nn\\
		&=& \hat x+x \label{8.6} \\ \nn\\
		\hat{D}^\dagger(x,p) \hat p \hat{D}(x,p) &=& \hat p+p \label{8.7}
	\end{eqnarray}
to write 
	\begin{eqnarray}
		&& \!\!\int\!\!\!\!\int\dd\tilde x_{}\,\dd\tilde p_{} \, \hat B^\dag(x_g,p_g;\tilde x_{},\tilde p_{}) U_{}^\dagger\!\left(\tfrac{\delta{}}2\right)\! \hat x U_{}\!\left(\tfrac{\delta{}}2\right)\! \hat B(x_g,p_g;\tilde x_{},\tilde p_{}) \nn\\
		&\!\!=& \!\!\! \int\!\!\!\!\int \frac{\dd\tilde x_{}\,\dd\tilde p_{}}{2\pi\hbar} \hat{D}(\tilde x,\tilde p) \sqrt{\hat \sigma}\left[   \hat x +\frac{1\!-\!\alpha}{1\!+\!\alpha}\tilde x + \frac{2\alpha x_g}{1\!+\!\alpha} + \frac{\delta{}}{2m}\!\left(  \hat p +\frac{1\!-\!\alpha}{1\!+\!\alpha}\tilde p + \frac{2 p_g}{1\!+\!\alpha}   \right)\!   \right]\sqrt{\hat \sigma}  \hat{D}^\dagger(\tilde x,\tilde p) \nn\\
		&\!\!=& \!\!\frac{2\alpha x_g}{1+\alpha} + \frac{1-\alpha}{1+\alpha}\hat x + \frac{\delta{}}{2m}\!\left(  \frac{2 p_g}{1+\alpha} +\frac{1-\alpha}{1+\alpha}\hat p  \right)\!.  \label{8.8}
	\end{eqnarray}
In the second step, \Eqref{O}, \Eqref{Ox}, and \Eqref{Op} from appendix A of this chapter were  used, as well as Tr$(\hat \sigma)=1$ and Tr$(\hat x\hat \sigma)=\mbox{Tr}(\hat p\hat \sigma)=0$. Substituting \Eqref{8.8} into \Eqref{8.4} and using \Eqref{6.16} we find
	\begin{eqnarray}
		 \mathcal L^\dagger \hat x   &=&  \frac{\hat p}{m}  -  \half(\hat x \hat R +  \hat R \hat x)  +   \int\!\!\!\!\int\dd x_g\,\dd p_g \frac{\mu_{T}(p_g)}{\delta{}} \nn\\  
		&\hspace{-10mm}\times&\hspace{-6mm}  U_{}\!\left(\tfrac{\delta{}}2\right)\!  \sqrt{  \hat \Gamma_\delta(x_g,p_g)}  \left[ \frac{2\alpha x_g}{1+\alpha} + \frac{1-\alpha}{1+\alpha}\hat x + \frac{\delta{}}{2m}\!\left(  \frac{2 p_g}{1+\alpha} +\frac{1-\alpha}{1+\alpha} \hat p \right)\! \right]  \sqrt{ \hat \Gamma_\delta(x_g,p_g)}   U_{}^\dagger\!\left(\tfrac{\delta{}}2\right)\!.  \nn\\ \label{8.9}
	\end{eqnarray}
With \Eqref{x} we write
	\begin{eqnarray}
		 \hat \Gamma_\delta(x_g,p_g)\hat x  &=&   \int\!\!\!\!\int \frac{\dd x\,\dd p}{2\pi\hbar} x 
		  \hat \Gamma_\delta(x_g,p_g)\ket{x,p}\!\bra{x,p}\nn\\
		 &\approx& \int\!\!\!\!\int_{0<\frac{x_{}-x_g}{\frac{p_g}{m_g}-\frac{p_{}}{m_{}}}<\delta{} }\frac{\dd x_{}\,\dd p_{}}{2\pi\hbar} x\ket{x_{},p_{}}\!\bra{x_{},p_{}} \label{8.10}
	\end{eqnarray}
where we also used relations~(\ref{6.12}) and~(\ref{6.13}). The same formula applies for $\hat x  \hat \Gamma_\delta(x_g,p_g)$ and $\sqrt{ \hat \Gamma_\delta(x_g,p_g)}\hat x \sqrt{ \hat \Gamma_\delta(x_g,p_g)}$ and therefore we can use
	\begin{equation}
		\hat x  \hat \Gamma_\delta(x_g,p_g) \,\approx\, \sqrt{ \hat \Gamma_\delta(x_g,p_g)}\hat x \sqrt{ \hat \Gamma_\delta(x_g,p_g)} \,\approx\,  \hat \Gamma_\delta(x_g,p_g)\hat x.
	\end{equation}
The same property, which is a consequence of $\hat \Gamma_\delta(x_g,p_g)$ being an approximate projection operator, also applies to the momentum operator and we can pull all $\hat R$ and $ \hat \Gamma_\delta(x_g,p_g)$  across position and momentum operators
	\begin{eqnarray}
		 \mathcal L^\dagger \hat x  &=&  \frac{\hat p}{m} -  \hat x  \hat R +   \int\!\!\!\!\int\dd x_g\,\dd p_g \frac{\mu_{T}(p_g)}{\delta{}} \nn\\  
		&\times&  U_{}\!\left(\tfrac{\delta{}}2\right)\!   \left[ \frac{2\alpha x_g}{1+\alpha} + \frac{1-\alpha}{1+\alpha}\hat x + \frac{\delta{}}{2m}\!\left(  \frac{2 p_g}{1+\alpha} +\frac{1-\alpha}{1+\alpha}\hat p  \right)\! \right]   \hat \Gamma_\delta(x_g,p_g)   U_{}^\dagger\!\left(\tfrac{\delta{}}2\right)\!. \qquad\qquad \label{8.12}
	\end{eqnarray}

Next we use \Eqref{6.23} and \Eqref{8.5} for
	\begin{eqnarray}
		U_{}^\dagger\!\left(\tfrac{\delta{}}2\right)\!   \hat x \hat R U_{}\!\left(\tfrac{\delta{}}2\right)\! &=& U_{}^\dagger\!\left(\tfrac{\delta{}}2\right)\!  \hat x U_{}\!\left(\tfrac{\delta{}}2\right)\! U_{}^\dagger\!\left(\tfrac{\delta{}}2\right)\! \hat R U_{}\!\left(\tfrac{\delta{}}2\right)\! \nn\\
		&=& \left(\hat x +\frac{\delta{}}{2m}\hat p\right)\hat R ,
	\end{eqnarray}
as well as $\hat R=\int\!\!\int \dd x_g\,\dd p_g\, \mu_T(p_g)\hat \Gamma_\delta(x_g,p_g)/\delta$, to find
	\begin{eqnarray}
		\mathcal L^\dagger \hat x  &=&  \frac{\hat p}{m}  +   U_{}\!\left(\tfrac{\delta{}}2\right)\!  \int\!\!\!\!\int\dd x_g\,\dd p_g \frac{\mu_{T}(p_g)}{\delta{}}    \left[ \frac{2\alpha (x_g-\hat x)}{1+\alpha}  + \frac{\delta{}}{m}\!\left(  \frac{p_g-\alpha \hat p}{1+\alpha}  \right)\! \right]   \hat \Gamma_\delta(x_g,p_g)   U_{}^\dagger\!\left(\tfrac{\delta{}}2\right)\!. \nn\\
		&=&\frac{\hat p}{m}  +   U_{}\!\left(\tfrac{\delta{}}2\right)\!  \int\!\!\!\!\int  \dd x_g\,\dd p_g\int\!\!\!\!\int\frac{\dd x\,\dd p}{2\pi\hbar} \frac{\mu_{T}(p_g)}{\delta{}}    \left[ \frac{2\alpha (x_g-x)}{1+\alpha}  + \frac{\delta{}}{m}\!\left(  \frac{p_g-\alpha p}{1+\alpha}  \right)\! \right]     \nn\\
		&&\times\; \ket{x,p}\!\bra{x,p}U_{}^\dagger\!\left(\tfrac{\delta{}}2\right)\!. \label{redf}
	\end{eqnarray}
In the second step, we used the definition \Eqref{6.11} of $\hat \Gamma_\delta(x_g,p_g)$, as well as \Eqref{8.10}. Upon integration over $x_g$, which is carried out in \Eqref{8.19} and \Eqref{8.19b} in the appendix B of this chapter, we find that all terms in the integrand cancel. Therefore, we are left with the expected
	\begin{eqnarray}
		 \mathcal L^\dagger \hat x  &=&  \frac{\hat p}{m}
	\end{eqnarray}
from which we deduce
	\begin{eqnarray}
		\frac{\dd \langle \hat x \rangle}{\dd t} &=& \frac{\langle\hat p\rangle}{m}. \label{8.21}
	\end{eqnarray}
That is, the mean position of the Brownian particle changes in time according to its mean velocity.

At this point we wish to mention that if we had used the continuity transition~(\ref{7.7}) instead of (\ref{7.9}), then, upon the integration over $x_g$, the integrand in \Eqref{redf} would only vanish in lowest order in $\delta$. The remaining part is of first order, and we would be left with a non-physical $\frac{\mbox{\scriptsize d} \langle \hat x \rangle}{\mbox{\scriptsize d} t} = \frac{\langle\hat p\rangle}{m} +\delta{} f\!\left( \frac{\langle\hat p\rangle}{m} \right)\!$ for some function $f$ involving Gauss and Error functions. See discussion in section~\ref{7.2} for more detail.

\section{Mean momentum}

Applying the adjoint Liouville operator \Eqref{8.4} on $\hat p$ results in
	\begin{eqnarray}
		 \mathcal L^\dagger \hat p   &=&  -  \half(\hat p \hat R +  \hat R\hat p)  +  \int\!\!\!\!\int\!\!\!\!\int\!\!\!\!\int\dd x_g\,\dd p_g\,\dd\tilde x_{}\,\dd\tilde p_{}  \nn\\
		&& \times\,  U_{}\!\left(\tfrac{\delta{}}2\right)\! \sqrt{  \hat R_\delta{}(x_g,p_g)} \hat B^\dag(x_g,p_g;\tilde x_{},\tilde p_{})  \hat p  \hat B(x_g,p_g;\tilde x_{},\tilde p_{}) \sqrt{  \hat R_\delta{}(x_g,p_g)} U_{}^\dagger\!\left(\tfrac{\delta{}}2\right)\!.  \qquad\qquad \label{8.22}
	\end{eqnarray}
Using a calculation analogous to \Eqref{8.8}, we find
	\begin{eqnarray}
		\int\!\!\!\!\int \dd\tilde x_{}\,\dd\tilde p_{}\, \hat B^\dag(x_g,p_g;\tilde x_{},\tilde p_{})  \hat p  \hat B(x_g,p_g;\tilde x_{},\tilde p_{}) &=& \frac{2p_g}{1+\alpha} + \frac{1-\alpha}{1+\alpha}\hat p,
	\end{eqnarray}
which we substitute into \Eqref{8.22}
	\begin{eqnarray}
		 \mathcal L^\dagger \hat p  &=&  -   \hat R\hat p   +   \int\!\!\!\!\int\dd x_g\,\dd p_g\,   U_{}\!\left(\tfrac{\delta{}}2\right)\! {  \hat R_\delta{}(x_g,p_g)} \left(\frac{2p_g}{1+\alpha} + \frac{1-\alpha}{1+\alpha}\hat p\right)  U_{}^\dagger\!\left(\tfrac{\delta{}}2\right)\nn\\
		 &=&   \int\dd p_g\,  {  \hat R(p_g)}\frac{2p_g-2\alpha \hat p}{1+\alpha} .  \qquad\quad \label{8.22}
	\end{eqnarray}
In the second step we used the definitions~(\ref{6.17}) and~(\ref{6.19b}) as well as $U(t) \hat R(p_g)\hat p U^\dagger(t)= \hat R(p_g)\hat p$ (see \Eqref{6.22}). 
Contrary to the previous section, here the contributions from $\hat p$ and $p_g$ do not cancel each other. This was to be expected as the gaseous environment has a damping effect on the momentum. Note that at this point there is no dependance on $\delta{}$.

Now we can use \Eqref{6.18} to arrive at the final result
	\begin{eqnarray}
		\frac{\dd\langle\hat p\rangle}{\dd t} &=& -\langle f_T(\hat p)\rangle  
	\end{eqnarray}
where
	\begin{eqnarray}
		f_T(\hat p) &=& \frac{2m_g}{1\!+\!\alpha}  \frac{n _g}{\sqrt{2\pi m_gk_B{T}}}\int\dd p_g \, \exp\!\left[\frac{-p_g^2}{2m_gk_B{T}}\right] \left| \frac{\hat p}m-\frac{p_g}{m_g} \right| \left( \frac{\hat p}m-\frac{p_g}{m_g} \right) \label{8.26}  \nn\\
		&=&  \frac{4m_g}{1\!+\!\alpha}  \frac{n _g}{\sqrt{2\pi m_gk_B{T}}} \! \Bigg[ \frac{2\hat pk_BT}{m}\exp\!\left(\frac{-\alpha \hat p^2}{2mk_BT}\right)  \nn\\
		&&\left. +  \int_0^{\alpha \hat p}\!\!\! \dd p_g \exp\!\left(\frac{-p^2_g}{2m_gk_BT}\right)\!  \left(\frac{\hat p^2}{m^2}+\frac{p_g^2}{m_g^2}\right)  \right]    \label{8.27}
	\end{eqnarray}
involves the error function. For the second line, results from the appendix of chapter~\ref{classmo} were used. This is the known result for classical Brownian motion, as derived in chapter~\ref{classmo}. 

For small velocities $p/m\ll\sqrt{k_BT/m_g}$ we can expand $f_T(\hat p)$ and write
	\begin{eqnarray}
		\frac{\dd\langle\hat p\rangle}{\dd t} &\approx& -\frac{n _g}{1+\alpha}\frac{\sqrt{2m_gk_BT}}{\sqrt\pi}\left(  4\frac{\langle\hat p\rangle}{m} + \frac{2m_g}{3k_BT}\frac{\langle\hat p^3\rangle}{m^3}  -  \frac{m_g^2}{30(k_BT)^2}\frac{\langle\hat p^5\rangle}{m^5}  \right). \qquad
	\end{eqnarray}
Therefore we find for linear friction 
	\begin{eqnarray}
		\frac{\dd\langle\hat p\rangle}{\dd t} &=& -f\frac{\langle \hat p\rangle}{m} \nn\\
		f &=& \frac{4n _g\sqrt{2m_gk_BT}}{(1+\alpha)\sqrt\pi}.
	\end{eqnarray}

Another interesting limit is the high velocity limit $p/m\gg\sqrt{k_BT/m_g}$, as it occurs e.g.\ in a cloud chamber. Then we can neglect the first term in the square brackets of \Eqref{8.27}, and extend the integration in the second term to infinity. We find
	\begin{eqnarray}
		\frac{\dd\langle\hat p\rangle}{\dd t} &=& -\frac{2n _gm_g}{1+\alpha} \left(\frac{\langle \hat p\, |\hat p|\rangle}{m^2}+\frac{k_BT\langle \hat p/ |\hat p|\rangle}{m_g}\right) ,
	\end{eqnarray}
which is the expected quadratic friction and can be explained as follows. In the high velocity limit, the velocity of the gas particles can be neglected and the momentum transfer per collision increases linear with the Brownian velocity. In addition, the collision rate increases linear with the velocity, and we end up with quadratic friction.

\section{Mean squared position\label{sec8.3}}

Applying the Liouville operator \Eqref{8.4} on $\hat x^2$ results in
	\begin{eqnarray}
		 \mathcal L^\dagger \hat x^2   &=&  \frac\imath{2\hbar m}\left[\hat p^2,\hat x^2\right]  -  \half\left(\hat x^2 \hat R +  \hat R\hat x^2\right)  +  \int\!\!\!\!\int\!\!\!\!\int\!\!\!\!\int\dd x_g\,\dd p_g\,\dd\tilde x_{}\,\dd\tilde p_{}  \nn\\
		& \hspace{-10mm}\times&\hspace{-6mm}  U_{}\!\left(\tfrac{\delta{}}2\right)\! \sqrt{  \hat R_\delta{}(x_g,p_g)} \hat B^\dag(x_g,p_g;\tilde x_{},\tilde p_{}) U_{}^\dagger\!\left(\tfrac{\delta{}}2\right)\! \hat x^2 U_{}\!\left(\tfrac{\delta{}}2\right)\! \hat B(x_g,p_g;\tilde x_{},\tilde p_{}) \sqrt{  \hat R_\delta{}(x_g,p_g)} U_{}^\dagger\!\left(\tfrac{\delta{}}2\right)\!.  \nn\\ \label{8.30b}
	\end{eqnarray}
We proceed in the same manner as in section~(\ref{sec8.1}) and find 
	\begin{eqnarray}
		&&\!\!\!\int\!\!\!\!\int\dd\tilde x_{}\, \dd\tilde p_{} \, \hat B^\dag(x_g,p_g;\tilde x_{},\tilde p_{}) U_{}^\dagger\!\left(\tfrac{\delta{}}2\right)\! \hat x^2 U_{}\!\left(\tfrac{\delta{}}2\right)\! \hat B(x_g,p_g;\tilde x_{},\tilde p_{})\nn\\
		&&\hspace{-7mm} =\, \int\!\!\!\!\int \frac{\dd\tilde x_{} \,\dd\tilde p_{}}{2\pi\hbar} \hat{D}(\tilde x,\tilde p) \sqrt{\hat \sigma}\left[   \hat x +\frac{1\!-\!\alpha}{1\!+\!\alpha}\tilde x + \frac{2\alpha x_g}{1\!+\!\alpha} + \frac{\delta{}}{2m}\!\left(  \hat p +\frac{1\!-\!\alpha}{1\!+\!\alpha}\tilde p + \frac{2 p_g}{1\!+\!\alpha}   \right)\!   \right]^2 \sqrt{\hat \sigma}  \hat{D}^\dagger(\tilde x,\tilde p) \nn\\
		&&\hspace{-7mm} = \,\frac{2\alpha }{(1\!+\!\alpha)^2}\left[W^2 +  \left(\frac{\delta{}\hbar}{2mW}\right)^{\!2}\right]  +  \left[ \frac{(1\!-\!\alpha)\hat x +2\alpha x_g}{1\!+\!\alpha} + \frac{\delta{}}{2m}\left(\frac{(1\!-\!\alpha)\hat p+2p_g}{1\!+\!\alpha}\right)   \right]^{2} \! .  \label{8.31}
	\end{eqnarray}
In the second step, we used formulas from appendix A of this chapter, as well as Tr$(\hat \sigma)=1$, Tr$(\hat x\hat \sigma)=\mbox{Tr}(\hat p\hat \sigma)=\mbox{Tr}[\hat \sigma(\hat x\hat p+\hat p\hat x)]=0$, Tr$[\hat \sigma\hat x^2]=W^2\frac{1+\alpha^2}{4\alpha}$, Tr$[\sqrt{\hat \sigma}\hat x\sqrt{\hat \sigma}\hat x]=W^2\frac{1-\alpha^2}{4\alpha}$, Tr$[\hat \sigma\hat p^2]=\frac{\hbar^2}{W^2}\frac{1+\alpha^2}{4\alpha}$, and Tr$[\sqrt{\hat \sigma}\hat p\sqrt{\hat \sigma}\hat p]=\frac{\hbar^2}{W^2}\frac{1-\alpha^2}{4\alpha}$.

With \Eqref{8.31} and $\imath[\hat p^2,\hat x^2]=2\hbar\{\hat x, \hat p\}$ we find for the adjoint Liouville operator applied on the squared position operator
	\begin{eqnarray}
		 \mathcal L^\dagger \hat x^2  &=& \frac{\{\hat x,\hat p\}}{m} - \hat R \hat x^2  + \frac{2\alpha }{(1+\alpha)^2}  \hat R \left[W^2 +  \left(\frac{\delta{}\hbar}{2mW}\right)^{\!2}\right]  + \int\!\!\!\!\int \dd x_g\,\dd p_g\, U\!\left(\tfrac{\delta{}}{2}\right)  \hat R_\delta{}(x_g,p_g)  \nn\\
		 &&  \times  \left[ \frac{(1-\alpha)\hat x +2\alpha x_g}{1+\alpha} + \frac{\delta{}}{2m}\left(\frac{(1-\alpha)\hat p+2p_g}{1+\alpha}\right)  \right]^2  U^\dagger\!\left(\tfrac{\delta{}}{2}\right) \nn\\
		 &=&  \frac{\{\hat x,\hat p\}}{m} + \frac{2\alpha }{(1+\alpha)^2}  \hat R \!\left[W^2 +  \left(\frac{\delta{}\hbar}{2mW}\right)^{\!2}\right] \! + \frac{4\alpha}{(1+\alpha)^2} U\!\left(\tfrac{\delta{}}{2}\right)\! \int\!\!\!\!\int \!\dd x_g\,\dd p_g\,\hat R_\delta(x_g,p_g) \nn\\
		 &&\times \left[ \alpha\!\left( x_g+\frac{\delta p_g}{2m_g}\right)^{\!2} \!+ (1-\alpha)\!\left(\hat x+\frac{\delta \hat p}{2m}\right)\!\left( x_g+\frac{\delta p_g}{2m_g}\right) \! - \! \left(\hat x+\frac{\delta \hat p}{2m}\right)^{\!2} \right]\! U^\dagger\!\left(\tfrac{\delta{}}{2}\right)\!\nn\\ \label{8.32}
	\end{eqnarray}
where we also used $\hat R\hat x^2 =U\!\left(\tfrac{\delta{}}{2}\right)  \hat R\left(\hat x+\frac{\delta{}}{2m}\hat p\right)^2 U^\dagger\!\left(\tfrac{\delta{}}{2}\right)$. Next, we use the definition of the rate operator \Eqref{6.16} and $\hat R=\int\!\!\int dx_g\,dp_g\, \hat R_\delta(x_g,p_g)$. Then, with \Eqref{x} to \Eqref{xp} we find relations like $\hat \Gamma_\delta(x_g,p_g)\hat x^2 = \int\!\!\int_{S_\delta(x_g,p_g)}\frac{d x \,d p}{2\pi\hbar} \left(x^2-W^2/2\right) \ket{x,p}\!\bra{x,p}$. All this substituted into \Eqref{8.32}, we proceed analog to section~\ref{sec8.1} and use the appendix B of this chapter to perform the integration over $x_g$
	\begin{eqnarray}
		 \mathcal L^\dagger \hat x^2  &=& \frac{\{\hat x,\hat p\}}{m} + \frac{4\alpha \hat R}{(1+\alpha)^2} \left[ W^2+\left(\frac{\delta{}\hbar}{2mW}\right)^2\right]   +  \frac{\delta{}^2\alpha^2}{3(1+\alpha)^2}  \int\dd p_g \,\mu_{T}(p_g) \left| \frac {\hat p}m-\frac{p_g}{m_g} \right|^3 \!. \nn\\ \label{8.34}
	\end{eqnarray}
Contrary to section~\ref{sec8.1}, not all contributions cancel out, but we are left with three terms, each representing position diffusion. The first term can be interpreted as the result of position measurements performed by the gas particles, because a measurement shifts the initial position towards the measured position. The second one is due the momentum shift of the momentum measurements, which over the coarse graining time $\delta$ also results in a position shift. Both these terms are multiplied by $\alpha$ because the measurements of light gas particles are weak ones. The third term is due to the (classical) collisional momentum exchange, which again, over a time interval $\delta$ gives rise to a position shift. 

We have seen in section~\ref{jumps} that position diffusion can not be a physical process induced by random collisions. Indeed, the second and third terms are clearly due to the coarse graining time approximation, while the first term is due to the assumption of instantaneous measurements, which in reality take the collision time $t_c$. Therefore, all position diffusion terms can be traced back to approximations we used to derive our master equation, and are fictitious. 

To show that these terms are small in the range of validity of the approximations used, we first discuss which one is the leading one. Here, we use the slow Brownian particle limit (noting that we need only qualitative estimates and that for fast Brownian particles the result is even more pronounced) to perform the integration
	\begin{equation}
		\int \dd p_g\,\mu_{T}(p_g)\left|\frac{p_g}{m_g}\right|^3 \;=\; \frac{n _g}{\sqrt\pi}\left(\frac{2k_BT}{m_g}\right)^{\frac32} \label{8.35}
	\end{equation}
and to use \Eqref{6.20} for the collision rate $\hat R$. Up to a factor $4\alpha n_g\sqrt{k_BT/m_g}$, we find $ W^2$ for the first term, $[\delta\hbar/(2mW)]^2$ for the second term, and $\delta^2k_BT/(6m)$ for the third term. From \Eqref{bbbb} we deduce that the third is large compared to the first, whereas according to \Eqref{7.18} the third term is also large compared to the first one.

Therefore, in the range of validity of our master equation, we can neglect the first two terms and arrive at
	\begin{eqnarray}
		\frac{\dd\left\langle\hat x^2\right\rangle}{\dd t} &=& \frac{\langle \{\hat x,\hat p\} \rangle}{m} + \delta{}^2 \langle g_T(\hat p) \rangle \nn\\
		g_T(\hat p) &=& \frac{\alpha^2}{3(1+\alpha)^2} \int\dd p_g \,\mu_{T}(p_g) \left| \frac {\hat p}m-\frac{p_g}{m_g} \right|^3  . \label{8.38}
	\end{eqnarray}
There is a class of states with $\langle \{\hat x,\hat p\} \rangle=0$ (e.g.\ the state $\ket{x=0,p=0}$). If the Brownian particle's state is one of these, then the position diffusion term in \Eqref{8.38} is the leading one. However, there are no such states for which $\langle \{\hat x,\hat p\} \rangle (t) \equiv0$ for all times. As we will see in section~\ref{solution}, for times longer than $\delta{}$ the first term will always become large compared to the second. Now we remember that our description of QBM is only valid on a time scale large compared to $\delta{}$, and therefore we can indeed neglect the position diffusion term. The result is
	\begin{eqnarray}
		\frac{\dd \left\langle\hat x^2\right\rangle}{\dd t} &=& \frac{\langle \{\hat x,\hat p\} \rangle}{m}. \label{8.39}
	\end{eqnarray}
	
We want to point out that the limit $\delta{}\to 0$ minimizes the artifically introduced position diffusion, although not to zero but to a value connected to the width $W$ of the wave packets used to define the approximate phase space projection operator in section \ref{ratedef}. Here we also wish to note that the same limit maximizes errors in the decoherence rates, because the rate operator then acts as a quantum-limited position-momentum measurement on the Brownian particle, and this is not actually performed. To describe all aspects of QBM, one should therefore choose some finite $\delta{}$ according to section \ref{refapprox}.

\section{Mean squared momentum}

Applying the adjoint Liouville operator \Eqref{8.4} on $\hat p^2$ results in
	\begin{eqnarray}
		 \mathcal L^\dagger \hat p^2   &=&  -  \half(\hat p^2 \hat R +  \hat R\hat p^2)  +  \int\!\!\!\!\int\!\!\!\!\int\!\!\!\!\int\dd x_g\,\dd p_g\,\dd\tilde x_{}\,\dd\tilde p_{}  \nn\\
		&& \times\,  U_{}\!\left(\tfrac{\delta{}}2\right)\! \sqrt{  \hat R_\delta{}(x_g,p_g)} \hat B^\dag(x_g,p_g;\tilde x_{},\tilde p_{})  \hat p^2  \hat B(x_g,p_g;\tilde x_{},\tilde p_{}) \sqrt{  \hat R_\delta{}(x_g,p_g)} U_{}^\dagger\!\left(\tfrac{\delta{}}2\right)  \qquad\quad\; \label{8.40a}
	\end{eqnarray}
where $[\hat p^2,\hat p^2]=0$ and $U(t)\hat p^2U^\dagger(t)=\hat p^2$ were used. Now we apply \Eqref{5.22} and~(\ref{8.7}) for a calculation analog to \Eqref{8.8} to find
	\begin{eqnarray}
		\int\!\!\!\!\int \dd\tilde x_{}\,\dd\tilde p_{}\, \hat B^\dag(x_g,p_g;\tilde x_{},\tilde p_{})  \hat p^2  \hat B(x_g,p_g;\tilde x_{},\tilde p_{})  \hspace{-50mm}&&\nn\\
		&=& \frac{2 +2\alpha^2}{(1+\alpha)^2} \mbox{Tr}(\hat \sigma \hat p^2) -  \frac{2 -2\alpha}{1+\alpha} \mbox{Tr}\left(\sqrt{\hat \sigma} \hat p \sqrt{\hat \sigma} \hat p\right) + \frac{[(1-\alpha)\hat p +2p_g ]^2}{(1+\alpha)^2}. \qquad\qquad
	\end{eqnarray}
Using $\mbox{Tr}(\sigma \hat p^2)=\mbox{Tr}\left(\sqrt{\hat \sigma} \hat p \sqrt{\hat \sigma} \hat p\right)=0$ in the limit $W\to\infty$\footnote{To be precise, one should include these terms as $W$ is finite. Then, we would find an extra momentum diffusion term, corresponding to the momentum measurements performed by the gas particles. At the end, this contribution would turn out to be negligible compared to the leading diffusion term. Because this is all analog to section~\ref{sec8.3}, we remove this term already here.}, we substitute into \Eqref{8.40a}
	\begin{eqnarray}
		 \mathcal L^\dagger \hat p^2   &=&  -   \hat R\hat p^2  +   \int\!\!\!\!\int\dd x_g\,\dd p_g \,   U_{}\!\left(\tfrac{\delta{}}2\right)\!   \hat R_\delta{}(x_g,p_g)\frac{[(1-\alpha)\hat p +2p_g ]^2}{(1+\alpha)^2} U_{}^\dagger\!\left(\tfrac{\delta{}}2\right)\nn\\
		 &=& \int\dd p_g \,     \hat R(p_g)\frac{4p_g^2+4(1-\alpha)p_g\hat p - 4\alpha\hat p^2}{(1+\alpha)^2}  .  \label{8.40}
	\end{eqnarray}
In the second step we used the definitions~(\ref{6.17}) and~(\ref{6.19b}) as well as $U(t) \hat R(p_g)\hat p^2 U^\dagger(t)= \hat R(p_g)\hat p^2$ (see \Eqref{6.23}). Now we can use \Eqref{6.18} to arrive at the final result
	\begin{eqnarray}
		\frac{\dd\left\langle\hat p^2\right\rangle}{\dd t} &=& -\langle h_T(\hat p)\rangle  \label{8.42}
	\end{eqnarray}
where
	\begin{eqnarray}
		h_T(\hat p) &=& \frac{4m_g}{(1\!+\!\alpha)^2}  \frac{n _g}{\sqrt{2\pi m_gk_B{T}}}\int\dd p_g \, \exp\!\left[\frac{-p_g^2}{2m_gk_B{T}}\right] \left| \frac{\hat p}m-\frac{p_g}{m_g} \right|  \left( \frac{\hat p}m-\frac{p_g}{m_g} \right)\!(\hat p+p_g) \nn\\
		&=& \frac{8n _g}{(1\!+\!\alpha)^2m_g\sqrt{2\pi m_gk_B{T}}}  \Bigg\{  \!\exp\!\left(\frac{-\alpha\hat p^2}{2mk_BT}\right)\left[2\alpha(1\!-\!\alpha)m_gk_BT\hat p^2 - 2(m_gk_BT)^2\right]    \nn\\
		&& + \left. \int_0^{\alpha\hat p}\dd p_g\, \exp\!\left(\frac{-p_g^2}{2m_gk_BT}\right) \left[ \alpha^2\hat p^3+(1-2\alpha)p_g^2\hat p\right] \right\} .  \nn\\
		 \label{8.43}
	\end{eqnarray}
For low velocities $|v|\ll \sqrt{k_BT/m_g}$ of the Brownian particle we can expand $h_T(\hat p)$ to find
	\begin{eqnarray}
		\frac{\dd\!\left\langle\hat p^2\right\rangle\!}{\dd t} &\!\approx\!& \frac{8n _g}{(1\!+\!\alpha)^2m_g\sqrt{2\pi m_gk_BT}} \!\left[  2(m_gk_BT)^2-\alpha(2\!-\!\alpha)m_gk_BT\!\left\langle \hat p^2\right\rangle-\frac{\alpha^3}{12}\left(4\!+\!\alpha\right)\! \left\langle \hat p^4\right\rangle \right]\!\!\nn\\
	\end{eqnarray}
	
Note that this equation gives the correct value for $\left\langle\hat p^2\right\rangle$ in the steady state limit only if $\alpha\ll 1$. The reason is, that, if the mass of the Brownian particle is comparable to the mass of the gas particles, then the Brownian particle can not be considered slow in the steady state, and therefore the approximation leading to the above equation is not satisfied anymore.

We therefore confirm the steady state $\rho\propto\exp\!\left(\frac{-\hat p^2}{2mk_BT}\right)$ for general mass ratios by substituting the assumed steady state in \Eqref{8.42}, and show that $\dd \!\left\langle\hat p^2\right\rangle/\dd t=0$. Disregarding all constants, we have to calculate
	\begin{eqnarray}
		\mbox{Tr}[\rho h_T(\hat p)] &\propto& \int\dd p \,\exp\!\left(\frac{-p^2}{2mk_BT}\right)h_T(p) \nn\\
		&\propto& \int\dd p \,\Bigg\{  \exp\!\left(\frac{-(1+\alpha) p^2}{2mk_BT}\right)\left[2\alpha(1-\alpha)m_gk_BT p^2 - 2(m_gk_BT)^2\right] \qquad   \nn\\
		&& + \left. \int_0^{\alpha p}\dd p_g\, \exp\!\left(\frac{-p_g^2-\alpha p^2}{2m_gk_BT}\right) \left[ \alpha^2 p^3+(1-2\alpha)p_g^2 p\right] \right\}.\label{8.44}
	\end{eqnarray}
While the first term is easily integrated to give
	\begin{equation}
		-\frac{4\alpha(m_gk_BT)^2\sqrt{2\pi mk_BT}}{(1+\alpha)^{3/2}}, \label{8.45}
	\end{equation}
the second term is not that straightforward as the integration over $p_g$ can not be carried out analytically. Therefore we will change the order of integration. The integration area splits up in to parts as seen in figure~\ref{fig5}.
\begin{figure}[htbp]
\begin{center} 
	\includegraphics[width=0.6\linewidth]{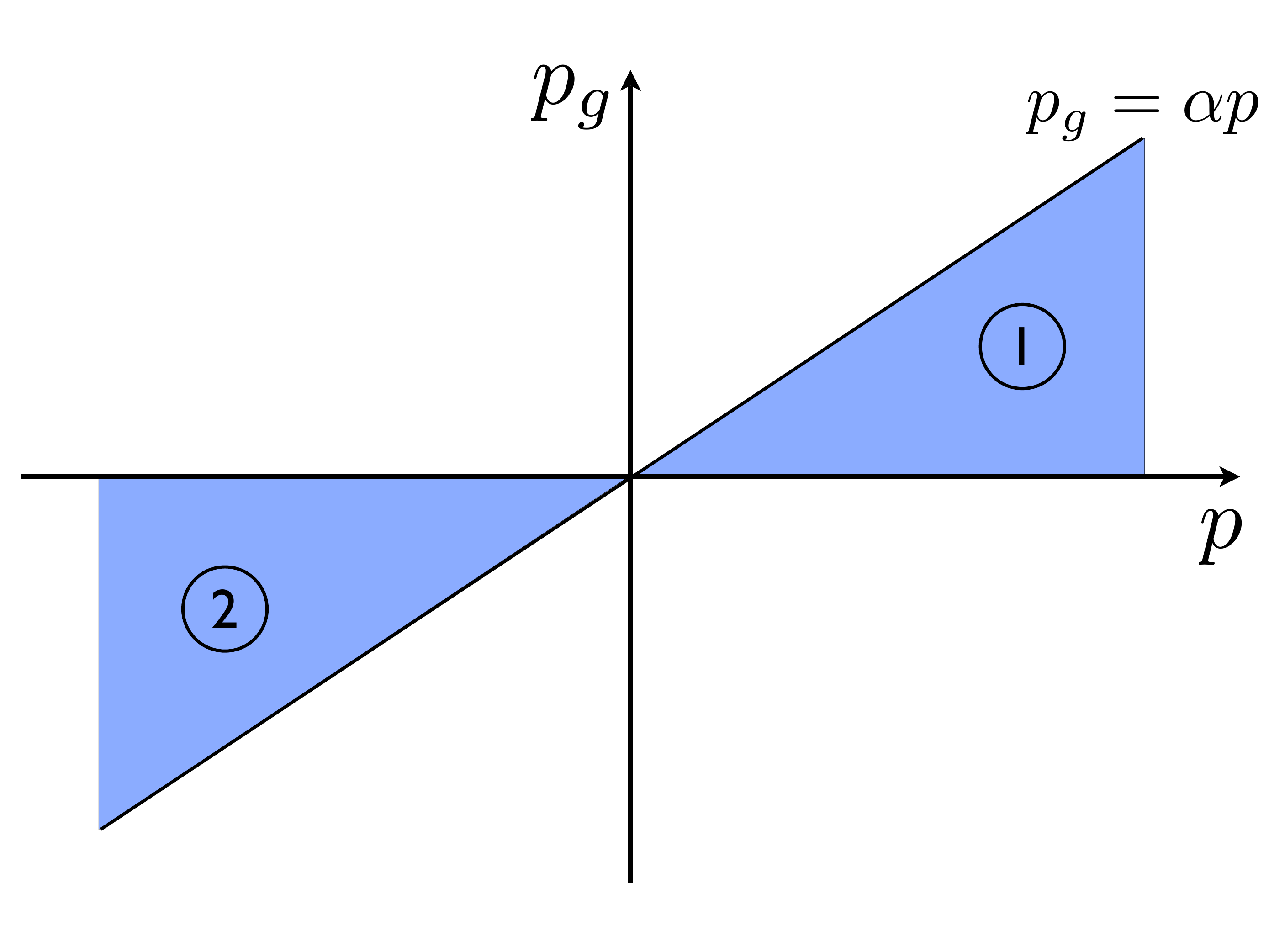}\vspace{-3mm}
\caption{\small Integration area of the second term in \Eqref{8.44}.\label{fig5}}
\end{center}
\end{figure}

For the first integration area, we write
	\begin{eqnarray}
		\int_0^\infty \dd p\int_0^{\alpha p}\dd p_g\,f(p,p_g) &=& \int_0^\infty\dd p_g\int_{p_g/\alpha}^\infty\dd p\, f(p,p_g)
	\end{eqnarray}
which is valid for arbitrary functions $f(p,p_g)$. For the second integration area we use
	\begin{eqnarray}
		\int_{-\infty}^0 \dd p\int_0^{\alpha p}\dd p_g\,f(p,p_g) &=& -\int_{-\infty}^0 \dd p\int_{\alpha p}^0  \dd p_g\,f(p,p_g) \nn\\
		&=& -\int_{-\infty}^0\dd p_g\int_{-\infty}^{p_g/\alpha} \dd p\,f(p,p_g) \nn\\
		&=& \int_{-\infty}^0\dd p_g\int^{\infty}_{p_g/\alpha} \dd p\,f(p,p_g)
	\end{eqnarray}
where the last equality is valid for functions $f(p,p_g)$ which are odd in $p$. Putting the two relations together,
	 \begin{eqnarray}
		\int \dd p\int_0^{\alpha p}\dd p_g\,f(p,p_g) &=& \int\dd p_g\int_{p_g/\alpha}^\infty\dd p\, f(p,p_g)
	\end{eqnarray}
can be applied to the second term in \Eqref{8.44}. The $p$-integration is then easily carried out by substitution of $p^2=u$ and integration by parts, and we find
	\begin{eqnarray}
		2\!\int\dd p_g\, \exp\!\left(\frac{-(1\!+\!\alpha)p_g^2}{2\alpha m_gk_BT}\right)\!  \left[(1\!-\!\alpha)p_g^2Mk_BT+\alpha^2(Mk_BT)^2 \right] &\!\!=\!\!& \frac{4\alpha(m_gk_BT)^2\sqrt{2\pi mk_BT}}{(1+\alpha)^{3/2}}\nn\\
	\end{eqnarray}
where the $p_g$-integration is now over the entire real axis. This cancels exactly with \Eqref{8.45} and we find
	\begin{eqnarray}
		\frac{\mbox{\dd}\left\langle\hat p^2\right\rangle}{\mbox{\dd} t}&=&0
	\end{eqnarray}
as required for a thermal state. Note that the derivation of this result is valid for all $\alpha$ and therefore also holds if the Brownian particle is smaller than the gas particles.

\section{Mean position times momentum\label{sec8.5}}

Applying the Liouville operator \Eqref{8.4} on $\{\hat x,\hat p\}$ results in
	\begin{eqnarray}
		 \mathcal L^\dagger \{\hat x,\hat p\}   &=&  \frac\imath{2\hbar m}\left[\hat p^2,\{\hat x,\hat p\}\right]  -  \half\left(\{\hat x,\hat p\} \hat R +  \hat R\{\hat x,\hat p\}\right)  +  \int\!\!\!\!\int\!\!\!\!\int\!\!\!\!\int\dd x_g\,\dd p_g\,\dd\tilde x_{}\,\dd\tilde p_{}  \nn\\
		& \hspace{-28mm}\times&\hspace{-16mm}  U_{}\!\left(\tfrac{\delta{}}2\right)\! \sqrt{  \hat R_\delta{}(x_g,p_g)} \hat B^\dag(x_g,p_g;\tilde x_{},\tilde p_{}) U_{}^\dagger\!\left(\tfrac{\delta{}}2\right)\! \{\hat x,\hat p\} U_{}\!\left(\tfrac{\delta{}}2\right)\! \hat B(x_g,p_g;\tilde x_{},\tilde p_{}) \sqrt{  \hat R_\delta{}(x_g,p_g)} U_{}^\dagger\!\left(\tfrac{\delta{}}2\right)\!.  \nn\\ \label{8.52}
	\end{eqnarray}
Similar to the sections before we find 
	\begin{eqnarray}
		\int\!\!\!\!\int\dd\tilde x_{}\,\dd\tilde p_{} \, \hat B^\dag(x_g,p_g;\tilde x_{},\tilde p_{}) U_{}^\dagger\!\left(\tfrac{\delta{}}2\right)\! \{\hat x,\hat p\} U_{}\!\left(\tfrac{\delta{}}2\right)\! \hat B(x_g,p_g;\tilde x_{},\tilde p_{}) \hspace{-77mm} &&\nn\\
		&=& \frac{1}{(1+\alpha)^2}  \left\{(1-\alpha)\hat x + 2\alpha x_g +\frac{\delta{}}{2m}\left[(1-\alpha)\hat p+2p_g^{\phantom 2} \right],(1-\alpha)\hat p+2p_g \right\}  \nn\\  \label{8.53}
	\end{eqnarray}
where we used Tr$\left(\sqrt{\hat \sigma} \hat p\sqrt{\hat \sigma}\hat x\right)=0$ and similar relations already used in section~\ref{sec8.3} as well as the limit $W\to\infty$. Together with $\left[\hat p^2,\{\hat x,\hat p\}\right] = -4\imath\hbar \hat p^2$ we get
	\begin{eqnarray}
		 \mathcal L^\dagger \{\hat x,\hat p\}   &=&  \frac{2\left\langle\hat p^2\right\rangle}{m}   -   \hat R\{\hat x,\hat p\}  + \frac{1}{(1+\alpha)^2} \int\!\!\!\!\int\dd x_g\,\dd p_g \,U_{}\!\left(\tfrac{\delta{}}2\right)\! {  \hat R_\delta{}(x_g,p_g)}  \nn\\
		&& \times  \left\{(1-\alpha)\hat x + 2\alpha x_g +\frac{\delta{}}{2m}\left[(1-\alpha)\hat p+2p_g^{\phantom 2} \right],(1-\alpha)\hat p+2p_g \right\}  U_{}^\dagger\!\left(\tfrac{\delta{}}2\right)\!.  \nn\\ \label{8.54}
	\end{eqnarray}
We continue with the help of
	\begin{eqnarray}
		  \hat R \{\hat x,\hat p\} &=&  2 \int\dd p_g\,\mu_{T}(p_g) \int\!\!\!\!\int  \frac{\dd x_{}\,\dd p_{}}{2\pi\hbar}\ket{x_{},p_{}}\!\bra{x_{},p_{}}  \left| \frac pm-\frac{p_g}{m_g}\right|  xp \quad
	\end{eqnarray}
and similar relations already used in section~\ref{sec8.3}, as well as \Eqref{xf(p)} to get the final result 
	\begin{eqnarray}
		\frac{\dd \langle\{\hat x,\hat p\}\rangle}{\dd t} &=& \frac{2\left\langle \hat p^2\right\rangle}{m} -  \langle\{\hat x,f_T(\hat p)\}\rangle
	\end{eqnarray}
with $f_T(\hat p)$ from \Eqref{8.27}. In the slow Brownian particle limit, we can again expand $f_T(\hat p)$ and find
	\begin{eqnarray}
		\frac{\dd \langle\{\hat x,\hat p\}\rangle}{\dd t} &\approx& \frac{2\left\langle \hat p^2\right\rangle}{m} - \frac{4n _g\sqrt{2 m_gk_BT}}{\sqrt{\pi}(m+m_g)} \langle\{\hat x,\hat p\}\rangle.
	\end{eqnarray}

\section{Solutions for expectation values\label{solution}}

In the limit of a slow Brownian particle $\left(|v|\ll \sqrt{k_BT/m_g}\right)$, the equations of motion derived in the previous subsections are exactly solvable. We will also use the limit of a heavy Brownian particle $1+\alpha\approx1$, to ensure the particle stays slow during the evolution. Then the equations of motion for the first two moments of position and momentum are
	\begin{eqnarray}
		\frac{\dd \langle \hat x \rangle}{\dd t} &=& \frac{\langle\hat p\rangle}{m}. \label{8.59} \\
		\frac{\dd\langle\hat p\rangle}{\dd t} &=& -\gamma \langle\hat p\rangle \label{ 8.60} \\
		\frac{\dd \left\langle\hat x^2\right\rangle}{\dd t} &=& \frac{\langle \{\hat x,\hat p\} \rangle}{m} + \frac{\alpha^2 n _g}{3\sqrt\pi}\left(\frac{2k_BT}{m_g}\right)^{\!3/2}  \delta{}^2  \label{8.61} \\
		\frac{\dd\left\langle\hat p^2\right\rangle}{\dd t} &=& 2\gamma \!\left[  mk_BT-\left\langle \hat p^2\right\rangle \right]  \label{8.62} \\
		\frac{\dd \langle\{\hat x,\hat p\}\rangle}{\dd t} &=& \frac{2\left\langle \hat p^2\right\rangle}{m} - \gamma \langle\{\hat x,\hat p\}\rangle , \label{8.63}
	\end{eqnarray}
where we defined
	\begin{equation}
		\gamma \;=\; \frac{4n _g\sqrt{2m_gk_BT}}{\sqrt\pi m}.
	\end{equation}
In the third equation we used \Eqref{8.35} to evaluate $g(\hat p)$ for slow Brownian velocities. We keep the position diffusion contribution, to show that it is negligible on time scales large compared to $\delta$. These are all linear first order differential  equations which are  easily solved using standard techniques. The second equation is immediately solved by
	\begin{eqnarray}
		\langle\hat p(t)\rangle &=& \exp\!\left(  -\gamma t  \right) \langle\hat p(0)\rangle .
	\end{eqnarray}
This in turn can be substituted into \Eqref{8.59} and integration results in
	\begin{eqnarray}
		\langle\hat x(t)\rangle &=& \langle\hat x(0)\rangle +  \frac{1}{m\gamma}  \left( 1-e^{-\gamma t} \right) \langle\hat p(0)\rangle .
	\end{eqnarray}
These are the classical motions for position and momentum of a linearly damped particle. We start the second moments with
	\begin{eqnarray}
		\left\langle\hat p^2(t)\right\rangle &=& mk_BT\left( 1-e^{-2\gamma t} \right) + e^{-2\gamma t} \left\langle\hat p^2(0)\right\rangle  ,\qquad
	\end{eqnarray}
 which we substitute into \Eqref{8.63} to find
	 \begin{eqnarray}
		\langle\{\hat x,\hat p\}(t)\rangle &=& \frac{2k_BT}{\gamma} \left( 1-e^{-\gamma t} \right)^2 + \frac{2\left\langle\hat p^2(0) \right\rangle }{m\gamma}   \left(e^{-\gamma t}-e^{-2\gamma t}\right)   +  \langle\{\hat x,\hat p\}(0)\rangle  e^{-\gamma t} , \nn\\
	\end{eqnarray}
 which in turn can be substituted in \Eqref{8.61} to yield
	\begin{eqnarray}
		\left\langle \hat x^2(t) \right\rangle &=& \left\langle \hat x^2(0) \right\rangle  +  \frac{\langle\{\hat x,\hat p\}(0)\rangle}{m\gamma}\left( 1-e^{-\gamma t} \right) + \frac{\left\langle\hat p^2(0) \right\rangle }{(m\gamma)^2}\left( 1-e^{-\gamma t} \right)^2 \nn\\
		&& - \frac{k_BT}{m\gamma^2}\left(  3-4e^{-\gamma t}+e^{-2\gamma t}  \right)   +t\left[  \frac{2k_BT}{m\gamma}+\delta{}^2\frac{\alpha^2 n _g}{3\sqrt\pi} \left(\frac{2k_BT}{m_g}\right)^{\!3/2}  \right]. \qquad\quad\label{dfsu}
	\end{eqnarray}

Now we only need to substitute $\gamma$ and to use \Eqref{eeee} to see that the term involving $\delta{}$ can be neglected compared to the other term in the square brackets. Therefore, the unphysical position diffusion is indeed small compared to the real physical position diffusion, which occurs on a time scale large compared to $\gamma^{-1}$. We also have to check how the unphysical position diffusion term behaves on short time scales. Expanding the exponentials in \Eqref{dfsu}, and substituting $\gamma$, we find for the second line
	\begin{eqnarray}
		\frac{n_g\sqrt{m_g}(2k_BT)^{3/2}}{3\sqrt{\pi}m^2}\left(4t^3+t\delta^2\right).
	\end{eqnarray}
Hence, for times large compared to the coarse graining time, the effect of the unphysical position diffusion is indeed small compared to the real position spreading, as promised in section~(\ref{sec8.3}). Because all our results are only valid on a coarse grained time scale, we can rightly claim that position diffusion is an artifact of the approximations used to derive the Markovian master equation. 

This is very reassuring. After all, this mysterious position diffusion could only be explained by random position jumps, and we already showed in section~\ref{jumps} that these are not possible in collisional quantum Brownian motion. 

 It is interesting to have a look at the long time limit $\gamma t\gg 1$ of these equations
	\begin{eqnarray}
		\langle\hat p(\infty)\rangle &=& 0 \\
		\langle\hat x(\infty)\rangle &=& \langle\hat x(0)\rangle +  \frac{\langle\hat p(0)\rangle}{m\gamma}    \label{8.71}\\
		\left\langle\hat p^2(\infty)\right\rangle &=&  mk_BT  \\
		\left\langle\{\hat x,\hat p\}(\infty)\right\rangle &=&  \frac{2k_BT}{\gamma}  \\
		\left\langle\hat x^2(\infty)\right\rangle &=& \left\langle \hat x^2(0) \right\rangle  +  \frac{\langle\{\hat x,\hat p\}(0)\rangle}{m\gamma} + \frac{\left\langle\hat p^2(0) \right\rangle }{(m\gamma)^2}  - \frac{3k_BT}{m\gamma^2} +  \frac{2k_BT}{m\gamma}t. \quad
	\end{eqnarray}
We find that the mean energy approaches $k_BT/2$ as required by the equipartition theorem of thermodynamics. Furthermore, the mean square displacement increases linearly in time, as is common in diffusive processes.

Also interesting is the short time limit $\gamma t\ll 1$. While the Brownian particle mean squared momentum increases linear in time, the mean squared displacement increases according to $2k_BT\gamma t^3/(3m)$. This is typical for the process of momentum diffusion.

To summarize, we find the expected momentum diffusion. On time scales large compared to the damping time $\gamma^{-1}$, this results in indirect position diffusion, because according to \Eqref{8.71},  each momentum kick $q$ can be associated with a position change of $q/(m\gamma)$ in the long term. Direct position diffusion (which is an artifact of our approximations) is found to be negligible on time scales large compared to the coarse graining time scale $\hbar/(k_BT)$.

\section{Standard form of QBM master equation}

A common master equation for QBM is of the form 
	\begin{eqnarray}
		\frac{\dd \rho_{}(t)}{\dd t} &=& -\frac\imath\hbar \left[ H_{},\rho_{}(t)\right] - \imath \frac{\gamma}{2\hbar}[\hat x,\{\hat p,\rho\}] - \frac{D_{pp}}{\hbar^2}[\hat x,[\hat x,\rho]] - \frac{D_{xx}}{\hbar^2}[\hat p,[\hat p,\rho]], \label{8.75} \qquad
	\end{eqnarray}
which we call the standard QBM master equation. To derive such an equation for collisional QBM \cite{Diosi,mainII,Hornberger3}, the limit of a small collisional momentum change of the Brownian particle has to be used. But also for non-collisional models, such as the Caldeira Leggett model, this form was put forward \cite{Breuer}. For this equation to be of Lindblad form,
	\begin{eqnarray}
		D_{xx}D_{pp} &\geqslant& (\hbar\gamma/4)^2 \label{8.76}
	\end{eqnarray}
is required. By comparing our equations of motions for the lowest order moments to the ones derived from the standard QBM master equation, we can find the parameters $\gamma$, $D_{xx}$, and $D_{pp}$ corresponding to our theory. We will see that this procedure is only possible in the heavy Brownian particle limit, and even then it is not certain whether the resulting master equation correctly describes other properties of QBM, like for example decoherence. However, for completeness and because of the simple form of the standard QBM master equation, we proceed with the task despite these problems.

The equations of motion for the first and second moments of position and momentum found from the standard QBM master equation are
	\begin{eqnarray}
		\frac{\dd \langle \hat x \rangle}{\dd t} &=& \frac{\langle\hat p\rangle}{m}. \nn \\
		\frac{\dd\langle\hat p\rangle}{\dd t} &=& -\gamma \langle\hat p\rangle \nn \\
		\frac{\dd \left\langle\hat x^2\right\rangle}{\dd t} &=& \frac{\langle \{\hat x,\hat p\} \rangle}{m} + 2D_{xx}  \nn\\
		\frac{\dd\left\langle\hat p^2\right\rangle}{\dd t} &=&  -2\gamma\left\langle \hat p^2\right\rangle + 2D_{pp}  \nn \\
		\frac{\dd \langle\{\hat x,\hat p\}\rangle}{\dd t} &=& \frac{2\left\langle \hat p^2\right\rangle}{m} - \gamma \langle\{\hat x,\hat p\}\rangle.
	\end{eqnarray}
These can be compared with the equations of motions, which we derived in sections~\ref{sec8.1} to \ref{sec8.5}, to find values for the parameters $\gamma,\;D_{pp}$, and $D_{xx}$. Interestingly, we can match these parameters only for a heavy Brownian particle ($\alpha\to0$) with small velocities $|v|\ll \sqrt{k_BT/m_g}$. That means, only in such circumstances the standard form of the QBM master equation might describe the physics correctly. We find
	\begin{eqnarray}
		\gamma &=& \frac{4n _g\sqrt{2m_gk_BT}}{\sqrt\pi m} \\
		D_{pp} &=& mk_BT\gamma \\
		D_{xx} &=& 0.
	\end{eqnarray}
Surely these parameters violate inequality~(\ref{8.76}) and therefore do not result in a completely positive master equation. However, had we used \Eqref{8.61} for the position variance and not neglected the term proportional to $\delta{}$, we had found
	\begin{eqnarray}
		D_{xx} &=& \frac{n _g}{6\sqrt\pi}\left(\frac{2k_BT}{m_g}\right)^{\frac32}  \delta{}^2. \label{yuh}
	\end{eqnarray}
Using \Eqref{7.19} it is straightforward to show that  this parameter fulfills the inequality~(\ref{8.76}). Therefore we find a completely positive master equation only if we keep contributions from unphysical position jumps. So what's going on here?

The problem is that QBM is not a strictly Markovian open system. Therefore, one needs approximations to produce a Markovian equation of motion. Here, as in most other approaches, that is the introduction of a coarse grained time scale. This approximation can result in a non-completly positive master equation even if the approximation is very well satisfied in the physical system as in QBM in the high temperature and low density limit.

Another prominent example of this sort of problems is the quantum optical master equation~\cite{Breuer}. The Born-Markov approximation is usually well satisfied, but it results in a master equation which is not completely positive. A second approximation is done, called the rotating wave approximation, to cast the master equation into Lindblad form.

In our approach to QBM the introduction of a coarse grained time scale does actually result in a completely positive master equation as is seen in \Eqref{7.13}. However, it introduces non-physical position jumps, and if these are removed by hand, it can not be expected that the master equation stays completely positive.

One should keep in mind that here \Eqref{8.75} is postulated, and although its parameters are chosen to produce the correct equations of motion for the first two moments of position and momentum, it is not necessarily the correct master equation describing QBM. For a correct treatment of QBM one has to refer to the more complicated master equation~(\ref{7.13}).

\section*{Appendix A}

Here we consider operators first shifted in phase space, and then integrated over phase space. For this purpose we will need 
	\begin{eqnarray}
		\mbox{Tr}(\hat O ) &=& \int \dd x \bra x \hat O \ket x \;=\, \int\!\!\!\!\int\!\!\!\!\int \dd x \frac{\dd x'\,\dd p'}{2\pi\hbar}\scalar{x}{x',p'}\!  \bra{x',p'} \hat O \ket x  \nn\\
		&=& \int\!\!\!\!\int\!\!\!\!\int \dd x \frac{\dd x'\,\dd p'}{2\pi\hbar}\bra{x',p'} \hat O \ket x \!\scalar{x}{x',p'} \;=\, \int\!\!\!\!\int \frac{\dd x'\,\dd p'}{2\pi\hbar}\bra{x',p'} \hat O \ket{x',p'}. \qquad\quad\;
	\end{eqnarray}
Then the relation
	\begin{eqnarray}
		\int\!\!\!\!\int \frac{\dd   x\,\dd  p}{2\pi\hbar} \hat{D}(  x,  p) \hat O \hat{D}^\dag(  x,  p) &=& \mbox{Tr}(\hat O) \one \label{O}
	\end{eqnarray}
is shown by its expectation value on a general Gaussian state $\ket{x',p'}$
	\begin{eqnarray}
		\bra{x',p'}\int\!\!\!\!\int \frac{\dd   x\,\dd  p}{2\pi\hbar} \hat{D}(  x,  p) \hat O \hat{D}^\dag(  x,  p) \ket{x',p'} &=& \int\!\!\!\!\int \frac{\dd x\,\dd p}{2\pi\hbar}\bra{x'\!-\!x,p'\!-\!p} \hat O \ket{x'\!-\!x,p'\!-\!p} \nn\\
		&=& \mbox{Tr}(\hat O), \nn
	\end{eqnarray}
which uniquely determines the operator. Applying \Eqref{O} to $\hat O\hat x$
	\begin{eqnarray}
		 \mbox{Tr}(\hat O \hat x) \one &=& \int\!\!\!\!\int \frac{\dd   x\,\dd  p}{2\pi\hbar} \hat{D}(  x,  p) \hat O D^\dag(  x,  p) D(  x,  p)\hat x \hat{D}^\dag(  x,  p) \nn\\
		 &=& \int\!\!\!\!\int \frac{\dd   x\,\dd  p}{2\pi\hbar} \hat{D}(  x,  p) \hat O D^\dag(  x,  p) (\hat x -x), \nn
	\end{eqnarray}
leads us to a related formula 
	\begin{eqnarray}
		\int\!\!\!\!\int \frac{\dd   x\,\dd  p}{2\pi\hbar}x \hat{D}(  x,  p) \hat O \hat{D}^\dag(  x,  p) &=& \mbox{Tr}(\hat O)\hat x -\mbox{Tr}(\hat O\hat x) \one. \label{Ox}
	\end{eqnarray}
In a similar manner, the following relations are found
	\begin{eqnarray}
		\int\!\!\!\!\int \frac{\dd   x\,\dd  p}{2\pi\hbar} p \hat{D}(  x,  p) \hat O \hat{D}^\dag\!(  x,  p) &\!=\!& \mbox{Tr}(\hat O)\hat p -\mbox{Tr}(\hat O\hat p) \one \label{Op} \\
		\int\!\!\!\!\int \frac{\dd   x\,\dd  p}{2\pi\hbar} x^2 \hat{D}(  x,  p) \hat O \hat{D}^\dag\!(  x,  p) &\!=\!& \mbox{Tr}(\hat O)\hat x^2 -2\mbox{Tr}(\hat O\hat x)\hat x + \mbox{Tr}(\hat O\hat x^2) \one \label{Oxx} \\
		\int\!\!\!\!\int \frac{\dd   x\,\dd  p}{2\pi\hbar} p^2 \hat{D}(  x,  p) \hat O \hat{D}^\dag\!(  x,  p) &\!=\!& \mbox{Tr}(\hat O)\hat p^2 -2\mbox{Tr}(\hat O\hat p)\hat p + \mbox{Tr}(\hat O\hat p^2) \one \label{Opp} \\
		\int\!\!\!\!\int\! \frac{\dd   x\,\dd  p}{2\pi\hbar} 2xp \hat{D}(  x,  p) \hat O \hat{D}^\dag\!(  x,  p) &\!=\!& \mbox{Tr}(\hat O)\{\hat x,\hat p\} - 2\mbox{Tr}(\hat O\hat x)\hat p - 2\mbox{Tr}(\hat O\hat p)\hat x + \mbox{Tr}(\hat O\{\hat x,\hat p\}) \one \nn\\ \label{Oxp} 
	\end{eqnarray}

\section*{Appendix B}

Here we evaluate several integrals over $x_g$, restricted to the region $S_\delta(x_g,p_g)$ defined in \Eqref{6.8}.
	\begin{eqnarray}
		\int \dd x_g \int\!\!\!\!\int_{0<\frac{x_{}-x_g}{\frac{p_g}{m_g}-\frac{p_{}}{m_{}}}<\delta{} } \dd x\,\dd p\, f(x,p) &=& \int\!\!\!\!  \int\!\!\!\!\int_{0<\frac{x_{}-x_g}{\frac{p_g}{m_g}-\frac{p_{}}{m_{}}}<\delta{} } \dd x_g\,\dd x\,\dd p\, f(x,p) \nn\\
		&\hspace{-30mm}=&\hspace{-15mm} \int\!\!\!\!  \int \dd x\,\dd p \,f(x,p) \int_{0<\frac{x_{}-x_g}{\frac{p_g}{m_g}-\frac{p_{}}{m_{}}}<\delta{} } \dd x_g \nn\\
		&\hspace{-30mm}=&\hspace{-15mm} \int\!\!\!\!  \int \dd x\,\dd p\, f(x,p) \left\{\!\!
			\begin{array}{lr}
				\int_{x\phantom{\frac pm}}^{x-\left( \frac{p_g}{m_g}-\frac{p}{m} \right)\delta} \dd x_g\; & \mbox{  for }\frac{p_g}{m_g}-\frac{p_{}}{m_{}}<0 \\
				\int_{x-\left( \frac{p_g}{m_g}-\frac{p_{}}{m_{}} \right)\delta}^{x\phantom{\frac pm}} \dd x_g  & \mbox{for }\frac{p_g}{m_g}-\frac{p_{}}{m_{}}>0 
			\end{array}\right.\!\nn\\
			&\hspace{-30mm}=&\hspace{-15mm} \delta\int\!\!\!\!\int \dd x\,\dd p \left| \frac{p_g}{m_g}-\frac{p_{}}{m_{}} \right| f(x,p) \label{8.19}
	\end{eqnarray}
	\begin{eqnarray}
		\int \dd x_g \left(x_g+\frac{\delta p_g}{2m_g}\right) \int\!\!\!\!\int_{0<\frac{x_{}-x_g}{\frac{p_g}{m_g}-\frac{p_{}}{m_{}}}<\delta{} } \dd x\,\dd p\, f(x,p) &&\nn\\
		&\hspace{-100mm}=&\hspace{-50mm} \int\!\!\!\!  \int \dd x\,\dd p \,f(x,p) \int_{0<\frac{x_{}-x_g}{\frac{p_g}{m_g}-\frac{p_{}}{m_{}}}<\delta{} } \dd x_g \left(x_g+\frac{\delta p_g}{2m_g}\right) \nn\\
		&\hspace{-100mm}=&\hspace{-50mm} \int\!\!\!\!  \int \dd x\,\dd p\, f(x,p) \left\{\!\!
			\begin{array}{lr}
				\int_{x+\frac{\delta p_g}{2m_g}}^{x-\frac{\delta p_g}{2m_g}+\frac{\delta p}{m}} \dd x_g\,x_g\; & \mbox{  for }\frac{p_g}{m_g}-\frac{p_{}}{m_{}}<0 \\
				\int^{x+\frac{\delta p_g}{2m_g}}_{x-\frac{\delta p_g}{2m_g}+\frac{\delta p}{m}} \dd x_g\,x_g\;  & \mbox{for }\frac{p_g}{m_g}-\frac{p_{}}{m_{}}>0
			\end{array}\right.\!\nn\\
			&\hspace{-100mm}=&\hspace{-50mm}\delta\int\!\!\!\!\int \dd x\,\dd p \left| \frac{p_g}{m_g}-\frac{p_{}}{m_{}} \right| \left( x+\frac{\delta p}{2m}\right)  f(x,p)\qquad \label{8.19b}
	\end{eqnarray}
	\begin{eqnarray}
		\int \dd x_g \left(x_g+\frac{\delta p_g}{2m_g}\right)^2 \int\!\!\!\!\int_{0<\frac{x_{}-x_g}{\frac{p_g}{m_g}-\frac{p_{}}{m_{}}}<\delta{} } \dd x\,\dd p\, f(x,p) &=& \nn\\
		&\hspace{-120mm}=&\hspace{-60mm} \int\!\!\!\!  \int \dd x\,\dd p\, f(x,p) \left\{\!\!
			\begin{array}{lr}
				\int_{x+\frac{\delta p_g}{2m_g}}^{x-\frac{\delta p_g}{2m_g}+\frac{\delta p}{m}} \dd x_g\,x_g^2\; & \mbox{  for }\frac{p_g}{m_g}-\frac{p_{}}{m_{}}<0 \\
				\int^{x+\frac{\delta p_g}{2m_g}}_{x-\frac{\delta p_g}{2m_g}+\frac{\delta p}{m}} \dd x_g\,x_g^2\;  & \mbox{for }\frac{p_g}{m_g}-\frac{p_{}}{m_{}}>0 
			\end{array}\right.\!\nn\\
			&\hspace{-120mm}=&\hspace{-60mm}\delta\!\int\!\!\!\!\int \dd x\,\dd p \left| \frac{p_g}{m_g}-\frac{p_{}}{m_{}} \right| \left[\! \left(x+\frac{\delta p}{2m}\right)^{\!2} \! + \frac{\delta^2}{12}\left(\frac{p_g}{m_g}-\frac pm\right)^{\!2} \right]\!  f(x,p) \qquad\qquad \label{8.19c}
	\end{eqnarray}

\chapter{Decoherence\label{final}}

Coherences of a quantum state are often described by the off-diagonals of the density matrix. This leads to the question: \emph{What basis should we use to examine decoherence?} The two bases which first come to mind are the position basis and the momentum basis, and indeed, these are the bases usually used in the literature~\cite{Joos,Gallis,Hornberger,review,Breuer2}. Nevertheless, they are not without problems. 

If one uses e.g.\ the momentum basis, one would be interested in how long a superposition of the form $(\ket{p_a}+\ket{p_b})/\sqrt2$ survives, or, equivalently, how fast the coherences $\ket{p_a}\!\bra{p_b}/2$ decrease. There is a major problem in this sort of question: A momentum eigenstate (or any state which is very localized in momentum) is itself a coherent superposition of widely separated position eigenstates, and there is no guarantee that these position coherences of each individual momentum eigenstate do not vanish on the same time scale, or even faster, as the momentum coherences of interest! In other words, by the time the coherence between $\ket{p_a}$ and $\ket{p_b}$ decreases, the Brownian particle might significantly change its momentum distribution. The same reasoning applies to using the position basis.

Our reservations are clearly related to the concept of a pointer basis, which we outlined in section~\ref{decointro}. Could we possibly single out a pointer basis from the measurement interpretation of a single collision? In chapter~\ref{meas} we saw that a colliding gas particle $\ket{x_g,p_g}_{W_g}$ performs a smeared out measurement on the Brownian particle in the basis $\ket{x,p}_W$, indicating that these states could be used as pointer basis. However, we also know from section~\ref{convex decomp}, that we have a choice in the width $W_g$ of the gas particle states. Does this mean, that we can use Gaussian states of any width $W=\sqrt{m/m_g}W_g$ as pointer basis, as long as $W_g$ satisfies the relation~(\ref{Wglim})
	\begin{eqnarray}
		 \frac{\hbar}{\sqrt{m_gk_BT}} \,\;\ll &W_g& \ll\;\, \frac1{n_g}, \label{Wglima}
	\end{eqnarray}
which we required for our treatment of a single collision?

As we mentioned in section~\ref{decointro}, there are still many open questions about the emergence of a pointer basis from the coupling to an environment. Nevertheless, because of the heuristic reasoning in the previous paragraph, and the lack of sensible alternatives, we will indeed use the Gaussian states $\ket{x,p}$ for the study of decoherence. In particular, we will discuss the vanishing of coherences of states of the form $\ket{x_a,p_a}+\ket{x_b,p_b}$, commonly referred to as \emph{cat states} in the literature.

Rather than using a density operator representation in terms of Gaussian states, we found it more graphic to use the Wigner function (see subsection~\ref{Wfunc}) to display coherences of Gaussian states. Decoherence can then be discussed in terms of the vanishing of the oscillatory behavior of the Wigner function.

We wish to mention that a Wigner function description of QMB was already put forward previously in a very heuristic derivation~\cite{Halliwell}, as well as in a more precise approach, but limited to states of the Brownian particle which are close to a thermal state~\cite{Hornberger3}. Both articles are concerned about the general form of a partial differential equation for the Wigner function, and do not study decoherence.

In the following, we will compare the Wigner function without a collision, to the Wigner function with a collision. This way, we will obtain the \emph{decoherence per collision}, which can be multiplied by the collision rate \Eqref{6.20}, $R\approx n_g\sqrt{2k_BT}/\sqrt{\pi m_g}$, to obtain the decoherence rate.

\section{Collisional decoherence for general gas particles\label{medi}}

In this chapter, we will frequently encounter density operators of the form
	\begin{eqnarray}
		\rho &=& \ket{x_a,p_a}\!\bra{x_a,p_a} + \ket{x_b,p_b}\!\bra{x_b,p_b} + ce^{\imath \varphi} \ket{x_b,p_b}\!\bra{x_a,p_a} + ce^{-\imath\varphi} \ket{x_a,p_a}\!\bra{x_b,p_b}.  \nn\\ \label{state1}
	\end{eqnarray}
Here, $c$ is bound between zero and one and is a measure for the strength of the coherences between $\ket{x_a,p_a}$ and $\ket{x_b,p_b}$, and $\varphi$ determines the relative phase between these states.  In particular, an incoming Brownian particle state $\ket{x_a,p_a}+\ket{x_b,p_b}$ before a collision corresponds to $c=1$ and $\varphi=0$.

Let us start by deriving a general analytic expression for the Wigner function of the state \Eqref{state1}. Because the Wigner function is linear in the density operator, it can be calculate term by term, using the position representation of the Gaussian states \Eqref{abel}, as well as the definition of the Wigner function \Eqref{wfunc}. The result is 
	\begin{eqnarray}
		W_\rho(x',p') &=& \frac1{\pi\hbar}\exp\!\left[ -\frac{(x'-x_a)^2}{W^2} \right] \exp\!\left[ -\frac{W^2(p'-p_a)^2}{\hbar^2} \right] \nn\\
		&+& \frac1{\pi\hbar}\exp\!\left[ -\frac{(x'-x_b)^2}{W^2} \right] \exp\!\left[ -\frac{W^2(p'-p_b)^2}{\hbar^2} \right] \nn\\
		&+& \frac{2c}{\pi\hbar} \exp\!\left[ -\frac{(x'-x_A)^2}{W^2} \right] \exp\!\left[ -\frac{W^2(p'-p_A)^2}{\hbar^2} \right] \nn\\
		&&\times\;\cos\!\left[\varphi + \frac{x_Ap_D-p_Ax_D}{2\hbar} + x_D\frac{p_A-p'}{\hbar} - p_D\frac{x_A-x'}{\hbar}  \right]\!, \qquad\quad \label{Wgeneral}
	\end{eqnarray}
where we have defined the average position $x_A=(x_a+x_b)/2$ and the difference $x_D=(x_a-x_b)$ (analogous for momenta).

We noted in subsection~\ref{Wfunc}, that the strength of coherences, indicated by oscillatory behavior of the Wigner function, do not change due to the unitary free evolution. Therefore, we will mostly use the interaction picture for what follows\footnote{The time evolved Wigner function is obtained by replacing $x'$ by $(x'-p't/m)$ in \Eqref{Wgeneral}. Because this would unnecessarily complicate matters, we will use the interaction picture Wigner function, i.e.\ without time evolution.}$^,$\footnote{We could also say, we study the collisional process according to figure~\ref{smatrix}~(b).}. We can then use the general formula \Eqref{Wgeneral}, with $c=1$ and $\varphi=0$, to plot the Wigner function of a state $\ket{\psi}=\ket{x_a,p_a}+\ket{x_b,p_b}$ without a collision in figure~\ref{Wcompare}~(a).

According to section~\ref{cat}, a collision with a gas particle state $\ket{x_g,p_g}$, results in
	\begin{eqnarray}
		\ket{x_{a/b},p_{a/b}} &\to& \ket{\bar x_{a/b},\bar p_{a/b}} \nn\\
		c=1 &\to& c=\exp\left[ -\frac\alpha{(1+\alpha)^2} \left(\frac{x_D^2}{W ^2} + \frac{W ^2p_D^2}{\hbar^2} \right)  \right]  \nn\\
		\varphi=0 &\to& \varphi=\frac{2\alpha(x_A p_D-x_D p_A) + (1-\alpha)p_gx_D - \alpha(1-\alpha)x_gp_D}{(1+\alpha)^2\hbar} ,\qquad \label{Wafter}
	\end{eqnarray}
where we used $\bar x_a=\bar x(x_g,x_a),\;\bar p_a=\dots,$ given by \Eqref{noname2}~-~(\ref{noname}). The Wigner function after the collision is plotted in figure~\ref{Wcompare}~(b).
\begin{figure}[htbp]
\begin{center}
	\includegraphics[width=\linewidth]{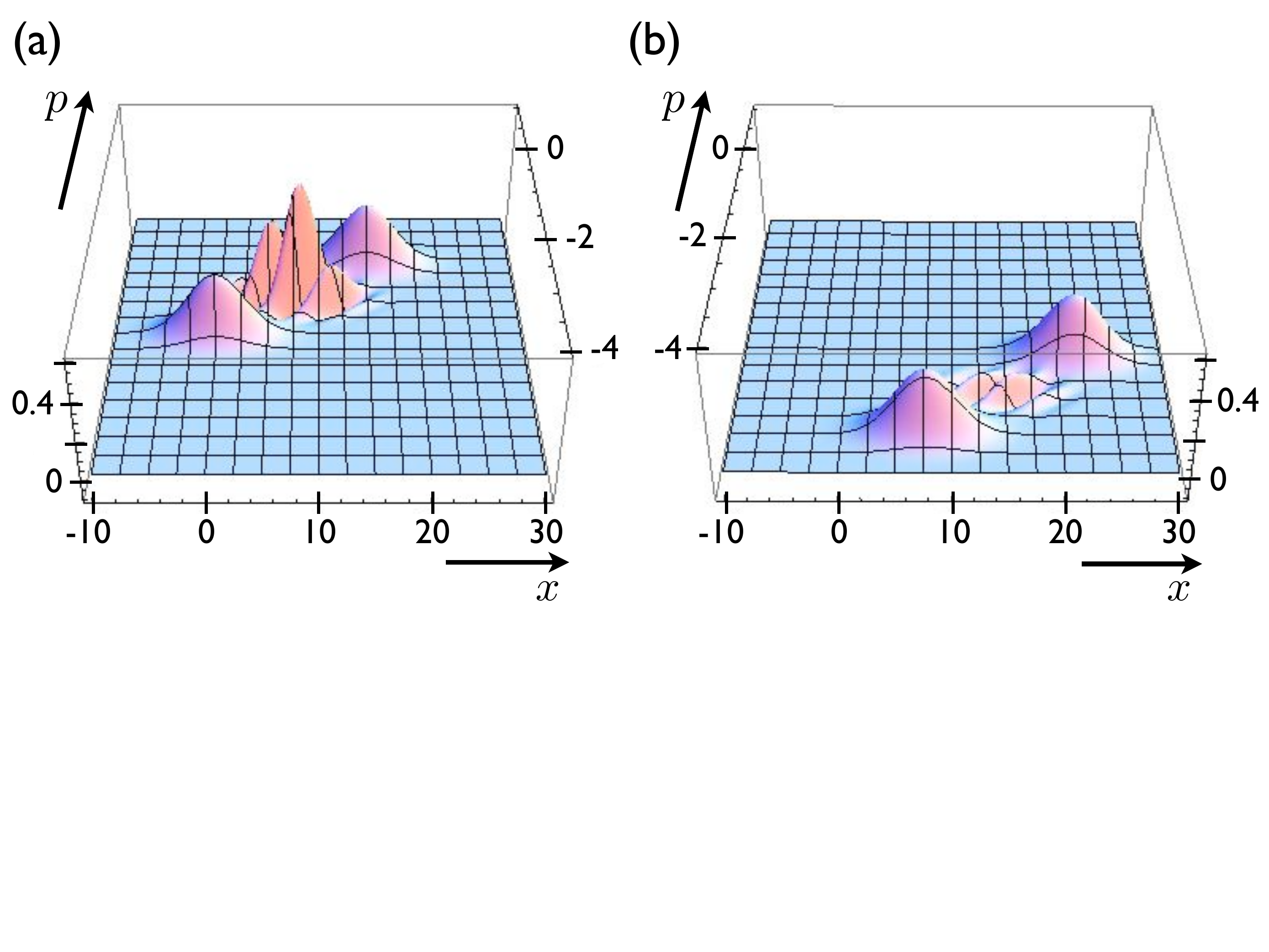}\vspace{-40mm}
\caption{\small The Wigner function of the initial coherent superposition $\ket{\psi}=\ket{x_a,p_a}+\ket{x_b,p_b}$~(a), and of the state resulting from a collision with $\ket{x_g,p_g}$~(b). Parameters are: $x_a=15,\;x_b=0,\;p_a=0,\;p_b=1.5,\;W=4,\;m=1,\;\hbar=1,\;x_g=100,\;p_g=-1,$ and $\alpha=0.04$.}
\label{Wcompare}
\end{center}
\end{figure}

It is quite astonishing, that, despite choosing a gas particle with only four percent of the mass of the Brownian particle, and, despite using a superposition of very close Gaussian wave functions, almost all coherences are lost after a single collision. If we had separated the initial Gaussians only slightly more, or had chosen only a slightly heavier gas particle, the coherences would be not visible at all, because $c$ in \Eqref{Wafter} decreases exponentially with these parameters. This observation is independent on the initial momentum and position of the colliding gas particle, as well as whether the Gaussian wave functions are separated predominantly in position or momentum. \textbf{It is therefore fair to say, that, unless the gas particle is much lighter than the Brownian particle, the decoherence rate equals the collision rate.} This result becomes even much more pronounced, if we average over different initial gas particle positions and momenta, and we will study this effect in the following section.

\section{Collisional decoherence for very light gas particles}

\begin{figure}[htbp]
\begin{center}
	\includegraphics[width=\linewidth]{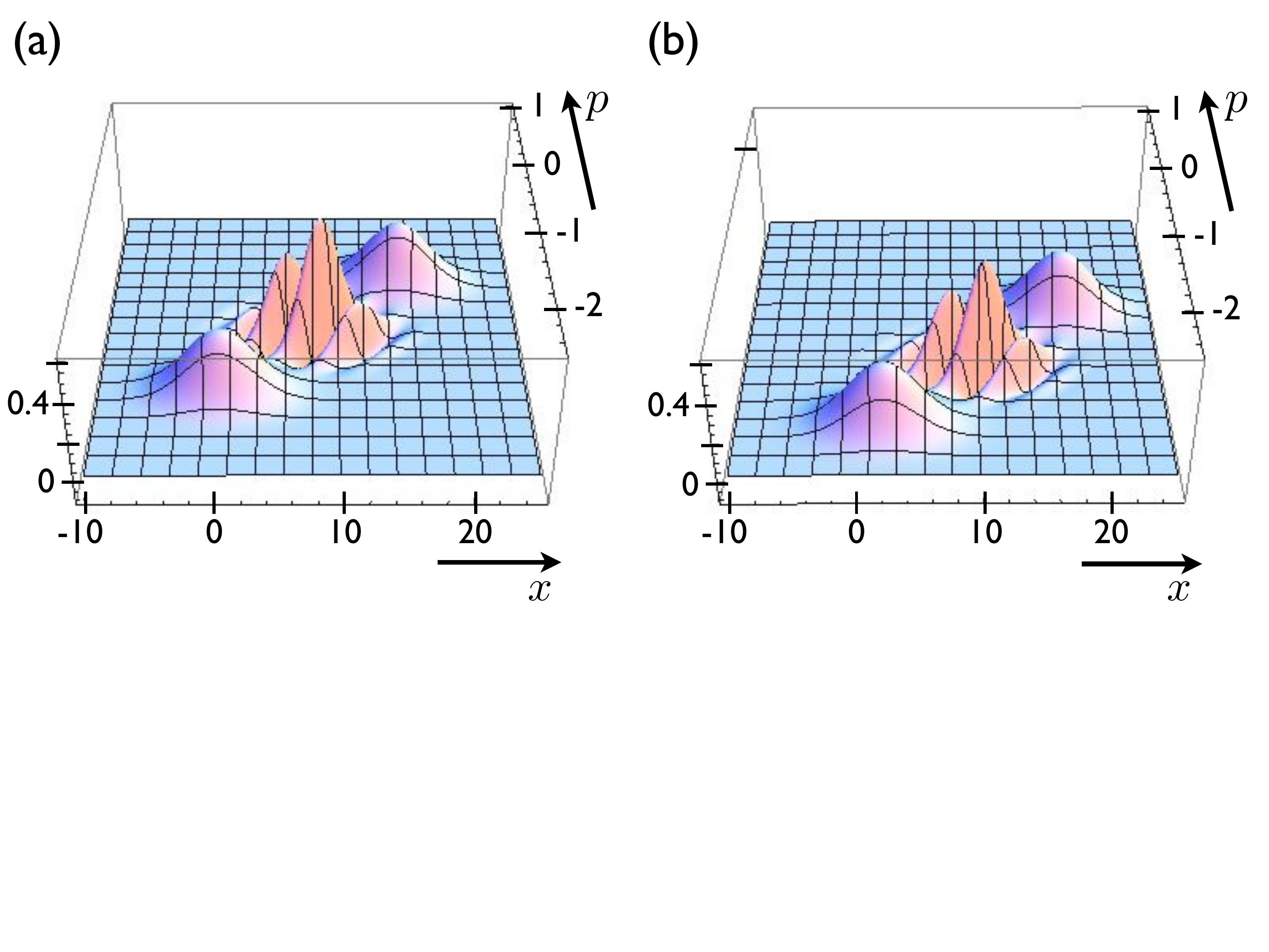}\vspace{-40mm}
\caption{\small As in figure~\ref{Wcompare}, but with different parameters for the gas particle: $p=-0.2,\;x_g=500,$ and $\alpha=0.002$.}
\label{Wcompare2}
\end{center}
\end{figure}

Figure~\ref{Wcompare2} shows how the Wigner function of a superposition state changes due to a collision with a very light gas particle. The coherences after the collision are still well pronounced. The reason is that the measurement, performed by a very light gas particle on the Brownian particle, is so imprecise that it can not distinguish between the two Gaussian wave functions of the Brownian particle.

The main effect of a collision with a very light gas particle is a shift of the entire Wigner distribution in phase space. That the oscillating part exactly shifts with the Gaussians, without obtaining an additional relative phase, might seem surprising, as this would certainly not be the case if we simply changed $\ket{x_{a/b},p_{a/b}}$ to $\ket{\bar x_{a/b},\bar p_{a/b}}$ (i.e.\ the relative height of the oscillations would be different). The reason can be found by looking at the argument of the cosine in \Eqref{Wgeneral} (and therefore applies also to heavy gas particle, where the effect is less easily observed, because of the strong suppression of coherences). In particular, the last two terms in the argument account for a phase shift according to the change of the average values $p_A\to\bar p_A$ and $x_A\to\bar x_A$. The first two terms in the argument account for an additional phase shift, but their sum does not change in a collision, because $(x_Ap_D-p_Ax_D)/(2\hbar)$ equals $(\bar x_A\bar p_D-\bar p_A\bar x_D)/(2\hbar)+\varphi$, with $\varphi$ taken from \Eqref{Wafter}.

Of course, if the gas particle state $\ket{x_g,p_g}$ is taken from a thermal gas, we have to average over all gas particle momenta $p_g$, weighted by the Maxwell-Boltzmann distribution $\mu_T(p_g)$, as well as over all gas particle positions $x_g$, which can reach the Brownian particle in a given time interval $(0,t)$. Because each possible combination of $x_g$ and $p_g$ results in a different  shift of the Wigner function in phase space, it is clear that this procedure strongly removes the oscillations. This is the decoherence effect we referred to `phase averaging' in section~\ref{cat}. It can still be very strong, even if the measurements which the gas particles perform are very weak. 

In the limit of a small mass ratio $\alpha$, the Brownian particle will be very localized compared to the gas particles, in both, position and velocity. Therefore, we do not have to use the full rate operator from  section~\ref{rate}, but can simply assume that a gas particle collides with the Brownian one during a time interval $(0,t)$, if $0<-x_gm_g/p_g<t$.

\subsection{Position decoherence}

To study position decoherence, we consider an initial Brownian particle state $\ket{x_a,p}+\ket{x_b,p}$. An example of the corresponding Wigner function in plotted in figure~\ref{Wpos}~(a). How the Wigner function after a collision with a gas particle state $\ket{x_g,p_g}$ is obtained, is  described in section~\ref{medi}, but now we also have to average over $x_g$ and $p_g$. To this end, we will randomly choose  pairs $(x_g,p_g)$ according to an appropriate distribution, and for each we will calculate the Wigner function. Finally, we will average over all of these to obtain the average Wigner function $W_{\rho_1}$ after one collision.

First, we need the normalized  collision probability distribution for $p_g$
	\begin{eqnarray}
		C(p_g)&=& \frac{R(p_g)}{\int\dd p_g \, R(p_g)} \nn\\
		&=& \frac{|p_g|}{2m_gk_BT}\exp\!\left(-\frac{p_g^2}{2m_gk_BT}\right)\! , \label{gjw}
	\end{eqnarray}
which we obtain from \Eqref{6.18} (or \Eqref{psq}) by assuming a slow Brownian particle. We will consider the cases $p_g\gtrless 0$ separately, starting with $p_g<0$. We therefore choose a random number $u_1\in(0,1/2)$, and solve
	\begin{eqnarray}
		u_1 &=& \int_{-\infty}^{p_g} \dd p_g'\, C(p_g') \nn\\ 
		\Rightarrow\quad p_g &=& -\sqrt{-2m_gk_BT\ln(2u_1)}
	\end{eqnarray}
to obtain a random gas particle momentum $p_g<0$. Because of symmetry of $C(p_g)$, we use $p_g = \sqrt{-2m_gk_BT\ln(2u_1)}$ for $p_g>0$ (after choosing a new random number $u_1$). For a given gas particle momentum $p_g<0$, the position $x_g$ has to be in the interval $x_g\in(0,p_gt/m)$. Therefore, we chose a second random number $u_2\in(0,1)$, and use
	\begin{eqnarray}
		x_g &=& -u_2\frac{p_g t}{m}
	\end{eqnarray}
as random gas particle position $x_g$.

We note at this point, that the position distribution of the colliding gas particle does not only depend on the gas temperature, but also on the time interval $(0,t)$. This is to be expected as the initial position of the gas particle is linearly related to the time of collision (see figure~\ref{smatrix}), and the time of collision can be anywhere within the interval $(0,t)$ of consideration.

\begin{figure}[htbp]\hspace{2mm}
	{\begin{rotate}{270} \includegraphics[width=1.26\linewidth]{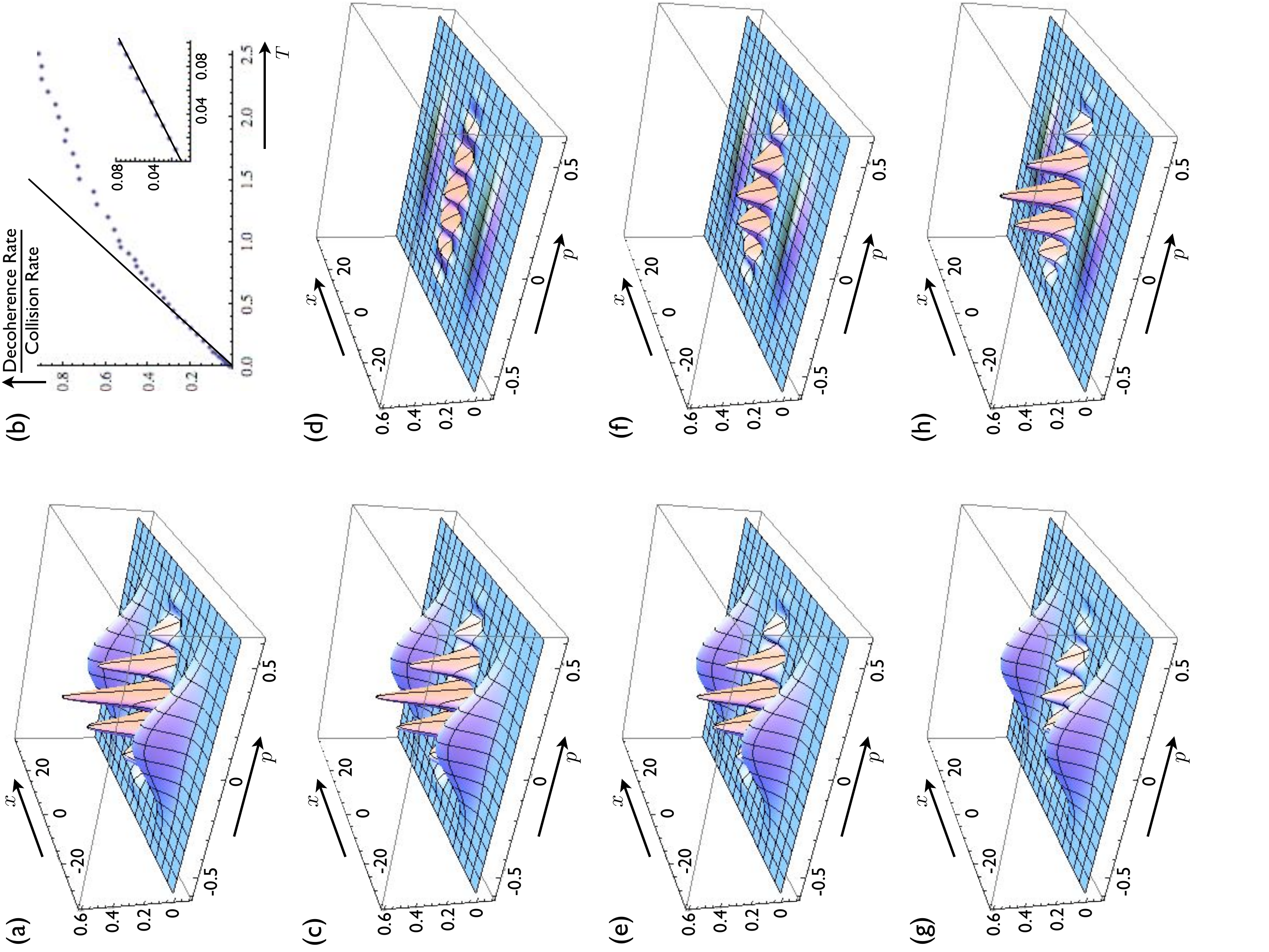}\end{rotate}}\vspace{17.5cm}
	\caption{\small (a): The Wigner function for the Brownian particle before a collision. (c), (e), (g): The average Wigner function after one collision with a gas particle at temperature $T=0.2,\;0.5$, and $1.5$, respectively. (d), (f), (h): The change of the Wigner function due to a collision at temperature $T=0.2,\;0.5$, and $1.5$, respectively. (b): The relative change of the Wigner function at the origin due to a collision. This serves as a quantitative measure of the decoherence per collision. The solid lines are the first order expansion in temperature. Parameters are: $x_a=20,\;x_b=-20,\;p_a=0,\;p_b=0,\;W=4,\;m=1,\;\hbar=1,\;k_B=1,\;t=20,$ and $\alpha=0.0001$.}
	\label{Wpos}
\end{figure}

The Wigner functions after a collision, averaged over 200 pairs $(x_g,p_g)$, are shown in figure~\ref{Wpos}~(c), (e), and (g), for the respective temperatures $T=0.2,\;0.5$, and $1.5$. The corresponding changes of the Wigner function from the initial one (figure~\ref{Wpos}~(a)) are shown figure~\ref{Wpos}~(d), (f), and (g). We use the maximum peak of the oscillations, relative to the maximum peak of the initial state's oscillations, as a quantitative measure of coherence. In particular, the relative change of the maximum peak serves as `decoherence per collision', and is plotted over the temperature $T$ in figure~\ref{Wpos}~(b). There is obviously a linear temperature dependence of the `decoherence per collision', as long as the momentum transfer in a collision is small. The small constant decoherence on top of the one linear in temperature is due to the very imprecise measurements, performed by the gas particles.

To explain this behavior, let us have a closer look at the oscillating term of the Wigner function \Eqref{Wgeneral} for the initial state $\ket{x_a,p}+\ket{x_b,p}$,
	\begin{eqnarray}
		\frac{2c}{\pi\hbar}\exp\!\left[ -\frac{W^2(p'-p)^2}{\hbar^2} \right]  \exp\!\left[ -\frac{(x'-x_A)^2}{W^2} \right]  \cos\!\left[\varphi - \frac{p\,x_D}{2\hbar} + x_D\frac{p-p'}{\hbar} \right]\quad  \label{Wosci}
	\end{eqnarray}
with $c=1$ and $\varphi=0$. According to section~\ref{medi}, a collision with a very light ($1+\alpha\approx 1$) gas particle results in a change of $c=\exp\!\left(-\alpha x_D^2/W^2\right)$, which is the responsible for a small decoherence independent of the temperature. Next, the position and momentum change to $\bar x_{a,b}=x_{a,b}+2\alpha x_g$ and $\bar p_{a,b}=p_{a,b}+2p_g$, respectively. This slightly broadens the exponentials in~(\ref{Wosci}) when averages over $x_g$ and $p_g$ are taken, but the effect is too small to be visible in the parameter range of figure~\ref{Wpos}. Further, we showed in section~\ref{medi}, that $[\varphi -px_D/(2\hbar)]$ does not change due to a collision. Therefore, the only significant effect is the change of $p$ to $p+2p_g$ in the cosine, which is a random phase shift corresponding to the random momentum of the colliding gas particle.

As we are interested in the reduction of the peak of the oscillations, we choose $p'$ such that $\cos\!\left[\varphi -p\,x_D/(2\hbar) + x_D(p-p')/\hbar \right]=1$, i.e.\ $\varphi -p\,x_D/(2\hbar) + x_D(p-p')/\hbar=0$. The same term after a collision at the very same $p'$ is then
	\begin{eqnarray}
		\cos\!\left(\varphi + \frac{-p\,x_D}{2\hbar} + x_D\frac{p+2p_g-p'}{\hbar} \right) &=& \cos\!\left(\frac{2x_D\,p_g}{\hbar}\right) \label{exjkl1} \\
		&\approx& 1-2\left(\frac{x_Dp_g}{\hbar}\right)^{\!2},\quad \label{exjkl}
	\end{eqnarray}
where we expanded the cosine around its maximum. For the `decoherence per collision', we have to average over the squared momenta of the colliding gas particle. Using \Eqref{gjw}, we find
	\begin{eqnarray}
		\left\langle p_g^2\right\rangle_{\mbox{\small coll}} &=& 2m_gk_BT.
	\end{eqnarray}
Note that the average squared momentum of a colliding gas particle is not the average squared momentum of a general gas particle, $\left\langle p_g^2\right\rangle=m_gk_BT$, obtained from the Maxwell-Boltzmann distribution (faster gas particles are more likely to collide). We finally deduce for the `decoherence per collision' for position superposition states
	\begin{eqnarray}
		\frac{\mbox{Decoherence}}{\mbox{Collision}}&=&\frac{4m_gk_BT}{\hbar^2}x_D^2. \label{qscz}
	\end{eqnarray}
Note that we would also have to add $\alpha x_D^2/W^2$ to account for the decrease of $c$, but according to relation~(\ref{7.18}), this term is small compared to \Eqref{qscz} in the range of validity of our work. This establishes, that the leading form of decoherence is phase averaging, and not information exchange between Brownian and colliding particles.\footnote{The non-zero y-intercept in the inset of figure~\ref{Wpos}~(b) is due to $c$. But the reduction of coherences from this information exchange is only dominating at temperatures so small, that relation~(\ref{7.18}) is violated and our approach is not valid anymore. Therefore, we can not eliminate the possibility that information exchange between the Brownian particle and the gas might not contribute to decoherence at all!}

The solid lines in figure~\ref{Wpos}~(b) are calculated from this formula, and agree very well with the numerical results as long as the temperature is low enough to justify the expansion \Eqref{exjkl}. Multiplying the last equation with the collision rate, we find the position decoherence rate
	\begin{eqnarray}
		D_x&=& \frac{8n _g\sqrt{m_g}(k_BT)^{3/2}}{\sqrt{2\pi}\hbar^2}x_D^2.
	\end{eqnarray}
This very nice result holds when $(2x_D)^2m_gk_BT\ll\hbar^2$ is valid, and agrees up to some constants with the decoherence rate obtained by Joos and Zeh {\small [Z. Phys. B: Condens. Matt. 59, 223 (1985)]}
\begin{figure}[htbp]\vspace{0cm}
	\includegraphics[width=\linewidth]{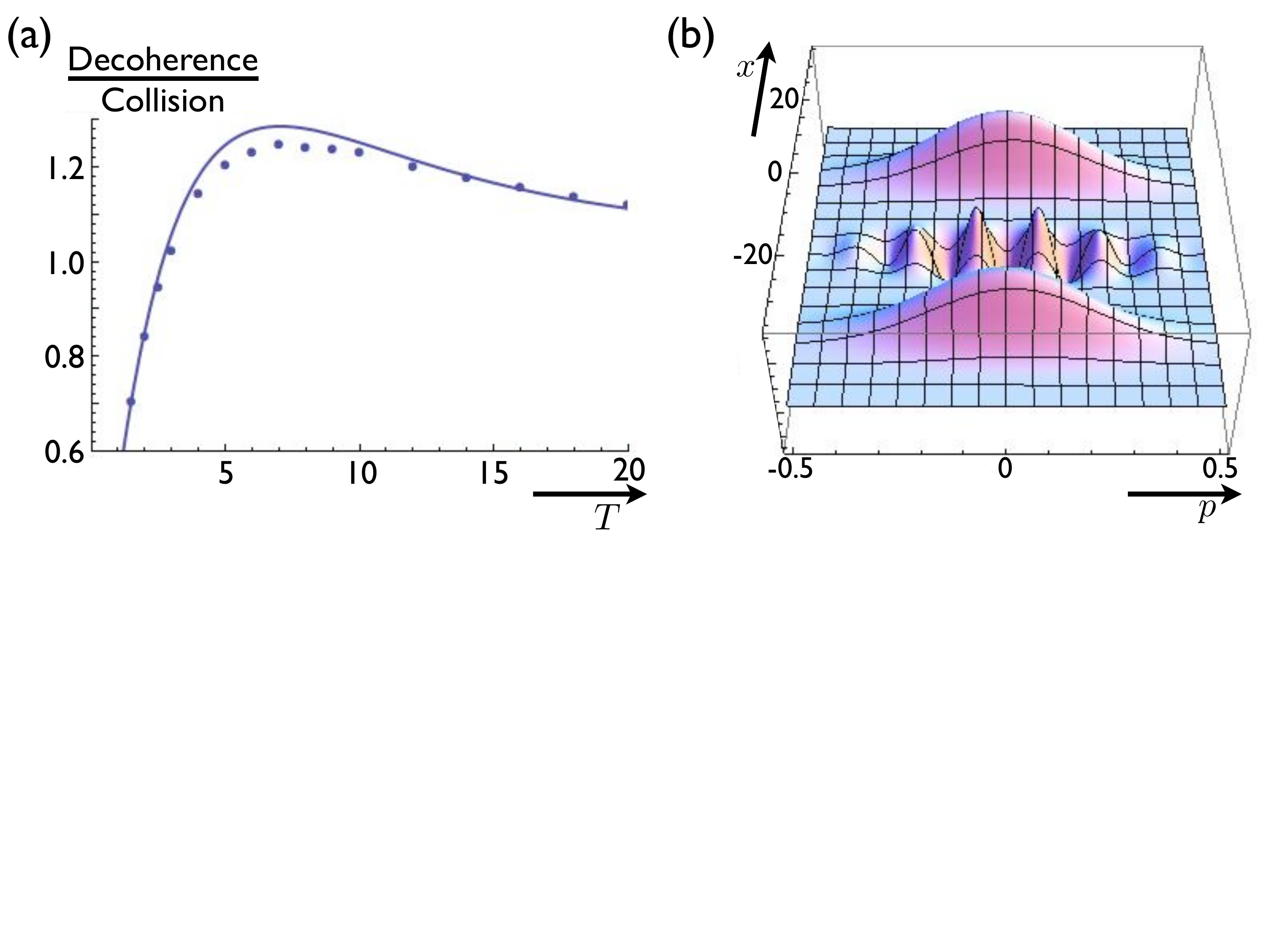} \vspace{-52mm}
	\caption{\small (a): The `decoherence per collision' plotted over the temperature, for parameters as in figure~\ref{Wpos}, but for larger temperatures. The dots are from the numerical Wigner function at the origin, and the line is from \Eqref{surprise}. The small discrepancy is due to the change of the Gaussians in \Eqref{Wosci}. \newline (b): The Wigner function after a collision at temperature $T=7$ shows indeed oscillations in opposite phase to the initial Wigner function (see figure~\ref{Wpos}~(a)). This leads to a decoherence rate which exceeds the collision rate.}
	\label{Wsurprise}
\end{figure}

We can generalize formula \Eqref{qscz} to high temperatures and large position separations, by directly averaging the cosine of \Eqref{exjkl1}, where we again use \Eqref{gjw}
	\begin{eqnarray}
		\left\langle  \cos\!\left(\frac{2x_D\,p_g}{\hbar}\right) \right\rangle_{\mbox{\small coll}} &=& \int_0^\infty \dd p_g\frac{p_g}{m_gk_BT} \exp\!\left( \frac{-p_g^2}{2m_gk_BT}\right) \cos\!\left(\frac{2x_D\,p_g}{\hbar}\right)\label{surprise1} \\
		&=& 1-\frac{2x_D}{\hbar} \int_0^\infty \dd p_g\, \exp\!\left( \frac{-p_g^2}{2m_gk_BT}\right) \sin\!\left(\frac{2x_D\,p_g}{\hbar}\right)\!. \qquad\quad \label{surprise}
	\end{eqnarray}
In the second step, we used integration by parts. We deduce for the `decoherence per collision' of position superposition states
	\begin{eqnarray}
		\frac{\mbox{Decoherence}}{\mbox{Collision}}&=& \frac{2x_D\sqrt{2m_gk_BT}}{\hbar} \int_0^\infty \dd u\, e^{-u^2} \sin\!\left(\frac{2x_D\sqrt{2m_gk_BT}}{\hbar}u\right)\!, \qquad\quad \label{surprise2}
	\end{eqnarray}
or for the decoherence rate
	\begin{eqnarray}
		D_x &=& \frac{4x_Dn_gk_BT}{\sqrt\pi \hbar} \int_0^\infty \dd u\, e^{-u^2} \sin\!\left(\frac{2x_D\sqrt{2m_gk_BT}}{\hbar}u\right)\!. \qquad\quad \label{surprise2}
	\end{eqnarray}
The `decoherence per collision' is plotted over temperature in figure~\ref{Wsurprise}~(a) for the same parameters as in figure~\ref{Wpos}~(b), but for higher temperatures. It might come as a surprise, that the decoherence rate exceeds the collision rate for $x_D\sqrt{2m_gk_BT}/\hbar \gtrsim 1$. The reason is that in this regime, the Wigner function after a collision shows oscillations, which are out of phase with the oscillations of the initial Wigner function, as shown in figure~\ref{Wsurprise}~(b). Therefore, if we write down the actual (interaction picture) Wigner function after some time $t$ as 
	\begin{eqnarray}
		W_{\rho(t)} &=& (1-Rt)W_{\rho_0}+RtW_{\rho_1},
	\end{eqnarray}
where $R$ is the total collision rate \Eqref{6.20}, and $\rho_0$ and $\rho_1$ are the density operators corresponding to either no or one collision, respectively, then the oscillations in $W_{\rho_0}$ and $W_{\rho_1}$ interfere destructively.

We note that it is often stated in the literature that, if the separation $x_D$ of two interfering wave packets is larger than the thermal wave length $\Lambda=\hbar/\sqrt{2\pi m_g k_BT}$ of the gas, then a colliding gas particle can distinguish between the two interfering wave packets, therefore removing their coherences. In finding that the decoherence rate is about the collision rate if $x_D \gtrsim \sqrt{\pi}\Lambda$, we confirm the latter part of this statement, but we also show that the loss of coherence is by no means due to the information exchange between the colliding particles, but due to phase averaging. This in turn results from a relative phase, which depends on the momentum of the colliding gas particle.

\subsection{Momentum decoherence}

Now, we consider the initial Brownian particle state $\ket{x,p_a}+\ket{x,p_b}$, whose Wigner function is plotted in figure~\ref{Wmom}~(a). Again, the main source of decoherence will be phase averaging. But contrary to the previous subsection, where the relative phase of the two Gaussians wave packets after a collision depended on the initial momentum of the colliding gas particle, for momentum decoherence this phase depends on the initial position of the colliding gas particle. The more variation of the initial position of a colliding gas particle, the more `decoherence per collision' we will find. For a given gas particle momenta $p_g$, the gas particle position can be anywhere in the interval $(-p_gt/m_g,0)$. Therefore, the decoherence per collision will not only depend on the temperature $T$ (for the distribution of $p_g$), but also on the considered time interval. In figure~\ref{Wmom}~(c) - (h), the effects of a collision on the Wigner function is shown for different temperatures and time intervals.  The dependence of the `decoherence per collision' for short times and/or low temperatures turns out to be linear in temperature and quadratic in time, as shown in~figure~\ref{Wmom}~(b). As a result, it is not possible to define a time independent decoherence rate for momentum superpositions. We will provide a physical interpretation at the end of this subsection, and first give a mathematical explanation of these results.

For this purpose, we consider again the oscillating term of the Wigner function \Eqref{Wgeneral}
	\begin{eqnarray}
		\frac{2c}{\pi\hbar} \exp\!\left[ -\frac{(x'-x)^2}{W^2} \right] \exp\!\left[ -\frac{W^2(p'-p_A)^2}{\hbar^2} \right] \cos\!\left[\varphi + \frac{x\,p_D}{2\hbar}  - p_D\frac{x-x'}{\hbar}  \right]\!, \label{tfcd}
	\end{eqnarray}
at its maximum position $\varphi +xp_D/(2\hbar)-p_D(x-x')/\hbar=0$. As discussed in section~\ref{medi}, a collision does not change the sum $\varphi +xp_D/(2\hbar)$ in the cosine, and we only have to consider the change of $x$ to $x+2\alpha x_g$ in the last term in the cosine. Therefore, we find the `decoherence per collision' by averaging over $\cos(2\alpha x_g\,p_D/\hbar)$. Here, we need the normalized distribution $C(x_g)$ of initial positions of colliding gas particle. The phase space distribution of gas particles is given by $\rho_g(x_g,p_g)$ in \Eqref{gaspp}. For a given $x_g<0$, a gas particle collides if its velocity is larger than $|x_g|/t$, which leads us to
	\begin{eqnarray}
		C(x_g) &\propto& \frac{n_g}{\sqrt{2\pi m_gk_BT}} \int_{|x_g|m_g/t}^\infty \dd p_g \exp\!\left(-\frac{p_g^2}{2m_gk_BT}\right)\!.
	\end{eqnarray}
The distribution is normalized either by integration over $x_g$, or directly by dividing by the total collision probability $Rt$. After substituting $u=p_g/\sqrt{2m_gk_BT}$ we find
	\begin{eqnarray}
		C(x_g) &=& \frac{\sqrt{m_g}}{t\sqrt{2k_BT}} \int_{\frac{|x_g|\sqrt{m_g}}{t\sqrt{2k_BT}}}^\infty \dd u\,e^{-u^2},
	\end{eqnarray}
and therefore
	\begin{eqnarray}
		\left\langle  \cos\!\left(\frac{2\alpha p_D\,x_g}{\hbar}\right) \right\rangle_{\mbox{\small coll}} &=& \frac{\sqrt{m_g}}{t\sqrt{2k_BT}} \int_{-\infty}^\infty \dd x_g\,\cos\!\left(\frac{2\alpha p_D\,x_g}{\hbar}\right) \int_{\frac{|x_g|\sqrt{m_g}}{t\sqrt{2k_BT}}}^\infty \dd u\,e^{-u^2} \nn\\
		&=&  \frac{2\sqrt{m_g}}{t\sqrt{2k_BT}} \int_0^\infty \dd u\,e^{-u^2} \int_0^{ut\sqrt{2k_BT}/\sqrt{m_g}} \dd x_g\,\cos\!\left(\frac{2\alpha p_D\,x_g}{\hbar}\right) \quad\nn\\
		&=& \frac{m\hbar}{t\sqrt{2m_gk_BT}p_D}\int_0^\infty \dd u\, e^{-u^2} \sin\!\left( \frac{2t\sqrt{2m_gk_BT}p_D}{m\hbar}u \right) \!,\nn\\ \label{pjrd}
	\end{eqnarray}
where we changed the order of integration (see figure~\ref{fig5} for detail) in the second line. As this functions represents the coherences after one collision, we have to subtract it from one to get the `decoherence per collision'
	\begin{eqnarray}
		\frac{\mbox{Decoherence}}{\mbox{Collision}} &=& 1- \frac{m\hbar}{t\sqrt{2m_gk_BT}p_D}\int_0^\infty \dd u\, e^{-u^2} \sin\!\left( \frac{2t\sqrt{2m_gk_BT}p_D}{m\hbar}u \right) \qquad\quad \label{kfur} \\
		&\approx& \frac{4m_gk_BT}{3\hbar^2}\left(\frac{tp_D}{m}\right)^2 .\label{kfur2}
	\end{eqnarray}
The approximation is obtained by expanding the sine to third order. The solid lines in figure~\ref{Wmom}~(b) are taken from \Eqref{kfur}, and agree with the data obtained by averaging over Wigner functions with random gas particle positions and momenta.\footnote{The non-zero y-intercept in the inset of figure~\ref{Wmom}~(b) is again due to $c$ (in addition to the spreading of the Gaussian in \Eqref{tfcd}). Similar to position decoherence, the reduction of coherences from this information exchange is only dominating at temperatures and times so small, that relation~(\ref{bbbb}) is violated, and our approach is not valid anymore.}

\begin{figure}[htbp]\hspace{2mm}
	{\begin{rotate}{270} \includegraphics[width=1.26\linewidth]{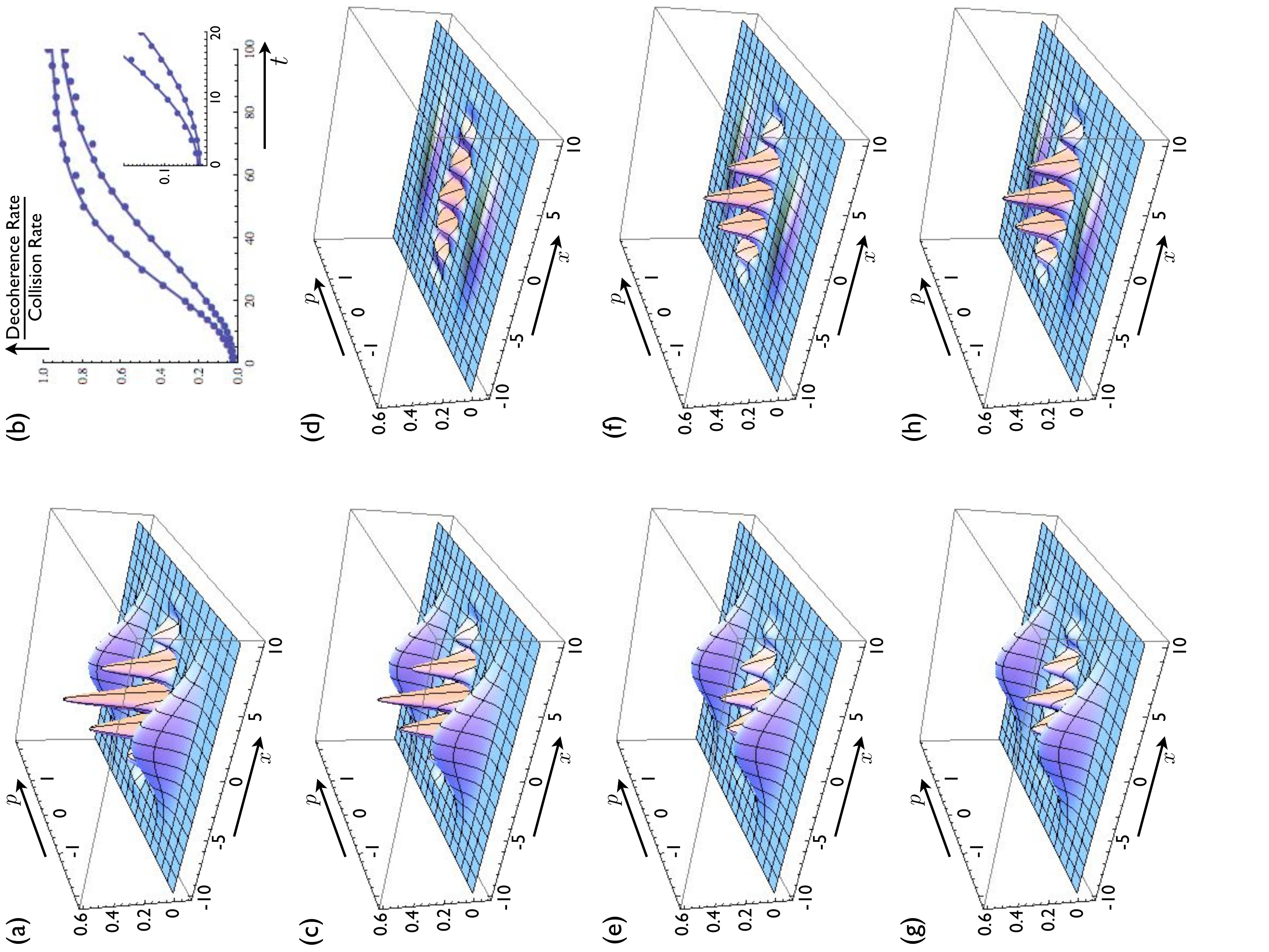} \end{rotate}}\vspace{17.3cm}
	\caption{\small (a): The Wigner function for the Brownian particle before a collision. (c), (e), (g): The average Wigner function after one collision with a gas particle within a time interval $(0,t)$ at temperature $T$, with $t=20,\;T=0.5$ in (c);  $t=50,\;T=0.5$ in (e);  $t=20,\;T=3$ in (g). (d), (f), (h): The corresponding change of the Wigner function due to a collision. (b): The relative change of the Wigner function at the origin due to a collision as function of the time interval for $T=0.5$ as well as $T=1$. Parameters are: $x_a=0,\;x_b=0,\;p_a=1.2,\;p_b=-1.2,\;W=4,\;m=1,\;\hbar=1,\;k_B=1,$ and $\alpha=0.0001$.}
	\label{Wmom}
\end{figure}

We recognize from \Eqref{kfur} and~(\ref{kfur2}), that for short time intervals, it seems as if collisions can not reduce the coherence of momentum superpositions states. The reason is that momentum decoherence is not a leading order process, but a higher order one, resulting from position decoherence due to the position separation $x_D(t) =tp_D/m$, which a Brownian particle with momentum separation $p_D$ acquires over time.


Let us investigate this proposition in detail, by splitting the time interval $(0,t)$ in many, but small intervals. Then, according to \Eqref{kfur2}, there should be no momentum decoherence at all, but the two wave packets will have a time dependent position separation $x_D(t')=t'p_D/m$. Therefore, if a collision happens during the time interval $(t',t'+\dd t)$, the coherence after the collision is given by substituting $x_D=t'p_D/m$ into the equation for position decoherence, \Eqref{surprise1}. Averaging over all possible $(t',t'+\dd t) \subset (0,t)$, we find for the coherence after a collision during the time interval $(0,t)$
	\begin{eqnarray}
		&&\frac1t\int_0^t \dd t'\,\int_0^\infty \dd u\,2u e^{-u^2} \cos\!\left(\frac{2t'\sqrt{2m_gk_BT}p_D}{m\hbar}u\right) \nn\\
		&=& \frac{m\hbar}{t\sqrt{2m_gk_BT}p_D} \int_0^\infty \dd u\, e^{-u^2} \sin\!\left(\frac{2t\sqrt{2m_gk_BT}p_D}{m\hbar}u\right)\!,\nn
	\end{eqnarray}
which is exactly the same as obtained by \Eqref{pjrd}. 

Therefore, we arrive at a very neat interpretation of the decoherence process of a superposition of two Gaussians wave packets. The decoherence due to a collision depends on the position separation $x_D$ of the two Gaussian wave packets at the time of the collision.\footnote{As the time of collision we understand here the time, when the mean value of the gas particle wave packet crosses over the mean value of the Brownian particle wave packet, such that a collision results in a change of $p_A$, but not of $x_A$.}
There is no direct decoherence due to the momentum separation $p_D$ of the two coherent wave packets. Over time, the momentum separation changes the position separation according to $x_D(t)=x_D+tp_D/m$. This leads to an indirect influence of $p_D$ on the decoherence rate, which, if there is no initial position separation, is described by \Eqref{kfur}.

\section{Conclusions from the study of decoherence}

We have seen in this chapter, that we can disregard the decoherences due to the measurements, which are performed by colliding gas particle. The reason is, that the measurement effects are small compared to phase averaging for very light gas particles; and that phase averaging destroys all coherences for slightly heavier gas particles, such that for both situations, the decoherence due to the measurements is not visible. Therefore, the entire dynamics of the Brownian particle due to a collision with a thermal gas particle seems to be captured, by changing the momentum in the Wigner function at the time of the collision according to
	\begin{eqnarray}
		W_\rho(x',p')&\to&W_\rho(x',p'-q),
	\end{eqnarray}
where the momentum change is $q=2(p_g-\alpha p')/(1-\alpha)$. This reminds very much of the change of a classical phase space probability distribution due to a collision, as described in section~\ref{rfko}. Noting that the Wigner function $W_\rho(x,p)$ behaves like a classical phase space probability distribution also during free evolution, it follows that it obeys the same master equation \Eqref{clasmas} as the phase space probability distribution $\rho(x,p)$ of a classical particle. If the momentum transfer is small compared to the variations of the Wigner function in momentum space, i.e.\ $W_\rho(x',p')\approx W_\rho(x',p'-q)$, then the Wigner function also obeys the classical Fokker-Planck equation \Eqref{fockerplanck2}, although the condition of small momentum transfer is more stringent than in classical dynamics, because of the possibly rapid oscillations of the Wigner functions.

After the huge effort of finding a very complicated master equation for the density operator of the Brownian particle, it finally seems possible that quantum Brownian motion could be much easier described, by applying the equations which govern the dynamics of the phase space probability distribution of a classical particle, to the Wigner function of a quantum particle. More work is needed to rigorously establish the described connection between classical and quantum Brownian motion. 

One should mention, that such a description of QBM would include all classical features like friction and momentum diffusion, but not the unphysical QPD, which all master equations of Lindblad form show. On the one hand, that is a very nice feature, but on the other hand, that also means that the dynamics described by such means, can not be completely positive. Therefore, we once again observe, that Markovian QBM is only an approximation to reality, which either leads to non-positive dynamics, or to a fictitious quantum contribution to position diffusion, but can not avoid both.

It is clear, that the process of a single collision with a gas particle might be quite different in three dimensions, including the measurement performed by a colliding gas particle. But if it turns out, that the decoherence effects due to the measurement is small compared to the phase averaging, as it is in one dimension, then also three dimensional quantum Brownian motion might be described by applying the classical equations to the Wigner function of the Brownian particle. 

The measurement performed by a colliding particle depends on the interaction potential. But if the measurement performed by a thermal particle is negligible compared by the decoherence effects due to the random momentum transfer, then it might be possible that also scattering of other particles, such as photons, could be described by such a Wigner function approach.

\chapter{General Conclusions}

We first applied the quantum measurement theory approach to the study of the coherent transfer of an electron along a rail of quantum dots. If used for the transfer of quantum information, we found, that the spin of an electron should be preferred as a qubit, rather than its position. Furthermore, we were able to specify the transfer fidelity for a given strength of Markovian and certain non-Markovian dephasing sources. This led us to conclude, that under certain circumstances, it is preferred to divide a long transfer distance into several smaller ones.

We then used quantum measurement theory, to study collisional quantum Brownian motion, a topic which has several controversial results. We first found, contrary to common belief, that there is no additional quantum contribution to position diffusion. Next we showed, that a colliding particle carries away information of the Brownian particle position \emph{and} momentum. Finally, this information transfer turned out \emph{not} to be the main source of decoherence of superposition states. Instead, phase averaging due to a random relative phase, which depends on the momentum of the colliding gas particle, was found to quickly and efficiently reduce coherences. 

The first statement of the above paragraph seems to be hard to test experimentally~\cite{review}, because the classical position diffusion would quickly dominate anyway. The second statement can not be observed in a thermal gas because of the third statement, although one might be able to measure the momentum of a particle by a collision with a smaller particle, if the latter is in a well prepared initial state. The last statement however, could be accessible to observation in e.g.\ double slit experiments, because it predicts, that, for certain experimental parameters, the decoherence rate can exceed the collision rate. This effect would be impossible, if information transfer between colliding particles were the source of decoherence.

Because this effect is predicted to occur at higher temperatures and with larger spatial separations of the superposition state, it might be more accessible to experiments, compared to the quadratic dependence in spatial positions separation for lower temperatures and lower spatial separation, which was already predicted previously.

Our results also let us to conjecture, that comparatively simple equations for the classical phase space probability distribution could be applied to the Wigner function of the Brownian particle, to successfully describe quantum Brownian motion, possibly even for much more general situations as considered in this thesis.

In conclusion, even in situations where the information transfer between system of interest and its environment does not play a significant role, the measurement approach was still able to lead us to considerable insight. However, instead of studying the effect of an environment on the system directly, but rather to first considers the effect of the system on the environment, and then to draw conclusions about the information transfer between both systems and eventually about the dynamics of the system itself, is somewhat indirect. This can result in considerable work if carried out rigorously, as especially the second part of this thesis shows. This possibly limits its value to situations, where other approaches failed to deliver conclusive results. 

On the other hand, a measurement interpretation of open quantum system might be a powerful tool if used heuristically, as in the first part of the thesis, where we did not consider the detailed interaction between electron and quantum point contacts. This might be true especially, if a first principle approach does not seem feasible.


\end{document}